\documentclass[11pt,graphicx,amsmath]{article}
\usepackage{amsmath}
\usepackage{graphicx}
\usepackage{subfigure}
\usepackage{CJK}
\usepackage{bm}
\usepackage{xcolor}
\usepackage[numbers,sort&compress]{natbib}
\numberwithin{equation}{section}

\def\be{\begin{equation}}
\def\ee{\end{equation}}
\def\ba{\begin{eqnarray}}
\def\ea{\end{eqnarray}}
\def\nn{\nonumber}
\def\bt{\bm{\theta}}
\def\cf{\mathcal{F}}
\def\ch{\mathcal{H}}
\def\bl#1\el{\begin{align}#1\end{align}}
\def\l{\left}
\def\r{\right}

\title{ Second-order cosmological perturbations.
III. Produced by scalar-scalar coupling
during radiation-dominated  stage}

\author{\small   Bo Wang  \thanks{ymwangbo@mail.ustc.edu.cn}
     \   \ and \ Yang  Zhang  \thanks{yzh@ustc.edu.cn}
           \\
 \small   Department of  Astronomy, Key Laboratory for Researches in Galaxies and Cosmology, \\
 \small    University of Science and Technology of China,   Hefei, Anhui, 230026,  China   }

 \date{}

\begin{document}
\maketitle

\def\bl#1\el{\begin{align}#1\end{align}}
\def\gsim{\;\rlap{\lower 2.5pt  \hbox{$\sim$}}\raise 1.5pt\hbox{$>$}\;}
\def\lsim{\;\rlap{\lower 2.5pt  \hbox{$\sim$}}\raise 1.5pt\hbox{$<$}\;}
\def\edth{\;\raise1.0pt\hbox{$'$}\hskip-6pt\partial\;}
\def\baredth{\;\overline{\raise1.0pt\hbox{$'$}\hskip-6pt \partial}\;}
\def\be{\begin{equation}}
\def\ee{\end{equation}}
\def\ba{\begin{eqnarray}}
\def\ea{\end{eqnarray}}
\def\nn{\nonumber}
\def\bt{\bm{\theta}}
\def\cf{\mathcal{F}}
\def\ch{\mathcal{H}}
\def\l{\left}
\def\r{\right}

\baselineskip=19truept
\Large


\begin{center}
\text{\large\bf Abstract}\\
\end{center}
We  study  the  2nd-order scalar, vector   and tensor   metric perturbations
in Robertson-Walker (RW) spacetime in synchronous coordinates during  the radiation dominated (RD) stage.
The dominant radiation is modeled by a relativistic fluid
described by a stress tensor $T_{\mu\nu}=(\rho+p)U_\mu U_\nu+g_{\mu\nu}p$
with $p= c^2_s \rho$,
and the 1st-order velocity is assumed to be curlless.
We analyze the solutions of 1st-order perturbations,
upon which the solutions of 2nd-order perturbation are based.
We show that the 1st-order tensor modes propagate at the speed of light
and are truly radiative, but the scalar and vector modes do not.
The 2nd-order perturbed Einstein equation
contains various couplings of 1st-order metric perturbations,
and the scalar-scalar coupling is considered in this paper.
We  decompose  the 2nd-order Einstein equation  into
the  evolution equations of 2nd-order scalar, vector, and tensor perturbations,
and the energy and momentum constraints.
The coupling terms and the stress tensor of the fluid
together serve as the effective source for
the 2nd-order metric perturbations.
The equation of covariant conservation of  stress tensor
is also needed to determine $\rho$ and $U^\mu$.
By solving this set of equations up to 2nd order analytically,
we obtain the 2nd-order integral
solutions of all the metric perturbations,
density contrast and  velocity.
To use  these solutions in applications,
one needs to carry out   seven types of the numerical integrals.
We perform the residual gauge transformations
between synchronous coordinates up to 2nd order,
and identify the gauge-invariant modes of  2nd-order solutions.

\section{Introduction}
\label{sec:introduction}

As the theoretical foundation of cosmology,
cosmological perturbation has been extensively studied to linear order
\cite{Lifshitz1946,LifshitzKhalatnikov1963,
   PressVishniac1980,Bardeen1980,KodamSasaki1984,Grishchuk1994},
and successfully applied to large-scale structure \cite{ Peebles1980},
cosmic microwave background radiation (CMB)
   \cite{BaskoPolnarev1984,Polnarev1985, MaBertschinger1995,Bertschinger,
   ZaldarriagaHarari1995,Kosowsky1996, ZhaoZhang2006,Baskaran}
and relic gravitational wave (RGW)
\cite{Grishchuk,Starobinsky,
    Allen1988, zhangyang05,ZhangWang2018}, etc.
Nonlinear effects beyond the linear perturbation
become  interesting in the era of precision cosmology
\cite{PyneCarroll1996,
JeongKomatsu2006,YangZhang2007,AnandaClarksonWands2007,Baumann2007,
Bartolo2010,Matarrese2007,Pietroni2008,Matsubara2008}.
The cosmic evolution in Big Bang cosmology
is very long, from inflation to the present accelerating stage,
and  the effects of nonlinear perturbation during expansion
might accumulate substantially and be significant for cosmology.
The metric perturbations in general relativity (GR) can be decomposed into
three irreducible parts: scalar, vector, and tensor,
giving rise to six types of couplings of metric perturbations:
scalar-scalar, scalar-vector, etc.,  at the 2nd-order level.
In a Robertson-Walker (RW) spacetime
the 2nd-order cosmological perturbation is the lowest order of nonlinear perturbations.
In the literature,  Tomita studied the 2nd-order perturbations in synchronous coordinates,
and analyzed the 2nd-order density contrast
for some special cases  \cite{Tomita1967}.
Gravitational instability was studied
in the 2nd-order perturbations in association
with large-scale structure \cite{MatarresePantanoSa'ez1994,Russ1996,Salopek}.
Ref.\cite{Baumann2012}
discussed the cosmological effects of small-scale nonlinearity.
In Lambda cold dark matter ($\Lambda$CDM) framework,
Ref.\cite{Brilenkov&Eingorm2017} calculated 2nd-order
scalar and vector perturbations in the Poisson gauge.
Gauge-invariant 2nd-order perturbations
were   studied in Refs.\cite{Nakamura2003,Domenech&Sasaki2017},
and in Ref.\cite{MalikWands2004}.
The 2nd-order density perturbation was studied  with the squared RGW as the source
in the Arnowitt-Deser-Misner (ADM) framework \cite{NohHwang2004,HwangNoh2012,HwangNoh2005}.

Matarrese {\it et al} studied the equations
of 2nd-order scalar and tensor perturbations for the scalar-scalar coupling
in Einstein-de Sitter model  filled with a pressureless dust \cite{Bruni97, Matarrese98},
the 2nd-order vector due to scalar-scalar coupling was explored
in Poisson gauge  \cite{MollerachHarariMatarrese2004,Lu2008}.
In Refs.\cite{WangZhang2017,ZhangQinWang2017}
we have performed a systematical study of 2nd-order perturbation
in the matter-dominated (MD) stage,
and for the scalar-scalar, scalar-tensor, tensor-tensor couplings,
we have derived all the analytical solutions of
2nd-order  scalar, vector, tensor metric perturbations,
and of the 2nd-order density contrast.

In the RD  stage
the scale factor $a(\tau)$ can increase up to  $\sim 10^{24}$ times,
much more than  in MD stage.
The 2nd-order perturbation during RD has not been
analytically explored
 in literature,
which motivates our study in this paper.
The dominant cosmic radiation driving
the RD expansion  can be modeled as  a relativistic fluid with
a stress tensor $T_{\mu\nu}=(\rho+p)U_\mu U_\nu+g_{\mu\nu}p$
with the pressure  $p= c^2_s \rho$ and $c_s^2=\frac13$,
which is  more complicated than
the dust model \cite{Bruni97, Matarrese98,WangZhang2017,ZhangQinWang2017},
as the velocity $U^\mu$ is involved.
For this, we have to solve the equation of
covariant conservation $T^{\mu\nu}\, _{;\nu} =0$
and determine $U^\mu$, as well as $\rho$,  up to the 2nd order.
The 2nd-order perturbed Einstein equation
will be decomposed into the equations of scalar, vector, and tensor,
by the procedures similar to
those for MD stage \cite{WangZhang2017,ZhangQinWang2017};
and for the case of  scalar-scalar coupling,
we derive the analytical solutions
of 2nd-order scalar, vector, and tensor perturbations  under general initial conditions.
Since the solutions  contain residual gauge  modes in synchronous coordinates,
we shall also calculate residual gauge transformations
from synchronous to synchronous up to the 2nd order,
and  identify the gauge-invariant modes
of the  2nd-order metric perturbations, the density contrast,  and the velocity.

In Sec. 2,
we give some basic setups,
including notations of metric perturbations,
and the relativistic fluid model.

In Sec. 3,
we derive the analytical solution of 1st-order perturbations,
and the 1st-order residual gauge transformations.
We analyze how the tensor modes  as dynamic degrees of freedom
(d.o.f.)
differ from the scalar and vector modes.

In Sec. 4,  for the case of scalar-scalar coupling,
we decompose the 2nd-order perturbed Einstein equation
into the equations of 2nd-order scalar, vector, tensor metric perturbations,
and also derive the equations of the 2nd-order density contrast and velocity
from the covariant conservation of $T_{\mu\nu}$.

In Sec. 5,
we derive the solutions of the 2nd-order metric perturbations,
density contrast, and velocity,
which involve time and momentum  integrals.

In Sec. 6,  we perform the synchronous-to-synchronous transformation,
and identify the residual gauge modes in these 2nd-order solutions.

 Section 7  is  the  conclusions and discussions.

Appendix A lists  the expressions of perturbed quantities used
in Einstein equations,
Appendix B gives the perturbed Einstein equations  up to 2nd order
   for a general RW spacetime,
and Appendix C gives  the  synchronous-to-synchronous
   gauge transformations of the metric perturbations,
   the density contrast, and the velocity up to 2nd order.
We use a unit with the speed of light $c=1$.

\section{ Basic setups}

A   flat Robertson-Walker metric in synchronous coordinates
is given by
\be \label{18q1}
ds^2=g_{\mu\nu}   dx^{\mu}dx^{\nu}=
a^2(\tau)[-d\tau^2
+\gamma_{ij}   dx^idx^j] ,
\ee
where $\tau$ is the  conformal time,
 the indices  $\mu,\nu = 0, 1, 2, 3$ and  $i,j= 1, 2, 3$,
and
\be\label{eq1}
\gamma_{ij} =\delta_{ij} + \gamma_{ij}^{(1)} + \frac{1}{2} \gamma_{ij}^{(2)}
\ee
with $\gamma_{ij}^{(1)}$ and  $\gamma_{ij}^{(2)}$
being  the 1st and 2nd-order metric perturbation, respectively.
Writing  $g^{ij}=a^{-2}\gamma^{ij}$,
one has
\be\label{metricUp2}
\gamma^{ij}=\delta^{ij} -\gamma^{(1)ij}
-\frac{1}{2}\gamma^{(2)ij}+\gamma^{(1)ik}\gamma^{(1)j}_{k} .
\ee
Raising and lowering
the 3-dim spatial indices,
such as $\gamma^{(1)ik}$ and  $\gamma^{(2)ik}$,
will be done by $\delta^{ij}$.
We use the same notations for metric perturbations
 as in Refs. \cite {Matarrese98,ZhangQinWang2017,WangZhang2017}.
The 1st- and 2nd-order metric perturbation
can be further written as
\be  \label{gqamma1}
\gamma^{(A)}_{ij}=-2\phi^{(A)}\delta_{ij}  +\chi_{ij}^{(A)},
~~\text{with}~~ A=1,2,
\ee
where  $\phi^{(A)}$ is the trace part of  scalar perturbation,
and $\chi_{ij}^{(A)}$ is  traceless and can be further decomposed into
the following
\be \label{xqiij1}
\chi_{ij}^{(A)} =D_{ij}\chi^{||(A)}
               +\chi^{\perp(A)}_{ij}
               +\chi^{\top(A)}_{ij},
~~\text{with}~~ A=1,2,
\ee
where $D_{ij} \equiv  \partial_i\partial_j-\frac{1}{3}\delta_{ij}\nabla^2 $
and $\chi^{||(A)}$ is a scalar function,
and  $D_{ij}\chi^{||(A)}$ is the traceless part of the scalar perturbation.
There are two scalar modes of metric perturbations,
$\phi^{(A)}$ and $D_{ij}\chi^{||(A)}$  of each order.
The vector metric perturbation  $\chi^{\perp(A) }_{ij}$
satisfies a condition
$\partial^i\partial^j  \chi^{\perp(A) }_{ij}=0$
and can be written as
\be\label{chiVec0}
\chi^{\perp(A) }_{ij}= \partial_i B^{(A)}_j+\partial_j B^{(A)}_i,
\,\,\,\,\,
\text{with}~~  \partial^i B^{(A)}_i =0,
~~ A=1,2 ,
\ee
where $B^{(A)}_i$ is a curl  vector,
and the vector metric perturbation also has  two independent modes.
The tensor metric perturbation  $\chi^{\top(A)}_{ij}$
satisfies the traceless and transverse  condition:
$\chi^{\top(A)i}\, _i=0$, $\partial^i\chi^{\top(A)}_{ij}=0$,
having two independent modes.

The RD stage of expansion is driven by a relativistic fluid  (without a shear stress),
whose  stress tensor is
$T_{\mu\nu}=(\rho+p)U_\mu U_\nu+g_{\mu\nu}p$,
where $\rho$ and $p$ are
the energy density and pressure measured by a comoving observer
in the locally inertial frame, respectively,
and $U^\mu=\frac{d x^\mu}{d\lambda}$ with $d\lambda^2 =-ds^2$
is the fluid  4-velocity
with a normalization condition $g_{\mu\nu}U^\mu U^\nu=-1$.
The  energy density $\rho$   is expanded up to 2nd order as the following
\be\label{rho}
\rho=\rho^{(0)} + \rho^{(1)}+\frac{1}{2}\rho^{(2)} ,
\ee
where $ \rho^{(0)}$ is the background density,
$ \rho^{(1)}$ is the 1st-order density perturbation, etc.
We introduce the  density contrast
\be\label{deltaA}
\delta^{(A)}\equiv\frac{\rho^{(A)}}{\rho^{(0)}}, ~~ A=1,2.
\ee
The pressure    is also expanded up to 2nd order
\be\label{p1}
p=p^{(0)}+p^{(1)}+\frac{1}{2}p^{(2)}.
\ee
For the fluid,
we introduce the 0th-, 1st-, and 2nd-order sound speeds as the following
\be \label{cn2}
c_s^2\equiv \frac {p^{(0)'}}{\rho^{(0)'}}= \frac {p^{(0)}}{\rho^{(0)}},
~~~~~~~~
c_L^2\equiv \frac {p^{(1)}}{\rho^{(1)}}
,
~~~~~~~~
c_N^2\equiv \frac {p^{(2)}}{\rho^{(2)}},
\ee
which can be taken to be  $c_s^2=\frac13$ and  $c_L^2=\frac13$
in applications \cite{Weinberg1972},
and  we assume $c_N^2=\frac13$ for computation convenience,
its actual value should be determined by future experiments.
The expansion of $U^\mu$ is expanded up to 2nd order
\be
U^\mu\equiv U^{(0)\mu}+U^{(1)\mu}+\frac{1}{2}U^{(2)\mu} .
\ee
The coordinate 3-velocity  \cite{Bertschinger}
$v^i \equiv \frac{d x^i}{d\tau} =\frac{U^i}{U^0} $
is expanded  up to the 2nd-order
\be\label{vi}
v^i=v^{(1)i}+\frac{1}{2}v^{(2)i}.
\ee
By $g_{\mu\nu}U^\mu U^\nu=-1$, the component $U^{0}$ is given  up to 2nd order
\be\label{U0element}
U^{(0)0}=a^{-1},
~~~~~~
U^{(1)0}=0,
~~~~~~
U^{(2)0}=a^{-1} v^{(1)k}v^{(1)}_{k} .
\ee
Here $m U^0$ is the total energy of a particle,
and $\frac{1}{2}m v^{(1)k}v^{(1)}_{k}$ is the  nonrelativistic kinetic energy
in the comoving coordinate in Newtonian mechanics.
By $U^i=v^i U^0$, one has
\be\label{Uielement}
U^{(0)i}=0,
~~~~~~
U^{(1)i}=a^{-1} v^{(1)i},
~~~~~~
U^{(2)i}= a^{-1}v^{(2)i} \, .
\ee
One then writes the covariant velocity  $U_\mu=g_{\mu\nu}U^\nu$ as the following
\be\label{Ulow0}
U_0=
-a \Big(1+\frac{1}{2}v^{(1)m}v^{(1)}_{m} \Big),
\ee
\be\label{Ulowi}
U_i=
a\Big(v^{(1)}_{i}  + \gamma_{ij}^{(1)}v^{(1)j} +\frac{1}{2}v^{(2)}_{i} \Big).
\ee
Note that  the lowest order of $U_i$ is   1st order.

The  Einstein equation is expanded up to 2nd-order of perturbations
\be\label{pertEinstein}
G^{(A)}_{\mu\nu}
\equiv R^{(A)}_{\mu\nu}-\frac{1}{2} \big[g_{\mu\nu}R\big]^{(A)}
=8\pi GT^{(A)}_{\mu\nu},
~~\text{with}~~ A= 0,1,2 .
\ee
For each  order of  (\ref{pertEinstein}),
the (00) component is the energy constraint,
 $(0i)$ components are the momentum constraints,
and  $(ij)$ components are the evolution equations.
The 0th-order Einstein equation gives the Friedmann equations
\be\label{Ein0th00}
\l(\frac{a'}{a}\r)^2
=\frac{8\pi G}{3} a^2\rho^{(0)},
\ee
\be\label{Ein0thij}
-2\frac{a''}{a}
 + \l(\frac{a'}{a} \r)^2
 = 8\pi G   a^2  p^{(0)}.
\ee
By   $c_s^2=\frac{1}{3}$ of    RD stage,
Eqs.(\ref{Ein0th00}) and (\ref{Ein0thij}) have a solution
$a(\tau)\propto\tau$ and
\be\label{rho0tau}
\rho^{(0)}(\tau)=\frac{3}{8\pi G}\frac{a'^{\,2}(\tau)}{a^4(\tau)}
\propto\tau^{-4}.
\ee
The   covariant conservation of the stress tensor is
\be\label{covcons}
T^{ \mu\nu}\,_{; \, \nu}=0 .
\ee
The dynamics of gravitational systems is
determined by (\ref{pertEinstein}) and (\ref{covcons}).
The component $\mu=0$ of (\ref{covcons}) gives the energy conservation
\bl\label{EnConsv2}
&g^{00}p_{,\,0}
+\partial_0\big[(\rho+p)U^0 U^0\big]
+\partial_i\big[(\rho+p)U^0 U^i\big]
+\Gamma^{0}_{00}(\rho+p)U^0 U^0
+\Gamma^{0}_{ij}(\rho+p)U^i U^j
\nn\\
&
+\Gamma^{0}_{00}(\rho+p)U^0 U^0
+\Gamma^{k}_{k0}(\rho+p)U^0 U^0
+\Gamma^{k}_{km}(\rho+p)U^0 U^m =0 ,
\el
and the component $\mu=i$ gives  the momentum conservation
\bl\label{MoConsv2}
&g^{ik}p_{,\,k}
+\partial_0\big[(\rho+p)U^i U^0\big]
+\partial_m\big[(\rho+p)U^i U^m\big]
+2\Gamma^{i}_{m0}(\rho+p)U^m U^0
+\Gamma^{i}_{ml}(\rho+p)U^m U^l
\nn\\
&
+\Gamma^{0}_{00}(\rho+p)U^i U^0
+\Gamma^{k}_{k0}(\rho+p)U^i U^0
+\Gamma^{k}_{km}(\rho+p)U^i U^m =0 ,
\el
where the nonvanishing Christopher symbols are  listed in
Eqs.~(\ref{Christopher000})--(\ref{Christopherijk}).
To each order of perturbation,
(\ref{EnConsv2}) and (\ref{MoConsv2}) will  determine $\rho$ and $U^\mu$.

\section{1st-order perturbations}
\label{sec:1st-order perturbed equations}

Although 1st-order perturbation in RD stage is known in the literature,
here we give  a detailed analysis of the equations,
solutions, and gauge modes,
because the 2nd-order solutions depend upon them.
In addition,
the 2nd-order equations and gauge transformations
have similar structures to the 1st-order ones.
Moreover, the 1st-order solutions will give an insight on
how the tensor perturbations as dynamic degrees of freedom
differ  from the scalar and vector   perturbations.

The $(00)$ component of 1st-order perturbed Einstein equation is
\be  \label{1st00RD}
 G^{(1)}_{00}\equiv
  R_{00}^{(1)}  -\frac{1}{2}g_{00}^{(0)}R^{(1)}
 = 8\pi G T_{00}^{(1)} .
\ee
The expressions of
the 1st-order perturbed Einstein tensors for a general RW spacetime
are listed in Appendix  A.
By  (\ref{G001})  and  (\ref{T001st})  for the RD stage,
the above  gives  the 1st-order energy constraint
\be\label{enCons1stRD}
-\frac{6}{\tau}\phi^{(1)'}
          +2\nabla^2\phi^{(1) }
          +\frac{1}{3}\nabla^2\nabla^2\chi^{||(1)}
=\frac{3}{\tau^2}\delta^{(1)} ,
\ee
which involves only the scalar modes $\phi^{(1)}$, $\chi^{||(1) }$
on the left-hand side (lhs),
and $ \delta^{(1)} $ on the right-hand side (rhs).
The  $(0i)$ component of   1st-order perturbed Einstein equation is
\be \label{cnmi19RD}
 G^{(1)}_{0i}=  R^{(1)}_{0i}=8\pi G T_{0i}^{(1)}.
\ee
By (\ref{R0i1st}) and (\ref{Ti01st}),
the above   for RD stage gives the 1st-order momentum constraint
\be \label{momentconstr1RD}
2\phi^{(1)' }_{,i} + \frac{1}{2} D_{ij}\chi^{||(1)',j  }
   + \frac{1}{2} \chi^{\perp(1)',j  }_{ij}
 = -\frac{4}{\tau^2} v^{(1)}_i ,
\ee
which involves  both  vector and scalar modes,
and the velocity $v^{(1)}_i$ on the rhs.
Both constraints equations (\ref{enCons1stRD}) and (\ref{momentconstr1RD})
 do not involve the tensor modes and
contain only first order time derivatives,
indicating that they are not   dynamical equations.
Now  decompose the 3-velocity as
$v^{(1)}_i=v^{\perp(1)}_i+v^{||(1)}_{,i}$
with $\partial^i v^{\perp(1)}_i=0$.
Applying $\nabla^{-2}\partial^i$  to  (\ref{momentconstr1RD})
leads to
the longitudinal momentum constraint
\be \label{moment1nonCurl}
2\phi^{(1)' }
+ \frac{1}{3} \nabla^2\chi^{||(1)'}
=
-\frac{4}{\tau^2}v^{||(1)},
\ee
which tells that the longitudinal velocity $v^{||(1)}$
is related to the scalar modes.
A combination
[$(\ref{momentconstr1RD})-\partial_i(\ref{moment1nonCurl})$]
gives the  transverse  momentum constraint
\be\label{moment1Curl}
\frac{1}{2} \chi^{\perp(1)',j  }_{ij} =
-\frac{4}{\tau^2} v^{\perp(1)}_i ;
\ee
i.e.,
the curl  velocity $v^{\perp(1)}_i$  is only related to
the 1st-order vector mode.
The $(ij)$ component of 1st-order perturbed Einstein equation is
\be\label{cnm29RD}
G^{(1)}_{ij}  \equiv
R^{(1)}_{ij}  -\frac{1}{2} \delta_{ij}a^2 R^{(1)}
    -\frac{1}{2}a^2\gamma^{(1)}_{ij}  R^{(0)}
 =8\pi G T^{(1)}_{ij},
\ee
which, by   (\ref{Gij1st}) and  (\ref{Tij1st}),
 gives the 1st-order evolution equation
\bl  \label{evoEq1stRD}
&
2\phi^{(1)''} \delta_{ij}
+\frac{4}{\tau}\phi^{(1)'}\delta_{ij}
+\phi^{(1) }_{,ij}
-\nabla^2\phi^{(1) }\delta_{ij}
\nn\\
&
+\frac{1}{2} D_{ij}\chi^{||(1)''}
+\frac{1}{\tau}D_{ij}\chi^{||(1)'}
+\frac{1}{6}\nabla^2D_{ij}\chi^{||(1)}
-\frac{1}{9}\delta_{ij}\nabla^2\nabla^2\chi^{||(1) }
\nn\\
&
+\frac{1}{2} \chi^{\perp(1)''}_{ij}
+\frac{1}{\tau}\chi^{\perp(1)'}_{ij}
\nn \\
&
+\frac{1}{2} \chi^{\top(1)''}_{ij}
+\frac{1}{\tau} \chi^{\top(1)'}_{ij}
- \frac{1}{2}\nabla^2\chi^{\top(1) }_{ij}
=
\frac{3 c_L^2}{\tau^2} \delta^{(1)} \delta_{ij}
\el
where
$\frac{1}{2}D^k_j\chi^{||(1)}_{,ik}
+\frac{1}{2}D^k_i\chi^{||(1)}_{,jk}
-\frac{1}{2}\delta_{ij}D_{kl}\chi^{||(1),kl }
=
\frac{2}{3}\nabla^2D_{ij}\chi^{||(1)}
-\frac{1}{9}\delta_{ij}\nabla^2\nabla^2\chi^{||(1) }$ and
$\chi^{\perp(1),\,k}_{k i,j}
+\chi^{\perp(1),\,k}_{k j,\,i}
    -\nabla^2\chi^{\perp(1)}_{i j}=0$
have been used.
The evolution equation (\ref{evoEq1stRD}) involves second order time derivatives of
all three types of the metric perturbations,
and gives  the dynamical equations.
We  decompose it  into the trace  and   traceless parts
\be\label{evoTrace1st}
\phi^{(1)''}
+\frac{2}{\tau}\phi^{(1)'}
-\frac{1}{3}\nabla^2\phi^{(1)}
-\frac{1}{18}\nabla^2\nabla^2\chi^{||(1) }
=   \frac{3 c_L^2}{2 \tau^2}\delta^{(1)},
\ee
\bl  \label{evoEq1stRDnontr}
&
D_{ij}\phi^{(1) }
+\frac{1}{2} D_{ij}\chi^{||(1)''}
+\frac{1}{\tau}D_{ij}\chi^{||(1)'}
+\frac{1}{6}\nabla^2D_{ij}\chi^{||(1)}
\nn\\
&
+\frac{1}{2} \chi^{\perp(1)''}_{ij}
+\frac{1}{\tau}\chi^{\perp(1)'}_{ij}
+\frac{1}{2} \chi^{\top(1)''}_{ij}
+\frac{1}{\tau} \chi^{\top(1)'}_{ij}
- \frac{1}{2}\nabla^2\chi^{\top(1) }_{ij}
=0.
\el
Applying $3 \nabla^{-2}\nabla^{-2}\partial^i\partial^j$
to Eq.(\ref{evoEq1stRDnontr}) gives the scalar mode equation
\be  \label{evoEq1stRDscalar}
 \chi^{||(1)''}
+\frac{2}{\tau}\chi^{||(1)'}
+\frac{1}{3}\nabla^2\chi^{||(1)} + 2\phi^{(1) }=0.
\ee
Note that the term $\frac{1}{3}\nabla^2\chi^{||(1)}$ in the above has a plus sign,
which is in contrast to the minus sign $-\frac{1}{3}\nabla^2\phi^{(1)}$
of the trace part equation (\ref{evoTrace1st}).
The scalar modes described by (\ref{evoTrace1st}) and (\ref{evoEq1stRDscalar})
behave  as a wave in RD stage,
as will be revealed by their solutions.
A combination
$\partial^i[(\ref{evoEq1stRDnontr})-\frac12 D_{ij}(\ref{evoEq1stRDscalar})]$
yields
  the vector mode equation
\be  \label{evoEq1stRDvec}
 \chi^{\perp(1)''}_{ij}
+\frac{2}{\tau}\chi^{\perp(1)'}_{ij}
=0  ,
\ee
where  an irrelevant $\bf x$-independent constant has been dropped.
Observe that (\ref{evoEq1stRDvec})
is not a hyperbolic type of partial differential equation,
implying that the vector modes  do not propagate in space  and are not a wave.
Besides,  (\ref{evoEq1stRDvec}) tells that the 1st-order vector modes
  is independent of the scalar and tensor modes,
but depends on the curl velocity by (\ref{moment1Curl}).
A combination
$[(\ref{evoEq1stRDnontr})-\frac12 D_{ij}(\ref{evoEq1stRDscalar})-(\ref{evoEq1stRDvec})]$,
yields  the   tensor mode equation
\be  \label{evoEq1stRDtensor}
 \chi^{\top(1)''}_{ij}
+\frac{2}{\tau} \chi^{\top(1)'}_{ij}
- \nabla^2\chi^{\top(1) }_{ij}
=0 ,
\ee
which is a hyperbolic  equation,
indicating that the tensor modes propagate at the speed of light
and are gravitational waves.
Also observe that (\ref{evoEq1stRDtensor})
is not related to the scalar nor the vector modes,
furthermore,
it is homogenous
and does not depend on
the perturbed stress tensor $T^{(1)}_{\mu\nu}$ of relativistic fluid.
We do not consider
the neutrino free-streaming during RD stage \cite{Weinberg2004,MiaoZhang2007}
that can give an anisotropic stress as a source of the tensor modes.
Therefore, at   1st order,
RGW is a free wave propagating in the  RW spacetime background.
So  far, the   1st-order evolution equation (\ref{evoEq1stRD})
has been decomposed  into three   equations,
(\ref{evoEq1stRDscalar}), (\ref{evoEq1stRDvec}),
and (\ref{evoEq1stRDtensor}),
for the three types of metric perturbations,
which are independent of each other.

To solve the equations of the scalars,
we need to know $\delta^{(1)}$ and $v^{(1)}_i$,
which serve as the source for the 1st-order metric perturbations.
For this, we resort to  the covariant conservation of stress tensor.
The 1st-order part of (\ref{EnConsv2})
gives the 1st-order equation of energy conservation
 (see also  (4.24) in Ref.\cite{Bertschinger})
\be\label{1stmaRD}
\rho^{(1)'}
+\frac{3}{\tau}(\rho^{(1)}+p^{(1)})
-3\phi^{(1)'}(\rho^{(0)}+p^{(0)})
+\partial_i\big[(\rho^{(0)}+p^{(0)})v^{(1)i}\big]  =0 ,
\ee
i.e., the continuity equation for a fluid.
By (\ref{deltaA}), the above is written in terms of the 1st-order density contrast
\be\label{enCons1st3}
\delta^{(1)'}
+\frac{3}{\tau}(c_L^2-\frac{1}{3})\delta^{(1)}
=
4\phi^{(1)'}
-\frac{4}{3} \nabla^2 v^{||(1)} ,
\ee
which  contains the scalar  $\phi^{(1)}$ and the longitudinal velocity  $ v^{||(1)}$.
The sound speed $c_L$ appears in the above 1st-order  equation.
Similarly,  the  1st order of  (\ref{MoConsv2})
gives the 1st-order  momentum conservation equation
(see  also  (4.25) in Ref.\cite{Bertschinger})
\be\label{MoConsv7RD}
c_L^2\rho^{(0)}\delta^{(1)}_{,\,i}
+\frac{4}{3}\rho^{(0)'}v^{(1)}_{i}
+\frac{4}{3}\rho^{(0)}v^{(1)'}_{\,i}
+\frac{16}{3\tau}\rho^{(0)}v^{(1)}_{i}  =0 ,
\ee
which is also written as
\be\label{MoConsv5}
c_L^2 \delta^{(1)}_{,\,i}
+\frac{4}{3}v^{(1)'}_{\,i}  =0,
\ee
i.e., the Euler equation for a fluid.
The vector equation  (\ref{MoConsv5}) can be decomposed into
 longitudinal and  transverse parts
\be\label{MoConsv6}
c_L^2\delta^{(1)}
+\frac{4}{3}v^{||(1)'} =0 ,
\ee
\be\label{MoConsv7curl}
v^{\perp(1)'}_i =0.
\ee

We find that the trace of evolution equation (\ref{evoTrace1st})
can be  given as the following combination
\be\label{relation1st1}
 (\ref{evoTrace1st})
=
-\frac{1}{6}(\ref{enCons1stRD})
-\frac{\tau}{6}\frac{d}{d\tau}(\ref{enCons1stRD})
+\frac{\tau}{6}\nabla^2(\ref{moment1nonCurl})
-\frac{1}{2\tau}(\ref{enCons1st3})
,
\ee
and so is the scalar part of
the traceless evolution equation (\ref{evoEq1stRDscalar})
as the following
\be\label{relation1st2}
(\ref{evoEq1stRDscalar})
=
\nabla^{-2}\Big[
(\ref{enCons1stRD})
+\tau\frac{d}{d\tau}(\ref{enCons1stRD})
-\tau\nabla^2(\ref{moment1nonCurl})
+3\frac{d}{d\tau}(\ref{moment1nonCurl})
+\frac{6}{\tau}(\ref{moment1nonCurl})
+\frac{3}{\tau}(\ref{enCons1st3})
-\frac{9}{\tau^2}(\ref{MoConsv6})
\Big].
\ee
(This also occurs for the 2nd-order scalar perturbations.)
This means that we can use the equations of constraints
and conservations  to solve the scalars,
and the solutions will satisfy the evolution equations automatically.

Now we solve for the 1st-order perturbations.
First,
the 1st-order tensor $\chi^{\top(1)}_{ij} $ in (\ref{evoEq1stRDtensor})
is written in terms of Fourier modes
\be  \label{Fourier}
\chi^{\top(1)}_{ij}  ( {\bf x},\tau)= \frac{1}{(2\pi)^{3/2}}
\int d^3k   e^{i \,\bf{k}\cdot\bf{x}}
\sum_{s={+,\times}} {\mathop \epsilon
\limits^s}_{ij}(k) ~ {\mathop h\limits^s}_k(\tau)
, \,\,\,\, {\bf k}=k\hat{k},
\ee
with two polarization tensors satisfying
\[
{\mathop \epsilon  \limits^s}_{ij}(k) \delta^{ij}=0,\,\,\,\,
{\mathop \epsilon  \limits^s}_{ij}(k)  k^i=0,\,\,\,
 {\mathop \epsilon  \limits^s}_{ij}(k)
{ \mathop \epsilon \limits^{s'}}{}^{ij}(k)
       =2\delta_{ss'}.
\]
For RGW  generated during inflation
 \cite{Grishchuk,Allen1988,zhangyang05},
the two polarization modes ${\mathop h\limits^s}_k(\tau)$ with  $s= {+,\times}$
are usually assumed to be statistically equivalent
and the superscript $s$ can be dropped.
During RD stage the mode is given by
\bl \label{GWmode}
h_k(\tau ) = &\frac{1}{a(\tau)}\sqrt{\frac{\pi}{2}}
   \sqrt{\frac{\tau}{2}}
     \big[b_1  H^{(1)}_{\frac{1}{2} } (k\tau )
          +b_2  H^{(2)}_{\frac{1}{2} } (k\tau ) \big] \nn \\
 = & \frac{1}{a(\tau)}
\frac{i}{ \sqrt{2 k}}
     \big[- b_1  e^{i k\tau}
          + b_2  e^{-i k\tau} \big],
\el
where  $b_1$, $b_2$
are $\bf k$-dependent coefficients,
to be determine by the initial condition during inflation,
or a possible subsequent reheating  stage \cite{Grishchuk,zhangyang05}.
There are cosmic  processes occurred during RD stage,
such QCD  transition,  and $e^+e^-$ annihilation \cite{WangZhang2008},
which  modify  only slightly the amplitude of RGW
and will be  neglected  in  this study.

Next, the vector mode equation (\ref{evoEq1stRDvec})
has a general solution
\be   \label{solutionChiperp}
\chi^{\perp(1)}_{ij}=
\frac{C_{1\,ij}(\mathbf x)}{\tau} +C_{2\,ij}(\mathbf x) ,
\ee
where $C_{1\,ij}$ and $C_{2\,ij}$ are arbitrary, time-independent functions,
and can be written in   a form
$C_{1\,ij}=\partial_jC^{\perp}_{1\,i}+\partial_iC^{\perp}_{1\,j}$
with $C^{\perp}_{1\,i}$ being a curl vector satisfying $\partial^i C^{\perp}_{1\,i}=0$,
and similar for $C_{2\,ij}$.
Substituting (\ref{solutionChiperp})
into the  transverse  momentum constraint  (\ref{moment1Curl})
yields the solution of transverse velocity
\be\label{vi1}
v^{\perp(1)}_i
=\frac{1}{8}C^{\,,j}_{1\,ij}(\mathbf x),
\ee
which is time independent,
consistent with the transverse  momentum conservation (\ref{MoConsv7curl}).
Neither vector nor tensor mode involves
the 1st-order sound speed $c_L$.

Then we solve for the scalars.
By the longitudinal momentum conservation   (\ref{MoConsv6}),
\be\label{delta1V1re}
\delta^{(1)}
= -\frac{4}{3 c_L^2} v^{||(1)'} ,
\ee
plugging it into the energy momentum conservation   (\ref{enCons1st3})
gives the  following:
\be\label{phiVrelate2}
\phi^{(1)'}
=  -\frac{1}{3c_L^2} v^{||(1)''}
-\frac{1}{\tau}\frac{(c_L^2-\frac{1}{3})}{c_L^2}v^{||(1)'}
+\frac{1}{3}\nabla^2v^{||(1)}.
\ee
Taking  $\frac{d}{d\tau}$ on the energy constraint (\ref{enCons1stRD})
and combing with the momentum constraint (\ref{moment1nonCurl})
and canceling the scalar $\chi^{||(1)}$ lead to
\be\label{Vdeltaphi}
 -6\frac{d}{d\tau}\Big[\frac{\phi^{(1)'}}{\tau}\Big]
 +\nabla^2 \Big[ -\frac{4}{\tau^2} v^{||(1)} \Big]
= 3\frac{d}{d\tau}\Big[\frac{1}{\tau^2}\delta^{(1)}\Big].
\ee
Plugging $\delta^{(1)}$ of (\ref{delta1V1re}) and $\phi^{(1)'}$ of
 (\ref{phiVrelate2}) into (\ref{Vdeltaphi}),
one arrives at a 3rd-order differential equation of longitudinal velocity
\be\label{v1eqCs1/3}
v^{||(1)'''}
+\frac{3c_L^2}{\tau}v^{||(1)''}
-\frac{6c_L^2+2}{\tau^2}v^{||(1)'}
-c_L^2\nabla^2v^{||(1)'}
-\frac{c_L^2}{\tau}\nabla^2v^{||(1)}
=0 ,
\ee
which can be solved directly.
By  the Fourier transformation
\be  \label{v||1Fourier}
v^{||(1)}( {\bf x},\tau)= \frac{1}{(2\pi)^{3/2}}
   \int d^3k   e^{i \,\bf{k}\cdot\bf{x}}
    v^{||(1)}_{ \bf k},
\ee
(\ref{v1eqCs1/3}) written in the $\bf k$-space is
\be\label{veq1}
v^{||(1)'''}_{ \bf k}
+\frac{3c_L^2}{\tau}v^{||(1)''}_{ \bf k}
+\l(c_L^2k^2-\frac{6c_L^2+2}{\tau^2}\r)v^{||(1)'}_{ \bf k}
+\frac{c_L^2 k^2}{\tau} v^{||(1)}_{\bf k}
=0,
\ee
which has a general   solution
\bl\label{v1||solgen}
v^{||(1)}_{\mathbf k}(\tau)=&
d_1  (\frac{c_L^{} k \tau}{2} )^{-\frac{3 c_L^2}{2}}
  \Gamma \left(\frac{3 c_L^2}{2}+1\right)
\left(J_{\frac{3 c_L^2}{2}}(c_L^{} k \tau)
+ (c_L^{} k \tau) J_{\frac{3 c_L^2}{2}+1}(c_L^{} k \tau)\right)
\nn\\
&
+  d_2   (\frac{c_L^{} k \tau}{2} )^3 \, _1F_2\left(2;\frac{5}{2},
\frac{3 c_L^2}{2}+\frac{5}{2}; -(\frac{c_L^{} k \tau}{2})^2\right)
\nn\\
&
 +d_3   (\frac{c_L^{} k \tau}{2})^{-3 c_L^2}  \,
   _1F_2\left(\frac{1}{2}-\frac{3 c_L^2}{2};
   -\frac{3 c_L^2}{2}-\frac{1}{2},1-\frac{3 c_L^2}{2};
   -(\frac{c_L^{} k \tau}{2})^2\right),
\el
where $d_1 ,d_2 ,d_3 $ are   $\bf k$-dependent integration constants,
$J_q(x)$ is the Bessel function,
and $\Gamma(x)$ is the Gamma function,
and
$_pF_q$
is the generalized hypergeometric function.
For the case of  the   sound speed $c_L^2 =\frac{1}{3}$  \cite{Grishchuk1994},
  the  solution (\ref{v1||solgen}) reduces to
\be\label{vsol}
v^{||(1)}_{\mathbf k}
=\frac{D_1}{k\tau}
+D_2\l(\frac{2}{k\tau}
    +\frac{i}{\sqrt3}\r)e^{-ik\tau/\sqrt3}
+D_3\l(\frac{2}{k\tau}
    -\frac{i}{\sqrt3}\r)e^{ik\tau/\sqrt3},
\ee
with $D_1$, $D_2$,  and $D_3$
being  $\bf k$-dependent  constants
with $D_1=3 \sqrt{3} d_2+2 \sqrt{3} d_3$,
$D_2=-\frac{3 \sqrt{3}}{4}d_2+\frac{i }{2} \sqrt{3} d_1$,
and $D_3 =-\frac{3 \sqrt{3}}{4}d_2+\frac{i }{2} \sqrt{3} d_1$.
Given the  solution  $v^{||(1)}$,
other 1st-order scalar perturbations follow straight forwardly.
The Fourier transform of  (\ref{delta1V1re}) gives the 1st-order density contrast
\bl\label{delta1sol}
\delta^{(1)}_{\mathbf k}
=&  -4v^{||(1)'}_{\mathbf k}
\nn\\
=&
\frac{4D_1}{k\tau^2}
+D_2\l(\frac{8}{k\tau^2}
+ \frac{8 \,i }{\sqrt3 \,\tau}
-\frac{4 k}{3}\r)e^{-ik\tau/\sqrt3}
+D_3\l(\frac{8}{k\tau^2}
-\frac{8\,i}{\sqrt3\,\tau}
-\frac{4 k}{3}\r)e^{ik\tau/\sqrt3}.
\el
Similarly,  time integration of the $k$-mode of (\ref{phiVrelate2})
yields
\bl\label{phi1sol}
\phi^{(1)}_{\mathbf k}
=&
-v^{||(1)'}_{\mathbf k}
-\frac{1}{3}k^2\int^\tau v^{||(1)}_{\mathbf k}d\tau   +D_4
\nn\\
=&
D_1\l(\frac{1}{k\tau^2}
-\frac{k\ln\tau }{3}\r)
+D_2\l(\frac{2}{k\tau^2}
+\frac{2 \,i }{\sqrt3 \,\tau}
    \r)e^{-ik\tau/\sqrt3}
+D_3\l(\frac{2}{k\tau^2}
-\frac{2\,i}{\sqrt3\,\tau}
    \r)e^{ik\tau/\sqrt3}
\nn\\
&
-\frac{2k}{3}\int^\tau\l[ D_2 e^{-ik\tau'/\sqrt3}
    +D_3e^{ik\tau'/\sqrt3}\r]\frac{d\tau'}{\tau'}
+D_4 \, ,
\el
where  $D_4$ is a   $\bf k$-dependent constant.
Equation (\ref{enCons1stRD}) in the $\bf k$-space gives
\bl\label{chi1sol}
\chi^{||(1)}_{\mathbf k}
=&
\frac{9}{k^4\tau^2}\delta^{(1)}_{\mathbf k}
+\frac{18}{k^4\tau}\phi^{(1)'}_{\mathbf k}
+\frac{6}{k^2}\phi^{(1) }_{\mathbf k}
\nn\\
=&
-D_1\frac{2\ln\tau }{k}
+D_2\frac{4\sqrt3\,i}{k^2\tau}e^{-ik\tau/\sqrt3}
-D_3\frac{4\sqrt3\,i}{k^2\tau}e^{ik\tau/\sqrt3}
+\frac{6 D_4}{k^2}
\nn\\
&
-\frac{4}{k}{}\int^\tau\l[ D_2 e^{-ik\tau'/\sqrt3}
    +D_3e^{ik\tau'/\sqrt3}\r]\frac{d\tau'}{\tau'}.
\el
We have checked that
these solutions
satisfy the scalar parts of the evolution equation,
(\ref{evoTrace1st}) and (\ref{evoEq1stRDscalar}).
The   constants  $D_1,D_2,D_3,D_4$
should be determined by the initial conditions,
say,  at the end of inflation.

The 1st-order solutions of
(\ref{solutionChiperp}) and (\ref{vsol})--(\ref{chi1sol})
contain coordinate-dependent gauge terms,
which need to be identified.
[See (\ref{gaugetrphi})--(\ref{gaugeGW}), (\ref{deltarho1Gen}),
and (\ref{v1TransGe})
in Appendix C
and also Ref.~\cite{WangZhang2017} for
the synchronous-synchronous transformation in an RW spacetime.]
For  RD stage   with $a(\tau)\propto\tau$,
one has the residual gauge transform of the metric perturbations
  as the following
\be\label{gaugetrphi1}
  \bar\phi^{(1)}  =   \phi^{(1)}
 +  \frac{1}{\tau^2}   A^{(1)}
 + \frac{ \nabla^2 A^{(1)}}{3} \ln\tau
 + \frac{1}{3}\nabla^2 C^{||(1)},
\ee
\be\label{gaugetrchi1}
 \bar \chi^{||(1)} =
 \chi^{||(1)}
 - 2 A^{(1)} \ln\tau
 -2C^{||(1)}
,
\ee
\be\label{gaugePerpchi1}
\bar\chi^{\perp(1)}_{ij}=\chi^{\perp(1)}_{ij}
-C^{\perp(1)}_{i,j}
-C^{\perp(1)}_{j,i},
\ee
\be\label{gaugeGW1}
\bar\chi^{\top(1)}_{ij}=\chi^{\top(1)}_{ij}  \, .
\ee
where $A^{(1)} $, $C\, ^{ ||(1) ,\,i} $
and  $C\, ^{ \bot(1) i}$ with $ \partial_i C^{ \bot(1)\, i}=0$
are small   $\bf x$-dependent functions
that describe the  transformation.
The transformation of 1st-order density contrast and  3-velocity are
\be \label{delta1transfRD}
\bar\delta^{(1)}
=   \delta^{(1)} + \frac{4}{\tau^2}A^{(1)},
\ee
\be \label{vGaugemodeNonCurl}
\bar v^{||(1)}
=  v^{||(1)}
+\frac{A^{(1)}}{\tau} ,
\ee
\be\label{vGaugemodeCurl}
\bar v^{\perp(1)i} =  v^{\perp(1)i}  .
\ee
Equation (\ref{vGaugemodeCurl}) states
that the transverse velocity is gauge invariant.
In the $\bf k $-space these  are
\be\label{gaugetrphi2}
  \bar\phi^{(1)}_{\mathbf k}  =   \phi^{(1)}_{\mathbf k}
 +   A^{(1)}_{\mathbf k} (\frac{1}{\tau^2} - \frac{ k^2}{3} \ln\tau)
 - \frac{k^2}{3} C^{||(1)}_{\mathbf k},
\ee
\be\label{gaugetrchi2}
 \bar \chi^{||(1)}_{\mathbf k} =  \chi^{||(1)}_{\mathbf k}
    - 2 A^{(1)}_{\mathbf k} \ln\tau   -2C^{||(1)}_{\mathbf k} ,
\ee
\be\label{gaugePerpchi1trans}
\bar\chi^{\perp(1)}_{ {\mathbf k} \, ij}=\chi^{\perp(1)}_{ {\mathbf k} \, ij}
  -C^{\perp(1)}_{ {\mathbf k} \, i,j}  -C^{\perp(1)}_{ {\mathbf k} \, j,i}  ,
\ee
\be\label{gaugeGW1trans}
\bar\chi^{\top(1)}_{ {\mathbf k} \, ij}=\chi^{\top(1)}_{ {\mathbf k}\, ij}  \, .
\ee
\be \label{delta1transf2}
\bar\delta^{(1)}_{\mathbf k}
=   \delta^{(1)}_{\mathbf k} + \frac{4}{\tau^2}A^{(1)}_{\mathbf k} ,
\ee
\be\label{vGaugemodeNonCurlk}
\bar v^{||(1)}_{\bf k}=  v^{||(1)}_{\bf k}+\frac{A^{(1)}_{\bf k}}{\tau} ,
\ee
\be\label{vGaugemodeCurltran}
\bar v^{\perp(1)i}_{\mathbf k}  =  v^{\perp(1)i}_{\mathbf k}   .
\ee
By these we can identify
the gauge modes in the 1st-order solutions.
First, the 1st-order tensor modes are gauge invariant  by (\ref{gaugeGW1}).
Next,
(\ref{gaugePerpchi1}) tells that the constant term $C_{2\,ij}$
in the vector  solution (\ref{solutionChiperp})
is a gauge term,
which can be removed by choosing the gauge transformation
function  $C\, ^{ \bot(1) i}$ to satisfy
$C^{\perp(1)}_{i,j} +C^{\perp(1)}_{j,i}=C_{2\, ij}$,
so that the gauge-invariant modes
in the vector solution (\ref{solutionChiperp})
are
\be\label{solutionChiperp1}
\chi^{\perp(1)}_{ij}=
\frac{C_{1\,ij}(\mathbf x)}{\tau} \, .
\ee
By the relation (\ref{Uielement}) and the solution (\ref{vi1}),
the transverse part of 1st-order velocity $U^{i}$ is
gauge invariant, given by
\be \label{trve}
a^{-1} v^{\perp(1)i}
=\frac{C^{\,ij}_{1\,,j}(\mathbf x)}{8\tau} .
\ee
Both (\ref{solutionChiperp1}) and (\ref{trve})
decay as $\tau^{-1}$ and can be neglected as an approximation.
i.e., we assume that the relativistic fluid is irrotational  during RD stage,
and  set
\be \label{notrsv}
v^{\perp(1)}_i= \chi^{\perp(1)}_{ij}= C_{1\,ij}= 0 \, .
\ee
This  approximation will simplify calculation of 2nd-order perturbations.

The transformations (\ref{gaugetrphi2}) and (\ref{gaugetrchi2})
state that $D_1$ and $D_4$ terms in the scalar solutions (\ref{phi1sol}) and (\ref{chi1sol})
are gauge terms and can be removed by choosing
\be\label{A1Trans}
 A^{(1)}_{\mathbf k}
=-\frac{ D_1}{k},
\ee
\be\label{C1perpTrans}
C^{||(1)}_{\mathbf k}=\frac{3 D_4}{k^2} ,
\ee
so that   the gauge-invariant modes of two scalars are
\bl
\phi^{(1)}_{\mathbf k}
=&
D_2\l(\frac{2}{k\tau^2}
+\frac{2 \,i }{\sqrt3 \,\tau}
    \r)e^{-ik\tau/\sqrt3}
+D_3\l(\frac{2}{k\tau^2}
-\frac{2\,i}{\sqrt3\,\tau}
    \r)e^{ik\tau/\sqrt3} \nn\\
& -\frac{2k}{3}\int^\tau\l[ D_2 e^{-ik\tau'/\sqrt3}
    +D_3e^{ik\tau'/\sqrt3}\r]\frac{d\tau'}{\tau'} ,
    \label{phi1sol2}
\\
\chi^{||(1)}_{\mathbf k}
 =&
D_2\frac{4\sqrt3\,i}{k^2\tau}e^{-ik\tau/\sqrt3}
-D_3\frac{4\sqrt3\,i}{k^2\tau}e^{ik\tau/\sqrt3}    \nn\\
& -\frac{4}{k}{}\int^\tau\l[ D_2 e^{-ik\tau'/\sqrt3}
    +D_3e^{ik\tau'/\sqrt3}\r]\frac{d\tau'}{\tau'},
    \label{chi1sol3}
\el
By  the  transformation (\ref{delta1transf2}),
 the $D_1$ term in the    density contrast  solution (\ref{delta1sol})
is a gauge term
and is removed by choosing the same $A^{(1)}_{\mathbf k}$ as (\ref{A1Trans}),
so that   the gauge-invariant  mode of density contrast is
\be\label{delta1sol3}
\delta^{(1)}_{\mathbf k}
=
D_2\l(\frac{8}{k\tau^2}
+ \frac{8 \,i }{\sqrt3 \,\tau}
-\frac{4 k}{3}\r)e^{-ik\tau/\sqrt3}
+D_3\l(\frac{8}{k\tau^2}
-\frac{8\,i}{\sqrt3\,\tau}
-\frac{4 k}{3}\r)e^{ik\tau/\sqrt3}
.
\ee
Finally,   (\ref{vGaugemodeNonCurlk})
shows that the $D_1$ term in the velocity (\ref{vsol})
is a gauge term,
and  the gauge-invariant  mode of velocity is
\be\label{vsol3}
 v^{||(1)}_{\mathbf k}
=
D_2\l(\frac{2}{k\tau}
    +\frac{i}{\sqrt3}\r)e^{-ik\tau/\sqrt3}
+D_3\l(\frac{2}{k\tau}
    -\frac{i}{\sqrt3}\r)e^{ik\tau/\sqrt3} \, .
\ee

It is revealing  to compare
the behaviors of the 1st-order perturbations in RD stage.
The scalar modes   (\ref{phi1sol2}) and (\ref{chi1sol3}),
the density contrast  (\ref{delta1sol3}),
and the longitudinal velocity (\ref{vsol3})
all contain a factor $e^{\pm ik\tau/\sqrt3}$,
so that they are waves propagating at the sound speed  $\frac{1}{\sqrt 3}$
of  the relativistic fluid.
On the other hand, as mentioned below (\ref{evoEq1stRDvec}),
the vector modes   (\ref{solutionChiperp1})
and the transverse velocity (\ref{trve})
are not a wave and do not propagate,
they  simply decrease with time.
In contrast,
the tensor modes (\ref{GWmode})  are waves
and  propagate at the speed of light.
It is also interesting to compare with
 the  MD stage \cite{Matarrese98,WangZhang2017,ZhangQinWang2017},
in which the scalar, vector, and density contrast
are not a wave and do not propagate,
only the tensor modes  propagate at the speed of light.
Therefore,
the tensor modes always propagate at the speed of light,
regardless of the spacetime background,
in contrast to  the scalar and vector modes.

So far,  we have obtained the gauge-invariant  1st-order solutions.
As we shall see in the next section,
the couplings (products) of the 1st-order solutions
will appear in the 2nd-order perturbed Einstein equation
and constitute a part of effective source for the 2nd-order metric perturbations.

\section{ 2nd-order perturbed equations by scalar-scalar couplings}

To study  the 2nd-order cosmological perturbations,
we need the 2nd-order perturbed Einstein equation,
which are  listed  in Appendix B
for a general RW spacetime.
Since
the vector mode and the curl velocity
are neglected in (\ref{notrsv}) as an approximation,
there remain  only three types of products of 1st-order perturbations:
scalar-scalar, scalar-tensor, and tensor-tensor.
In this paper
we consider the scalar-scalar coupling.

The  $(00)$ component of 2nd-order perturbed Einstein equation
is given by Eq.(\ref{Ein2th003}),
which for RD stage   is   written as
\be  \label{Ein2th003RD}
-\frac{6}{\tau} \phi^{(2)'}_S
+2\nabla^2\phi^{(2)}_S
+\frac{1}{3}\nabla^2\nabla^2\chi^{||(2)}_{S}
=
\frac {3}{\tau^2}\delta^{(2)}_S +E_S  .
\ee
where a subscript ``S" in $\phi^{(2)}_S$, etc.,
indicates the scalar-scalar coupling;
 $\delta^{(2)}_S $ is  the 2nd-order density contrast, and
\bl\label{ES1RD}
E_S
=
&
\frac{8}{\tau^2}v^{||(1),\,k}v^{||(1)}_{,\,k}
+\frac{24}{\tau}\phi^{(1)'}\phi^{(1)}
-6\phi^{(1) '}\phi^{(1) '}
-6\phi^{(1)}_{,\,k}\phi^{(1),\,k}
-16\phi^{(1)}\nabla^2\phi^{(1)}
\nn\\
    &
-\frac{8}{3}\phi^{(1)}\nabla^2\nabla^2\chi^{||(1)}
+2\phi^{(1),\,kl}\chi^{||(1)}_{,\,kl}
-\frac{2}{3}\nabla^2\phi^{(1)}\nabla^2\chi^{||(1)}
+\frac{1}{4} \chi^{||(1)',\,kl}\chi^{||(1)'}_{,\,kl}
            \nn\\
            &
-\frac{1}{12} \nabla^2\chi^{||(1)'}\nabla^2\chi^{||(1)'}
+\frac{2}{\tau} \chi^{||(1),\,kl}\chi^{||(1)'}_{,\,kl}
-\frac{2}{3\tau}\nabla^2\chi^{||(1)}\nabla^2\chi^{||(1)'}
+\frac{1}{3}\chi^{||(1),\,kl}\nabla^2\chi^{||(1)}_{,\,kl}
\nn\\
&
-\frac{1}{9}\nabla^2\chi^{||(1)}\nabla^2\nabla^2\chi^{||(1)}
-\frac{1}{4}\chi^{||(1),\,klm}\chi^{||(1)}_{,\,klm}
+\frac{5}{12}\nabla^2\chi^{||(1)}_{,\,k}\nabla^2\chi^{||(1),\,k}
\el
which consists
the coupling terms of 1st-order scalar metric perturbations,
as well as the longitudinal velocity $v^{||(1)}$
which is absent in the dust model as for the MD stage \cite{Matarrese98,WangZhang2017}.
Equation (\ref{Ein2th003RD})
is formally similar to the 1st-order equation (\ref{enCons1stRD}),
except for $E_S$ as a part of the effective source on the rhs.

The  $(0i)$ component of the 2nd-order perturbed Einstein equation
(\ref{MoConstr2ndv3}) for RD stage is given by
\be\label{MoConstr2ndv3RD}
2 \phi^{(2)'} _{S,\,i}
+\frac{1}{2}D_{ij}\chi^{||(2)',\,j}_S
+\frac{1}{2}\chi^{\perp(2)',\,j}_{S\,ij}
=
-\frac{4}{\tau^2}v^{(2)}_{S\,i} +M_{S\,i},
\ee
where
\bl\label{MSi1RD}
M_{S\,i}
\equiv&
\frac{16}{\tau^2}\phi^{(1)}v^{||(1)}_{,i}
-\frac{8}{\tau^2}v^{||(1),k}\chi^{||(1)}_{,ki}
+\frac{8}{3\tau^2}v^{||(1)}_{,\,i}\nabla^2\chi^{||(1)}
-\frac{6}{\tau^2}(1+c_L^2)\delta^{(1)}v^{||(1)}_{,i}
\nn\\
&
-8\phi^{(1)'} \phi^{(1) } _{,\,i}
-8\phi^{(1) } \phi^{(1) '} _{,\,i}
-\frac{4}{3}\phi^{(1) }\nabla^2\chi^{||(1)'}_{,\,i}
-\frac{4}{3}\phi^{(1)'}\nabla^2\chi^{||(1)}_{ ,\,i}
\nn\\
&
+ \phi^{(1) ,\,k }\chi^{||(1)' }_{,\,ki}
-\frac{1}{3}\phi^{(1) }_{,\,i }\nabla^2\chi^{||(1)' }
-2\phi^{(1)' ,\,k }\chi^{||(1) }_{,\,ki}
+\frac{2}{3}\phi^{(1)'}_{ ,\,i }\nabla^2\chi^{||(1) }
\nn\\
&
+\frac{2}{3}\chi^{||(1)' }_{,\,ki}\nabla^2\chi^{||(1),\,k}
-\frac{1}{18}\nabla^2\chi^{||(1)}_{,\,i}\nabla^2\chi^{||(1)' }
-\frac{1}{3}\chi^{||(1)  }_{,\,ki}\nabla^2\chi^{||(1)',\,k}
\nn\\
&
+\frac{1}{9}\nabla^2\chi^{||(1)  }\nabla^2\chi^{||(1)'}_{,\,i}
- \frac{1}{2}\chi^{||(1)',\,kl }\chi^{||(1) }_{,\,kli} .
\el
Equation (\ref{MoConstr2ndv3RD}) is similar to the 1st-order equation (\ref{momentconstr1RD}),
except for $M_{S\,i}$ in the effective source,
which contains
the 1st-order  pressure perturbation $c_L^2 \delta^{(1)} \propto p^{(1)}$
and the velocity $v^{||(1)}$ from $T^{(2)}_{0i}$,
as well as  the scalar-scalar metric products.
Equation (\ref{MoConstr2ndv3RD}) can be further decomposed into two parts,
 similar  to the 1st-order case  in Section 3.
Writing  the 3-velocity as $v^{(2)}_i=v^{\perp(2)}_i+v^{||(2)}_{,i}$
with $\partial^i v^{\perp(2)}_i=0$ and
applying $\nabla^{-2}\partial^i$ on (\ref{MoConstr2ndv3RD}) lead to
the longitudinal part of momentum constraint
\be \label{MoConstr2ndv3RD2}
2\phi^{(2)'}_{S}
+\frac{1}{3}\nabla^2\chi^{||(2)'}_S
=
-\frac{4}{\tau^2}v^{||(2)}_{S}
+\nabla^{-2}M_{S\,l}^{,\,l},
\ee
where
\bl\label{MSkkScalarRD}
\nabla^{-2}M_{S\,l}^{,l}
&=
-8\phi^{(1)'} \phi^{(1) }
-\frac{1}{3}\phi^{(1) }\nabla^2\chi^{||(1)' }
-\frac{4}{3}\phi^{(1)'}\nabla^2\chi^{||(1)}
-\frac{1}{18}\nabla^2\chi^{||(1)}\nabla^2\chi^{||(1)' }
\nn\\
&
-\frac{1}{6}\chi^{||(1)}_{,\,k}\nabla^2\chi^{||(1)',\,k}
+\nabla^{-2}\Big[
\frac{16}{\tau^2}\phi^{(1),k}v^{||(1)}_{,\,k}
+\frac{16}{\tau^2}\phi^{(1)}\nabla^2v^{||(1)}
-\frac{8}{\tau^2}v^{||(1),kl}\chi^{||(1)}_{,kl}
\nn\\
&
-\frac{16}{3\tau^2}v^{||(1),k}\nabla^2\chi^{||(1)}_{,k}
+\frac{8}{3\tau^2}\nabla^2v^{||(1)}\nabla^2\chi^{||(1)}
-\frac{6}{\tau^2}(1+c_L^2)\delta^{(1)}\nabla^2v^{||(1)}
\nn\\
&
-\frac{6}{\tau^2}(1+c_L^2)\delta^{(1)}_{,\,k}v^{||(1),\,k}
+ \phi^{(1) ,\,kl }\chi^{||(1)' }_{,\,kl}
-\phi^{(1) }\nabla^2\nabla^2\chi^{||(1)'}
+2\nabla^2\phi^{(1)'}\nabla^2\chi^{||(1) }
\nn\\
&
-2\phi^{(1)' ,\,kl}\chi^{||(1) }_{,\,kl}
+\frac{1}{6}\chi^{||(1)}_{,\,k}\nabla^2\nabla^2\chi^{||(1)',\,k}
+\frac{1}{6}\nabla^2\chi^{||(1)  }\nabla^2\nabla^2\chi^{||(1)'}
\nn\\
&
+\frac{1}{6}\chi^{||(1)'}_{,\,kl}\nabla^2\chi^{||(1),\,kl}
+\frac{2}{3}\nabla^2\chi^{||(1)' }_{,\,k}\nabla^2\chi^{||(1),\,k}
- \frac{1}{2}\chi^{||(1)',\,klm }\chi^{||(1) }_{,\,klm}
\Big]
.
\el
A combination
[(\ref{MoConstr2ndv3RD})-$\partial_i$(\ref{MoConstr2ndv3RD2})]
gives the transverse part  of momentum constraint
\be \label{MoCons2ndCurlRD1}
\frac{1}{2}\chi^{\perp(2)',\,j}_{S\,ij}
=
-\frac{4}{\tau^2}v^{\perp(2)}_{S\,i}
+\l(
M_{S\,i}
-\partial_i\nabla^{-2}M_{S\,l}^{,\,l}
\r)
\ ,
\ee
where
{\allowdisplaybreaks
\bl\label{MSiCurlRD}
&
\l( M_{S\,i}-\partial_i\nabla^{-2}M_{S\,l}^{,l} \r)
\nn\\
=&
\frac{16}{\tau^2}\phi^{(1)}v^{||(1)}_{,i}
-\frac{8}{\tau^2}v^{||(1),k}\chi^{||(1)}_{,ki}
+\frac{8}{3\tau^2}v^{||(1)}_{,\,i}\nabla^2\chi^{||(1)}
-\frac{6}{\tau^2}(1+c_L^2)\delta^{(1)}v^{||(1)}_{,i}
\nn\\
&
-\phi^{(1) }\nabla^2\chi^{||(1)' }_{,\,i}
+2\phi^{(1)'}_{ ,\,i }\nabla^2\chi^{||(1) }
+ \phi^{(1) ,\,k }\chi^{||(1)' }_{,\,ki}
-2\phi^{(1)' ,\,k }\chi^{||(1) }_{,\,ki}
+\frac{2}{3}\chi^{||(1)' }_{,\,ki}\nabla^2\chi^{||(1),\,k}
\nn\\
&
-\frac{1}{6}\chi^{||(1)}_{,\,ki}\nabla^2\chi^{||(1)',\,k}
+\frac{1}{6}\chi^{||(1)}_{,\,k}\nabla^2\chi^{||(1)',\,k}_{,\,i}
+\frac{1}{6}\nabla^2\chi^{||(1)  }\nabla^2\chi^{||(1)'}_{,\,i}
- \frac{1}{2}\chi^{||(1)',\,kl }\chi^{||(1) }_{,\,kli}
\nn\\
&
+\partial_i\nabla^{-2}\Big[
-\frac{16}{\tau^2}\phi^{(1),k}v^{||(1)}_{,\,k}
-\frac{16}{\tau^2}\phi^{(1)}\nabla^2v^{||(1)}
+\frac{8}{\tau^2}v^{||(1),kl}\chi^{||(1)}_{,kl}
+\frac{16}{3\tau^2}v^{||(1),k}\nabla^2\chi^{||(1)}_{,k}
\nn\\
&
-\frac{8}{3\tau^2}\nabla^2v^{||(1)}\nabla^2\chi^{||(1)}
+\frac{6}{\tau^2}(1+c_L^2)\delta^{(1)}\nabla^2v^{||(1)}
+\frac{6}{\tau^2}(1+c_L^2)\delta^{(1)}_{,\,k}v^{||(1),\,k}
\nn\\
&
- \phi^{(1) ,\,kl }\chi^{||(1)' }_{,\,kl}
+\phi^{(1) }\nabla^2\nabla^2\chi^{||(1)'}
-2\nabla^2\phi^{(1)'}\nabla^2\chi^{||(1) }
+2\phi^{(1)' ,\,kl}\chi^{||(1) }_{,\,kl}
\nn\\
&
 -\frac{1}{6}\chi^{||(1)}_{,\,k}\nabla^2\nabla^2\chi^{||(1)',\,k}
-\frac{1}{6}\nabla^2\chi^{||(1)  }\nabla^2\nabla^2\chi^{||(1)'}
-\frac{1}{6}\chi^{||(1)'}_{,\,kl}\nabla^2\chi^{||(1),\,kl}
\nn\\
&
-\frac{2}{3}\nabla^2\chi^{||(1)' }_{,\,k}\nabla^2\chi^{||(1),\,k}
+ \frac{1}{2}\chi^{||(1)',\,klm }\chi^{||(1) }_{,\,klm}
\Big]  .
\el
}
Equations (\ref{MoConstr2ndv3RD2}) and (\ref{MoCons2ndCurlRD1})
are similar to the 1st-order equations
(\ref{moment1nonCurl}) and (\ref{moment1Curl}), respectively.

The $(ij)$ component of 2nd-order perturbed Einstein equation
 (\ref{Evo2ndSs1}) for the RD stage is given by
\bl\label{Evo2ndSs1RD}
&
2\phi^{(2)''}_S \delta_{ij}
+\frac{4}{\tau}\phi^{(2)'}_S \delta_{ij}
+\phi^{(2)}_{S,\,ij}
-\nabla^2\phi^{(2)}_S \delta_{ij}
\nn\\
&
+\frac{1}{2}D_{ij}\chi^{||(2)''}_S
+\frac{1}{\tau}D_{ij}\chi^{||(2)'}_S
+\frac{1}{6}\nabla^2D_{ij}\chi^{||(2)}_S
-\frac{1}{9}\delta_{ij}\nabla^2\nabla^2\chi^{||(2) }_S
\nn\\
&
+\frac{1}{2}\chi^{\perp(2)''}_{S\,ij}
+\frac{1}{\tau}\chi^{\perp(2)'}_{S\,ij}
\nn\\&
+\frac{1}{2}\chi^{\top(2)''}_{S\,ij}
+\frac{1}{\tau}\chi^{\top(2)'}_{S\,ij}
-\frac{1}{2}\nabla^2\chi^{\top(2)}_{S\,ij}
=
\frac{3 c_N^2}{\tau^2}\delta^{(2)}_{S}\delta_{ij}
+S_{S\,ij} .
\el
The 2nd-order pressure perturbation $c_N^2\delta^{(2)}_S \propto p^{(2)}$
appears on the rhs.
This 2nd-order evolution
 equation is similar to the 1st-order   (\ref{evoEq1stRD}),
except for the extra part in the effective source
{\allowdisplaybreaks
\bl\label{Ss2ndij2RD}
S_{S\,ij}
&=
\frac{8}{\tau^2}v^{||(1)}_{,i}v^{||(1)}_{,j}
-\frac{12}{\tau^2}c_L^2\delta^{(1)}\phi^{(1)}\delta_{ij}
+\frac{6}{\tau^2}c_L^2\delta^{(1)}\chi^{||(1)}_{,\,ij}
-\frac{2}{\tau^2}c_L^2\delta^{(1)}\nabla^2\chi^{||(1)}\delta_{ij}
-6\phi^{(1)}_{,\,i}\phi^{(1)}_{,\,j}
\nn\\
&
+4\phi^{(1),\,k}\phi^{(1)}_{,\,k}\delta_{ij}
+4\phi^{(1)}\nabla^2\phi^{(1)}\delta_{ij}
-4\phi^{(1)}\phi^{(1)}_{,\,ij}
-2\phi^{(1)'}\phi^{(1)'}\delta_{ij}
-\frac{12}{\tau}\phi^{(1)'}\chi^{||(1)}_{,\,ij}
\nn\\
&
+\frac{4}{\tau}\phi^{(1)'}\nabla^2\chi^{||(1)}\delta_{ij}
-\phi^{(1)'}\chi^{||(1)'}_{,\,ij}
+\frac{1}{3}\phi^{(1)'}\nabla^2\chi^{||(1)'}\delta_{ij}
-6\phi^{(1)''}\chi^{||(1)}_{,\,ij}
+2\phi^{(1)''}\nabla^2\chi^{||(1)}\delta_{ij}
\nn\\
&
-\phi^{(1)}_{,\,j}\nabla^2\chi^{||(1)}_{,\,i}
-\phi^{(1)}_{,\,i}\nabla^2\chi^{||(1)}_{,\,j}
+\phi^{(1)}_{,\,k}\chi^{||(1),\,k}_{,\,ij}
-\frac{2}{3}\phi^{(1)}\nabla^2\chi^{||(1)}_{,\,ij}
+\frac{1}{3}\phi^{(1)}_{,\,k}\nabla^2\chi^{||(1),\,k}\delta_{ij}
\nn\\
&
+\frac{2}{3}\phi^{(1)}\nabla^2\nabla^2\chi^{||(1)}\delta_{ij}
+4\chi^{||(1)}_{,\,ij}\nabla^2\phi^{(1)}
-\frac{4}{3}\nabla^2\phi^{(1)}\nabla^2\chi^{||(1)}\delta_{ij}
-2\phi^{(1)}_{,\,ki}\chi^{||(1),\,k}_{,\,j}
\nn\\
&
-2\phi^{(1)}_{,\,kj}\chi^{||(1),\,k}_{,\,i}
+\frac{4}{3}\phi^{(1)}_{,\,ij}\nabla^2\chi^{||(1)}
+\frac{1}{4}\chi^{||(1),\,klm}\chi^{||(1)}_{,\,klm}\delta_{ij}
-\frac{11}{36}\nabla^2\chi^{||(1),\,k}\nabla^2\chi^{||(1)}_{,\,k}\delta_{ij}
\nn\\
&
-\frac{2}{9}\nabla^2\chi^{||(1)}\nabla^2\nabla^2\chi^{||(1)}\delta_{ij}
-\frac{1}{6}\nabla^2\chi^{||(1)}_{,\,i}\nabla^2\chi^{||(1)}_{,\,j}
-\frac{1}{3}\chi^{||(1),\,k}_{,\,j}\nabla^2\chi^{||(1)}_{,\,ik}
-\frac{1}{3}\chi^{||(1),\,k}_{,\,i}\nabla^2\chi^{||(1)}_{,\,jk}
\nn\\
&
+\frac{2}{3}\chi^{||(1)}_{,\,ij}\nabla^2\nabla^2\chi^{||(1)}
+\frac{2}{9}\nabla^2\chi^{||(1)}\nabla^2\chi^{||(1)}_{,\,ij}
-\frac{1}{2}\chi^{||(1),\,kl}_{,\,i}\chi^{||(1)}_{,\,klj}
+\frac{2}{3}\chi^{||(1)}_{,\,kij}\nabla^2\chi^{||(1),\,k}
\nn\\
&
-\frac{2}{\tau}\chi^{||(1),\,kl}\chi^{||(1)'}_{,\,kl}\delta_{ij}
+\frac{2}{3\tau}\nabla^2\chi^{||(1)}\nabla^2\chi^{||(1)'}\delta_{ij}
-\frac{3}{4}\chi^{||(1)',\,kl}\chi^{||(1)'}_{,\,kl}\delta_{ij}
-\chi^{||(1),\,kl}\chi^{||(1)''}_{,\,kl}\delta_{ij}
\nn\\
&
+\frac{1}{3}\nabla^2\chi^{||(1)}\nabla^2\chi^{||(1)''}\delta_{ij}
+\frac{13}{36}\nabla^2\chi^{||(1)'}\nabla^2\chi^{||(1)'}\delta_{ij}
+\chi^{||(1)',\,k}_{,\,i}\chi^{||(1)'}_{,\,kj}
-\frac{2}{3}\chi^{||(1)'}_{,\,ij}\nabla^2\chi^{||(1)'}
 ,
\el
}
which contains the scalar-scalar metric products,
as well as
the terms of  $v^{||(1)}$ and $c_L^2 \delta^{(1)} \propto p^{(1)}$
that are absent in the dust model.

We also   decompose the evolution equation (\ref{Evo2ndSs1RD}),
by similar procedures to the 1st-order case.
First,
the trace part of  (\ref{Evo2ndSs1RD}) is  the following
\be\label{Evo2ndSsTr2RD}
2\phi^{(2)''}_S
+\frac{4}{\tau}\phi^{(2)'}_S
-\frac{2}{3}\nabla^2\phi^{(2)}_S
-\frac{1}{9}\nabla^2\nabla^2\chi^{||(2) }_S
=
\frac{3c_N^2}{\tau^2}\delta^{(2)}_{S}
+\frac{1}{3}S_{S\,k}^k
\,,
\ee
where
\bl\label{SijTraceRD}
S_{S\,k}^k
=&
\frac{8}{\tau^2} v^{||(1)}_{,\,k}v^{||(1),\,k}
-\frac{36}{\tau^2} c_L^2\delta^{(1)}\phi^{(1)}
+6\phi^{(1)}_{,\,k}\phi^{(1),\,k}
+8\phi^{(1)}\nabla^2\phi^{(1)}
-6\phi^{(1)'}\phi^{(1)'}
\nn\\
&
-4\phi^{(1)}_{,\,kl}\chi^{||(1),\,kl}
+\frac{4}{3}\phi^{(1)}\nabla^2\nabla^2\chi^{||(1)}
+\frac{4}{3}\nabla^2\phi^{(1)}\nabla^2\chi^{||(1)}
+\frac{2}{9}\nabla^2\chi^{||(1)}\nabla^2\nabla^2\chi^{||(1)}
\nn\\
&
-\frac{2}{3}\chi^{||(1),\,kl}\nabla^2\chi^{||(1)}_{,\,kl}
-\frac{5}{12}\nabla^2\chi^{||(1),\,k}\nabla^2\chi^{||(1)}_{,\,k}
+\frac{1}{4}\chi^{||(1),\,klm}\chi^{||(1)}_{,\,klm}
-\frac{6}{\tau}\chi^{||(1),\,kl}\chi^{||(1)'}_{,\,kl}
\nn\\
&
+\frac{2}{\tau}\nabla^2\chi^{||(1)}\nabla^2\chi^{||(1)'}
-\frac{5}{4}\chi^{||(1)',\,kl}\chi^{||(1)'}_{,\,kl}
-3\chi^{||(1),\,kl}\chi^{||(1)''}_{,\,kl}
+\nabla^2\chi^{||(1)}\nabla^2\chi^{||(1)''}
\nn\\
&
+\frac{5}{12}\nabla^2\chi^{||(1)'}\nabla^2\chi^{||(1)'}
 .
\el
The traceless part of (\ref{Evo2ndSs1RD}) is
\bl\label{Evo2ndSsNoTr1RD}
&
D_{ij}\phi^{(2)}_{S}
+\frac{1}{2}D_{ij}\chi^{||(2)''}_S
+\frac{1}{\tau} D_{ij}\chi^{||(2)'}_S
+\frac{1}{6}\nabla^2D_{ij}\chi^{||(2)}_S
\nn\\
&
+\frac{1}{2}\chi^{\perp(2)''}_{S\,ij}
+\frac{1}{\tau} \chi^{\perp(2)'}_{S\,ij}
\nn\\
&
+\frac{1}{2}\chi^{\top(2)''}_{S\,ij}
+\frac{1}{\tau} \chi^{\top(2)'}_{S\,ij}
-\frac{1}{2}\nabla^2\chi^{\top(2)}_{S\,ij}
=
\bar S_{S\,ij}
,
\el
where
{
\bl\label{SijTracelessRD}
\bar S_{S\,ij}
\equiv&
S_{S\,ij}-\frac{1}{3}S_{S\,k}^k\delta_{ij}
\nn\\
=&
\frac{8}{\tau^2}v^{||(1)}_{,i}v^{||(1)}_{,j}
-\frac{8}{3\tau^2}v^{||(1)}_{,\,k}v^{||(1),\,k}\delta_{ij}
+\frac{6}{\tau^2}c_L^2\delta^{(1)}\chi^{||(1)}_{,\,ij}
-\frac{2}{\tau^2}c_L^2\delta^{(1)}\nabla^2\chi^{||(1)}\delta_{ij}
-6\phi^{(1)}_{,\,i}\phi^{(1)}_{,\,j}
\nn\\
&
+2\phi^{(1),\,k}\phi^{(1)}_{,\,k}\delta_{ij}
-4\phi^{(1)}\phi^{(1)}_{,\,ij}
+\frac{4}{3}\phi^{(1)}\nabla^2\phi^{(1)}\delta_{ij}
-\frac{12}{\tau}\phi^{(1)'}\chi^{||(1)}_{,\,ij}
+\frac{4}{\tau}\phi^{(1)'}\nabla^2\chi^{||(1)}\delta_{ij}
\nn\\
&
-\phi^{(1)'}\chi^{||(1)'}_{,\,ij}
+\frac{1}{3}\phi^{(1)'}\nabla^2\chi^{||(1)'}\delta_{ij}
-6\phi^{(1)''}\chi^{||(1)}_{,\,ij}
+2\phi^{(1)''}\nabla^2\chi^{||(1)}\delta_{ij}
-\phi^{(1)}_{,\,j}\nabla^2\chi^{||(1)}_{,\,i}
\nn\\
&
-\phi^{(1)}_{,\,i}\nabla^2\chi^{||(1)}_{,\,j}
+\phi^{(1)}_{,\,k}\chi^{||(1),\,k}_{,\,ij}
+\frac{1}{3}\phi^{(1)}_{,\,k}\nabla^2\chi^{||(1),\,k}\delta_{ij}
-\frac{2}{3}\phi^{(1)}\nabla^2\chi^{||(1)}_{,\,ij}
\nn\\
&
+\frac{2}{9}\phi^{(1)}\nabla^2\nabla^2\chi^{||(1)}\delta_{ij}
+4\chi^{||(1)}_{,\,ij}\nabla^2\phi^{(1)}
-\frac{4}{3}\nabla^2\phi^{(1)}\nabla^2\chi^{||(1)}\delta_{ij}
-2\phi^{(1)}_{,\,ki}\chi^{||(1),\,k}_{,\,j}
\nn\\
&
-2\phi^{(1)}_{,\,kj}\chi^{||(1),\,k}_{,\,i}
+\frac{4}{3}\phi^{(1)}_{,\,kl}\chi^{||(1),\,kl}\delta_{ij}
+\frac{4}{3}\phi^{(1)}_{,\,ij}\nabla^2\chi^{||(1)}
-\frac{4}{9}\nabla^2\phi^{(1)}\nabla^2\chi^{||(1)}\delta_{ij}
\nn\\
&
+\frac{2}{3}\chi^{||(1)}_{,\,kij}\nabla^2\chi^{||(1),\,k}
-\frac{1}{6}\nabla^2\chi^{||(1)}_{,\,i}\nabla^2\chi^{||(1)}_{,\,j}
-\frac{1}{6}\nabla^2\chi^{||(1),\,k}\nabla^2\chi^{||(1)}_{,\,k}\delta_{ij}
-\frac{1}{3}\chi^{||(1),\,k}_{,\,j}\nabla^2\chi^{||(1)}_{,\,ik}
\nn\\
&
-\frac{1}{3}\chi^{||(1),\,k}_{,\,i}\nabla^2\chi^{||(1)}_{,\,jk}
+\frac{2}{9}\chi^{||(1),\,kl}\nabla^2\chi^{||(1)}_{,\,kl}\delta_{ij}
+\frac{2}{3}\chi^{||(1)}_{,\,ij}\nabla^2\nabla^2\chi^{||(1)}
+\frac{2}{9}\nabla^2\chi^{||(1)}\nabla^2\chi^{||(1)}_{,\,ij}
\nn\\
&
-\frac{8}{27}\nabla^2\chi^{||(1)}\nabla^2\nabla^2\chi^{||(1)}\delta_{ij}
-\frac{1}{2}\chi^{||(1),\,kl}_{,\,i}\chi^{||(1)}_{,\,klj}
+\frac{1}{6}\chi^{||(1),\,klm}\chi^{||(1)}_{,\,klm}\delta_{ij}
+\chi^{||(1)',\,k}_{,\,i}\chi^{||(1)'}_{,\,kj}
\nn\\
&
-\frac{1}{3}\chi^{||(1)',\,kl}\chi^{||(1)'}_{,\,kl}\delta_{ij}
-\frac{2}{3}\chi^{||(1)'}_{,\,ij}\nabla^2\chi^{||(1)'}
+\frac{2}{9}\nabla^2\chi^{||(1)'}\nabla^2\chi^{||(1)'}\delta_{ij}
\ .
\el
}
Applying
$3\nabla^{-2}\nabla^{-2}\partial_i\partial_j$ to (\ref{Evo2ndSsNoTr1RD})
gives the   equation for the scalar $\chi^{||(2)}_S$
as the following:
\bl\label{Evo2ndSsChi1RD}
&
\chi^{||(2)''}_S
+\frac{2}{\tau}\chi^{||(2)'}_S
+\frac{1}{3}\nabla^2\chi^{||(2)}_S
+2\phi^{(2)}_{S}
=
3\nabla^{-2}\nabla^{-2}\bar S_{S\,kl}^{\, ,\,kl}
,
\el
where
\bl\label{SijPijbarRD}
\bar S_{S\,kl}^{\,,kl}
=&
\frac{2}{3}\nabla^2\nabla^2\Big[
-\frac{9}{4}\phi^{(1)}\phi^{(1)}
-\frac{2}{3}\phi^{(1)}\nabla^2\chi^{||(1)}
-\frac{1}{18}\nabla^2\chi^{||(1)}\nabla^2\chi^{||(1)}
-\frac{1}{16}\chi^{||(1)}_{,\,kl}\chi^{||(1),\,kl}
\Big]
\nn\\
&
+\nabla^2\Big[
\frac{4}{3\tau^2} v^{||(1)}_{,\,k}v^{||(1),\,k}
+\frac{4}{\tau^2}c_L^2\delta^{(1)}\nabla^2\chi^{||(1)}
-\frac{5}{3}\phi^{(1)}\nabla^2\phi^{(1)}
\nn\\
&
-\frac{8}{\tau}\phi^{(1)'}\nabla^2\chi^{||(1)}
-\frac{2}{3}\phi^{(1)'}\nabla^2\chi^{||(1)'}
-4\phi^{(1)''}\nabla^2\chi^{||(1)}
-\frac{4}{9}\phi^{(1),k}\nabla^2\chi^{||(1)}_{,\,k}
+\frac{4}{3}\phi^{(1)}_{,\,kl}\chi^{||(1),kl}
\nn\\
&
+\frac{1}{54} \nabla^2\chi^{||(1),k}\nabla^2\chi^{||(1)}_{,\,k}
+\frac{11}{36}\chi^{||(1)}_{,\,kl}\nabla^2\chi^{||(1),kl}
+\frac{1}{6}\chi^{||(1)',kl}\chi^{||(1)'}_{,\,kl}
-\frac{1}{9}\nabla^2\chi^{||(1)'}\nabla^2\chi^{||(1)'}
\Big]
\nn\\
&
+\frac{8}{\tau^2}\nabla^2v^{||(1)}\nabla^2v^{||(1)}
+\frac{8}{\tau^2}v^{||(1)}_{,\,k}\nabla^2v^{||(1),\,k}
-\frac{6}{\tau^2}c_L^2\nabla^2\delta^{(1)}\nabla^2\chi^{||(1)}
+\frac{6}{\tau^2}c_L^2\delta^{(1),\,kl}\chi^{||(1)}_{,\,kl}
\nn\\
&
+2\phi^{(1),\,k}\nabla^2\phi^{(1)}_{,\,k}
+2\phi^{(1)}\nabla^2\nabla^2\phi^{(1)}
-\frac{12}{\tau}\phi^{(1)',\,kl}\chi^{||(1)}_{,\,kl}
+\frac{12}{\tau}\nabla^2\phi^{(1)'}\nabla^2\chi^{||(1)}
\nn\\
&
-\phi^{(1)',\,kl}\chi^{||(1)'}_{,\,kl}
+\nabla^2\phi^{(1)'}\nabla^2\chi^{||(1)'}
-6\phi^{(1)'',\,kl}\chi^{||(1)}_{,\,kl}
+6\nabla^2\phi^{(1)''}\nabla^2\chi^{||(1)}
\nn\\
&
+\frac{4}{3}\nabla^2\nabla^2\chi^{||(1)}\nabla^2\phi^{(1)}
+\frac{11}{3}\nabla^2\chi^{||(1)}_{,\,k}\nabla^2\phi^{(1),k}
+\frac{2}{3}\phi^{(1),\,k}\nabla^2\nabla^2\chi^{||(1)}_{,\,k}
-3\phi^{(1)}_{,\,klm}\chi^{||(1),klm}
\nn\\
&
+\frac{7}{9}\nabla^2\chi^{||(1)}_{,\,k}\nabla^2\nabla^2\chi^{||(1),k}
+\frac{5}{18}\nabla^2\nabla^2\chi^{||(1)}\nabla^2\nabla^2\chi^{||(1)}
-\frac{1}{2}\chi^{||(1)}_{,\,klm}\nabla^2\chi^{||(1),\,klm}
\nn\\
&
+\frac{1}{3}\chi^{||(1)'}_{,\,kl}\nabla^2\chi^{||(1)',\,kl}
+\frac{1}{3}\nabla^2\chi^{||(1)'}_{,\,k}\nabla^2\chi^{||(1)',\,k}
\ .
\el
By a combination
$\partial^i$\big[(\ref{Evo2ndSsNoTr1RD})$ -\frac12 D_{ij}$(\ref{Evo2ndSsChi1RD})\big],
one has
\be\label{Evo2ndSsVec1}
\frac{1}{2}\chi^{\perp(2)'',\,k}_{S\,kj}
+\frac{1}{\tau} \chi^{\perp(2)',\,k}_{S\,kj}
=
\bar S_{S\,kj}^{\,,k}
-\partial_j\nabla^{-2}\bar S_{S\,kl}^{\, ,\,kl}
.
\ee
By  $\nabla^{-2}\big[\partial_i(\ref{Evo2ndSsVec1})
+(i\leftrightarrow j)\big]$ and by (\ref{chiVec0}),
one gets the vector mode  equation
\be\label{Evo2ndSsVec2RD}
\frac{1}{2}\chi^{\perp(2)''}_{S\,ij}
+\frac{1}{\tau} \chi^{\perp(2)'}_{S\,ij}
= V_{S\,ij}
~,
\ee
where  the rhs is the effective source [see (\ref{SourceCurl1})]
{
\allowdisplaybreaks
\bl\label{SourceCurl1RD}
V_{S\,ij}
\equiv
&
\nabla^{-2}\bar S_{S\,kj,\,i}^{,\,k}
+\nabla^{-2}\bar S_{S\,ki,j}^{,\,k}
-2\nabla^{-2}\nabla^{-2}\bar S_{S\,kl,\,ij}^{\, ,\,kl}
\nn\\
=&
\partial_i\Big[\frac{1}{3}\phi^{(1)}\nabla^2\chi^{||(1)}_{,\,j}
+2\phi^{(1),\,k}\chi^{||(1)}_{,\,kj}
+\frac{1}{3}\chi^{||(1)}_{,\,kj}\nabla^2\chi^{||(1),\,k}
\Big]
\nn\\
&
-\partial_i\partial_j\nabla^{-2}\Big[
\frac{6}{\tau^2}c_L^2\delta^{(1)}\nabla^2\chi^{||(1)}
-\frac{12}{\tau}\phi^{(1)'}\nabla^2\chi^{||(1)}
-\phi^{(1)'}\nabla^2\chi^{||(1)'}
-6\phi^{(1)''}\nabla^2\chi^{||(1)}
\nn\\
&
+\frac{1}{3}\phi^{(1)}\nabla^2\nabla^2\chi^{||(1)}
+2\phi^{(1),\,kl}\chi^{||(1)}_{,\,kl}
+\frac{1}{12}\nabla^2\chi^{||(1)}_{,\,k}\nabla^2\chi^{||(1),\,k}
+\frac{1}{3}\chi^{||(1)}_{,\,kl}\nabla^2\chi^{||(1),\,kl}
\Big]
\nn\\
&
+\partial_i\nabla^{-2}\Big[
\frac{8}{\tau^2}v^{||(1)}_{,j}\nabla^2v^{||(1)}
+\frac{6}{\tau^2}c_L^2\delta^{(1)}\nabla^2\chi^{||(1)}_{,\,j}
+\frac{6}{\tau^2}c_L^2\delta^{(1),\,k}\chi^{||(1)}_{,\,kj}
\nn\\
&
+2\phi^{(1)}\nabla^2\phi^{(1)}_{,\,j}
-\frac{12}{\tau}\phi^{(1)',\,k}\chi^{||(1)}_{,\,kj}
-\frac{12}{\tau}\phi^{(1)'}\nabla^2\chi^{||(1)}_{,\,j}
-\phi^{(1)'}\nabla^2\chi^{||(1)'}_{,\,j}
-\phi^{(1)',\,k}\chi^{||(1)'}_{,\,kj}
\nn\\
&
-6\phi^{(1)''}\nabla^2\chi^{||(1)}_{,\,j}
-6\phi^{(1)'',\,k}\chi^{||(1)}_{,\,kj}
+\frac{4}{3}\nabla^2\chi^{||(1)}_{,\,j}\nabla^2\phi^{(1)}
-\frac{5}{3}\phi^{(1),\,k}\nabla^2\chi^{||(1)}_{,\,kj}
\nn\\
&
-3\phi^{(1)}_{,\,kl}\chi^{||(1),\,kl}_{,\,j}
+\frac{5}{18}\nabla^2\chi^{||(1)}_{,\,j}\nabla^2\nabla^2\chi^{||(1)}
-\frac{1}{2}\chi^{||(1)}_{,\,klj}\nabla^2\chi^{||(1),\,kl}
+\frac{1}{3}\chi^{||(1)'}_{,\,kj}\nabla^2\chi^{||(1)',\,k}
\Big]
\nn\\
&
-\partial_i\partial_j\nabla^{-2}\nabla^{-2}\Big[
\frac{8}{\tau^2}\nabla^2v^{||(1)}\nabla^2v^{||(1)}
+\frac{8}{\tau^2}v^{||(1)}_{,\,k}\nabla^2v^{||(1),\,k}
-\frac{6}{\tau^2}c_L^2\nabla^2\delta^{(1)}\nabla^2\chi^{||(1)}
\nn\\
&
+\frac{6}{\tau^2}c_L^2\delta^{(1),\,kl}\chi^{||(1)}_{,\,kl}
+2\phi^{(1),\,k}\nabla^2\phi^{(1)}_{,\,k}
+2\phi^{(1)}\nabla^2\nabla^2\phi^{(1)}
-\frac{12}{\tau}\phi^{(1)',\,kl}\chi^{||(1)}_{,\,kl}
\nn\\
&
+\frac{12}{\tau}\nabla^2\phi^{(1)'}\nabla^2\chi^{||(1)}
-\phi^{(1)',\,kl}\chi^{||(1)'}_{,\,kl}
+\nabla^2\phi^{(1)'}\nabla^2\chi^{||(1)'}
-6\phi^{(1)'',\,kl}\chi^{||(1)}_{,\,kl}
\nn\\
&
+6\nabla^2\phi^{(1)''}\nabla^2\chi^{||(1)}
+\frac{4}{3}\nabla^2\nabla^2\chi^{||(1)}\nabla^2\phi^{(1)}
+\frac{11}{3}\nabla^2\chi^{||(1)}_{,\,k}\nabla^2\phi^{(1),k}
+\frac{2}{3}\phi^{(1),k}\nabla^2\nabla^2\chi^{||(1)}_{,\,k}
\nn\\
&
-3\phi^{(1)}_{,\,klm}\chi^{||(1),\,klm}
+\frac{7}{9}\nabla^2\chi^{||(1)}_{,\,k}\nabla^2\nabla^2\chi^{||(1),\,k}
-\frac{1}{2}\chi^{||(1)}_{,\,klm}\nabla^2\chi^{||(1),\,klm}
\nn\\
&
+\frac{5}{18}\nabla^2\nabla^2\chi^{||(1)}\nabla^2\nabla^2\chi^{||(1)}
+\frac{1}{3}\chi^{||(1)'}_{,\,kl}\nabla^2\chi^{||(1)',\,kl}
+\frac{1}{3}\nabla^2\chi^{||(1)'}_{,\,k}\nabla^2\chi^{||(1)',\,k}
\Big]
\nn\\
&
+(i\leftrightarrow j)
\ ,
\el
}
which consists of the scalar-scalar metric products,
as well as  $v^{||(1)}$ and $c_L^2 \delta^{(1)} \propto p^{(1)}$.
Equation (\ref{Evo2ndSsVec2RD}) is similar to
the 1st-order equation (\ref{evoEq1stRDvec}) aside from the source $ V_{S\,ij}$.
For consistency, we have checked that  $V_{S\, ij}$
satisfies the condition (\ref{chiVec0}).
Although  the 1st-order vector is vanishing by assumption,
the 2nd-order  vector  is
 generated by the  coupling of 1st-order scalar perturbations.

Finally,
[(\ref{Evo2ndSsNoTr1RD})$-\frac12 D_{ij}$(\ref{Evo2ndSsChi1RD})
$-$(\ref{Evo2ndSsVec2RD})] gives the 2nd-order tensor mode equation
\be \label{Evo2ndSsTen1RD}
\frac{1}{2}\chi^{\top(2)''}_{S\,ij}
+\frac{1}{\tau} \chi^{\top(2)'}_{S\,ij}
-\frac{1}{2}\nabla^2\chi^{\top(2)}_{S\,ij}
= J_{S\,ij} .
\ee
Unlike  the 1st-order equation (\ref{evoEq1stRDtensor}),
it is inhomogeneous with the rhs being the effective   source
[see (\ref{2ndTensorSource})]
{
\allowdisplaybreaks
\bl\label{2ndTensorSourceRD}
J_{S\,ij}
&
\equiv
\bar S_{S\,ij}
-\frac{3}{2}D_{ij}\nabla^{-2}\nabla^{-2}\bar S_{S\,kl}^{\, ,\,kl}
-\nabla^{-2}\bar S_{S\,kj,\,i}^{,\,k}
-\nabla^{-2}\bar S_{S\,ki,j}^{,\,k}
+2\nabla^{-2}\nabla^{-2}\bar S_{S\,kl,\,ij}^{\, ,\,kl}
\nn\\
&=
D_{ij}\Big[
-\frac{3}{4}\phi^{(1)}\phi^{(1)}
-\frac{1}{3}\phi^{(1)}\nabla^2\chi^{||(1)}
-2\phi^{(1)}_{,\,k}\chi^{||(1),\,k}
+\frac{1}{6}\nabla^2\chi^{||(1)}\nabla^2\chi^{||(1)}
\nn\\
&
+\frac{1}{16}\chi^{||(1)}_{,\,kl}\chi^{||(1),\,kl}
-\frac{1}{3}\chi^{||(1)}_{,\,k}\nabla^2\chi^{||(1),\,k}
\Big]
+\frac{8}{\tau^2}v^{||(1)}_{,i}v^{||(1)}_{,j}
-\frac{8}{3\tau^2} v^{||(1)}_{,\,k}v^{||(1),\,k}\delta_{ij}
\nn\\
&
+\frac{6}{\tau^2}c_L^2\delta^{(1)}\chi^{||(1)}_{,\,ij}
-\frac{2}{\tau^2}c_L^2\delta^{(1)}\nabla^2\chi^{||(1)}\delta_{ij}
+2\phi^{(1)}\phi^{(1)}_{,\,ij}
-\frac{2}{3}\phi^{(1)}\nabla^2\phi^{(1)}\delta_{ij}
\nn\\
&
-\frac{12}{\tau}\phi^{(1)'}\chi^{||(1)}_{,\,ij}
+\frac{4}{\tau}\phi^{(1)'}\nabla^2\chi^{||(1)}\delta_{ij}
-\phi^{(1)'}\chi^{||(1)'}_{,\,ij}
+\frac{1}{3}\phi^{(1)'}\nabla^2\chi^{||(1)'}\delta_{ij}
-6\phi^{(1)''}\chi^{||(1)}_{,\,ij}
\nn\\
&
+2\phi^{(1)''}\nabla^2\chi^{||(1)}\delta_{ij}
+\frac{7}{3}\phi^{(1)}_{,\,ij}\nabla^2\chi^{||(1)}
-\frac{7}{9}\nabla^2\phi^{(1)}\nabla^2\chi^{||(1)}\delta_{ij}
+\frac{1}{3}\phi^{(1)}\nabla^2\chi^{||(1)}_{,\,ij}
\nn\\
&
-\frac{1}{9}\phi^{(1)}\nabla^2\nabla^2\chi^{||(1)}\delta_{ij}
+3\phi^{(1)}_{,\,k}\chi^{||(1),\,k}_{,\,ij}
-\phi^{(1)}_{,\,k}\nabla^2\chi^{||(1),\,k}\delta_{ij}
+2\chi^{||(1),\,k}\phi^{(1)}_{,\,kij}
\nn\\
&
-\frac{2}{3}\chi^{||(1),\,k}\nabla^2\phi^{(1)}_{,\,k}\delta_{ij}
+4\chi^{||(1)}_{,\,ij}\nabla^2\phi^{(1)}
-\frac{4}{3}\nabla^2\phi^{(1)}\nabla^2\chi^{||(1)}\delta_{ij}
+\frac{2}{3}\chi^{||(1)}_{,\,ij}\nabla^2\nabla^2\chi^{||(1)}
\nn\\
&
-\frac{2}{9}\nabla^2\chi^{||(1)}\nabla^2\nabla^2\chi^{||(1)}\delta_{ij}
-\frac{7}{18}\nabla^2\chi^{||(1)}_{,\,i}\nabla^2\chi^{||(1)}_{,\,j}
+\chi^{||(1)}_{,\,kij}\nabla^2\chi^{||(1),\,k}
\nn\\
&
-\frac{11}{54}\nabla^2\chi^{||(1),\,k}\nabla^2\chi^{||(1)}_{,\,k}\delta_{ij}
-\frac{1}{9}\chi^{||(1),\,k}\nabla^2\nabla^2\chi^{||(1)}_{,\,k}\delta_{ij}
+\frac{1}{3}\chi^{||(1)}_{,\,k}\nabla^2\chi^{||(1),\,k}_{,\,ij}
\nn\\
&
-\frac{1}{2}\chi^{||(1),\,kl}_{,\,i}\chi^{||(1)}_{,\,klj}
+\frac{1}{6}\chi^{||(1),\,klm}\chi^{||(1)}_{,\,klm}\delta_{ij}
+\chi^{||(1)',\,k}_{,\,i}\chi^{||(1)'}_{,\,kj}
-\frac{1}{3}\chi^{||(1)',\,kl}\chi^{||(1)'}_{,\,kl}\delta_{ij}
\nn\\
&
-\frac{2}{3}\chi^{||(1)'}_{,\,ij}\nabla^2\chi^{||(1)'}
+\frac{2}{9}\nabla^2\chi^{||(1)'}\nabla^2\chi^{||(1)'}\delta_{ij}
    \nn\\
    &
-\frac{3}{2}D_{ij}\nabla^{-2}\Big[
\frac{4}{3\tau^2}v^{||(1)}_{,\,k}v^{||(1),\,k}
+\frac{4}{\tau^2}c_L^2\delta^{(1)}\nabla^2\chi^{||(1)}
-\frac{5}{3}\phi^{(1)}\nabla^2\phi^{(1)}
\nn\\
&
-\frac{8}{\tau}\phi^{(1)'}\nabla^2\chi^{||(1)}
-\frac{2}{3}\phi^{(1)'}\nabla^2\chi^{||(1)'}
-4\phi^{(1)''}\nabla^2\chi^{||(1)}
-\frac{4}{9}\phi^{(1),k}\nabla^2\chi^{||(1)}_{,\,k}
+\frac{4}{3}\phi^{(1)}_{,\,kl}\chi^{||(1),kl}
\nn\\
&
+\frac{1}{54} \nabla^2\chi^{||(1),k}\nabla^2\chi^{||(1)}_{,\,k}
+\frac{11}{36}\chi^{||(1)}_{,\,kl}\nabla^2\chi^{||(1),kl}
+\frac{1}{6}\chi^{||(1)',kl}\chi^{||(1)'}_{,\,kl}
-\frac{1}{9}\nabla^2\chi^{||(1)'}\nabla^2\chi^{||(1)'}
\Big]
\nn\\
&
-\frac{3}{2}D_{ij}\nabla^{-2}\nabla^{-2}
\Big[
\frac{8}{\tau^2}\nabla^2v^{||(1)}\nabla^2v^{||(1)}
+\frac{8}{\tau^2}v^{||(1)}_{,\,k}\nabla^2v^{||(1),\,k}
-\frac{6}{\tau^2}c_L^2\nabla^2\delta^{(1)}\nabla^2\chi^{||(1)}
\nn\\
&
+\frac{6}{\tau^2}c_L^2\delta^{(1),\,kl}\chi^{||(1)}_{,\,kl}
+2\phi^{(1),\,k}\nabla^2\phi^{(1)}_{,\,k}
+2\phi^{(1)}\nabla^2\nabla^2\phi^{(1)}
-\frac{12}{\tau}\phi^{(1)',\,kl}\chi^{||(1)}_{,\,kl}
\nn\\
&
+\frac{12}{\tau}\nabla^2\phi^{(1)'}\nabla^2\chi^{||(1)}
-\phi^{(1)',\,kl}\chi^{||(1)'}_{,\,kl}
+\nabla^2\phi^{(1)'}\nabla^2\chi^{||(1)'}
-6\phi^{(1)'',\,kl}\chi^{||(1)}_{,\,kl}
\nn\\
&
+6\nabla^2\phi^{(1)''}\nabla^2\chi^{||(1)}
+\frac{4}{3}\nabla^2\nabla^2\chi^{||(1)}\nabla^2\phi^{(1)}
+\frac{11}{3}\nabla^2\chi^{||(1)}_{,\,k}\nabla^2\phi^{(1),k}
+\frac{2}{3}\phi^{(1),\,k}\nabla^2\nabla^2\chi^{||(1)}_{,\,k}
\nn\\
&
-3\phi^{(1)}_{,\,klm}\chi^{||(1),klm}
+\frac{7}{9}\nabla^2\chi^{||(1)}_{,\,k}\nabla^2\nabla^2\chi^{||(1),k}
+\frac{5}{18}\nabla^2\nabla^2\chi^{||(1)}\nabla^2\nabla^2\chi^{||(1)}
\nn\\
&
-\frac{1}{2}\chi^{||(1)}_{,\,klm}\nabla^2\chi^{||(1),\,klm}
+\frac{1}{3}\chi^{||(1)'}_{,\,kl}\nabla^2\chi^{||(1)',\,kl}
+\frac{1}{3}\nabla^2\chi^{||(1)'}_{,\,k}\nabla^2\chi^{||(1)',\,k}
\Big]
\nn
\\
    &
-\partial_i\Big[
\frac{1}{3}\phi^{(1)}\nabla^2\chi^{||(1)}_{,\,j}
+2\phi^{(1),\,k}\chi^{||(1)}_{,\,kj}
+\frac{1}{3}\chi^{||(1)}_{,\,kj}\nabla^2\chi^{||(1),\,k}
\Big]
\nn\\
&
-\partial_j\Big[
\frac{1}{3}\phi^{(1)}\nabla^2\chi^{||(1)}_{,\,i}
+2\phi^{(1),\,k}\chi^{||(1)}_{,\,ki}
+\frac{1}{3}\chi^{||(1)}_{,\,ki}\nabla^2\chi^{||(1),\,k}
\Big]
\nn\\
&
+\partial_i\partial_j\nabla^{-2}\Big[
\frac{12}{\tau^2}c_L^2\delta^{(1)}\nabla^2\chi^{||(1)}
-\frac{24}{\tau}\phi^{(1)'}\nabla^2\chi^{||(1)}
-2\phi^{(1)'}\nabla^2\chi^{||(1)'}
-12\phi^{(1)''}\nabla^2\chi^{||(1)}
\nn\\
&
+\frac{2}{3}\phi^{(1)}\nabla^2\nabla^2\chi^{||(1)}
+4\phi^{(1),\,kl}\chi^{||(1)}_{,\,kl}
+\frac{1}{6}\nabla^2\chi^{||(1)}_{,\,k}\nabla^2\chi^{||(1),\,k}
+\frac{2}{3}\chi^{||(1)}_{,\,kl}\nabla^2\chi^{||(1),\,kl}
\Big]
\nn\\
&
-\partial_i\nabla^{-2}\Big[
\frac{8}{\tau^2}v^{||(1)}_{,j}\nabla^2v^{||(1)}
+\frac{6}{\tau^2}c_L^2\delta^{(1)}\nabla^2\chi^{||(1)}_{,\,j}
+\frac{6}{\tau^2}c_L^2\delta^{(1),\,k}\chi^{||(1)}_{,\,kj}
\nn\\
&
+2\phi^{(1)}\nabla^2\phi^{(1)}_{,\,j}
-\frac{12}{\tau}\phi^{(1)',\,k}\chi^{||(1)}_{,\,kj}
-\frac{12}{\tau}\phi^{(1)'}\nabla^2\chi^{||(1)}_{,\,j}
-\phi^{(1)'}\nabla^2\chi^{||(1)'}_{,\,j}
-\phi^{(1)',\,k}\chi^{||(1)'}_{,\,kj}
\nn\\
&
-6\phi^{(1)''}\nabla^2\chi^{||(1)}_{,\,j}
-6\phi^{(1)'',\,k}\chi^{||(1)}_{,\,kj}
+\frac{4}{3}\nabla^2\chi^{||(1)}_{,\,j}\nabla^2\phi^{(1)}
-\frac{5}{3}\phi^{(1),\,k}\nabla^2\chi^{||(1)}_{,\,kj}
-3\phi^{(1)}_{,\,kl}\chi^{||(1),\,kl}_{,\,j}
\nn\\
&
+\frac{5}{18}\nabla^2\chi^{||(1)}_{,\,j}\nabla^2\nabla^2\chi^{||(1)}
-\frac{1}{2}\chi^{||(1)}_{,\,klj}\nabla^2\chi^{||(1),\,kl}
+\frac{1}{3}\chi^{||(1)'}_{,\,kj}\nabla^2\chi^{||(1)',\,k}
\Big]
\nn\\
&
-\partial_j\nabla^{-2}\Big[
\frac{8}{\tau^2}v^{||(1)}_{,i}\nabla^2v^{||(1)}
+\frac{6}{\tau^2}c_L^2\delta^{(1)}\nabla^2\chi^{||(1)}_{,\,i}
+\frac{6}{\tau^2}c_L^2\delta^{(1),\,k}\chi^{||(1)}_{,\,ki}
\nn\\
&
+2\phi^{(1)}\nabla^2\phi^{(1)}_{,\,i}
-\frac{12}{\tau}\phi^{(1)',\,k}\chi^{||(1)}_{,\,ki}
-\frac{12}{\tau}\phi^{(1)'}\nabla^2\chi^{||(1)}_{,\,i}
-\phi^{(1)'}\nabla^2\chi^{||(1)'}_{,\,i}
-\phi^{(1)',\,k}\chi^{||(1)'}_{,\,ki}
\nn\\
&
-6\phi^{(1)''}\nabla^2\chi^{||(1)}_{,\,i}
-6\phi^{(1)'',\,k}\chi^{||(1)}_{,\,ki}
+\frac{4}{3}\nabla^2\chi^{||(1)}_{,\,i}\nabla^2\phi^{(1)}
-\frac{5}{3}\phi^{(1),\,k}\nabla^2\chi^{||(1)}_{,\,ki}
-3\phi^{(1)}_{,\,kl}\chi^{||(1),\,kl}_{,\,i}
\nn\\
&
+\frac{5}{18}\nabla^2\chi^{||(1)}_{,\,i}\nabla^2\nabla^2\chi^{||(1)}
-\frac{1}{2}\chi^{||(1)}_{,\,kli}\nabla^2\chi^{||(1),\,kl}
+\frac{1}{3}\chi^{||(1)'}_{,\,ki}\nabla^2\chi^{||(1)',\,k}
\Big]
\nn\\
&
+\partial_i\partial_j\nabla^{-2}\nabla^{-2}\Big[
\frac{16}{\tau^2}\nabla^2v^{||(1)}\nabla^2v^{||(1)}
+\frac{16}{\tau^2}v^{||(1)}_{,\,k}\nabla^2v^{||(1),\,k}
\nn\\
&
-\frac{12}{\tau^2}c_L^2\nabla^2\delta^{(1)}\nabla^2\chi^{||(1)}
+\frac{12}{\tau^2}c_L^2\delta^{(1),\,kl}\chi^{||(1)}_{,\,kl}
+4\phi^{(1),\,k}\nabla^2\phi^{(1)}_{,\,k}
+4\phi^{(1)}\nabla^2\nabla^2\phi^{(1)}
\nn\\
&
-\frac{24}{\tau}\phi^{(1)',\,kl}\chi^{||(1)}_{,\,kl}
+\frac{24}{\tau}\nabla^2\phi^{(1)'}\nabla^2\chi^{||(1)}
-2\phi^{(1)',\,kl}\chi^{||(1)'}_{,\,kl}
+2\nabla^2\phi^{(1)'}\nabla^2\chi^{||(1)'}
\nn\\
&
-12\phi^{(1)'',\,kl}\chi^{||(1)}_{,\,kl}
+12\nabla^2\phi^{(1)''}\nabla^2\chi^{||(1)}
+\frac{8}{3}\nabla^2\nabla^2\chi^{||(1)}\nabla^2\phi^{(1)}
+\frac{22}{3}\nabla^2\chi^{||(1)}_{,\,k}\nabla^2\phi^{(1),\,k}
\nn\\
&
+\frac{4}{3}\phi^{(1),\,k}\nabla^2\nabla^2\chi^{||(1)}_{,\,k}
-6\phi^{(1)}_{,\,klm}\chi^{||(1),\,klm}
+\frac{14}{9}\nabla^2\chi^{||(1)}_{,\,k}\nabla^2\nabla^2\chi^{||(1),\,k}
-\chi^{||(1)}_{,\,klm}\nabla^2\chi^{||(1),\,klm}
\nn\\
&
+\frac{5}{9}\nabla^2\nabla^2\chi^{||(1)}\nabla^2\nabla^2\chi^{||(1)}
+\frac{2}{3}\chi^{||(1)'}_{,\,kl}\nabla^2\chi^{||(1)',\,kl}
+\frac{2}{3}\nabla^2\chi^{||(1)'}_{,\,k}\nabla^2\chi^{||(1)',\,k}
\Big]
\ ,
\el
}
which
consists of the scalar-scalar couplings only.

So far the 2nd-order perturbed Einstein equation has been decomposed
into the equations for
the 2nd-order scalar, vector, and tensor, respectively.
To solve them,
we need to specify the 2nd-order density contrast and   velocity,
and resort to the 2nd-order energy-momentum conservation
(\ref{enCons2ndGeneral}) and (\ref{MoCons2ndGeneral}).
For RD stage (\ref{enCons2ndGeneral}) for the scalar-scalar coupling
gives the 2nd-order energy conservation
\bl
&
\delta^{(2)'}_S
+\frac{3}{\tau}\l(c_N^2-\frac{1}{3}\r)\delta^{(2)}_S
+\frac{4}{3}\nabla^2v^{||(2)}_S
-4\phi^{(2)'}_S
\nn\\
&
+\frac{16}{3}v^{||(1)',k}v^{||(1)}_{,k}
+2(1+c_L^2)\delta^{(1)}_{,k}v^{||(1),k}
+2(1+c_L^2)\delta^{(1)}\nabla^2v^{||(1)}
\nn\\
&
-6(1+c_L^2)\delta^{(1)}\phi^{(1)'}
-16 \phi^{(1)'}\phi^{(1)}
-\frac{4}{3}D_{kl}\chi^{||(1)'}D^{kl}\chi^{||(1)}
-8\phi^{(1)}_{,\,k}v^{||(1),k}
=0.
\el
Moving the coupling terms to the rhs,
this is also written as
\be\label{enCons2ndRD2}
\delta^{(2)'}_S
+\frac{3}{\tau}\l(c_N^2-\frac{1}{3}\r)\delta^{(2)}_S
=
-\frac{4}{3}\nabla^2v^{||(2)}_S
+4\phi^{(2)'}_S
+A_S ,
\ee
which is similar to the 1st-order (\ref{enCons1st3}),
where
\bl\label{AS}
A_S\equiv
&
-\frac{16}{3}v^{||(1)',k}v^{||(1)}_{,k}
-2(1+c_L^2)\delta^{(1)}_{,k}v^{||(1),k}
-2(1+c_L^2)\delta^{(1)}\nabla^2v^{||(1)}
+6(1+c_L^2)\delta^{(1)}\phi^{(1)'}
\nn\\
&
+16 \phi^{(1)'}\phi^{(1)}
+\frac{4}{3}\chi^{||(1)'}_{,kl}\chi^{||(1),kl}
-\frac{4}{9}\nabla^2\chi^{||(1)'}\nabla^2\chi^{||(1)}
+8\phi^{(1)}_{,\,k}v^{||(1),k}
\el
is part of the effective source.
The 2nd-order pressure perturbation $c_N^2\delta^{(2)}_S \propto p^{(2)}$
appears in Eq.(\ref{enCons2ndRD2}).
From (\ref{MoCons2ndGeneral}) follows the  2nd-order momentum conservation
\be\label{MoCons2ndRD2}
c_N^2 \delta^{(2)}_{S,\,i}
+\frac{4}{3} v^{(2)'}_{S\,i}
=
F_{S\,i}
,
\ee
which is similar to the 1st-order (\ref{MoConsv5}),
 where
\bl\label{FS}
F_{S\,i}\equiv
&
-2(1+c_L^2) \delta^{(1)'}v^{||(1)}_{,i}
-2(1+c_L^2) \delta^{(1)}v^{||(1)'}_{,i}
-\frac{8}{3}v^{||(1)}_{,ik} v^{||(1),k}
\nn\\
&
-\frac{8}{3} v^{||(1)}_{,i}\nabla^2v^{||(1)}
-4c_L^2\delta^{(1)}_{,\,i} \phi^{(1)}
+\frac{40}{3}v^{||(1)}_{,i}\phi^{(1)'}
+2c_L^2 \delta^{(1),\,k}\chi^{||(1)}_{,ki}
\nn\\
&
-\frac{2}{3}c_L^2 \delta^{(1)}_{,\,i}\nabla^2\chi^{||(1)}
-\frac{8}{3}v^{||(1),k}\chi^{||(1)'}_{,ki}
+\frac{8}{9}v^{||(1)}_{,\,i}\nabla^2\chi^{||(1)'}
.
\el
To proceed, by [$\nabla^{-2}\partial^i$(\ref{MoCons2ndRD2})],
(\ref{MoCons2ndRD2}) is decomposed into a longitudinal part
\be\label{MoC2ndNonCurl}
c_N^2 \delta_S^{(2)}
+\frac{4}{3} v_S^{||(2)'}
=
F^{||}_S
\ee
with
\bl\label{FSnoncu}
F^{||}_S
\equiv
&
\nabla^{-2}\partial^iF_{S\,i}
\nn\\
=&
-\frac{4}{3}v^{||(1)}_{,k} v^{||(1),k}
+\nabla^{-2}\Big[
-2(1+c_L^2) \delta^{(1)'}_{,k}v^{||(1),k}
-2(1+c_L^2) \delta^{(1)'}\nabla^2v^{||(1)}
\nn\\
&
-2(1+c_L^2) \delta^{(1),k}v^{||(1)'}_{,k}
-2(1+c_L^2) \delta^{(1)}\nabla^2v^{||(1)'}
-\frac{8}{3} v^{||(1)}_{,k}\nabla^2v^{||(1),k}
-\frac{8}{3} \nabla^2v^{||(1)}\nabla^2v^{||(1)}
\nn\\
&
-4c_L^2\delta^{(1)}_{,\,k} \phi^{(1),k}
-4c_L^2 \phi^{(1)}\nabla^2\delta^{(1)}
+\frac{40}{3}v^{||(1),k}\phi^{(1)'}_{,k}
+\frac{40}{3}\phi^{(1)'}\nabla^2v^{||(1)}
+2c_L^2 \delta^{(1),\,kl}\chi^{||(1)}_{,kl}
\nn\\
&
+\frac{4}{3}c_L^2 \delta^{(1),\,k}\nabla^2\chi^{||(1)}_{,k}
-\frac{2}{3}c_L^2 \nabla^2\delta^{(1)}\nabla^2\chi^{||(1)}
-\frac{8}{3}v^{||(1),kl}\chi^{||(1)'}_{,kl}
\nn\\
&
-\frac{16}{9}v^{||(1),k}\nabla^2\chi^{||(1)'}_{,k}
+\frac{8}{9}\nabla^2v^{||(1)}\nabla^2\chi^{||(1)'}
\Big] .
\el
By [(\ref{MoCons2ndRD2})-$\partial_i$(\ref{MoC2ndNonCurl})],
one gets the  transverse part
\be\label{MoC2ndCurl}
\frac{4}{3} v^{\perp(2)'}_{S\,i}
=
F^\perp_{S\,i}
\ee
with the effective source
{
\allowdisplaybreaks
\bl\label{FScurl}
F^\perp_{S\,i}
\equiv
&
F_{S\,i}-\partial_i F^{||}_{S}
\nn\\
=&
-2(1+c_L^2) \delta^{(1)'}v^{||(1)}_{,i}
-2(1+c_L^2) \delta^{(1)}v^{||(1)'}_{,i}
-\frac{8}{3} v^{||(1)}_{,i}\nabla^2v^{||(1)}
-4c_L^2\delta^{(1)}_{,\,i} \phi^{(1)}
\nn\\
&
+\frac{40}{3}v^{||(1)}_{,i}\phi^{(1)'}
+2c_L^2 \delta^{(1),\,k}\chi^{||(1)}_{,ki}
-\frac{2}{3}c_L^2 \delta^{(1)}_{,\,i}\nabla^2\chi^{||(1)}
-\frac{8}{3}v^{||(1),k}\chi^{||(1)'}_{,ki}
+\frac{8}{9}v^{||(1)}_{,\,i}\nabla^2\chi^{||(1)'}
        \nn\\
    &
+\partial_i\nabla^{-2}\Big[
2(1+c_L^2) \delta^{(1)'}_{,k}v^{||(1),k}
+2(1+c_L^2) \delta^{(1)'}\nabla^2v^{||(1)}
+2(1+c_L^2) \delta^{(1),k}v^{||(1)'}_{,k}
\nn\\
&
+2(1+c_L^2) \delta^{(1)}\nabla^2v^{||(1)'}
+\frac{8}{3} v^{||(1)}_{,k}\nabla^2v^{||(1),k}
+\frac{8}{3} \nabla^2v^{||(1)}\nabla^2v^{||(1)}
+4c_L^2\delta^{(1)}_{,\,k} \phi^{(1),k}
\nn\\
&
+4c_L^2 \phi^{(1)}\nabla^2\delta^{(1)}
-\frac{40}{3}v^{||(1),k}\phi^{(1)'}_{,k}
-\frac{40}{3}\phi^{(1)'}\nabla^2v^{||(1)}
-2c_L^2 \delta^{(1),\,kl}\chi^{||(1)}_{,kl}
\nn\\
&
-\frac{4}{3}c_L^2 \delta^{(1),\,k}\nabla^2\chi^{||(1)}_{,k}
+\frac{2}{3}c_L^2 \nabla^2\delta^{(1)}\nabla^2\chi^{||(1)}
+\frac{8}{3}v^{||(1),kl}\chi^{||(1)'}_{,kl}
+\frac{16}{9}v^{||(1),k}\nabla^2\chi^{||(1)'}_{,k}
\nn\\
&
-\frac{8}{9}\nabla^2v^{||(1)}\nabla^2\chi^{||(1)'}
\Big]
\el
}
being a transverse vector function.

Similar to the 1st-order relations (\ref{relation1st1})
and (\ref{relation1st2}),
we find that the trace of the 2nd-order evolution equation (\ref{Evo2ndSsTr2RD})
can be formed by a combination
\be\label{relation1}
 (\ref{Evo2ndSsTr2RD})
 =
 -\frac{1}{3}(\ref{Ein2th003RD})
-\frac{\tau}{3}\frac{d}{d\tau}(\ref{Ein2th003RD})
+\frac{\tau}{3}\nabla^2(\ref{MoConstr2ndv3RD2})
-\frac{1}{\tau}(\ref{enCons2ndRD2})
,
\ee
and the scalar part of
the 2nd-order traceless evolution equation (\ref{Evo2ndSsChi1RD})
can be formed by a  combination
\be\label{relation3}
(\ref{Evo2ndSsChi1RD})
=
\nabla^{-2}\Big[
(\ref{Ein2th003RD})
+\tau\frac{d}{d\tau}(\ref{Ein2th003RD})
-\tau\nabla^2(\ref{MoConstr2ndv3RD2})
+3\frac{d}{d\tau}(\ref{MoConstr2ndv3RD2})
+\frac{6}{\tau}(\ref{MoConstr2ndv3RD2})
+\frac{3}{\tau}(\ref{enCons2ndRD2})
-\frac{9}{\tau^2}(\ref{MoC2ndNonCurl})
\Big]  .
\ee
Thus,  we can  use the equations of constraints
and conservations
to solve the scalars,
and the solutions will satisfy the evolution equations automatically.

\section{Solutions of the 2nd-order perturbations}

Now we  solve the 2nd-order equations given in the last section.
First, the 2nd-order tensor   equation (\ref{Evo2ndSsTen1RD})
has the   general   solution
\be\label{solgw}
\chi^{\top(2)}_{S\,ij}({\bf x},\tau)
=\int  \frac{d^3k}{(2\pi)^{\frac{3}{2}}}  e^{i{\bf k\cdot x}}
\Big(
\bar I_{S\, ij}({\bf k}, \tau) +
\sum_{s={+,\times}} {\mathop \epsilon
\limits^s}_{ij}({\bf k})
\Big[-b_{1}^s \sqrt{\frac{2}{\pi}}\,\frac{i\, e^{i  k\tau}}{ k\tau}
+b_{2}^s \sqrt{\frac{2}{\pi}}\,\frac{i\, e^{-i  k\tau}}{ k\tau} \Big]
\Big),
\ee
where $b_{1}^s$ and  $b_{2}^s $ are  polarization-dependent
and ${\bf k}$-dependent coefficients,
to be determined by    initial conditions,
their  associated term is   the  homogeneous solution of  (\ref{Evo2ndSsTen1RD})
and has   the same form as   (\ref{GWmode}).
The integrand of the inhomogenous solution in (\ref{solgw}) is given by
\be\label{GW2ndStepI}
\bar I_{S\,ij}({\bf k},\tau)\equiv
\frac{ie^{-ik\tau}}{k\tau}\int^{\tau}\tau'e^{ik\tau'}\bar J_{S\, ij}({\bf k},\tau')d\tau'
-\frac{ie^{ik\tau}}{k\tau}\int^{\tau}\tau'e^{-ik\tau'}\bar J_{S\, ij}({\bf k},\tau')d\tau'
,
\ee
with $\bar J_{S\, ij}$ being the Fourier transform of the source
$J_{S\,ij}$ in (\ref{2ndTensorSourceRD})
consisting of  products of 1st-order solutions.

Next,
the vector mode equation  (\ref{Evo2ndSsVec2RD}) has the  general solution
\be\label{chiVecSol}
\chi^{\perp(2)}_{S\,ij}({\bf x},\tau)
=c_{1ij}({\bf x})
+\frac{c_{2ij}({\bf x})}{\tau}
+ 2 \int^\tau \frac{d\tau'}{\tau^{'2}}
 \int^{\tau'} \tau^{''2}\,V_{S\,ij}({\bf x},\tau'')d\tau''
\,,
\ee
with $V_{S\,ij}$ given by (\ref{SourceCurl1RD}),
and
$c_{1ij}$ and $c_{2ij}$ are  two time-independent functions
and correspond to the homogenous solution
to be determined by initial values.
(Actually $c_{1ij}$ is a gauge mode as shall be seen in the next section.)
Plugging the solution (\ref{chiVecSol}) into (\ref{MoCons2ndCurlRD1})
gives  the solution of transverse  2nd-order velocity
\bl\label{Vperp2ndSol}
v^{\perp(2)}_{S\,i}
=&
\frac{c_{2ij}^{,j}({\bf x})}{8}
+\frac{\tau^2}{4}\l(
M_{S\,i}
-\partial_i\nabla^{-2}M_{S\,k}^{,\,k}
\r)
- \frac{1}{4}
    \int^{\tau}\tau^{'2}\,V_{S\,ij}^{,j}({\bf x},\tau')d\tau'
\ ,
\el
where $(M_{S\,i}-\partial_i\nabla^{-2}M_{S\,k}^{,k})$
has been defined  in (\ref{MSiCurlRD}).
This solution can also be derived from the integration of
the transverse part of the momentum conservation (\ref{MoC2ndCurl}),
as we have checked.
Although the 1st-order curl vector $v^{\perp(1)}_{i}$ is vanishing by assumption,
nevertheless, the 2nd-order $v^{\perp(2)}_{i}$
is generated according to (\ref{Vperp2ndSol}).

Next,
we solve the 2nd-order scalars by similar procedures to the 1st-order case.
From the longitudinal part of 2nd-order momentum conservation
(\ref{MoC2ndNonCurl}),
\be \label{deltaV2}
 \delta_S^{(2)}
=
-\frac{4}{3c_N^2} v_S^{||(2)'}
+\frac{1}{c_N^2}F^{||}_S
\ .
\ee
Plugging this  $\delta^{(2)}_S$ into
the energy conservation (\ref{enCons2ndRD2})
gives $\phi^{(2)'}_S$ in terms of $v_S^{||(2)}$,
 as the following
\be\label{Vphi2}
\phi^{(2)'}_S
=
-\frac{1}{3 c_N^2} v_S^{||(2)''}
-\frac{1}{\tau}\frac{c_N^2-\frac{1}{3}}{c_N^2} v_S^{||(2)'}
+\frac{1}{3}\nabla^2v^{||(2)}_S
+\frac{1}{4 c_N^2}F^{||'}_S
+\frac{3}{4\tau}\frac{c_N^2-\frac{1}{3}}{c_N^2}F^{||}_S
-\frac{1}{4}A_S
.
\ee
Taking [$\frac{d}{d\tau}$(\ref{Ein2th003RD})] gives
\be
-\frac{6}{\tau} \phi^{(2)''}_S
+\frac{6}{\tau^2} \phi^{(2)'}_S
+\nabla^2 \Big[
2\phi^{(2)'}_S
+\frac{1}{3}\nabla^2\chi^{||(2)'}_{S}
\Big]
=
\frac{3}{\tau^2}\delta^{(2)'}_S
-\frac{6}{\tau^3}\delta^{(2)}_S
+E^{\,'}_S
.
\ee
Plugging the longitudinal momentum constraint (\ref{MoConstr2ndv3RD2}) into the above
gives
\be \label{3dt}
-\frac{6}{\tau} \phi^{(2)''}_S
+\frac{6}{\tau^2} \phi^{(2)'}_S
-\frac{4}{\tau^2}\nabla^2v^{||(2)}_{S}
+M_{S\,l}^{,\,l}
=
\frac{3}{\tau^2}\delta^{(2)'}_S
-\frac{6}{\tau^3}\delta^{(2)}_S
+E^{\,'}_S
.
\ee
Then,
plugging $\delta^{(2)}_S$  of  (\ref{deltaV2}) and
$\phi^{(2)'}_S $ of  (\ref{Vphi2}) into (\ref{3dt})
yields a third order  differential equation of $v^{||(2)}_{S}$
as
\be\label{V2ndeq1}
v_S^{||(2)'''}
+\frac{3c_N^2}{\tau} v_S^{||(2)''}
-\frac{6c_N^2+2}{\tau^2}v_S^{||(2)'}
-\frac{c_N^2}{\tau}\nabla^2v^{||(2)}_{S}
-c_N^2\nabla^2v^{||(2)'}_S
=Z_S
,
\ee
where the effective source is
\ba \label{ZScn1}
Z_S & \equiv &
\frac{3}{4}F^{||''}_S
+\frac{9c_N^2}{4\tau}F^{||'}_S
-\frac{9c_N^2+3}{2\tau^2}F^{||}_S
-\frac{3c_N^2}{4}A_S^{\,'}
+\frac{3c_N^2}{4\tau}A_S
  \nn \\
&& -\frac{\tau}{2}c_N^2M_{S\,l}^{,\,l}
+\frac{\tau}{2}c_N^2E^{\,'}_S
\,.
\ea
Written in  the $\bf k$-space,  (\ref{V2ndeq1}) and (\ref{ZScn1}) are
\be\label{V2eqFourier1}
v_{S\mathbf k}^{||(2)'''}
+\frac{3c_N^2}{\tau} v_{S\mathbf k}^{||(2)''}
+\l(
c_N^2k^2
-\frac{6c_N^2+2}{\tau^2}
\r)v_{S\mathbf k}^{||(2)'}
+\frac{c_N^2}{\tau}k^2v_{S\mathbf k}^{||(2)}
=Z_{S\mathbf k}
,
\ee
\ba
Z_{S\mathbf k}  & \equiv &
\frac{3}{4}F^{||''}_{S\mathbf k}
+\frac{9c_N^2}{4\tau}F^{||'}_{S\mathbf k}
-\frac{9c_N^2+3}{2\tau^2}F^{||}_{S\mathbf k}
-\frac{3c_N^2}{4}A_{S\mathbf k}^{\,'}
+\frac{3c_N^2}{4\tau}A_{S\mathbf k}  \nn \\
&& -\frac{\tau}{2}c_N^2M_{S\mathbf k\,l}^{,\,l}
+\frac{\tau}{2}c_N^2E^{\,'}_{S\mathbf k}
\, .
\ea
For  a general 2nd-order sound speed  $c_N$,
the homogeneous solution of (\ref{V2eqFourier1})
is similar to the 1st-order solution  (\ref{v1||solgen})
with a replacement of $c_L\rightarrow c_N$,
the inhomogeneous solution of (\ref{V2eqFourier1}) is complicated.
For the case $c_N^2 =\frac{1}{3}$ and a general $c_L$,
(\ref{V2eqFourier1}) becomes
\be\label{V2eqFourier2}
v_{S\mathbf k}^{||(2)'''}
+\frac{1}{\tau} v_{S\mathbf k}^{||(2)''}
+\l(\frac{k^2}{3}
-\frac{4}{\tau^2}\r)v_{S\mathbf k}^{||(2)'}
+\frac{k^2}{3\tau}v_{S\mathbf k}^{||(2)}
  =Z_{S\mathbf k}(\tau)  .
\ee
The explicit expression of $Z_S $ in (\ref{ZScn1}) is
{\allowdisplaybreaks
\bl\label{ZSall}
Z_S ({\bf x},\tau) &
 =
\nabla^{2}\bigg[
\frac{4\tau}{3}\phi^{(1)'}\phi^{(1)}
+\frac{\tau}{18}\phi^{(1)}\nabla^2\chi^{||(1)'}
+\frac{2\tau}{9}\phi^{(1)'}\nabla^2\chi^{||(1)}
+\frac{\tau}{108}\nabla^2\chi^{||(1)}\nabla^2\chi^{||(1)'}
\nn\\
&
+\frac{\tau}{36}\chi^{||(1)}_{,\,l}\nabla^2\chi^{||(1)',\,l}
\bigg]
+\frac{4}{3\tau^2}v^{||(1)}_{,\,l}v^{||(1),\,l}
-\frac{8}{3\tau}v^{||(1)',\,l}v^{||(1)}_{,\,l}
+\frac{1}{2\tau}(1+c_L^2)\delta^{(1)}_{,\,l}v^{||(1),\,l}
\nn\\
&
+\frac{1}{2\tau}(1+c_L^2)\delta^{(1)}\nabla^2v^{||(1)}
-\frac{2}{3\tau}\phi^{(1),\,l}v^{||(1)}_{,\,l}
-\frac{8}{3\tau}\phi^{(1)}\nabla^2v^{||(1)}
+\frac{4}{3\tau}v^{||(1),\,lm}\chi^{||(1)}_{,\,lm}
\nn\\
&
+\frac{8}{9\tau}v^{||(1),\,l}\nabla^2\chi^{||(1)}_{,\,l}
-\frac{4}{9\tau}\nabla^2v^{||(1)}\nabla^2\chi^{||(1)}
+\frac{3}{2\tau}(1+c_L^2)\delta^{(1)}\phi^{(1)'}
-\frac{8}{3}v^{||(1)',\,l}v^{||(1)'}_{,\,l}
\nn\\
&
-\frac{8}{3}v^{||(1)'',\,l}v^{||(1)}_{,\,l}
+\frac{1}{2}(1+c_L^2)\delta^{(1)'}_{,\,l}v^{||(1),\,l}
+\frac{1}{2}(1+c_L^2)\delta^{(1)}_{,\,l}v^{||(1)',\,l}
\nn\\
&
+\frac{1}{2}(1+c_L^2)\delta^{(1)'}\nabla^2v^{||(1)}
+\frac{1}{2}(1+c_L^2)\delta^{(1)}\nabla^2v^{||(1)'}
-\frac{3}{2}(1+c_L^2)\delta^{(1)'}\phi^{(1)'}
\nn\\
&
-\frac{3}{2}(1+c_L^2)\delta^{(1)}\phi^{(1)''}
-2\phi^{(1)'}_{,\,l}v^{||(1),\,l}
-2\phi^{(1)}_{,\,l}v^{||(1)',\,l}
-2\tau\phi^{(1)'}\phi^{(1)''}
-2\tau\phi^{(1)}_{,\,l}\phi^{(1)',\,l}
\nn\\
&
-\frac{8\tau}{3}\phi^{(1)'}\nabla^2\phi^{(1)}
-\frac{8\tau}{3}\phi^{(1)}\nabla^2\phi^{(1)'}
-\frac{4\tau}{9}\phi^{(1)'}\nabla^2\nabla^2\chi^{||(1)}
-\frac{5\tau}{18}\phi^{(1)}\nabla^2\nabla^2\chi^{||(1)'}
\nn\\
&
+\frac{2\tau}{3}\phi^{(1)',\,lm}\chi^{||(1)}_{,\,lm}
+\frac{\tau}{6}\phi^{(1),\,lm}\chi^{||(1)'}_{,\,lm}
-\frac{4\tau}{9}\nabla^2\phi^{(1)'}\nabla^2\chi^{||(1)}
-\frac{\tau}{9}\nabla^2\phi^{(1)}\nabla^2\chi^{||(1)'}
\nn\\
&
+\frac{\tau}{12}\chi^{||(1)',\,lm}\chi^{||(1)''}_{,\,lm}
-\frac{\tau}{36}\nabla^2\chi^{||(1)'}\nabla^2\chi^{||(1)''}
+\frac{\tau}{36}\chi^{||(1)',\,lm}\nabla^2\chi^{||(1)}_{,\,lm}
\nn\\
&
+\frac{\tau}{18}\chi^{||(1),\,lm}\nabla^2\chi^{||(1)'}_{,\,lm}
-\frac{\tau}{54}\nabla^2\chi^{||(1)'}\nabla^2\nabla^2\chi^{||(1)}
-\frac{5\tau}{108}\nabla^2\chi^{||(1)}\nabla^2\nabla^2\chi^{||(1)'}
\nn\\
&
+\frac{\tau}{36}\nabla^2\chi^{||(1)}_{,\,l}\nabla^2\chi^{||(1)',\,l}
-\frac{\tau}{36}\chi^{||(1)}_{,\,l}\nabla^2\nabla^2\chi^{||(1)',\,l}
\nn\\
&
+\nabla^{-2}\bigg[
-\frac{3}{2}(1+c_L^2)\delta^{(1)'''}_{,\,l}v^{||(1),\,l}
-\frac{3}{2}(1+c_L^2)\delta^{(1)'''}\nabla^2v^{||(1)}
-\frac{9}{2}(1+c_L^2)\delta^{(1)'',\,l}v^{||(1)'}_{,\,l}
\nn\\
&
-\frac{9}{2}(1+c_L^2)\delta^{(1)''}\nabla^2v^{||(1)'}
-\frac{9}{2}(1+c_L^2)\delta^{(1)',\,l}v^{||(1)''}_{,\,l}
-\frac{9}{2}(1+c_L^2)\delta^{(1)'}\nabla^2v^{||(1)''}
\nn\\
&
-\frac{3}{2}(1+c_L^2)\delta^{(1),\,l}v^{||(1)'''}_{,\,l}
-\frac{3}{2}(1+c_L^2)\delta^{(1)}\nabla^2v^{||(1)'''}
+4v^{||(1)'}_{,\,lm}v^{||(1)',\,lm}
\nn\\
&
+4v^{||(1)}_{,\,lm}v^{||(1)'',\,lm}
-4\nabla^2v^{||(1)'}\nabla^2v^{||(1)'}
-4\nabla^2v^{||(1)}\nabla^2v^{||(1)''}
-3c_L^2\delta^{(1)''}_{,\,l}\phi^{(1),\,l}
\nn\\
&
-3c_L^2\phi^{(1)}\nabla^2\delta^{(1)''}
-6c_L^2\delta^{(1)'}_{,\,l}\phi^{(1)',\,l}
-6c_L^2\phi^{(1)'}\nabla^2\delta^{(1)'}
-3c_L^2\phi^{(1)''}\nabla^2\delta^{(1)}
\nn\\
&
-3c_L^2\delta^{(1)}_{,\,l}\phi^{(1)'',\,l}
+10v^{||(1)'',\,l}\phi^{(1)'}_{,\,l}
+10\phi^{(1)'}\nabla^2v^{||(1)''}
+20v^{||(1)',\,l}\phi^{(1)''}_{,\,l}
\nn\\
&
+20\phi^{(1)''}\nabla^2v^{||(1)'}
+10v^{||(1),\,l}\phi^{(1)'''}_{,\,l}
+10\phi^{(1)'''}\nabla^2v^{||(1)}
+\frac{3}{2}c_L^2\delta^{(1)'',\,lm}\chi^{||(1)}_{,lm}
\nn\\
&
+c_L^2\delta^{(1)'',\,l}\nabla^2\chi^{||(1)}_{,\,l}
-\frac{1}{2}c_L^2\nabla^2\delta^{(1)''}\nabla^2\chi^{||(1)}
+3c_L^2\delta^{(1)',\,lm}\chi^{||(1)'}_{,lm}
+2c_L^2\delta^{(1)',\,l}\nabla^2\chi^{||(1)'}_{,\,l}
\nn\\
&
-c_L^2\nabla^2\delta^{(1)'}\nabla^2\chi^{||(1)'}
+\frac{3}{2}c_L^2\delta^{(1),\,lm}\chi^{||(1)''}_{,lm}
+c_L^2\delta^{(1),\,l}\nabla^2\chi^{||(1)''}_{,\,l}
-\frac{1}{2}c_L^2\nabla^2\delta^{(1)}\nabla^2\chi^{||(1)''}
\nn\\
&
-2v^{||(1)'',lm}\chi^{||(1)'}_{,lm}
-\frac{4}{3}v^{||(1)'',\,l}\nabla^2\chi^{||(1)'}_{,\,l}
+\frac{2}{3}\nabla^2v^{||(1)''}\nabla^2\chi^{||(1)'}
-4v^{||(1)',lm}\chi^{||(1)''}_{,lm}
\nn\\
&
-\frac{8}{3}v^{||(1)',\,l}\nabla^2\chi^{||(1)''}_{,\,l}
+\frac{4}{3}\nabla^2v^{||(1)'}\nabla^2\chi^{||(1)''}
-2v^{||(1),lm}\chi^{||(1)'''}_{,lm}
-\frac{4}{3}v^{||(1),\,l}\nabla^2\chi^{||(1)'''}_{,\,l}
\nn\\
&
+\frac{2}{3}\nabla^2v^{||(1)}\nabla^2\chi^{||(1)'''}
-\frac{3}{2\tau}(1+c_L^2)\delta^{(1)''}_{,\,l}v^{||(1),\,l}
-\frac{3}{2\tau}(1+c_L^2)\delta^{(1)''}\nabla^2v^{||(1)}
\nn\\
&
-\frac{3}{2\tau}(1+c_L^2)\delta^{(1),\,l}v^{||(1)''}_{,\,l}
-\frac{3}{2\tau}(1+c_L^2)\delta^{(1)}\nabla^2v^{||(1)''}
-\frac{3}{\tau}(1+c_L^2)\delta^{(1)',\,l}v^{||(1)'}_{,\,l}
\nn\\
&
-\frac{3}{\tau}(1+c_L^2)\delta^{(1)'}\nabla^2v^{||(1)'}
+\frac{4}{\tau}v^{||(1)}_{,lm}v^{||(1)',lm}
-\frac{4}{\tau}\nabla^2v^{||(1)}\nabla^2v^{||(1)'}
-\frac{3}{\tau}c_L^2\delta^{(1)'}_{,\,l}\phi^{(1),\,l}
\nn\\
&
-\frac{3}{\tau}c_L^2\delta^{(1)}_{,\,l}\phi^{(1)',\,l}
-\frac{3}{\tau}c_L^2\phi^{(1)'}\nabla^2\delta^{(1)}
-\frac{3}{\tau}c_L^2\phi^{(1)}\nabla^2\delta^{(1)'}
+\frac{10}{\tau}v^{||(1)',\,l}\phi^{(1)'}_{,\,l}
\nn\\
&
+\frac{10}{\tau}v^{||(1),\,l}\phi^{(1)''}_{,\,l}
+\frac{10}{\tau}\phi^{(1)''}\nabla^2v^{||(1)}
+\frac{10}{\tau}\phi^{(1)'}\nabla^2v^{||(1)'}
+\frac{3}{2\tau}c_L^2\delta^{(1)',\,lm}\chi^{||(1)}_{,lm}
\nn\\
&
+\frac{3}{2\tau}c_L^2\delta^{(1),\,lm}\chi^{||(1)'}_{,\,lm}
+\frac{1}{\tau}c_L^2\delta^{(1)',\,l}\nabla^2\chi^{||(1)}_{,\,l}
+\frac{1}{\tau}c_L^2\delta^{(1),l}\nabla^2\chi^{||(1)'}_{,\,l}
-\frac{1}{2\tau}c_L^2\nabla^2\delta^{(1)'}\nabla^2\chi^{||(1)}
\nn\\
&
-\frac{1}{2\tau}c_L^2\nabla^2\delta^{(1)}\nabla^2\chi^{||(1)'}
-\frac{2}{\tau}v^{||(1)',lm}\chi^{||(1)'}_{,lm}
-\frac{2}{\tau}v^{||(1),\,lm}\chi^{||(1)''}_{,lm}
-\frac{4}{3\tau}v^{||(1)',\,l}\nabla^2\chi^{||(1)'}_{,\,l}
\nn\\
&
-\frac{4}{3\tau}v^{||(1),\,l}\nabla^2\chi^{||(1)''}_{,\,l}
+\frac{2}{3\tau}\nabla^2v^{||(1)'}\nabla^2\chi^{||(1)'}
+\frac{2}{3\tau}\nabla^2v^{||(1)}\nabla^2\chi^{||(1)''}
\nn\\
&
+\frac{6}{\tau^2}(1+c_L^2)\delta^{(1)'}_{,\,l}v^{||(1),\,l}
+\frac{6}{\tau^2}(1+c_L^2)\delta^{(1)'}\nabla^2v^{||(1)}
+\frac{6}{\tau^2}(1+c_L^2)\delta^{(1),\,l}v^{||(1)'}_{,\,l}
\nn\\
&
+\frac{6}{\tau^2}(1+c_L^2)\delta^{(1)}\nabla^2v^{||(1)'}
+\frac{8}{\tau^2}v^{||(1)}_{,\,l}\nabla^2v^{||(1),l}
+\frac{8}{\tau^2}\nabla^2v^{||(1)}\nabla^2v^{||(1)}
+\frac{12}{\tau^2}c_L^2\delta^{(1)}_{,\,l}\phi^{(1),\,l}
\nn\\
&
+\frac{12}{\tau^2}c_L^2\phi^{(1)}\nabla^2\delta^{(1)}
-\frac{40}{\tau^2}v^{||(1),\,l}\phi^{(1)'}_{,\,l}
-\frac{40}{\tau^2}\phi^{(1)'}\nabla^2v^{||(1)}
-\frac{6}{\tau^2}c_L^2\delta^{(1),\,lm}\chi^{||(1)}_{,lm}
\nn\\
&
-\frac{4}{\tau^2}c_L^2\delta^{(1),\,l}\nabla^2\chi^{||(1)}_{,\,l}
+\frac{2}{\tau^2}c_L^2\nabla^2\delta^{(1)}\nabla^2\chi^{||(1)}
+\frac{8}{\tau^2}v^{||(1),lm}\chi^{||(1)'}_{,lm}
\nn\\
&
+\frac{16}{3\tau^2}v^{||(1),\,l}\nabla^2\chi^{||(1)'}_{,\,l}
-\frac{8}{3\tau^2}\nabla^2v^{||(1)}\nabla^2\chi^{||(1)'}
\bigg]
\,,
\el
}
and its Fourier transform is $Z_{S\mathbf k}(\tau)$ can be given,
which is a lengthy expression.
Here,  as an illustration,  one typical term of (\ref{ZSall})
has  the Fourier transformation  as the  following
\bl\label{onetm}
&
\frac1{(2\pi)^{3/2}}\int d^3x
\l[
\frac{12}{\tau^2}c_L^2\phi^{(1)}\nabla^2\delta^{(1)}
\r]
e^{-i{\bf k\cdot x}}
\nn\\
=&
\frac{12 c_L^2}{(2\pi)^{9/2}\tau^2}\int d^3x
\l(\int d^3k_1\phi^{(1)}_{\bf k_1}e^{i{\bf k_1\cdot x}} \r)
\l(\int d^3k_2\delta^{(1)}_{\bf k_2}e^{i{\bf k_2\cdot x}} (-|{\bf k_2}|^2) \r)
e^{-i{\bf k\cdot x}}
\nn\\
=&
-\frac{12 c_L^2}{(2\pi)^{3}\,\tau^2}\int d^3k_1\int d^3k_2\,
|{\bf k_2}|^2
\phi^{(1)}_{\bf k_1}\delta^{(1)}_{\bf k_2}
\delta^{(3)}(\bf k_1+k_2- k)
\nn\\
=&
-\frac{12 c_L^2}{(2\pi)^{3}\,\tau^2}\int d^3k_1\,
\big|{\bf k -k_1}\big|^2\,
\phi^{(1)}_{\bf k_1}\,
\delta^{(1)}_{(\bf k -k_1)}
,
\el
where the 1st-order  $\phi^{(1)}_{\bf k}$ and $\delta^{(1)}_{\bf k}$
are known in (\ref{phi1sol2}) and (\ref{delta1sol3})
for  $c_L^2=\frac{1}{3}$.
[For a general $c_L$,
$\delta^{(1)}_{\bf k}$ and $\phi^{(1)}_{\bf k}$
are given by (\ref{delta1V1re}) and (\ref{phiVrelate2}),
using the solution  $v^{||(1)}$ (\ref{v1||solgen}).]
Other terms in (\ref{ZSall}) can be calculated similarly to  (\ref{onetm}).
Once  $Z_{S\mathbf k}$ is obtained,
the general  solution of (\ref{V2eqFourier2}) follows
\bl\label{v2ndsol}
v^{||(2)}_{S\,\mathbf k}
&=
\frac{G_1}{k\tau}
+G_2\l(\frac{2}{k\tau}
    +\frac{i}{\sqrt3}\r)e^{-ik\tau/\sqrt3}
+G_3\l(\frac{2}{k\tau}
    -\frac{i}{\sqrt3}\r)e^{ik\tau/\sqrt3}
\nn\\
&
-\left(
\frac{2}{k\tau} \cos (\frac{k \tau}{\sqrt{3}})
+\frac{1}{\sqrt{3}}\sin (\frac{k \tau}{\sqrt{3}})
\right)
\int^\tau
\bigg(
9\cos (\frac{k \tau'}{\sqrt{3}})
+3\sqrt{3}k \tau' \sin (\frac{k \tau'}{\sqrt{3}})
\bigg)
\frac{  Z_{S\mathbf k}(\tau') }{ k^3 \tau'} \, d\tau'
\nn\\
&
-\left(
\frac{ 2}{k\tau} \sin (\frac{k \tau}{\sqrt{3}})
-\frac{1}{\sqrt{3}} \cos (\frac{k\tau}{\sqrt{3}})
\right)
    \int^\tau
    \bigg( 9 \sin (\frac{k \tau'}{\sqrt{3}})
- 3\sqrt{3} k \tau' \cos (\frac{k \tau'}{\sqrt{3}})\bigg)
    \frac{Z_{S\mathbf k}(\tau')}{k^3 \tau'}d\tau'
    \nn\\
&
+\frac{1}{ k\tau}
\int^\tau \frac{
    3\left(k^2 \tau'^2+6\right)}{k^3 \tau'} Z_{S\mathbf k}(\tau')d\tau'
,
\el
where $G_1 $, $G_2 $, $G_3 $
are $\bf k$-dependent constants,
and correspond to the homogeneous solution, having
the same form as the 1st-order solution (\ref{vsol}).
(The $G_1 $ term is a gauge mode as shall be seen in the next section.)

 The general   solution of 2nd-order density contrast  in  $\bf k$-space
follows  from  (\ref{deltaV2}) as the following
{\allowdisplaybreaks
\bl\label{deltaV2solu}
 \delta_{S\,{\bf k}}^{(2)}=&
-4 v_{S\,{\bf k}}^{||(2)'}
+3 F^{||}_{S\,{\bf k}}
\nn\\
=&
\frac{4 G_1}{k\tau^2}
+G_2
\l(
\frac{8}{k\tau^2}
+\frac{8 i}{\sqrt3\,\tau}
-\frac{4 k}{3}
\r)e^{-ik\tau/\sqrt3}
\nn\\
    &
+G_3 \l(
\frac{8}{k\tau^2}
-\frac{8 i}{\sqrt3\,\tau}
-\frac{4 k}{3}
\r)e^{ik\tau/\sqrt3}
+3 F^{||}_{S\,{\bf k}}
        \nn\\
        &
+\bigg(
-\frac{8}{k\tau^2} \cos (\frac{k \tau}{\sqrt{3}})
-\frac{8}{\sqrt3\,\tau} \sin (\frac{k \tau}{\sqrt{3}})
+\frac{4k}{3}\cos (\frac{k \tau}{\sqrt{3}})
\bigg)
\int^\tau
\bigg(
9\cos (\frac{k \tau'}{\sqrt{3}})
\nn\\
&
+3\sqrt{3}k \tau' \sin (\frac{k \tau'}{\sqrt{3}})
\bigg)
\frac{  Z_{S\mathbf k}(\tau') }{ k^3 \tau'} \, d\tau'
\nn\\
&
+\bigg(
-\frac{ 8}{k\tau^2} \sin (\frac{k \tau}{\sqrt{3}})
+\frac{ 8}{\sqrt{3}\,\tau} \cos (\frac{k \tau}{\sqrt{3}})
+\frac{4k}{3} \sin (\frac{k\tau}{\sqrt{3}})
\bigg)
    \int^\tau
    \bigg( 9 \sin (\frac{k \tau'}{\sqrt{3}})
\nn\\
&
- 3\sqrt{3} k \tau' \cos (\frac{k \tau'}{\sqrt{3}})\bigg)
    \frac{Z_{S\mathbf k}(\tau')}{k^3 \tau'}d\tau'
   +\frac{1}{ k\tau^2}
\int^\tau \frac{
    12(k^2 \tau'^2+6)}{k^3 \tau'} Z_{S\mathbf k}(\tau')d\tau'
.
\el
}
The general  solution of scalar $\phi^{(2)}_{S\,{\bf k}}$
is given by integrating (\ref{Vphi2}),
{\allowdisplaybreaks
\bl\label{phi2SkSol}
\phi^{(2)}_{S\,{\bf k}}
&=
-v_{S\,{\bf k}}^{||(2)'}
-\int^\tau \frac{k^2}{3}v^{||(2)}_{S\,{\bf k}} d\tau'
+\frac{3}{4}F^{||}_{S\,{\bf k}}
- \frac{1}{4}\int^\tau A_{S\,{\bf k}} d\tau'
+G_4
\nn\\
=&
G_1 \l( \frac{1}{k\tau^2} -\frac{k\ln\tau}{3} \r)
+G_2 \l(
\frac{2}{k\tau^2} +\frac{2 i}{\sqrt3\,\tau}\r) e^{-ik\tau/\sqrt3}
+G_3 \l(\frac{2}{k\tau^2}-\frac{2 i}{\sqrt3\,\tau}\r) e^{ik\tau/\sqrt3}
   +G_4        \nn\\
&
-\frac{2 k}{3}\int^\tau
\l[
G_2 e^{-ik\tau'/\sqrt3}
+G_3 e^{ik\tau'/\sqrt3}
\r]
\frac{d\tau'}{\tau'}
+\frac{3}{4}F^{||}_{S\,{\bf k}}(\tau)
- \frac{1}{4}\int^\tau A_{S\,{\bf k}}(\tau') d\tau'
\nn\\
&
+\int^\tau\frac{(k^2 \tau'^2+6)\ln\tau'+3}{k^2 \tau'} Z_{S\mathbf k}(\tau')d\tau'
+\l(
\frac{1}{ k\tau^2}
-\frac{ k\ln\tau}{3}
\r)\int^\tau \frac{
    3(k^2 \tau'^2+6)}{k^3 \tau'} Z_{S\mathbf k}(\tau')d\tau'
\nn\\
&
-\bigg(
\frac{2}{k\tau^2} \cos (\frac{k \tau}{\sqrt{3}})
+\frac{2}{\sqrt3\,\tau} \sin (\frac{k \tau}{\sqrt{3}})
\bigg)
\int^\tau
\bigg(
9\cos (\frac{k \tau'}{\sqrt{3}})
+3\sqrt{3}k \tau' \sin (\frac{k \tau'}{\sqrt{3}})
\bigg)
\frac{  Z_{S\mathbf k}(\tau') }{ k^3 \tau'} \, d\tau'
\nn\\
&
-\bigg(
\frac{ 2}{k\tau^2} \sin (\frac{k \tau}{\sqrt{3}})
-\frac{ 2}{\sqrt{3}\,\tau} \cos (\frac{k \tau}{\sqrt{3}})
\bigg)
    \int^\tau
    \bigg( 9 \sin (\frac{k \tau'}{\sqrt{3}})
- 3\sqrt{3} k \tau' \cos (\frac{k \tau'}{\sqrt{3}})\bigg)
    \frac{Z_{S\mathbf k}(\tau')}{k^3 \tau'}d\tau'
\nn\\
&
+\int^\tau
\bigg[
\frac{2}{k\tau''} \cos (\frac{k \tau''}{\sqrt{3}})
\int^{\tau''}
\bigg( 3\cos (\frac{k \tau'}{\sqrt{3}})
 +\sqrt{3}k \tau' \sin (\frac{k \tau'}{\sqrt{3}})
\bigg)
\frac{  Z_{S\mathbf k}(\tau') }{ k \tau'} \, d\tau'
\bigg]d\tau''
\nn\\
&
+\int^\tau
\bigg[
\frac{ 2}{k\tau''} \sin (\frac{k \tau''}{\sqrt{3}})
    \int^{\tau''}    \bigg(    3 \sin (\frac{k \tau'}{\sqrt{3}})
      - \sqrt{3} k \tau' \cos (\frac{k \tau'}{\sqrt{3}})\bigg)
    \frac{Z_{S\mathbf k}(\tau')}{k \tau'}d\tau'
\bigg]d\tau''
.
\el
}
where $G_4 $ is a $\bf k$-dependent  constant,
which is also a gauge term, as shall be seen in the next section.
Finally, plugging (\ref{deltaV2solu}) and (\ref{phi2SkSol}) into
(\ref{Ein2th003RD}) in $\bf k$-space,
one obtains  the general solution for scalar $\chi^{||(2)}_{S\,{\bf k}}$
as the following:
{\allowdisplaybreaks
\bl\label{chi2Ssol}
\chi^{||(2)}_{S\,{\bf k}}
=&
\frac{18}{k^4\tau} \phi^{(2)'}_{S\,{\bf k}}
+\frac{6}{k^2}\phi^{(2)}_{S\,{\bf k}}
+\frac{9}{k^4\tau^2}\delta^{(2)}_{S\,{\bf k}}
+\frac{3}{k^4} E_{S\,{\bf k}}
\nn\\
=&
-G_1 \frac{2\ln\tau}{k}
+G_2 \frac{4\sqrt3 \,i}{k^2\tau}e^{-ik\tau/\sqrt3}
-G_3 \frac{4\sqrt3 \,i}{k^2\tau}e^{ik\tau/\sqrt3}
+\frac{6 G_4 }{k^2}
        \nn\\
        &
-\frac{4}{k}\int^\tau
\l[
G_2 e^{-ik\tau'/\sqrt3}
+G_3 e^{ik\tau'/\sqrt3}
\r]
\frac{d\tau'}{\tau'}
\nn\\
&
-\frac{2 \ln\tau}{k }\int^\tau \frac{
    3(k^2 \tau'^2+6)}{k^3 \tau'} Z_{S\mathbf k}(\tau')d\tau'
+\int^\tau\frac{6(k^2 \tau'^2+6)\ln\tau'+18}{k^4 \tau'} Z_{S\mathbf k}(\tau')d\tau'
    \nn\\
&
-\frac{4\sqrt3}{k^2\tau} \sin (\frac{k \tau}{\sqrt{3}})
\int^\tau
\bigg(
9\cos (\frac{k \tau'}{\sqrt{3}})
+3\sqrt{3}k \tau' \sin (\frac{k \tau'}{\sqrt{3}})
\bigg)
\frac{  Z_{S\mathbf k}(\tau') }{ k^3 \tau'} \, d\tau'
\nn\\
&
+\frac{4\sqrt3}{k^2\tau} \cos (\frac{k \tau}{\sqrt{3}})
    \int^\tau
    \bigg( 9 \sin (\frac{k \tau'}{\sqrt{3}})
- 3\sqrt{3} k \tau' \cos (\frac{k \tau'}{\sqrt{3}})\bigg)
    \frac{Z_{S\mathbf k}(\tau')}{k^3 \tau'}d\tau'
\nn\\
&
+\int^\tau
\bigg[
\frac{12}{k^3\tau''} \cos (\frac{k \tau''}{\sqrt{3}})
\int^{\tau''}
\bigg(
3\cos (\frac{k \tau'}{\sqrt{3}})
+\sqrt{3}k \tau' \sin (\frac{k \tau'}{\sqrt{3}})
\bigg)
\frac{  Z_{S\mathbf k}(\tau') }{ k \tau'} \, d\tau'
\bigg]d\tau''
\nn\\
&
+\int^\tau
\bigg[
\frac{12}{k^3\tau''} \sin (\frac{k \tau''}{\sqrt{3}})
    \int^{\tau''}
    \bigg(
    3 \sin (\frac{k \tau'}{\sqrt{3}})
- \sqrt{3} k \tau' \cos (\frac{k \tau'}{\sqrt{3}})\bigg)
    \frac{Z_{S\mathbf k}(\tau')}{k \tau'}d\tau'
\bigg]d\tau''
        \nn\\
        &
+\frac{3}{k^4} E_{S\,{\bf k}}
+\frac{27}{2k^4\tau}F^{||\, '}_{S\,{\bf k}}
+\frac{27}{k^4\tau^2} F^{||}_{S\,{\bf k}}
+\frac{9}{2k^2}F^{||}_{S\,{\bf k}}
        \nn\\
        &
- \frac{9}{2k^4\tau} A_{S\,{\bf k}}
-\frac{3}{2k^2}\int^\tau A_{S\,{\bf k}}(\tau') d\tau'
.
\el
}
We have checked that the scalar solutions
(\ref{v2ndsol}), (\ref{deltaV2solu}), (\ref{phi2SkSol}), and (\ref{chi2Ssol})
satisfy the scalar parts of the evolution equation
(\ref{Evo2ndSsTr2RD}) and (\ref{Evo2ndSsChi1RD}).
Thus  far, all the solutions of the 2nd-order perturbations have  been  given.

The above 2nd-order solutions  involve many  terms of integrals of
the scalar-scalar coupling terms,
each of which contains  $\int d^3 k$  integrations
and time integrations $\int d\tau$.
In the $\bf k$-integrations,
two functions $D_2({\bf k})$ and  $D_3({\bf k})$ appear,
which depend upon the concrete initial condition at the beginning of RD stage
and should be practically determined by the precedent inflation
or reheating stages.
Various models of inflation will give different
$D_2(\bf k)$,  $D_3(\bf k)$.
Moreover, in actually doing integration,
one should avoid     IR and UV divergences  \cite{ZhangWang2018}
which may arise from the lower and upper limits of  $\int d^3 k$,
so that  $D_2(\bf k)$, $D_3(\bf k)$ may be required to satisfy certain  conditions.
As an illustration,
suppose $D_2({\bf k}) \propto k^{N_1}$ and $D_3({\bf k}) \propto k^{N_2}$.
Then we shall have the following typical integration terms:
\[
\int  d^3k_1\,
k_1^{n_1} \big|{\bf k -k_1}\big|^{n_2}
\propto  \int_{K_1}^{K_2} dk_1
k_1^{n_1+2}
\frac{ (k+k_1)^{ n_2+2}
- |k-k_1| ^{ n_2+2 } }{k k_1  },
\]
where $n_1$ and $n_2$ are linearly related to $N_1$ and $N_2$ ,
and the integration limits $K_1$ and $K_2$
are  introduced as possible cutoffs to ensure IR and UV convergence.
All other terms can be treated similarly.

\begin{figure}
\subfigure[]{
\begin{minipage}[t]{0.5\linewidth}
\centering
\includegraphics[width=7cm]{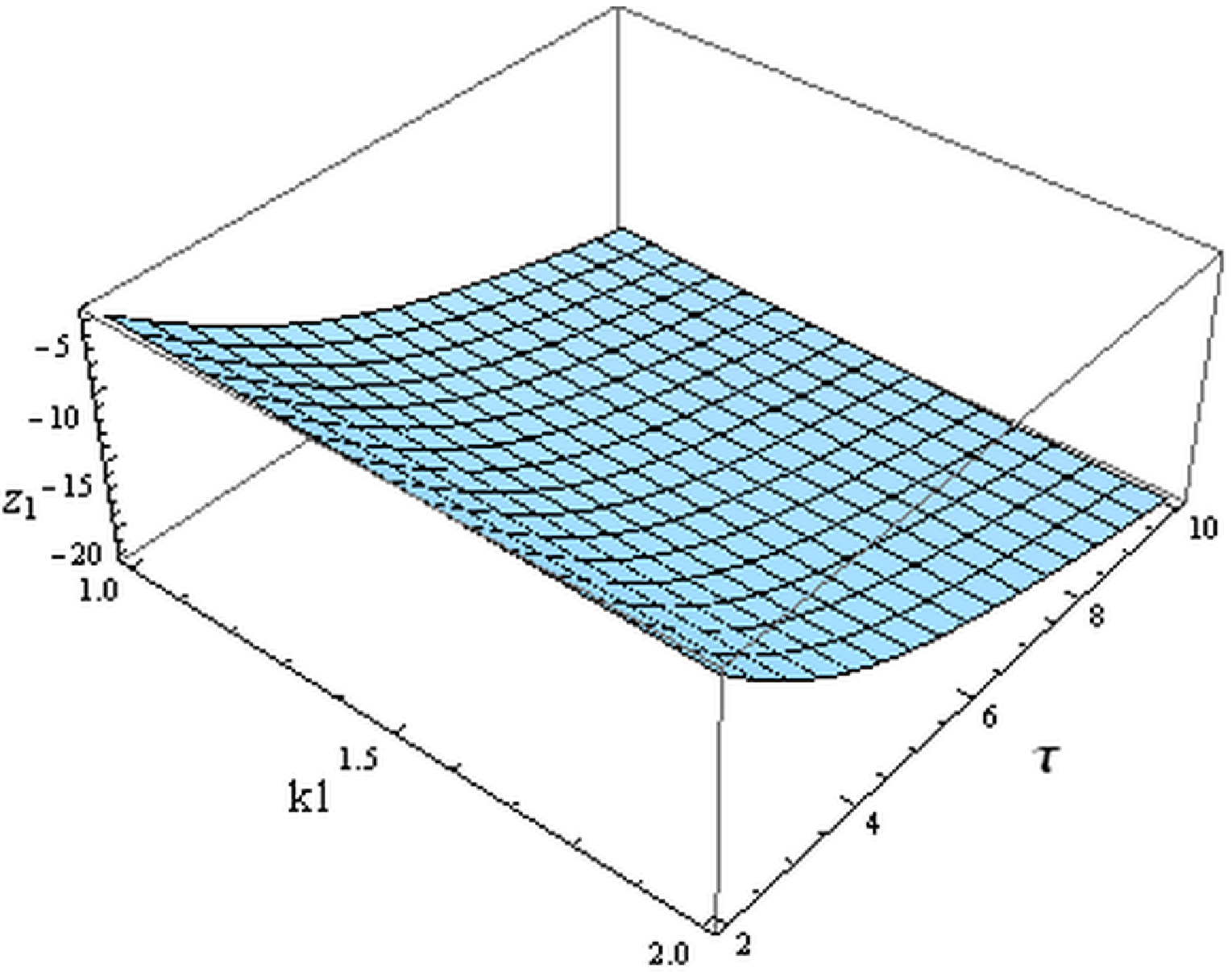}
\end{minipage}%
}
\subfigure[]{
\begin{minipage}[t]{0.5\linewidth}
\centering
\includegraphics[width=7cm]{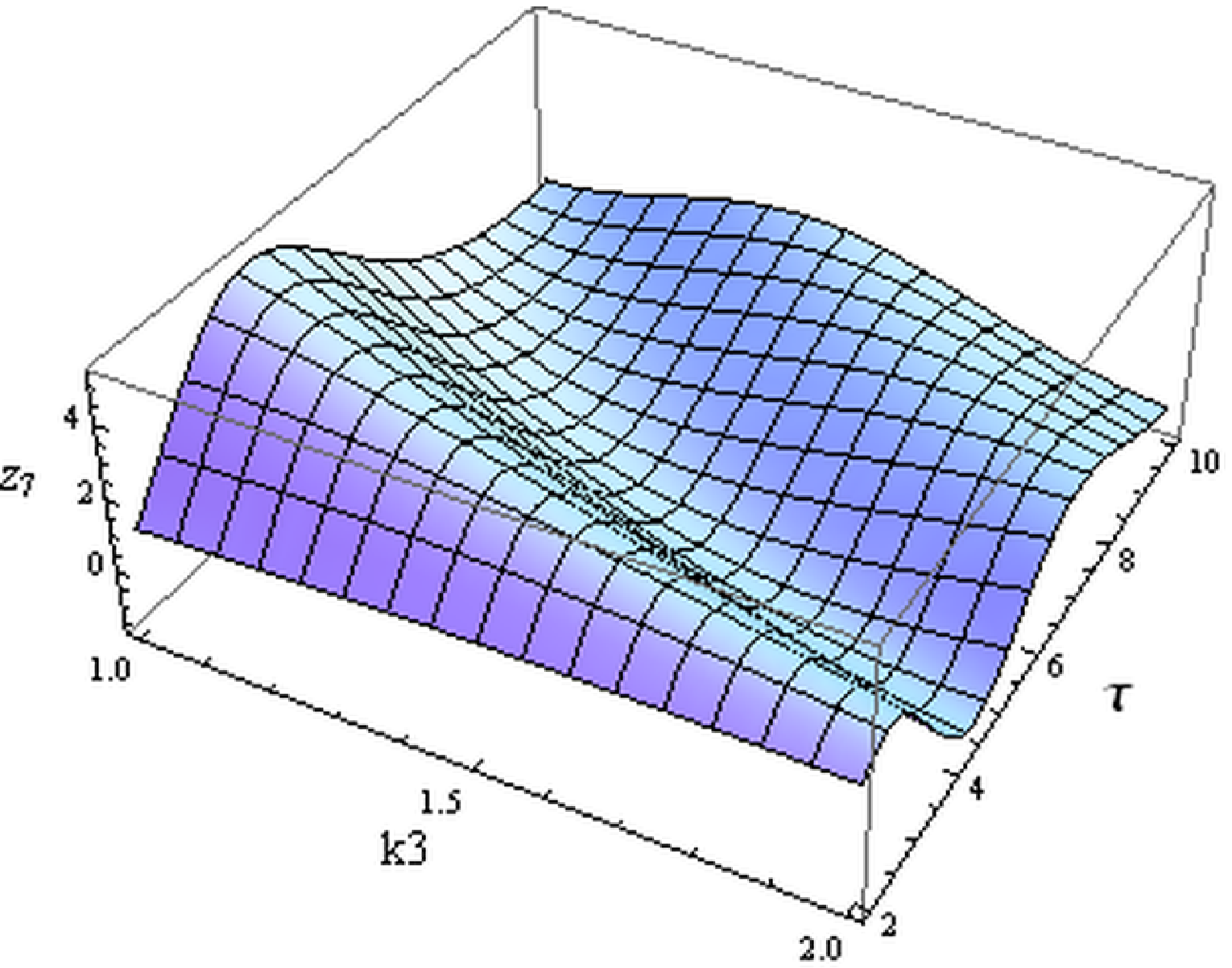}
\end{minipage}
}
\caption{
Left:
real part of $z_1(\tau; 1; k_1,k_2)$ at fixed $k_2$;
right: real part of $z_7(\tau; 1; k_1,k_2,k_3,k_4)$ with fixed
$k_1$, $k_2$, and $k_4$.
}
\label{z1z7}
\end{figure}

The time integrations can also be carried out.
The types of time integrations of $A_{S {\bf k}}$ in (\ref{phi2SkSol}) and (\ref{chi2Ssol})
are  already  contained in those of $Z_{S {\bf k}}(\tau)$,
the latter has four types of  terms:
$ \frac{1}{\tau^n}$, $ \frac{1}{\tau^n} e^{ -i \frac{ k\tau}{\sqrt{3}} }$,
$\int^{\tau}  \frac{ d\tau'}{\tau'} e^{ -i \frac{ k\tau'}{\sqrt{3}}  }$,
and $(\int^{\tau}  \frac{ d\tau'}{\tau'} e^{ -i \frac{ k_1\tau'}{\sqrt{3}} } )
( \int^{\tau}  \frac{ d\tau''}{\tau''} e^{ -i \frac{ k_2\tau''}{\sqrt{3}} }) $.
The single time integrations of $Z_{S{\bf k}}(\tau)$
have the following nontrivial terms:
\[
 \int^\tau\frac{d\tau'}{\tau'^n} e^{ -i \frac{ k\tau'}{\sqrt{3}} }
 \propto  k ^{n-1} \Gamma (1-n,\frac{i k  \tau }{\sqrt{3}} ),
\]
\[
 \int^\tau\frac{d\tau'}{\tau'^n} \int^{\tau' }
 \frac{ d\tau''}{\tau''} e^{ -i \frac{ k\tau''}{\sqrt{3}}  }
\propto
   \tau ^{1-n}   \text{Ei}(-\frac{i k \tau }{\sqrt{3}})
 +  3^{\frac{1}{2}-\frac{n}{2}}
(i k )^{n-1} \Gamma (1-n,\frac{i k \tau }{\sqrt{3}}),
\]
\[
 \int^\tau\frac{d\tau'}{\tau'^n}
 (\int^{\tau'}  \frac{ d\tau''}{\tau''} e^{ -i \frac{ k_1\tau''}{\sqrt{3}}  })
( \int^{\tau '} \frac{ d\tau'''}{\tau'''} e^{ -i \frac{ k_2\tau'''}{\sqrt{3}} } )
  \equiv z_1 (\tau;n;k_1,k_2),
\]
\[
\int^\tau
\frac{d\tau'}{\tau^{'n}}
e^{-i \frac{ k_1\tau'}{\sqrt{3}}}
\int^{\tau'}
\frac{d\tau''}{\tau^{''}}
e^{-i \frac{ k_2\tau''}{\sqrt{3}}} \equiv z_2 (\tau;n; k_1,k_2),
\]
\[
\int^{\tau}  \frac{ d\tau'}{\tau'^n} e^{ i \frac{ k_3\tau'}{\sqrt{3}}  }
(\int^{\tau'}  \frac{ d\tau''}{\tau''} e^{ -i \frac{ k_1\tau''}{\sqrt{3}}  })
( \int^{\tau'}  \frac{ d\tau'''}{\tau'''} e^{ -i \frac{ k_2\tau'''}{\sqrt{3}} })
 \equiv  z_3 (\tau;n; k_1,k_2,k_3),
\]
\bl
 \int^\tau \frac{d\tau'}{\tau'^n}\ln\tau' e^{ -i \frac{ k\tau'}{\sqrt{3}} } \propto\
 &
    3^{\frac{1-n}{2}}  (i k   )^{n-1}  \ln\tau
     \Gamma  (1-n,0,\frac{i k \tau }{\sqrt{3}} ) \nn \\
&   -(1-n)^{-2} \tau ^{1-n}
       \, _2F_2  (1-n,1-n;2-n,2-n;-\frac{i k \tau }{\sqrt{3}} )    ,
\nn
\el
\bl
\int^\tau \frac{d\tau'}{\tau'^n}
\ln\tau'
&
\int^{\tau'}
\frac{ d\tau''}{\tau''} e^{ -i \frac{ k\tau''}{\sqrt{3}}  } \propto
\nn\\
&
-   (n-1)^{-3}  \tau^{1-n} \, _2F_2 (1-n,1-n;2-n,2-n;-\frac{i k \tau }{\sqrt{3}} )
\nn\\
&
-(n-1)^{-2}  \tau^{1-n}  \big(1+ (n-1) \ln\tau   \big) \text{Ei} (-\frac{i k \tau }{\sqrt{3}} )
\nn\\
&
+(n-1)^{-2} 3^{\frac12 -\frac{n}{2} }
    i^{n+1}    k^{n-1}
\left(\Gamma  (1-n,\frac{i k \tau }{\sqrt{3}} )-(n-1) \ln\tau
\Gamma  (1-n,0,\frac{i k \tau }{\sqrt{3}} ) \right)   ,
\nn
\el

\[
\int^\tau \frac{d\tau'}{\tau'^n}\ln\tau'
 (\int^{\tau'}  \frac{ d\tau''}{\tau''} e^{ -i \frac{ k_1\tau''}{\sqrt{3}}  })
( \int^{\tau'}  \frac{ d\tau'''}{\tau'''} e^{ -i \frac{ k_2\tau'''}{\sqrt{3}}  })
 \equiv  z_4 (\tau;n; k_1,k_2 ).
\hspace{1.5cm}
\]
The double time integrations of $Z_{S{}\bf k}$ have the following nontrivial terms:
\[
 \int^{\tau}  \frac{ d\tau'}{\tau'} e^{ i \frac{ k_1\tau'}{\sqrt{3}}  }
 \int^{\tau'}  \frac{ d\tau''}{\tau''^n} e^{i \frac{ k_2\tau''}{\sqrt{3}}}
 \equiv  z_5 (\tau;n; k_1,k_2 ),
\]
\[
\int^{\tau}  \frac{ d\tau'}{\tau'} e^{ i \frac{ k_3\tau'}{\sqrt{3}}  }
 \int^{\tau'}  \frac{ d\tau''}{\tau''^n} e^{i \frac{ k_1\tau''}{\sqrt{3}}  }
\int^{\tau''}  \frac{ d\tau'''}{\tau'''} e^{ -i \frac{ k_2\tau'''}{\sqrt{3}}}
 \equiv  z_6 (\tau;n; k_1,k_2,k_3 ),
\]
\[
\int^{\tau}  \frac{ d\tau'}{\tau'} e^{ i \frac{ k_3\tau'}{\sqrt{3}}  }
 \int^{\tau'}  \frac{ d\tau''}{\tau''^n} e^{i \frac{ k_4\tau''}{\sqrt{3}}  }
(\int^{\tau''}  \frac{ d\tau'''}{\tau'''} e^{ -i \frac{ k_1\tau'''}{\sqrt{3}}  })
( \int^{\tau''}  \frac{ d\tau''''}{\tau''''} e^{ -i \frac{ k_2\tau''''}{\sqrt{3}}})
\equiv  z_7 (\tau;n; k_1,k_2,k_3,k_4 ).
\]
In actual computing,
 $z_1$, ...,  $z_7$ in the  above can be defined as functions
and recalled.
In our test computing, the triple time integrals,  $z_6$ and $z_7$,
take more computing  time than $z_1$, ... $z_5$.
As an illustration, we plot real parts of  $z_1$ and $z_7$ in Fig.\ref{z1z7}.

We mention that in the MD model \cite{WangZhang2017,ZhangQinWang2017},
the scalar modes are not a wave
and do not contain the oscillating factors $e^{\pm \frac{i k \tau}{\sqrt{3}}}$,
so that the time integrations consist of powers of time $\tau$,
which are simpler than the RD model.

\section{2nd-order residual gauge modes}

The 2nd order perturbation solutions in the last section
have the residual degrees of gauge freedom.
A general 2nd-order  coordinate transformation can involve
a 2nd-order transformation vector $\xi^{(2)\mu}$,
and  the square of a 1st-order transformation vector$\xi^{(1)\mu}$ as well.
For synchronous-to-synchronous coordinate transformations
in a general RW spacetime,
we list  $\xi^{(1)\mu}$ in  (\ref{xi0trans}) and (\ref{gi0}),
 $\xi^{(2)\mu}$ in (\ref{alpha2_3}) and (\ref{xi2ge}) in Appendix C
 (see also Ref.\cite{WangZhang2017}).
For the RD stage and  the scalar-scalar coupling,
 $\xi^{(2)\mu}$  is given  in (\ref{alpha2RD1}), (\ref{beta2RD1}),
 and (\ref{d2RD1}).
The 2nd-order transformations of the 2nd-order metric perturbations
for a general RW spacetime are given by
(\ref{phi2TransGen}),
(\ref{chi||2transF2}),
(\ref{chiPerp2TransF}),
and (\ref{chiT2transF2}),
which,  for the RD stage and  the scalar-scalar coupling, reduce to
{\allowdisplaybreaks
\bl\label{phi2TransRD}
\bar \phi^{(2)}_S  = &
\phi^{(2)}_S
+\frac{1}{\tau^2} \bigg [
-\frac{2}{3}A^{(1)}\nabla^2A^{(1)}
-\frac{1}{3}A^{(1),\,l}A^{(1)}_{,\,l}
-4\phi^{(1)}A^{(1)}  \bigg ]
-\frac{2}{\tau}\phi^{(1)'}A^{(1)}
\nn\\
&
+\frac{\ln\tau}{\tau^2}
\bigg[
-\frac{4}{3}A^{(1)} \nabla^2  A^{(1)}
- A^{(1)}_{,\,l}A^{(1),\,l}
\bigg]
-\frac{(\ln\tau )^2}{3}A^{(1)}_{,\,lm} A^{(1),\,lm}
\nn\\
&
+ \ln\tau
\bigg[
\frac{2}{3} A^{(1),\,lm}C^{||(1)}_{,\,lm}
+ \frac{2}{3}A^{(1),\,l}\nabla^2C^{||(1)}_{,\,l}
-\frac{4}{3} \phi^{(1)} \nabla^2  A^{(1)}
-2\phi^{(1)}_{,\,l}A^{(1),\,l}
\nn\\
&
+\frac{2}{3}A^{(1),\,lm}D_{\,lm}\chi^{||(1)}
\bigg]
+  \nabla^2A^{(1)}  \int^\tau\frac{4\phi^{(1)}(\tau',{\bf x})}{3\tau'}d\tau'
+A^{(1)}_{,\,l}\int^\tau\frac{4\phi^{(1)}(\tau',{\bf x})^{,\,l}}{3\tau'}d\tau'
\nn\\
&
-A^{(1),\,lm}\int^\tau\frac{2D_{\,lm}\chi^{||(1)}(\tau',{\bf x})}{3\tau'}d\tau'
-A^{(1)}_{,\,l}\int^\tau\frac{4\nabla^2\chi^{||(1)}(\tau',{\bf x})^{,\,l}}{9\tau'}d\tau'
\nn\\
&
+\frac{A^{(2)}}{\tau^2}
+\frac{\ln\tau}{3}\nabla^2A^{(2)}
+\frac{1}{3}\nabla^2C^{||(2)}
,
\el
}
where    $A^{(2)}$ and $C^{||(2)}$  in the last line are due to the vector $\xi^{(2)\mu}$,
and
{\allowdisplaybreaks
\bl\label{chi||2transRD}
 \bar\chi^{||(2)}_S
&=
 \chi^{||(2)}_S
+ \frac{1}{\tau^2}   \bigg[
A^{(1)}A^{(1)}
+\nabla^{-2}\big(2 A^{(1)}\nabla^2A^{(1)}
+ 8 A^{(1)}\nabla^2C^{||(1)}\big)
\nn\\
&
+\nabla^{-2}\nabla^{-2}\big(
3A^{(1),\,lm}A^{(1)}_{,\,lm}
-3\nabla^2A^{(1)}\nabla^2A^{(1)}
-4A^{(1)}\nabla^2\nabla^2\chi^{||(1)}
-6A^{(1),\,lm}D_{\,lm}\chi^{||(1)}
\nn\\
&
-8 A^{(1),\,l}\nabla^2\chi^{||(1)}_{,\,l}
+12A^{(1),\,lm}C^{||(1)}_{,\,lm}
-12\nabla^2A^{(1)}\nabla^2C^{||(1)}\big)
\bigg]
\nn\\
&
+\frac{4\ln\tau}{\tau^2}
  \bigg[
2\nabla^{-2}\big(A^{(1)}\nabla^2A^{(1)}\big)
+3\nabla^{-2}\nabla^{-2}\big(
A^{(1),\,lm}A^{(1)}_{,\,lm}
-\nabla^2A^{(1)}\nabla^2A^{(1)}\big)
\bigg]
\nn\\
&
-  \frac{1}{\tau}   \nabla^{-2} \nabla^{-2}\bigg[
2A^{(1)}\nabla^2\nabla^2\chi^{||(1)'}
+3A^{(1),\,lm}D_{\,lm}\chi^{||(1)'}
+4A^{(1),\,m}\nabla^2\chi^{(1)'}_{,\,m} \bigg]
\nn \\
&
+(\ln\tau )^2 \bigg[
\nabla^{-2}\big(
2A^{(1)}_{,\,lm}A^{(1),\,lm}
\big)
+3\nabla^{-2}\nabla^{-2}\big(
\nabla^2A^{(1),\,l}\nabla^2A^{(1)}_{,\,l}
-A^{(1),\,lmn}A^{(1)}_{,\,lmn}\big)
\bigg]
\nn\\
&
+  \ln\tau
 \bigg[
2A^{(1),\,l}C^{||(1)}_{,\,l}
+2\nabla^{-2}\big(4\phi^{(1)}\nabla^2A^{(1)}
+A^{(1),\,lm}D_{\,lm}\chi^{||(1)}
-2A^{(1),\,l}\nabla^2C^{||(1)}_{,\,l}
\big)
\nn\\
&
+\nabla^{-2}\nabla^{-2}\big(
-2A^{(1)}_{,l}\nabla^2\nabla^2\chi^{||(1),\,l}
-9A^{(1),\,lmn}D_{\,lm}\chi^{||(1)}_{,n}
-8A^{(1),\,lm}\nabla^2\chi^{||(1)}_{,lm}
\nn\\
&
-4\nabla^2\chi^{||(1)}_{,\,l}\nabla^2A^{(1),\,l}
-6D_{\,lm}\chi^{||(1)}\nabla^2A^{(1),\,lm}
+12\phi^{(1),\,lm}A^{(1)}_{,\,lm}
-12\nabla^2\phi^{(1)}\nabla^2A^{(1)}
\nn\\
&
+6\nabla^2A^{(1),\,l}\nabla^2C^{||(1)}_{,\,l}
-6A^{(1),\,lmn}C^{||(1)}_{,\,lmn}
\big)
\bigg]
+   \bigg[
2C^{||(1)}_{,\,l}C^{||(1),\,l}
+\nabla^{-2}\big(8\phi^{(1)}\nabla^2C^{||(1)}
\nn\\
&
+2 C^{||(1),\,lm}D_{\,lm}\chi^{||(1)}
-2 C^{||(1),\,l}\nabla^2C^{||(1)}_{,\,l}\big)
+\nabla^{-2}\nabla^{-2}\big(
12\phi^{(1),\,lm}C^{||(1)}_{,\,lm}
\nn\\
&
-12\nabla^2\phi^{(1)}\nabla^2C^{||(1)}
-2C^{||(1)}_{,\,l}\nabla^2\nabla^2\chi^{||(1),\,l}
-9 C^{||(1),\,lmn}D_{\,lm}\chi^{||(1)}_{,n}
-8 C^{||(1),\,lm}\nabla^2\chi^{||(1)}_{,\,lm}
\nn\\
&
-4\nabla^2\chi^{||(1)}_{,\,l}\nabla^2C^{||(1),\,l}
-6D_{\,lm}\chi^{||(1)}\nabla^2C^{||(1),\,lm}
+3\nabla^2C^{||(1),\,l}\nabla^2C^{||(1)}_{,\,l}
\nn \\
&
-3C^{||(1),\,lmn}C^{||(1)}_{,\,lmn}
\big)
\bigg]
-4 \nabla^{-2}\bigg[
\nabla^2A^{(1)} \int^\tau\frac{2\phi^{(1)}(\tau',{\bf x})}{\tau'}d\tau'
+A^{(1)}_{,\,l} \int^\tau\frac{2\phi^{(1)}(\tau',{\bf x})^{,\,l}}{\tau'}d\tau'
\nn\\
&
- A^{(1),\,lm} \int^\tau \frac{D_{\,lm}\chi^{||(1)}(\tau',{\bf x})}{\tau'}d\tau'
- A^{(1)}_{,\,l}  \int^\tau \frac{2\nabla^2\chi^{||(1)}(\tau',{\bf x})^{,\,l}}{3\tau'}d\tau'
\bigg]
\nn\\
& -2    A^{(2)} \ln\tau  -2C^{||(2)}       ,
\el
}
and
{
\allowdisplaybreaks
\bl\label{chiPerp2TransRD}
\bar\chi^{\perp(2)}_{S\,ij}
&=
\chi^{\perp(2)}_{S\,ij}
+ \frac{1}{\tau^2} \bigg[
2\partial_i\nabla^{-2}\big(
- A^{(1)}_{,j}\nabla^2A^{(1)}
+4A^{(1),\,l}C^{||(1)}_{,\,lj}
-4A^{(1)}_{,j}\nabla^2C^{||(1)}
-\frac{4}{3}A^{(1)}\nabla^2\chi^{||(1)}_{,j}
\nn\\
&
-2A^{(1),\,l} D_{\,lj}\chi^{||(1)} \big)
+\partial_i\partial_j\nabla^{-2}\big(
A^{(1),\,l}A^{(1)}_{,\,l}\big)
+2\partial_i\partial_j\nabla^{-2}\nabla^{-2}\big(
-A^{(1),\,lm}A^{(1)}_{,\,lm}
\nn\\
&
+\nabla^2A^{(1)}\nabla^2A^{(1)}
+4\nabla^2A^{(1)}\nabla^2C^{||(1)}
-4A^{(1),\,lm}C^{||(1)}_{,\,lm}
+\frac{4}{3}A^{(1)}\nabla^2\nabla^2\chi^{||(1)}
\nn\\
&
+2A^{(1),\,lm}D_{\,lm}\chi^{||(1)}
+\frac{8}{3}A^{(1),m}\nabla^2\chi^{||(1)}_{,m}
\big)
\bigg]
+\frac{4\ln\tau}{\tau^2}\bigg[
-2\partial_i\nabla^{-2}\big( A^{(1)}_{,j}\nabla^2A^{(1)} \big)
\nn\\
&
+\partial_i\partial_j\nabla^{-2}\big(
A^{(1),\,l}A^{(1)}_{,\,l}
\big)
-2\partial_i\partial_j\nabla^{-2}\nabla^{-2}\big(
A^{(1),\,lm}A^{(1)}_{,\,lm}
-\nabla^2A^{(1)}\nabla^2A^{(1)}
\big)
\bigg]
\nn\\
&
- \frac{1}{\tau} \bigg[
2\partial_i\nabla^{-2}\big(
\frac{2}{3}A^{(1)}\nabla^2\chi^{||(1)'}_{,j}
+A^{(1),\,l} D_{\,lj}\chi^{||(1)'}\big)
-2\partial_i\partial_j\nabla^{-2}\nabla^{-2}\big(
\frac{2}{3}A^{(1)}\nabla^2\nabla^2\chi^{||(1)'}
\nn \\
&
+A^{(1),\,lm}D_{\,lm}\chi^{||(1)'}
+\frac{4}{3}A^{(1),m}\nabla^2\chi^{||(1)'}_{,m}
\big)
\bigg]
+ \Big[\ln\tau\Big]^2\bigg[
2\partial_i\nabla^{-2}\big( A^{(1)}_{,\,lj}\nabla^2A^{(1),\,l} \big)
\nn \\
&
-\partial_i\partial_j\nabla^{-2}\big(
A^{(1),\,lm}A^{(1)}_{,\,lm}
\big)
+2\partial_i\partial_j\nabla^{-2}\nabla^{-2}\big(
A^{(1),\,lmn}A^{(1)}_{,\,lmn}
-\nabla^2A^{(1),\,l}\nabla^2A^{(1)}_{,\,l}
\big)
\bigg]
\nn\\
&
-\ln\tau \bigg[
2\partial_i\nabla^{-2}\big(
4\phi^{(1)}_{,j}\nabla^2A^{(1)}
-4\phi^{(1),\,l}A^{(1)}_{,\,lj}
+\frac{2}{3}A^{(1),\,l}\nabla^2\chi^{||(1)}_{,\,lj}
+2 A^{(1),\,lm}D_{\,lj}\chi^{||(1)}_{,m}
\nn \\
&
+\frac{2}{3}A^{(1),\,l}_{,j}\nabla^2\chi^{||(1)}_{,\,l}
+A^{(1),\,lm}_{,j}D_{\,lm}\chi^{||(1)}
+D_{\,lj}\chi^{||(1)}\nabla^2A^{(1),\,l}
+2A^{(1),\,lm}C^{||(1)}_{,\,lmj}
\nn\\
&
-2A^{(1)}_{,\,lj}\nabla^2C^{||(1),\,l}
\big)
-2\partial_i\partial_j\nabla^{-2}\nabla^{-2}\big(
4\nabla^2\phi^{(1)}\nabla^2A^{(1)}
-4\phi^{(1),\,lm}A^{(1)}_{,\,lm}
\nn\\
&
+\frac{2}{3}A^{(1),\,l}\nabla^2\nabla^2\chi^{||(1)}_{,\,l}
+3A^{(1),\,lmn}D_{\,lm}\chi^{||(1)}_{,n}
+\frac{8}{3}A^{(1),\,lm}\nabla^2\chi^{||(1)}_{,\,lm}
+\frac{4}{3}\nabla^2\chi^{||(1)}_{,m}\nabla^2A^{(1),m}
\nn\\
&
+2D_{\,lm}\chi^{||(1)}\nabla^2A^{(1),\,lm}
+2A^{(1),\,lmn}C^{||(1)}_{,\,lmn}
-2\nabla^2A^{(1),\,l}\nabla^2C^{||(1)}_{,\,l}
\big)
\bigg]
\nn\\
&
+\bigg[
-2\partial_i\nabla^{-2}\big(
4\phi^{(1)}_{,j}\nabla^2C^{||(1)}
-4\phi^{(1),\,l}C^{||(1)}_{,\,lj}
+\frac{2}{3}C^{||(1),\,l}\nabla^2\chi^{||(1)}_{,\,lj}
+2C^{||(1),\,lm}D_{\,lj}\chi^{||(1)}_{,m}
\nn\\
&
+\frac{2}{3}C^{||(1),\,l}_{,j}\nabla^2\chi^{||(1)}_{,\,l}
+C^{||(1),\,lm}_{,j}D_{\,lm}\chi^{||(1)}
+D_{\,lj}\chi^{||(1)}\nabla^2C^{||(1),\,l}
+C^{||(1),\,ln}C^{||(1)}_{,\,lnj}
\nn\\
&
-C^{||(1)}_{,\,lj}\nabla^2C^{||(1),\,l}
\big)
+2\partial_i\partial_j\nabla^{-2}\nabla^{-2}\big(
4\nabla^2\phi^{(1)}\nabla^2C^{||(1)}
-4\phi^{(1),\,lm}C^{||(1)}_{,\,lm}
\nn\\
&
+\frac{2}{3}C^{||(1),\,l}\nabla^2\nabla^2\chi^{||(1)}_{,\,l}
+\frac{8}{3}C^{||(1),lm}\nabla^2\chi^{||(1)}_{,\,lm}
+3C^{||(1),lmn}D_{lm}\chi^{||(1)}_{,n}
+\frac{4}{3}\nabla^2\chi^{||(1)}_{,m}\nabla^2C^{||(1),m}
\nn\\
&
+2D_{\,lm}\chi^{||(1)}\nabla^2C^{||(1),\,lm}
+C^{||(1),\,lmn}C^{||(1)}_{,\,lmn}
-\nabla^2C^{||(1),\,l}\nabla^2C^{||(1)}_{,\,l}
\big)
\bigg]
\nn\\
&
+\partial_i\bigg[
A^{(1),\,l}\int^\tau \frac{2D_{\,lj}\chi^{||(1)}(\tau',{\bf x})}{\tau'}d\tau'
-A^{(1)}_{,j}\int^\tau \frac{4\phi^{(1)}(\tau',{\bf x})}{\tau'}d\tau'
\bigg]
\nn \\
&
+\partial_i\partial_j\nabla^{-2}\bigg[
\nabla^2A^{(1)} \int^\tau \frac{4\phi^{(1)}(\tau',{\bf x})}{\tau'}d\tau'
+A^{(1)}_{,\,l}\int^\tau \frac{4\phi^{(1)}(\tau',{\bf x})^{,\,l}}{\tau'}d\tau'
\nn\\
&
-A^{(1),\,lm}\int^\tau \frac{2D_{\,lm}\chi^{||(1)}(\tau' ,{\bf x})}{\tau' }d\tau
-A^{(1)}_{,\,l}\int^\tau \frac{4\nabla^2\chi^{||(1)}(\tau' ,{\bf x})^{,\,l}}{3\tau'}d\tau'
       \bigg]
\nn\\
&
  -C^{\perp(2)}_{i,j}
+(i \leftrightarrow j ) \ ,
\el
}
and
{\allowdisplaybreaks
\bl\label{chiT2transRD}
\bar\chi^{\top(2)}_{S\,ij}
&=
\chi^{\top(2)}_{S\,ij}
+\frac{1}{\tau^2}
\bigg[
-\delta_{ij}\nabla^{-2}\big(
-A^{(1),\,lm}A^{(1)}_{,\,lm}
+\nabla^2A^{(1)}\nabla^2A^{(1)}
+\frac{4}{3}A^{(1)}\nabla^2\nabla^2\chi^{||(1)}
\nn\\
&
+2A^{(1),\,lm}D_{\,lm}\chi^{||(1)}
+\frac{8}{3}A^{(1),m}\nabla^2\chi^{||(1)}_{,m}
+4\nabla^2A^{(1)}\nabla^2C^{||(1)}
-4A^{(1),\,lm}C^{||(1)}_{,\,lm}
\big)
\nn\\
&
-4A^{(1)}D_{\,ij}\chi^{||(1)}
+4\partial_i\nabla^{-2}\big(
\frac{2}{3}A^{(1)}\nabla^2\chi^{||(1)}_{,j}
+A^{(1),\,l}D_{\,lj}\chi^{||(1)}
\big)
\nn\\
&
+4\partial_j\nabla^{-2}\big(
\frac{2}{3}A^{(1)}\nabla^2\chi^{||(1)}_{,i}
+A^{(1),\,l}D_{\,li}\chi^{||(1)}
\big)
+4\nabla^{-2}\big(
A^{(1)}_{,ij}\nabla^2A^{(1)}
-A^{(1),\,l}_{,i}A^{(1)}_{,\,lj}
\nn\\
&
+2A^{(1)}_{,ij}\nabla^2C^{||(1)}
+2C^{||(1)}_{,ij}\nabla^2A^{(1)}
-2A^{(1),\,l}_{,i}C^{||(1)}_{,\,lj}
-2A^{(1),\,l}_{,j}C^{||(1)}_{,\,li}
\big)
\nn\\
&
-\partial_i\partial_j\nabla^{-2}\nabla^{-2}\big(
-A^{(1),\,lm}A^{(1)}_{,\,lm}
+\nabla^2A^{(1)}\nabla^2A^{(1)}
+\frac{4}{3}A^{(1)}\nabla^2\nabla^2\chi^{||(1)}
\nn\\
&
+2A^{(1),\,lm}D_{\,lm}\chi^{||(1)}
+\frac{8}{3}A^{(1),m}\nabla^2\chi^{||(1)}_{,m}
+4A^{(1),\,lm}C^{||(1)}_{,\,lm}
-4\nabla^2A^{(1)}\nabla^2C^{||(1)}
\big)
\bigg]
\nn\\
&
+\frac{4\ln\tau}{\tau^2}\bigg[
\delta_{ij}\nabla^{-2}\big(
A^{(1),\,lm}A^{(1)}_{,\,lm}
-\nabla^2A^{(1)}\nabla^2A^{(1)}
\big)
+4\nabla^{-2}\big(
A^{(1)}_{,ij}\nabla^2A^{(1)}
\nn\\
&
-A^{(1),\,l}_{,i}A^{(1)}_{,\,lj}
\big)
+\partial_i\partial_j\nabla^{-2}\nabla^{-2}\big(
A^{(1),\,lm}A^{(1)}_{,\,lm}
-\nabla^2A^{(1)}\nabla^2A^{(1)}
\big)
\bigg]
\nn\\
&
- \frac{1}{\tau}
\bigg[
\delta_{ij}\nabla^{-2}\big(
\frac{2}{3}A^{(1)}\nabla^2\nabla^2\chi^{||(1)'}
+A^{(1),\,lm}D_{\,lm}\chi^{||(1)'}
+\frac{4}{3}A^{(1),m}\nabla^2\chi^{||(1)'}_{,m}
\big)
\nn\\
&
+2A^{(1)}D_{\,ij}\chi^{||(1)'}
-2\partial_i\nabla^{-2}
\big(
\frac{2}{3}A^{(1)}\nabla^2\chi^{||(1)'}_{,j}
+A^{(1),\,l}D_{\,lj}\chi^{||(1)'}
\big)
\nn\\
&
-2\partial_j\nabla^{-2}
\big(
\frac{2}{3}A^{(1)}\nabla^2\chi^{||(1)'}_{,i}
+A^{(1),\,l}D_{\,li}\chi^{||(1)'}
\big)
+\partial_i\partial_j\nabla^{-2}\nabla^{-2}\big(
\frac{2}{3}A^{(1)}\nabla^2\nabla^2\chi^{||(1)'}
\nn\\
&
+A^{(1),\,lm}D_{\,lm}\chi^{||(1)'}
+\frac{4}{3}A^{(1),m}\nabla^2\chi^{||(1)'}_{,m}
\big) \bigg]
\nn\\
&
-(\ln\tau )^2
\bigg[
\delta_{ij}\nabla^{-2}\big(
A^{(1),\,lmn}A^{(1)}_{,\,lmn}
-\nabla^2A^{(1),\,l}\nabla^2A^{(1)}_{,\,l}
\big)
+2\nabla^{-2}\big(-2A^{(1),\,lm}_{,i}A^{(1)}_{,\,lmj}
\nn\\
&
+2A^{(1)}_{,\,lij}\nabla^2A^{(1),\,l}
\big)
+\partial_i\partial_j\nabla^{-2}\nabla^{-2}\big(
A^{(1),\,lmn}A^{(1)}_{,\,lmn}
-\nabla^2A^{(1),\,l}\nabla^2A^{(1)}_{,\,l}
\big)
\bigg]
\nn\\
&
+ \ln\tau
\bigg[
\delta_{ij}\nabla^{-2}\big(
4\phi^{(1),\,lm}A^{(1)}_{,\,lm}
-4\nabla^2\phi^{(1)}\nabla^2A^{(1)}
-\frac{2}{3}A^{(1),\,l}\nabla^2\nabla^2\chi^{||(1)}_{,\,l}
\nn\\
&
-\frac{8}{3}A^{(1),\,lm}\nabla^2\chi^{||(1)}_{,\,lm}
+A^{(1),\,lmn}D_{\,lm}\chi^{||(1)}_{,n}
-\frac{4}{3}\nabla^2\chi^{||(1)}_{,m}\nabla^2A^{(1),m}
+2 A^{(1),\,lm}\nabla^2D_{\,lm}\chi^{||(1)}
\nn\\
&
-2A^{(1),\,lmn}C^{||(1)}_{,\,lmn}
+2\nabla^2A^{(1),\,l}\nabla^2C^{||(1)}_{,\,l}
\big)
-2 A^{(1),\,l}D_{\,ij}\chi^{||(1)}_{,\,l}
-2A^{(1),\,l}_{,j}D_{\,li}\chi^{||(1)}
\nn\\
&
-2A^{(1),\,l}_{,i}D_{\,lj}\chi^{||(1)}
-4\nabla^{-2}\big(
2\phi^{(1),\,l}_{,i}A^{(1)}_{,\,lj}
+2\phi^{(1),\,l}_{,j}A^{(1)}_{,\,li}
-2\phi^{(1)}_{,ij}\nabla^2A^{(1)}
-2A^{(1)}_{,ij}\nabla^2\phi^{(1)}
\nn\\
&
-A^{(1),\,lm}_{,i}C^{||(1)}_{,\,lmj}
-A^{(1),\,lm}_{,j}C^{||(1)}_{,\,lmi}
+A^{(1)}_{,\,lij}\nabla^2C^{||(1),\,l}
+C^{||(1)}_{,\,lij}\nabla^2A^{(1),\,l}
\big)
    \nn\\
    &
+2\partial_i\nabla^{-2}\big(
\frac{2}{3}A^{(1),m}\nabla^2\chi^{||(1)}_{,mj}
+2A^{(1),\,lm}D_{\,lj}\chi^{||(1)}_{,m}
+\frac{2}{3}A^{(1),\,l}_{,j}\nabla^2\chi^{||(1)}_{,\,l}
+A^{(1),\,lm}_{,j}D_{\,lm}\chi^{||(1)}
\nn\\
&
+D_{\,lj}\chi^{||(1)}\nabla^2A^{(1),\,l}
\big)
+2\partial_j\nabla^{-2}\big(
\frac{2}{3}A^{(1),m}\nabla^2\chi^{||(1)}_{,mi}
+2A^{(1),\,lm}D_{\,li}\chi^{||(1)}_{,m}
    \nn\\
    &
+\frac{2}{3}A^{(1),\,l}_{,i}\nabla^2\chi^{||(1)}_{,\,l}
+A^{(1),\,lm}_{,i}D_{\,lm}\chi^{||(1)}
+D_{\,li}\chi^{||(1)}\nabla^2A^{(1),\,l}
\big)
\nn\\
&
+\partial_i\partial_j\nabla^{-2}\nabla^{-2}\big(
4\phi^{(1),\,lm}A^{(1)}_{,\,lm}
-4\nabla^2\phi^{(1)}\nabla^2A^{(1)}
-\frac{2}{3}A^{(1),\,l}\nabla^2\nabla^2\chi^{||(1)}_{,\,l}
\nn\\
&
-\frac{8}{3}A^{(1),\,lm}\nabla^2\chi^{||(1)}_{,\,lm}
-7 A^{(1),\,lmn}D_{\,lm}\chi^{||(1)}_{,n}
-\frac{4}{3}\nabla^2\chi^{||(1)}_{,m}\nabla^2A^{(1),m}
-4D_{\,lm}\chi^{||(1)}\nabla^2A^{(1),\,lm}
\nn\\
&
-2 A^{(1),\,lm}\nabla^2D_{\,lm}\chi^{||(1)}
-2A^{(1),\,lmn}C^{||(1)}_{,\,lmn}
+2\nabla^2A^{(1),\,l}\nabla^2C^{||(1)}_{,\,l}
\big)
\bigg]
\nn\\
&
+\bigg[
\delta_{ij}\nabla^{-2}\big(
4\phi^{(1),\,lm}C^{||(1)}_{,\,lm}
-4\nabla^2\phi^{(1)}\nabla^2C^{||(1)}
-\frac{2}{3}C^{||(1),\,l}\nabla^2\nabla^2\chi^{||(1)}_{,\,l}
\nn\\
&
-\frac{8}{3}C^{||(1),\,lm}\nabla^2\chi^{||(1)}_{,\,lm}
+C^{||(1),\,lmn}D_{\,lm}\chi^{||(1)}_{,n}
-\frac{4}{3}\nabla^2\chi^{||(1)}_{,m}\nabla^2C^{||(1),m}
\nn\\
&
+2C^{||(1),\,lm}\nabla^2D_{\,lm}\chi^{||(1)}
+\nabla^2C^{||(1),\,l}\nabla^2C^{||(1)}_{,\,l}
-C^{||(1),\,lmn}C^{||(1)}_{,\,lmn}
\big)
\nn\\
&
-2C^{||(1),\,l}D_{\,ij}\chi^{||(1)}_{,\,l}
-2C^{||(1),\,l}_{,j}D_{\,li}\chi^{||(1)}
-2C^{||(1),\,l}_{,i}D_{\,lj}\chi^{||(1)}
-4\nabla^{-2}\big(2\phi^{(1),\,l}_{,i}C^{||(1)}_{,\,lj}
\nn\\
&
+2\phi^{(1),\,l}_{,j}C^{||(1)}_{,\,li}
-2\phi^{(1)}_{,ij}\nabla^2C^{||(1)}
-2C^{||(1)}_{,ij}\nabla^2\phi^{(1)}
-C^{||(1),\,lm}_{,i}C^{||(1)}_{,\,lmj}
+C^{||(1)}_{,\,lij}\nabla^2C^{||(1),\,l}
\big)
\nn\\
&
+2 \partial_i\nabla^{-2}\big(
\frac{2}{3}C^{||(1),m}\nabla^2\chi^{||(1)}_{,mj}
+2 C^{||(1),\,lm}D_{\,lj}\chi^{||(1)}_{,m}
+\frac{2}{3}C^{||(1),\,l}_{,j}\nabla^2\chi^{||(1)}_{,\,l}
\nn\\
&
+C^{||(1),\,lm}_{,j}D_{\,lm}\chi^{||(1)}
+D_{\,lj}\chi^{||(1)}\nabla^2C^{||(1),\,l}
\big)
+2 \partial_j\nabla^{-2}\big(
\frac{2}{3}C^{||(1),m}\nabla^2\chi^{||(1)}_{,mi}
\nn\\
&
+2C^{||(1),\,lm}D_{\,li}\chi^{||(1)}_{,m}
+\frac{2}{3}C^{||(1),\,l}_{,i}\nabla^2\chi^{||(1)}_{,\,l}
+C^{||(1),\,lm}_{,i}D_{\,lm}\chi^{||(1)}
+D_{\,li}\chi^{||(1)}\nabla^2C^{||(1),\,l}
\big)
\nn\\
&
-\partial_i\partial_j\nabla^{-2}\nabla^{-2}\big(
4\nabla^2\phi^{(1)}\nabla^2C^{||(1)}
-4\phi^{(1),\,lm}C^{||(1)}_{,\,lm}
+\frac{2}{3}C^{||(1),\,l}\nabla^2\nabla^2\chi^{||(1)}_{,\,l}
\nn\\
&
+\frac{8}{3}C^{||(1),\,lm}\nabla^2\chi^{||(1)}_{,\,lm}
+7C^{||(1),\,lmn}D_{\,lm}\chi^{||(1)}_{,n}
+\frac{4}{3}\nabla^2\chi^{||(1)}_{,\,l}\nabla^2C^{||(1),\,l}
\nn\\
&
+4D_{\,lm}\chi^{||(1)}\nabla^2C^{||(1),\,lm}
+2 C^{||(1),\,lm}\nabla^2D_{\,lm}\chi^{||(1)}
+C^{||(1),\,lmn}C^{||(1)}_{,\,lmn}
\nn\\
&
-\nabla^2C^{||(1),\,l}\nabla^2C^{||(1)}_{,\,l}
\big)
\bigg]
   .
\el
}

The  2nd-order transformation of the 2nd-order density, density contrast
and velocity for a general RW spacetime
are give in
(\ref{rho2Trans}), (\ref{delta2TransGe}),
 (\ref{v2iTransGe}), (\ref{v||2TransGe}), and (\ref{vperp2TransGe}).
For the RD stage and  the scalar-scalar coupling,
 (\ref{delta2TransGe}) reduces to
 the following transformation of the density contrast:
\bl\label{delta2TransRD}
\bar\delta^{(2)}_S
=&
\delta^{(2)}_S
+\frac{24}{\tau^4}A^{(1)}A^{(1)}
-\frac{4\ln\tau}{\tau^2}A^{(1)}_{,\, l}A^{(1),\,l}
+\frac{1}{\tau^2}
\bigg[
-4A^{(1)}_{,\, l}C\, ^{ ||(1) ,\,l}
+8\delta^{(1)}A^{(1)}
\bigg]
\nn\\
&
-\frac{2}{\tau}\delta^{(1)'}A^{(1)}
-2\delta^{(1)}_{,\, l}C\, ^{ ||(1) ,\,l}
-2 \delta^{(1)}_{,\, l}A^{(1),\,l} \ln\tau
+\frac{4}{\tau^2}A^{(2)}
,
\el
and   (\ref{v||2TransGe}) and (\ref{vperp2TransGe}) reduce to
the following   transformation of  velocity
\bl\label{v||2TransRD}
\bar v^{||(2)}_S
=&
v^{||(2)}_S
+\frac{2}{\tau^3} A^{(1)}A^{(1)}
+\frac{2 }{\tau^2}\nabla^{-2}
\bigg[
v^{||(1),\,l}A^{(1)}_{,\,l}
+A^{(1)}\nabla^2v^{||(1)}
\bigg]
\nn\\
&
+\frac{1}{\tau}
\bigg[
A^{(1)}_{,\,l}C^{||(1),\,l}
+\nabla^{-2}
\big(
-2C^{||(1),\,l}\nabla^2A^{(1)}_{,\,l}
+2A^{(1)}_{,\,l}\nabla^2C^{||(1),\,l}
+4 A^{(1),\,l}\phi^{(1)}_{,\,l}
\nn\\
&
+4\phi^{(1)}\nabla^2A^{(1)}
-2 A^{(1)}_{,\,lm}D^{\,lm}\chi^{||(1)}
-\frac{4}{3} A^{(1)}_{,\,l}\nabla^2\chi^{||(1),\,l}
\big)
\bigg]
+\frac{\ln\tau}{\tau} A^{(1)}_{,\,l}A^{(1),\,l}
\nn\\
&
+\bigg[
2 v^{||(1),\,l}C^{||(1)}_{,\,l}
+\nabla^{-2}
\big(
-4 v^{||(1)}_{,\,lm}C^{||(1),\,lm}
-4 C^{||(1),\,l}\nabla^2v^{||(1)}_{,\,l}
\big)
\bigg]
\nn\\
&
+  \ln\tau
\bigg[
2 v^{||(1),\,l}A^{(1)}_{,\,l}
+\nabla^{-2}
\big(
-4 v^{||(1)}_{,\,lm}A^{(1),\,lm}
-4A^{(1),\,l}\nabla^2v^{||(1)}_{,\,l}
\big)
\bigg]
+\frac{A^{(2)}}{\tau}
\ ,
\el
\bl\label{vperp2TransRD}
\bar v^{\perp(2)}_{S\,i}
=&
v^{\perp(2)}_{S\,i}
+\frac{2 }{\tau^2}
\bigg[
v^{||(1)}_{,i}A^{(1)}
+\partial_i\nabla^{-2}
\big(
-v^{||(1),\,l}A^{(1)}_{,\,l}
-A^{(1)}\nabla^2v^{||(1)}
\big)
\bigg]
\nn\\
&
+\frac{1}{\tau}
\bigg[
-2A^{(1)}_{,\,l i}C^{||(1),\,l}
+2A^{(1),\,l}C^{||(1)}_{,\,li}
+4 A^{(1)}_{,i}\phi^{(1)}
-2 A^{(1),\,l}D_{\,li}\chi^{||(1)}
\nn\\
&
+\partial_i\nabla^{-2}
\big(
2C^{||(1),\,l}\nabla^2A^{(1)}_{,\,l}
-2A^{(1)}_{,\,l}\nabla^2C^{||(1),\,l}
-4 A^{(1),\,l}\phi^{(1)}_{,\,l}
-4\phi^{(1)}\nabla^2A^{(1)}
\nn\\
&
+2 A^{(1)}_{,\,lm}D^{\,lm}\chi^{||(1)}
+\frac{4}{3} A^{(1)}_{,\,l}\nabla^2\chi^{||(1),\,l}
\big)
\bigg]
\nn\\
&
+\bigg[
-4 v^{||(1)}_{,\,l i}C^{||(1),\,l}
+\partial_i\nabla^{-2}
\big(
4 v^{||(1)}_{,\,lm}C^{||(1),\,lm}
+4 C^{||(1),\,l}\nabla^2v^{||(1)}_{,\,l}
\big)
\bigg]  \nn\\
&
+  \ln\tau
\bigg[
-4 v^{||(1)}_{,\,l i}A^{(1),\,l}
+\partial_i\nabla^{-2}
\big(
4 v^{||(1)}_{,\,lm}A^{(1),\,lm}
+4A^{(1),\,l}\nabla^2v^{||(1)}_{,\,l}
\big)
\bigg]
.
\el
The above  synchronous-to-synchronous transformations are general,
in the sense that
two vector fields $\xi^{(1)\mu}$ and $\xi^{(2)\mu}$ are involved simultaneously.
These expressions are lengthy due to
the  parameters $A^{(1)}$,   $C^{||(1)}$, and $C^{\perp(1)}_i$
of the 1st-order vector $\xi^{(1)\mu}$,
and only a few terms are due to the parameters
$A^{(2)}$, $C^{||(2)}$, and $C^{\perp(2)}_i$  of $\xi^{(2)\mu}$.
In particular,
(\ref{chiT2transRD}) and (\ref{vperp2TransRD}) tell that
the transformation of 2nd-order tensor and curl velocity
involve only  $\xi^{(1)\mu}$,
independent of  $A^{(2)}$,   $C^{||(2)}$, and $C^{\perp(2)}_i $.
Recall that the transformations are similar
in the dust model for MD stage \cite{WangZhang2017},
but there is no velocity.

However, distinctions should be made
between  $\xi^{(2)\mu}$ and   $\xi^{(1)\mu}$,
as pointed out in Ref.\cite{WangZhang2017}.
In applications
we are often interested in the following case:
the 2nd-order solutions  are transformed \cite{Gleiser1996}
at the same time the 1st-order solutions are fixed.
That is, we just transform the 2nd-order solutions
without altering   the 1st-order ones.
This is referred to as the effective 2nd-order transformations,
which requires
\be\label{effecttrs}
\xi^{(1)\mu}=0, ~~  {\text{but}} ~~ \xi^{(2)\mu} \ne 0 .
\ee
Contrarily,
if one would set $\xi^{(2)\mu}=0$  \cite{Abramo1997,HwangNoh2012},
only  $\xi^{(1)\mu}  $ remains,
one would have no freedom to make  $\bar g^{(2)}_{00}=0$
and  $\bar g^{(2)}_{0i}=0$ anymore,
because  $\xi^{(1)\mu}$ has been already fixed
in ensuring $\bar g^{(1)}_{00}=0$, $\bar g^{(1)}_{0i}=0$
and keeping the obtained 1st-order solutions as gauge-invariant.

For the effective 2nd-order transformations,
(\ref{alpha2RD1}), (\ref{beta2RD1}), and (\ref{d2RD1})  reduce to
\be\label{alpha2RD2}
\alpha^{(2)}(\tau,\mathbf x)
=\frac{ A^{(2)}(\mathbf x)}{\tau} \, ,
\ee
\be\label{beta2RD2}
\beta^{(2)}(\tau,\mathbf x)=
 A^{(2)}({\bf x})  \ln\tau
+C^{||(2)} ({\bf x})
\,,
\ee
\be
\label{d2RD2}
d^{(2)}_i(\mathbf x)
=C^{\perp(2)}_i ({\bf x})   ,
\ee
and
(\ref{phi2TransRD}) --- (\ref{vperp2TransRD})
reduce to
\be\label{phi2TransRD2}
\bar \phi^{(2)}_S (\tau,\mathbf x) =
\phi^{(2)}_S(\tau,\mathbf x)
+\frac{A^{(2)}({\bf x})}{\tau^2}
+\frac{\ln\tau}{3}\nabla^2A^{(2)}({\bf x})
+\frac{1}{3}\nabla^2C^{||(2)}({\bf x})
,
\ee
\be\label{chi||2transRD2}
 \bar\chi^{||(2)}_S(\tau,\mathbf x)
=
 \chi^{||(2)}_S(\tau,\mathbf x)
 -2   A^{(2)}({\bf x}) \ln\tau
-2C^{||(2)}({\bf x})
    ,
\ee
\be\label{chiPerp2TransRD2}
\bar\chi^{\perp(2)}_{S\,ij}(\tau,\mathbf x)
=
\chi^{\perp(2)}_{S\,ij}(\tau,\mathbf x)
 -\partial_{j}C^{\perp(2)}_{i}({\bf x})
-\partial_{i}C^{\perp(2)}_{j}({\bf x})\ ,
\ee
\bl\label{chiT2transRD2}
\bar\chi^{\top(2)}_{S\,ij}(\tau,\mathbf x)
=
\chi^{\top(2)}_{S\,ij}(\tau,\mathbf x),
\el
\be\label{delta2TransRD2}
\bar\delta^{(2)}_S(\tau,\mathbf x)
=
\delta^{(2)}_S(\tau,\mathbf x)
+\frac{4}{\tau^2}A^{(2)}({\bf x})
.
\ee
\be\label{v||2TransRD2}
\bar v^{||(2)}_S(\tau,\mathbf x)
=
v^{||(2)}_S(\tau,\mathbf x)
+\frac{A^{(2)}({\bf x})}{\tau}
,
\ee
\be\label{vperp2TransRD2}
\bar v^{\perp(2)}_{S\,i}(\tau,\mathbf x)
=
v^{\perp(2)}_{S\,i}(\tau,\mathbf x)
,
\ee
which has the same structure as the 1st-order residual transformations
(\ref{gaugetrphi1})-- (\ref{vGaugemodeNonCurl}).
From  (\ref{chiT2transRD2}) and (\ref{vperp2TransRD2}) we see
that the 2nd-order tensor and curl velocity are invariant
under the 2nd-order  transformation  within synchronous coordinates,
so the solution $\chi^{\top(2)}_{S\,ij}$ of (\ref{solgw})
and $v^{\perp(2)}_{S\,i}$ of (\ref{Vperp2ndSol}) are gauge-invariant modes.
Equation (\ref{chiPerp2TransRD2}) tells us
that  $c_{1ij}$  in the solution $\chi^{\perp(2)}_{S\,ij}$ of (\ref{chiVecSol})
is a gauge term and can be eliminated,
so that the gauge-invariant vector mode  is
\be\label{chiVecSolphy}
\chi^{\perp(2)}_{S\,ij}({\bf x},\tau)
=\frac{c_{2ij}({\bf x})}{\tau}
 + 2\int^\tau \frac{d\tau'}{\tau^{'2}}
    \int^{\tau'}\tau^{''2}\,V_{S\,ij}({\bf x},\tau'')d\tau''
\,,
\ee
where
$V_{S\,ij}$ is in (\ref{SourceCurl1RD})
and does not change under this effective 2nd-order
residual transformation with
$\xi^{(2)\mu} \ne 0$ but $\xi^{(1)\mu}=0$.

To identify the residual gauge modes in the 2nd-order scalar solutions,
we write (\ref{phi2TransRD2}), (\ref{chi||2transRD2}),
(\ref{delta2TransRD2}), and (\ref{v||2TransRD2})
 in $\bf k$-space
\be\label{phi2TransRD2F}
\bar \phi^{(2)}_{S\,{\bf k}} (\tau) =
\phi^{(2)}_{S\,{\bf k}} (\tau)
+A^{(2)}_{\bf k}\l(
\frac{1}{\tau^2}
-\frac{k^2}{3}\ln\tau
\r)
-\frac{k^2}{3}C^{||(2)}_{\bf k}
,
\ee
\be\label{chi||2transRD2F}
 \bar\chi^{||(2)}_{S\,{\bf k}} (\tau)
=
 \chi^{||(2)}_{S\,{\bf k}} (\tau)
 -2A^{(2)}_{\bf k}\ln\tau
-2C^{||(2)}_{\bf k}
    ,
\ee
\be\label{delta2TransRD2F}
\bar\delta^{(2)}_{S\,{\bf k}} (\tau)
=
\delta^{(2)}_{S\,{\bf k}} (\tau)
+\frac{4}{\tau^2}A^{(2)}_{\bf k}
,
\ee
\be\label{v||2TransRD2F}
\bar v^{||(2)}_{S\,{\bf k}} (\tau)
=
v^{||(2)}_{S\,{\bf k}} (\tau)
+\frac{A^{(2)}_{\bf k}}{\tau}
.
\ee
Comparing them
with the solutions (\ref{phi2SkSol}),  (\ref{chi2Ssol}),
(\ref{deltaV2solu}), and (\ref{v2ndsol}) respectively,
tells us that $G_1$ and $G_4$ terms
in  the solutions are gauge terms,
which can be removed simultaneously by choosing
\be\label{A2Transk}
 A^{(2)}_{\mathbf k}
=-\frac{ G_1}{k},
\ee
\be\label{C2||Transk}
C^{||(2)}_{\mathbf k}=\frac{3 G_4}{k^2} .
\ee
Thus,  the gauge-invariant  modes of the 2nd-order scalar
perturbations are
{\allowdisplaybreaks
\bl\label{phi2SkSolphy}
\phi^{(2)}_{S\,{\bf k}}
&=
G_2
\l(
\frac{2}{k\tau^2}
+\frac{2 i}{\sqrt3\,\tau}
\r)e^{-ik\tau/\sqrt3}
+G_3\l(
\frac{2}{k\tau^2}
-\frac{2 i}{\sqrt3\,\tau}
\r)e^{ik\tau/\sqrt3}
        \nn\\
        &
-\frac{2 k}{3}\int^\tau
\l[
G_2 e^{-ik\tau'/\sqrt3}
+G_3 e^{ik\tau'/\sqrt3}
\r]
\frac{d\tau'}{\tau'}
+\frac{3}{4}F^{||}_{S\,{\bf k}}(\tau)
- \frac{1}{4}\int^\tau A_{S\,{\bf k}}(\tau') d\tau'
\nn\\
&
+\int^\tau\frac{(k^2 \tau'^2+6)\ln\tau'+3}{k^2 \tau'} Z_{S\mathbf k}(\tau')d\tau'
+\l(
\frac{1}{ k\tau^2}
-\frac{ k\ln\tau}{3}
\r)\int^\tau \frac{
    3(k^2 \tau'^2+6)}{k^3 \tau'} Z_{S\mathbf k}(\tau')d\tau'
\nn\\
&
-\bigg(
\frac{2}{k\tau^2} \cos (\frac{k \tau}{\sqrt{3}})
+\frac{2}{\sqrt3\,\tau} \sin (\frac{k \tau}{\sqrt{3}})
\bigg)
\int^\tau
\bigg(
9\cos (\frac{k \tau'}{\sqrt{3}})
+3\sqrt{3}k \tau' \sin (\frac{k \tau'}{\sqrt{3}})
\bigg)
\frac{  Z_{S\mathbf k}(\tau') }{ k^3 \tau'} \, d\tau'
\nn\\
&
-\bigg(
\frac{ 2}{k\tau^2} \sin (\frac{k \tau}{\sqrt{3}})
-\frac{ 2}{\sqrt{3}\,\tau} \cos (\frac{k \tau}{\sqrt{3}})
\bigg)
    \int^\tau
    \bigg( 9 \sin (\frac{k \tau'}{\sqrt{3}})
- 3\sqrt{3} k \tau' \cos (\frac{k \tau'}{\sqrt{3}})\bigg)
    \frac{Z_{S\mathbf k}(\tau')}{k^3 \tau'}d\tau'
\nn\\
&
+\int^\tau
\bigg[
\frac{2}{k\tau''} \cos (\frac{k \tau''}{\sqrt{3}})
\int^{\tau''}
\bigg(
3\cos (\frac{k \tau'}{\sqrt{3}})
+\sqrt{3}k \tau' \sin (\frac{k \tau'}{\sqrt{3}})
\bigg)
\frac{  Z_{S\mathbf k}(\tau') }{ k \tau'} \, d\tau'
\bigg]d\tau''
\nn\\
&
+\int^\tau
\bigg[
\frac{ 2}{k\tau''} \sin (\frac{k \tau''}{\sqrt{3}})
    \int^{\tau''}
    \bigg(
    3 \sin (\frac{k \tau'}{\sqrt{3}})
- \sqrt{3} k \tau' \cos (\frac{k \tau'}{\sqrt{3}})\bigg)
    \frac{Z_{S\mathbf k}(\tau')}{k \tau'}d\tau'
\bigg]d\tau''
,
\el
\bl\label{chi2Ssolphy}
\chi^{||(2)}_{S\,{\bf k}}
=&
G_2 \frac{4\sqrt3 \,i}{k^2\tau}e^{-ik\tau/\sqrt3}
-G_3 \frac{4\sqrt3 \,i}{k^2\tau}e^{ik\tau/\sqrt3}
        \nn\\
        &
-\frac{4}{k}\int^\tau
\l[
G_2 e^{-ik\tau'/\sqrt3}
+G_3 e^{ik\tau'/\sqrt3}
\r]
\frac{d\tau'}{\tau'}
\nn\\
&
-\frac{2 \ln\tau}{k }\int^\tau \frac{
    3(k^2 \tau'^2+6)}{k^3 \tau'} Z_{S\mathbf k}(\tau')d\tau'
+\int^\tau\frac{6(k^2 \tau'^2+6)\ln\tau'+18}{k^4 \tau'} Z_{S\mathbf k}(\tau')d\tau'
    \nn\\
&
-\frac{4\sqrt3}{k^2\tau} \sin (\frac{k \tau}{\sqrt{3}})
\int^\tau
\bigg(
9\cos (\frac{k \tau'}{\sqrt{3}})
+3\sqrt{3}k \tau' \sin (\frac{k \tau'}{\sqrt{3}})
\bigg)
\frac{  Z_{S\mathbf k}(\tau') }{ k^3 \tau'} \, d\tau'
\nn\\
&
+\frac{4\sqrt3}{k^2\tau} \cos (\frac{k \tau}{\sqrt{3}})
    \int^\tau
    \bigg( 9 \sin (\frac{k \tau'}{\sqrt{3}})
- 3\sqrt{3} k \tau' \cos (\frac{k \tau'}{\sqrt{3}})\bigg)
    \frac{Z_{S\mathbf k}(\tau')}{k^3 \tau'}d\tau'
\nn\\
&
+\int^\tau
\bigg[
\frac{12}{k^3\tau''} \cos (\frac{k \tau''}{\sqrt{3}})
\int^{\tau''}
\bigg(
3\cos (\frac{k \tau'}{\sqrt{3}})
+\sqrt{3}k \tau' \sin (\frac{k \tau'}{\sqrt{3}})
\bigg)
\frac{  Z_{S\mathbf k}(\tau') }{ k \tau'} \, d\tau'
\bigg]d\tau''
\nn\\
&
+\int^\tau
\bigg[
\frac{12}{k^3\tau''} \sin (\frac{k \tau''}{\sqrt{3}})
    \int^{\tau''}
    \bigg(
    3 \sin (\frac{k \tau'}{\sqrt{3}})
- \sqrt{3} k \tau' \cos (\frac{k \tau'}{\sqrt{3}})\bigg)
    \frac{Z_{S\mathbf k}(\tau')}{k \tau'}d\tau'
\bigg]d\tau''
        \nn\\
        &
+\frac{3}{k^4} E_{S\,{\bf k}}
+\frac{27}{2k^4\tau}F^{||'}_{S\,{\bf k}}(\tau)
+\frac{27}{k^4\tau^2} F^{||}_{S\,{\bf k}}
+\frac{9}{2k^2}F^{||}_{S\,{\bf k}}(\tau)
        \nn\\
        &
- \frac{9}{2k^4\tau} A_{S\,{\bf k}}(\tau)
-\frac{3}{2k^2}\int^\tau A_{S\,{\bf k}}(\tau') d\tau'
,
\el
\bl\label{deltaV2soluphy}
\delta_{S\,{\bf k}}^{(2)}
=&
G_2
\l(
\frac{8}{k\tau^2}
+\frac{8 i}{\sqrt3\,\tau}
-\frac{4 k}{3}
\r)e^{-ik\tau/\sqrt3}
+G_3 \l(
\frac{8}{k\tau^2}
-\frac{8 i}{\sqrt3\,\tau}
-\frac{4 k}{3}
\r)e^{ik\tau/\sqrt3}
        \nn\\
        &
+3 F^{||}_{S\,{\bf k}}
+\bigg(
-\frac{8}{k\tau^2} \cos (\frac{k \tau}{\sqrt{3}})
-\frac{8}{\sqrt3\,\tau} \sin (\frac{k \tau}{\sqrt{3}})
+\frac{4k}{3}\cos (\frac{k \tau}{\sqrt{3}})
\bigg)
\int^\tau
\bigg(
9\cos (\frac{k \tau'}{\sqrt{3}})
\nn\\
&
+3\sqrt{3}k \tau' \sin (\frac{k \tau'}{\sqrt{3}})
\bigg)
\frac{  Z_{S\mathbf k}(\tau') }{ k^3 \tau'} \, d\tau'
\nn\\
&
+\bigg(
-\frac{ 8}{k\tau^2} \sin (\frac{k \tau}{\sqrt{3}})
+\frac{ 8}{\sqrt{3}\,\tau} \cos (\frac{k \tau}{\sqrt{3}})
+\frac{4k}{3} \sin (\frac{k\tau}{\sqrt{3}})
\bigg)
    \int^\tau
    \bigg( 9 \sin (\frac{k \tau'}{\sqrt{3}})
\nn\\
&
- 3\sqrt{3} k \tau' \cos (\frac{k \tau'}{\sqrt{3}})\bigg)
    \frac{Z_{S\mathbf k}(\tau')}{k^3 \tau'}d\tau'
    \nn\\
&
+\frac{1}{ k\tau^2}
\int^\tau \frac{
    12(k^2 \tau'^2+6)}{k^3 \tau'} Z_{S\mathbf k}(\tau')d\tau'
,
\el
\bl\label{v2ndsolphy}
v^{||(2)}_{\mathbf k}
&=
G_2 \l(\frac{2}{k\tau}
    +\frac{i}{\sqrt3}\r)e^{-ik\tau/\sqrt3}
+G_3 \l(\frac{2}{k\tau}
    -\frac{i}{\sqrt3}\r)e^{ik\tau/\sqrt3}
\nn\\
&
-\left(
\frac{2}{k\tau} \cos (\frac{k \tau}{\sqrt{3}})
+\frac{1}{\sqrt{3}}\sin (\frac{k \tau}{\sqrt{3}})
\right)
\int^\tau
\bigg(
9\cos (\frac{k \tau'}{\sqrt{3}})
+3\sqrt{3}k \tau' \sin (\frac{k \tau'}{\sqrt{3}})
\bigg)
\frac{  Z_{S\mathbf k}(\tau') }{ k^3 \tau'} \, d\tau'
\nn\\
&
-\left(
\frac{ 2}{k\tau} \sin (\frac{k \tau}{\sqrt{3}})
-\frac{1}{\sqrt{3}} \cos (\frac{k\tau}{\sqrt{3}})
\right)
    \int^\tau
    \bigg( 9 \sin (\frac{k \tau'}{\sqrt{3}})
- 3\sqrt{3} k \tau' \cos (\frac{k \tau'}{\sqrt{3}})\bigg)
    \frac{Z_{S\mathbf k}(\tau')}{k^3 \tau'}d\tau'
    \nn\\
&
+\frac{1}{ k\tau}
\int^\tau \frac{
    3\left(k^2 \tau'^2+6\right)}{k^3 \tau'} Z_{S\mathbf k}(\tau')d\tau'
,
\el
}
where
$Z_S$, $E_S$, $A_S$, $F^{||}_S$ are in
(\ref{ZSall}), (\ref{ES1RD}), (\ref{AS}), and (\ref{FSnoncu}),
which do not change
under this effective 2nd-order transformation
induced by $\xi^{(2)\mu}$.
So far,
we have obtained all the gauge invariant solutions of
the 2nd-order perturbations in
Eqs.  (\ref{solgw}), (\ref{Vperp2ndSol}),  (\ref{chiVecSolphy}),
(\ref{phi2SkSolphy}), (\ref{chi2Ssolphy}),
(\ref{deltaV2soluphy}), and (\ref{v2ndsolphy}).

\section{   Conclusion and discussion}
\label{sec:Conclusion}

We present a systematic study  of the  2nd-order cosmological  perturbations
in RD stage in synchronous coordinates
based on the Einstein equation.
The dominant  radiation is modeled by a relativistic fluid
with the energy density $\rho$,
the pressure $p= c^2_s \rho$ with $c_s^2=\frac13$ and the velocity  $U^\mu$,
and we assume that there is no shear for the fluid
and  the 1st-order velocity is curlless.
The model has  more complications than
the pressureless dust model
in the MD stage \cite{Matarrese98,WangZhang2017,ZhangQinWang2017}
since the spatial components  $T_{ij}$  of stress tensor are nonvanishing
and have to be specified nontrivially.

We  give a detailed analysis of the structure
and the 1st-order  solutions and   residual gauge transformations,
on which the 2nd-order perturbations are based.
We have demonstrated that,   during RD stage,
the 1st-order scalar modes, density contrast and longitudinal velocity
all propagate at the sound speed  $\frac{1}{\sqrt 3}$,
instead of the speed of light.
The 1st-order vector modes and curl velocity
are not a wave and do not propagate,
and  they simply decrease with time.
In contrast,
the tensor modes are waves and propagate at the speed of light.
Compare this with the MD stage
during which
only the tensor modes  propagate at the speed of light,
whereas all other perturbations are not a wave and do not propagate.
Hence, we conclude
that the tensor modes are truly radiative as dynamic degrees of freedom,
regardless of the background matter,
but the scalar and vector modes are not.
Sometimes in the literature
all metric perturbations were misleadingly
referred to as six gravitational waves (GW).
Our analysis suggests that the term gravitational waves
should be reserved for the tensor modes only.

The 2nd-order perturbed Einstein equation
contains various couplings of 1st-order metric perturbations
derived.
The case of scalar-scalar coupling has been considered in this paper.
The 1st-order vector metric perturbations is taken to be vanishing,
consistent with the curlless 1st-order velocity.
When these coupling terms are moved to the rhs of the Einstein equation,
they together with $T_{\mu\nu}$ of the fluid
serve as the effective source for the 2nd-order metric perturbations.
The resulting 2nd-order Einstein equation
has a similar structure to the 1st-order  Einstein equation,
except that the effective source is now more complicated.
The $(00)$ component of Einstein equation
         is the energy constraint,
the $(0i)$ components are the momentum constraints
which are  decomposed into longitudinal and transverse parts,
and the $(ij)$ components are decomposed
into the respective evolution equations of
2nd-order scalar, vector, and tensor metric perturbations.
Moreover, to specify $\rho$, $p$,  and $U^\mu$
that appear in the source,
the equation of covariant conservation $T^{\mu\nu}\, _{;\nu} =0$
up to the 2nd-order needs to be solved.
The presence of velocity $U^\mu$
makes the calculations algebraically more involved than the dust model.
We have solved the set of equations for the 2nd-order
metric perturbations, density contrast and velocity analytically,
and obtained all the  2nd-order solutions  in the integral form.
They consist of the homogeneous part for general initial conditions
and the inhomogeneous part due to the effective source.
We have analyzed also the general 2nd-order residual gauge transformations
in synchronous coordinates,
which involve both the 2nd-order vector $\xi^{(2)\mu}$ and
the 1st-order vector $\xi^{(1)\mu}$.
We have obtained the explicit expressions of transformations
for all the 2nd-order perturbations,
which are lengthy and have many terms due to  the square of  $\xi^{(1)\mu}$
and the products of 1st-order perturbations with $\xi^{(1)\mu}$.
In particular,
we also have distinguished
the transformations due to the 2nd-order vector $\xi^{(2)\mu}$
from those due to the combinations of the 1st-order vector $\xi^{(1)\mu}$.
The 1st-order solutions are often fixed in actual applications,
only those transformations due to $\xi^{(2)\mu}$ are effective.
In this case the effective transformation of the 2nd-order perturbations
have similar structure to the 1st-order  perturbations.
After this analysis,
we have identified the gauge-invariant modes of the 2nd-order solutions.

These  2nd-order analytical results of RD stage,
in conjunction with the 2nd-order results
of MD stage \cite{WangZhang2017,ZhangQinWang2017},
can be used to study
nonlinear effects of cosmological perturbations.
As an advantage of analytical results,
one will be able to focus on
certain aspect of the nonlinearity of 2nd-order perturbation.
For example, one can study  separately the tensor modes, as well as the vector modes,
besides  the scalar modes.
And  one can examine the individual contribution of each $k$-mode
to the nonlinearity,
the transfer of perturbation power among different  modes,
the interference of positive and negative frequency modes
represented by $D_2$ and $D_3$,
 the influences by initial conditions,
the separate influence by the growing and decaying modes.
Furthermore, one can  pick up a particular period  of evolution
and estimate the dynamic behavior there.
These aspects can provide possible advantages that other methods often lack.
To use these solutions in specific applications,
one needs to carry out numerical integrals,
such as $z_1$, ..., $z_7$,
and  choose  the appropriate initial conditions.
These possible applications will need more work.

To improve the above work,
one can calculate the couplings involving the 1st-order tensor mode,
which may have effects comparable to the scala-scalar.
As possible extensions,
one can study 2nd-order perturbations for inflationary models.
These will be left for future studies.

\

\textbf{Acknowledgements}

Y. Zhang is supported by
NSFC Grant No. 11421303, 11675165, 11633001
SRFDP, and CAS, the Strategic Priority Research Program
``The Emergence of Cosmological Structures"
of the Chinese Academy of Sciences, No. XDB09000000.

\newpage

\appendix

\section{Perturbed quantities}

The quantities listed  in this appendix
are  valid for a general flat RW spacetime in synchronous coordinate.
By  (\ref{18q1}) and (\ref{eq1}),
the nonvanishing  Christopher symbols
up to 2nd order are
\be\label{Christopher000}
\Gamma^{0}_{00}=\frac{a'}{a},
\ee
\be\label{Christopher0ij}
\Gamma^{0}_{ij}=\frac{a'}{a}\delta_{ij}
+\bigg(\frac{a'}{a}\gamma^{(1)}_{ij}
+\frac{1}{2}\gamma^{(1)'}_{ij}\bigg)
+\bigg(\frac{a'}{2a}\gamma^{(2)}_{ij}
+\frac{1}{4}\gamma^{(2)'}_{ij}\bigg),
\ee
\be\label{Christopherij0}
\Gamma^i_{j0}=\frac{a'}{a}\delta^i_j
+\bigg(\frac{1}{2}\gamma^{(1)'i}_{j}\bigg)
+\bigg(\frac{1}{4}\gamma^{(2)'i}_j
-\frac{1}{2}\gamma^{(1)ik}\gamma^{(1)'}_{jk}\bigg),
\ee
\bl\label{Christopherijk}
\Gamma^{i}_{jk}=&
\bigg(\frac{1}{2}\gamma^{(1)i}_{j,k}
+\frac{1}{2}\gamma^{(1)i}_{k,j}
-\frac{1}{2}\gamma^{(1),i}_{jk}\bigg)
+\bigg(
-\frac{1}{2}\gamma^{(1)im}\gamma^{(1)}_{mj,k}
-\frac{1}{2}\gamma^{(1)im}\gamma^{(1)}_{mk,j}
+\frac{1}{2}\gamma^{(1)im}\gamma^{(1)}_{jk,m}
\nn\\
&
+\frac{1}{4}\gamma^{(2)i}_{j,k}
+\frac{1}{4}\gamma^{(2)i}_{k,j}
-\frac{1}{4}\gamma^{(2),i}_{jk}
\bigg).
\el
The Ricci tensor  are
$R _{ \mu\nu} =\Gamma^{\alpha}_{ \mu\nu,\alpha}-\Gamma^{\alpha}_{\nu\alpha,\mu}
+\Gamma^{\alpha}_{\lambda\alpha}\Gamma^{\lambda}_{\mu\nu}
-\Gamma^{\alpha}_{\lambda\nu}\Gamma^{\lambda}_{\alpha\mu}$.
One calculates  the  0th-order Ricci tensors are
$R^{(0)}_{00}=-\frac{3a''}{a}+3\l(\frac{a'}{a}\r)^2$,
$R^{(0)}_{0i}=0$,
$R^{(0)}_{ij}=\delta_{ij} \l(\frac{a''}{a}+ (\frac{a'}{a} )^2\r)$,
$R^{(0)}=\frac{6}{a^2} \frac{a''}{a}$,
and   the 1st-order perturbed Ricci tensor
\be
R^{(1) }_{00}=   3\phi^{(1)''} +3\frac{a'}{a}\phi^{(1)'} ,
\ee
\be\label{R0i1st}
R^{(1)}_{0i}= 2\phi^{(1)' }_{,i}
+ \frac{1}{2} D_{ij}\chi^{||(1)',j  }
+ \frac{1}{2} \chi^{\perp(1)',j  }_{ij},
\ee
\bl \label{cnm11R}
R^{(1)}_{ij}=&-  5\frac{a'}{a}\phi^{(1)'}\delta_{ij}
-2\frac{a''}{a}\phi^{(1)}\delta_{ij}-2(\frac{a'}{a})^2\phi^{(1)}\delta_{ij}
- \phi^{(1)''}\delta_{ij}+\nabla^2 \phi^{(1) }\delta_{ij}+\phi^{(1) }_{,ij}
 \nn\\
&+\frac{1}{2} D_{ij}\chi^{||(1)''}+\frac{a'}{a}D_{ij}\chi^{||(1)'} +\frac{a''}{a}D_{ij}\chi^{||(1)}
+(\frac{a'}{a})^2D_{ij}\chi^{||(1)}
-\frac{1}{2}\nabla^2D_{ij}\chi^{||(1) }    \nn\\
& +\frac{1}{2}D^k_j\chi^{||(1)}_{,ik}+\frac{1}{2}D^k_i\chi^{||(1)}_{,jk} \nn\\
&
+\frac{1}{2} \chi^{\perp(1)''}_{ij}
+\frac{a'}{a}\chi^{\perp(1)'}_{ij}
 +\frac{a''}{a}\chi^{\perp(1)}_{ij}
+(\frac{a'}{a})^2\chi^{\perp(1)}_{ij}
+\frac{1}{2}\l[
\chi^{\perp(1),k}_{kj,i}
+\chi^{\perp(1),k}_{ki,j}
-\nabla^2\chi^{\perp(1) }_{ij}
\r]
\nn \\
&+\frac{a'}{a} \chi^{\top(1)'}_{ij}
 +\frac{a''}{a} \chi^{\top(1)}_{ij} +(\frac{a'}{a})^2 \chi^{\top(1) }_{ij}
 +\frac{1}{2} \chi^{\top(1)''}_{ij}
 - \frac{1}{2}\nabla^2\chi^{\top(1) }_{ij} .
\el
The 1st-order Ricci scalar is
\be\label{RicciScalar1st}
R^{(1)}   =  \frac{1}{a^2}(-6\phi^{(1)''} -18\frac{a'}{a}\phi^{(1)'}
          +4\nabla^2\phi^{(1) }+D_{ij}\chi^{||(1),ij  }) ,
\ee
which is independent of the 1st-order vector and tensor.

The 1st-order Einstein tensor is
\be\label{G001}
 G^{(1)}_{00}\equiv
  R_{00}^{(1)}  -\frac{1}{2}g_{00}^{(0)}R^{(1)}
=
-6\frac{a'}{a}\phi^{(1)'}
          +2\nabla^2\phi^{(1) }
          +\frac{1}{3}\nabla^2\nabla^2\chi^{||(1)} ,
\ee
\be \label{G0i1st}
 G^{(1)}_{0i}\equiv  R^{(1)}_{0i} ,
\ee
\bl\label{Gij1st}
G^{(1)}_{ij}
\equiv&
R^{(1)}_{ij}  -\frac{1}{2} \delta_{ij}a^2 R^{(1)}  -\frac{1}{2}a^2\gamma^{(1)}_{ij}  R^{(0)}
    \nn\\
    =&
2\phi^{(1)''} \delta_{ij}
+4\frac{a'}{a}\phi^{(1)'}\delta_{ij}
+\phi^{(1) }_{,ij}
-\nabla^2\phi^{(1) }\delta_{ij}
+\l[4\frac{a''}{a}-2(\frac{a'}{a})^2\r]\phi^{(1)}\delta_{ij}
\nn\\
&
+\frac{1}{2} D_{ij}\chi^{||(1)''}
+\frac{a'}{a}D_{ij}\chi^{||(1)'}
+\l[(\frac{a'}{a})^2-2\frac{a''}{a}\r]D_{ij}\chi^{||(1)}
\nn\\
&
+\l[\frac{1}{2}D^k_j\chi^{||(1)}_{,ik}
+\frac{1}{2}D^k_i\chi^{||(1)}_{,jk}
-\frac{1}{2}\delta_{ij}D_{kl}\chi^{||(1),kl }\r]
-\frac{1}{2}\nabla^2D_{ij}\chi^{||(1) }
\nn\\
&
+\frac{1}{2} \chi^{\perp(1)''}_{ij}
+\frac{a'}{a}\chi^{\perp(1)'}_{ij}
+\l[(\frac{a'}{a})^2-2\frac{a''}{a}\r]\chi^{\perp(1)}_{ij}
+\frac{1}{2}\l[
\chi^{\perp(1),k}_{kj,i}
+\chi^{\perp(1),k}_{ki,j}
-\nabla^2\chi^{\perp(1) }_{ij}
\r]
\nn \\
&
+\frac{1}{2} \chi^{\top(1)''}_{ij}
+\frac{a'}{a} \chi^{\top(1)'}_{ij}
+\l[(\frac{a'}{a})^2 -2\frac{a''}{a}\r]\chi^{\top(1)}_{ij}
- \frac{1}{2}\nabla^2\chi^{\top(1) }_{ij} .
\el

The 2nd-order Ricci tensors are
{\allowdisplaybreaks
 \bl\label{cnm12}
R^{(2) }_{00}=& \frac{3a'}{2a} \phi^{(2)'}
 +\frac{3}{2} \phi^{(2)''} +6\frac{a'}{a}\phi^{(1) }  \phi^{(1) '}
+6 \phi^{(1) }  \phi^{(1)''} +3\phi^{(1) '}\phi^{(1) '}
+\frac{1}{2}D^{ij}\chi^{||(1)}D_{ij}\chi^{||(1)''}
\nn\\
&
+\frac{1}{4}D^{ij}\chi^{||(1)'}D_{ij}\chi^{||(1)'}
+\frac{a'}{2a}D^{ij}\chi^{||(1)}D_{ij}\chi^{||(1)'}
+\frac{1}{2} \chi^{\top(1)ij} \chi^{\top(1)''}_{ij}
+\frac{1}{4} \chi^{\top(1)'ij} \chi^{\top (1)'}_{ij}
\nn\\
&
+\frac{a'}{2a} \chi^{\top(1)ij} \chi^{\top (1)'}_{ij}
+\frac{1}{2}\chi^{\top(1)ij}D_{ij}\chi^{||(1)''}
+\frac{1}{2}\chi^{\top(1)'ij}D_{ij}\chi^{||(1)'}
+\frac{a'}{2a}\chi^{\top(1)ij}D_{ij}\chi^{||(1)'}
\nn\\&
+\frac{1}{2} \chi^{\top(1)''}_{ij}D^{ij}\chi^{||(1)}
 +\frac{a'}{2a}\chi^{\top (1)'}_{ij}D^{ij}\chi^{||(1)} ,
\el
}
{\allowdisplaybreaks
\bl  \label{cnm13}
R^{(2) }_{0i}=&  \phi^{(2)'} _{,\,i}
+\frac{1}{4}D_{ij}\chi^{||(2)',\,j}
+\frac{1}{4}\chi^{\perp(2)',\,j}_{ij}
+4\phi^{(1)'} \phi^{(1) } _{,\,i}+4\phi^{(1) } \phi^{(1) '} _{,\,i}
+\phi^{(1) }D_{ij}\chi^{||(1)',\,j}
\nn\\
&
+\phi^{(1)'}D_{ij}\chi^{||(1) ,\,j}
-\frac{1}{2} \phi^{(1) ,\,j }D_{ij}\chi^{||(1)' }
+\phi^{(1)' ,\,j }D_{ij}\chi^{||(1) }
-\frac{1}{2} \phi^{(1) ,\,j }\chi^{\top(1)' }_{ij}
\nn\\
&
    +\phi^{(1)' ,\,j } \chi^{\top(1) }_{ij}
-\frac{1}{2} D^k_{j}\chi^{||(1),\,j}D_{ik}\chi^{||(1)' }
-\frac{1}{2} D^k_{j}\chi^{||(1)  }D_{ik}\chi^{||(1)',\,j}
  + \frac{1}{2} D^{jk}\chi^{||(1)'}_{,\,i}D_{jk}\chi^{||(1)  }
   \nn\\
&  + \frac{1}{4} D^{jk}\chi^{||(1)' }D_{jk}\chi^{||(1) }_{,\,i}
 -\frac{1}{2}\chi^{\top(1) k }_{j}D_{ik}\chi^{||(1)',\,j}
+ \frac{1}{2}\chi^{\top(1)j k} D_{jk}\chi^{||(1)' }_{,\,i}
    + \frac{1}{4} \chi^{\top(1)  jk}_{,\,i} D_{jk}\chi^{||(1)'}
    \nn\\
&    -\frac{1}{2} \chi^{\top(1)',\,j}_{ik}D^k_{j}\chi^{||(1)}
    + \frac{1}{2}\chi^{\top(1)' }_{jk,  \, i}D^{jk}\chi^{||(1)}
    + \frac{1}{4}\chi^{\top(1)'}_{jk }D^{jk}\chi^{||(1) }_{,\,i}
-\frac{1}{2} \chi^{\top (1)' }_{ik}D^k_{j}\chi^{||(1),\,j}
\nn\\
& -\frac{1}{2} \chi^{\top(1) k }_{j} \chi^{\top(1)',\,j}_{ik}
+ \frac{1}{2}\chi^{\top(1)j k} \chi^{\top(1)' }_{jk,i}
  + \frac{1}{4}  \chi^{\top(1)  jk}_{ ,\,i} \chi^{\top(1)'}_{jk },
\el
}
{\allowdisplaybreaks
\bl\label{cnm14}
R_{ij}^{(2)}=&
\delta_{ij}\Big[
-\frac52\frac{{a'}}{a}\phi^{(2)'}
-( \frac{a'}a )^2 \phi^{(2)}
-\frac{a''}{a}\phi^{(2)}
-\frac{1}{2}\phi^{(2)''}
+\frac{1}{2}\nabla^2\phi^{(2)}
+(\phi^{(1)'})^2
+\phi^{(1),\,k}\phi^{(1)}_{,\,k}
\nn\\
&
+2\phi^{(1)}\nabla^2\phi^{(1)}
-\phi^{(1)}_{,\,k}D^{lk}\chi^{||(1)}_{,\,l}
-\phi^{(1)}_{,\,kl}D^{kl}\chi^{||(1)}
-\phi^{(1)}_{,\,kl}\chi^{\top(1)kl}
-\frac{a'}{2a}D^{kl}\chi^{||(1)}D_{kl}\chi^{||(1)'}
\nn\\
&-\frac{a'}{2a}\chi^{\top(1)'}_{kl}D^{kl}\chi^{||(1)}
-\frac{a'}{2a}\chi^{\top(1)kl}D_{kl}\chi^{||(1)'}
-\frac{a'}{2a}\chi^{\top(1)kl}\chi^{\top(1)'}_{kl}
\Big] +\frac{1}{2}\phi^{(2)}_{,\,ij}    \nn
\\
&+\frac{1}{4}D_{ij}\chi^{||(2)''}
 +\frac{1}{2}\l[(\frac{a'}a )^2
    +\frac{a''}a \r]D_{ij}\chi^{||(2)}
+\frac{a'}{2a} D_{ij}\chi^{||(2)'}
+\frac{1}{4}D^k_j\chi^{||(2)}_{,\,ki}
+\frac{1}{4}D^k_i\chi^{||(2)}_{,\,kj}
\nn\\
&
-\frac{1}{4}\nabla^2D_{ij}\chi^{||(2)}
+\frac{1}{4}\chi^{\perp(2)''}_{ij}
+\frac{1}{2}\l[(\frac{a'}a )^2
    +\frac{a''}a \r]\chi^{\perp(2)}_{ij}
+\frac{a'}{2a} \chi^{\perp(2)'}_{ij}
+\frac{1}{4}\chi^{\perp(2),\,k}_{kj,\,i}
\nn\\
&
+\frac{1}{4}\chi^{\perp(2),\,k}_{ki,\,j}
-\frac{1}{4}\nabla^2\chi^{\perp(2)}_{ij}
+\frac{1}{4}\chi^{\top(2)''}_{ij}
+\frac{1}{2}\l[(\frac{a'}a )^2
    +\frac{a''}a \r]\chi^{\top(2)}_{ij}
+\frac{a'}{2a} \chi^{\top(2)'}_{ij}
-\frac{1}{4}\nabla^2\chi^{\top(2)}_{ij}
\nn\\
&
+3\phi^{(1)}_{,\,i}\phi^{(1)}_{,\,j}
+2\phi^{(1)}\phi^{(1)}_{,\,ij}
-\frac{3a'}a \phi^{(1)'}D_{ij}\chi^{||(1)}
+\frac{1}{2}\phi^{(1)'}D_{ij}\chi^{||(1)'}
\nn\\
&
+\frac{1}{2}\phi^{(1)}_{,\,k}D^k_j\chi^{||(1)}_{,\,i}
+\frac{1}{2}\phi^{(1)}_{,\,k}D^k_i\chi^{||(1)}_{,\,j}
-\frac32\phi^{(1)}_{,\,k}D_{ij}\chi^{||(1),\,k}
+\phi^{(1)}D^k_j\chi^{||(1)}_{,\,ik}
+\phi^{(1)}D^k_i\chi^{||(1)}_{,\,jk}
\nn
\\
&
-\phi^{(1)}\nabla^2D_{ij}\chi^{||(1)}
+\phi^{(1)}_{,\,i}D^k_j\chi^{||(1)}_{,\,k}
+\phi^{(1)}_{,\,j}D^k_i\chi^{||(1)}_{,\,k}
+\phi^{(1)}_{,\,ki}D^k_j\chi^{||(1)}
+\phi^{(1)}_{,\,kj}D^k_i\chi^{||(1)}
\nn\\
&
-\frac{3a'}a \phi^{(1)'}\chi^{\top(1)}_{ij}
+\frac{1}{2}\phi^{(1)'}\chi^{\top(1)'}_{ij}
+\frac{1}{2}\phi^{(1)}_{,\,k}\chi^{\top(1)k}_{j,\,i}
+\frac{1}{2}\phi^{(1)}_{,\,k}\chi^{\top(1)k}_{i,\,j}
-\frac32\phi^{(1)}_{,\,k}\chi^{\top(1),\,k}_{ij}
\nn
\\
&
-\phi^{(1)}\nabla^2\chi^{\top(1)}_{ij}
+\phi^{(1)}_{,\,ki}\chi^{\top(1)k}_{j}
+\phi^{(1)}_{,\,kj}\chi^{\top(1)k}_{i}
-\frac{1}{2}D^k_i\chi^{||(1)'}D_{kj}\chi^{||(1)'}
-\frac{1}{2}D^{kl}\chi^{||(1)}_{,\,l}D_{kj}\chi^{||(1)}_{,\,i}
\nn\\
&
-\frac{1}{2}D^{kl}\chi^{||(1)}_{,\,l}D_{ki}\chi^{||(1)}_{,\,j}
+\frac{1}{2}D^{lk}\chi^{||(1)}_{,\,l}D_{ij}\chi^{||(1)}_{,\,k}
-\frac{1}{2}D^{kl}\chi^{||(1)}D_{lj}\chi^{||(1)}_{,\,ik}
-\frac{1}{2}D^{kl}\chi^{||(1)}D_{li}\chi^{||(1)}_{,\,jk}
\nn
\\
&
+\frac{1}{2}D^{kl}\chi^{||(1)}D_{ij}\chi^{||(1)}_{,\,kl}
+\frac{1}{2}D^{kl}\chi^{||(1)}D_{kl}\chi^{||(1)}_{,\,ij}
+\frac{1}{4} D^{kl}\chi^{||(1)}_{,\,i}D_{kl}\chi^{||(1)}_{,\,j}
-\frac{1}{2}D^l_i\chi^{||(1)}_{,\,k}D^k_j\chi^{||(1)}_{,\,l}
\nn\\
&
+\frac{1}{2}D^k_i\chi^{||(1)}_{,\,l}D_{kj}\chi^{||(1),\,l}
-\frac{1}{2}\chi^{\top(1)'}_{kj}D^k_i\chi^{||(1)'}
-\frac{1}{2}\chi^{\top(1)'k}_{i}D_{kj}\chi^{||(1)'}
-\frac{1}{2}\chi^{\top(1)}_{kj,\,i}D^{kl}\chi^{||(1)}_{,\,l}
\nn\\
&
-\frac{1}{2}\chi^{\top(1)}_{ki,\,j}D^{kl}\chi^{||(1)}_{,\,l}
+\frac{1}{2}\chi^{\top(1)}_{ij,\,k}D^{lk}\chi^{||(1)}_{,\,l}
-\frac{1}{2}\chi^{\top(1)}_{lj,\,ik}D^{kl}\chi^{||(1)}
-\frac{1}{2}\chi^{\top(1)kl}D_{lj}\chi^{||(1)}_{,\,ik}
\nn
\\
&
-\frac{1}{2}\chi^{\top(1)}_{li,\,jk}D^{kl}\chi^{||(1)}
-\frac{1}{2}\chi^{\top(1)kl}D_{li}\chi^{||(1)}_{,\,jk}
+\frac{1}{2}\chi^{\top(1)}_{ij,\,kl}D^{kl}\chi^{||(1)}
+\frac{1}{2}\chi^{\top(1)kl}D_{ij}\chi^{||(1)}_{,\,kl}
\nn
\\
&
+\frac{1}{2}\chi^{\top(1)}_{kl,\,ij}D^{kl}\chi^{||(1)}
+\frac{1}{2}\chi^{\top(1)kl}D_{kl}\chi^{||(1)}_{,\,ij}
+\frac{1}{4}\chi^{\top(1)}_{kl,\,j}D^{kl}\chi^{||(1)}_{,\,i}
+\frac{1}{4}\chi^{\top(1)kl}_{,\,i}D_{kl}\chi^{||(1)}_{,\,j}
\nn
\\
&
-\frac{1}{2}\chi^{\top(1)k}_{j,\,l}D^l_i\chi^{||(1)}_{,\,k}
-\frac{1}{2}\chi^{\top(1)l}_{i,\,k}D^k_j\chi^{||(1)}_{,\,l}
+\frac{1}{2}\chi^{\top(1),\,l}_{kj}D^k_i\chi^{||(1)}_{,\,l}
+\frac{1}{2}\chi^{\top(1)k}_{i,\,l}D_{kj}\chi^{||(1),\,l}
\nn\\
&
-\frac{1}{2}\chi^{\top(1)'k}_{i}\chi^{\top(1)'}_{kj}
-\frac{1}{2}\chi^{\top(1)kl}\chi^{\top(1)}_{lj,\,ik}
-\frac{1}{2}\chi^{\top(1)kl}\chi^{\top(1)}_{li,\,jk}
+\frac{1}{2}\chi^{\top(1)kl}\chi^\top{(1)}_{ij,\,kl}
\nn
\\
&
+\frac{1}{2}\chi^{\top(1)kl}\chi^{\top(1)}_{kl,\,ij}
+\frac{1}{4}\chi^{\top(1)kl}_{,\,i}\chi^{\top(1)}_{kl,\,j}
-\frac{1}{2}\chi^{\top(1)l}_{i,\,k}\chi^{\top(1)k}_{j,\,l}
+\frac{1}{2}\chi^{\top(1)k}_{i,\,l}\chi^{\top(1),\,l}_{kj}
.
\el
}
The above expression of $R_{ij}^{(2)}$ corrects
some typos in (B3) of Ref.~\cite{WangZhang2017}
where  a factor $\frac{1}{2}$ was missed
in the three terms of $\l[(\frac{a'}a )^2+\frac{a''}a \r]$.
The 2nd-order Ricci scalar is
{\allowdisplaybreaks
\bl\label{cnm18}
R^{(2) } =&\frac{1}{a^2}\Bigg[
2\nabla^2\phi^{(2)}
-9\frac{a'}a \phi^{(2)'}
-3\phi^{(2)''}
+\frac{1}{2}D^{kl}\chi^{||(2)}_{,\,kl}
-12\phi^{(1)}\phi^{(1)''}
-36\frac{a'}a\phi^{(1)'}\phi^{(1)}
\nn
\\
&
+6\phi^{(1)}_{,\,k}\phi^{(1),\,k}
+16\phi^{(1)}\nabla^2\phi^{(1)}
+4\phi^{(1)}D^{kl}\chi^{||(1)}_{,\,kl}
-2\phi^{(1)}_{,\,kl}D^{kl}\chi^{||(1)}
-2\phi^{(1)}_{,\,kl}\chi^{\top(1)kl}
\nn\\
    &
-D^{kl}\chi^{||(1)}D_{kl}\chi^{||(1)''}
-\frac{3}{4}D^{kl}\chi^{||(1)'}D_{kl}\chi^{||(1)'}
    -3\frac{a'}a D^{kl}\chi^{||(1)}D_{kl}\chi^{||(1)'}
    \nn
    \\
    &
    -2D_{ml}\chi^{||(1),\,l}_{,\,k}D^{km}\chi^{||(1)}
+D^{km}\chi^{||(1)}\nabla^2D_{km}\chi^{||(1)}
-D^{km}\chi^{||(1)}_{,\,k}D_{ml}\chi^{||(1),\,l}
\nn\\
&
    +\frac{3}{4}D^{km}\chi^{||(1)}_{,\,l}D_{km}\chi^{||(1),\,l}
-\frac{1}{2}D^{km}\chi^{||(1)}_{,\,l}D^l_k\chi^{||(1)}_{,\,m}
-D^{kl}\chi^{||(1)}\chi^{\top(1)''}_{kl}
-\chi^{\top(1)kl}D_{kl}\chi^{||(1)''}
\nn
\\
&
-\frac32\chi^{\top(1)'}_{kl}D^{kl}\chi^{||(1)'}
-3\frac{a'}a \chi^{\top(1)'}_{kl}D^{kl}\chi^{||(1)}
-3\frac{a'}a \chi^{\top(1)kl}D_{kl}\chi^{||(1)'}
-2\chi^{\top(1)km}D_{ml}\chi^{||(1),\,l}_{,\,k}
    \nn\\
    &
+\chi^{\top(1)km}\nabla^2D_{km}\chi^{||(1)}
+D^{km}\chi^{||(1)}\nabla^2\chi^{\top(1)}_{km}
+\frac32\chi^{\top(1),\,l}_{km}D^{km}\chi^{||(1)}_{,\,l}
-\chi^{\top(1)l}_{k,\,m}D^{km}\chi^{||(1)}_{,\,l}
    \nn
    \\
    &
-\chi^{\top(1)kl}\chi^{\top(1)''}_{kl}
-\frac{3}{4}\chi^{\top(1)'kl}\chi^{\top(1)'}_{kl}
    -3\frac{a'}a \chi^{\top(1)kl}\chi^{\top(1)'}_{kl}
+\chi^{\top(1)km}\nabla^2\chi^{\top(1)}_{km}
    \nn
    \\
    &
    +\frac{3}{4}\chi^{\top(1)km}_{,\,l}\chi^{\top(1),\,l}_{km}
    -\frac{1}{2}\chi^{\top(1)km}_{,\,l}\chi^{\top(1)l}_{k,\,m}
\Bigg] .
\el
}
The 2nd-order perturbed Einstein tensors are
{\allowdisplaybreaks
\bl \label{G002}
G_{00}^{(2)} \equiv &
         R_{00}^{(2)} -\frac{1}{2}g^{(0)}_{00}R^{(2)} \nn \\
 = &
\nabla^2\phi^{(2)}
-\frac{3a'}{a} \phi^{(2)'}
+\frac{1}{4}D^{kl}\chi^{||(2)}_{,\,kl}
\nn\\
&
-12\frac{a'}a\phi^{(1)'}\phi^{(1)}
+3\phi^{(1) '}\phi^{(1) '}
+3\phi^{(1)}_{,\,k}\phi^{(1),\,k}
+8\phi^{(1)}\nabla^2\phi^{(1)}
+2\phi^{(1)}D^{kl}\chi^{||(1)}_{,\,kl}
\nn\\
    &
-\phi^{(1)}_{,\,kl}D^{kl}\chi^{||(1)}
-\frac{1}{8} D^{kl}\chi^{||(1)'}D_{kl}\chi^{||(1)'}
-\frac{a'}a D^{kl}\chi^{||(1)}D_{kl}\chi^{||(1)'}
            \nn\\
            &
-D_{ml}\chi^{||(1),\,l}_{,\,k}D^{km}\chi^{||(1)}
+\frac{1}{2}D^{km}\chi^{||(1)}\nabla^2D_{km}\chi^{||(1)}
-\frac{1}{2}D^{km}\chi^{||(1)}_{,\,k}D_{ml}\chi^{||(1),\,l}
\nn\\
&
    +\frac{3}{8}D^{km}\chi^{||(1)}_{,\,l}D_{km}\chi^{||(1),\,l}
-\frac{1}{4}D^{km}\chi^{||(1)}_{,\,l}D^l_k\chi^{||(1)}_{,\,m}
-\phi^{(1)}_{,\,kl}\chi^{\top(1)kl}
\nn\\
&
-\frac{1}{4}\chi^{\top(1)'}_{kl}D^{kl}\chi^{||(1)'}
-\frac{a'}a \chi^{\top(1)kl}D_{kl}\chi^{||(1)'}
-\frac{a'}a \chi^{\top(1)'}_{kl}D^{kl}\chi^{||(1)}
    \nn\\
    &
-\chi^{\top(1)km}D_{ml}\chi^{||(1),\,l}_{,\,k}
+\frac{1}{2}\chi^{\top(1)km}\nabla^2D_{km}\chi^{||(1)}
+\frac{1}{2}D^{km}\chi^{||(1)}\nabla^2\chi^{\top(1)}_{km}
    \nn\\
    &
+\frac{3}{4}\chi^{\top(1),\,l}_{km}D^{km}\chi^{||(1)}_{,\,l}
-\frac{1}{2}\chi^{\top(1)l}_{k,\,m}D^{km}\chi^{||(1)}_{,\,l}
-\frac{1}{8}\chi^{\top(1)'kl}\chi^{\top(1)'}_{kl}
 -\frac{a'}a \chi^{\top(1)kl}\chi^{\top(1)'}_{kl}
    \nn
    \\
    &
+\frac{1}{2}\chi^{\top(1)km}\nabla^2\chi^{\top(1)}_{km}
+\frac{3}{8}\chi^{\top(1)km}_{,\,l}\chi^{\top(1),\,l}_{km}
       -\frac{1}{4}\chi^{\top(1)km}_{,\,l}\chi^{\top(1)l}_{k,\,m},
\el
}
\be
 G_{0i}^{(2)} \equiv   R_{0i}^{(2)},
\ee
{\allowdisplaybreaks
\bl\label{EinsteinTensor2f}
G^{(2)}_{ij}
&=
R_{ij}^{(2)}
-\frac{1}{2}a^2(\frac{1}{2}\gamma_{ij}^{(2)})R^{(0)}
-\frac{1}{2}a^2\gamma_{ij}^{(1)}R^{(1)}
-\frac{1}{2}a^2\delta_{ij}R^{(2)}
\nn\\
=&
\delta_{ij}\Big[
-\frac{1}{2}\nabla^2\phi^{(2)}
+\phi^{(2)''}
+2\frac{{a'}}{a}\phi^{(2)'}
+\l[2\frac{a''}{a}-(\frac{a'}{a})^2\r]\phi^{(2)}
-\frac{1}{4}D^{kl}\chi^{||(2)}_{,\,kl}
\nn\\
&
-2\phi^{(1),\,k}\phi^{(1)}_{,\,k}
-2\phi^{(1)}\nabla^2\phi^{(1)}
+\l(\phi^{(1)'}\r)^2
-\phi^{(1)}_{,\,k}D^{lk}\chi^{||(1)}_{,\,l}
-\phi^{(1)}D^{kl}\chi^{||(1)}_{,\,kl}
\nn\\
&
+\frac{a'}{a}D^{kl}\chi^{||(1)}D_{kl}\chi^{||(1)'}
+D_{ml}\chi^{||(1),\,l}_{,\,k}D^{km}\chi^{||(1)}
-\frac{1}{2} D^{km}\chi^{||(1)}\nabla^2D_{km}\chi^{||(1)}
\nn
\\
&
-\frac38D^{km}\chi^{||(1),\,l}D_{km}\chi^{||(1)}_{,\,l}
+\frac{1}{4}D^{km}\chi^{||(1),\,l}D_{kl}\chi^{||(1)}_{,\,m}
+\frac{1}{2}D^{kl}\chi^{||(1)}_{,\,l}D^m_k\chi^{||(1)}_{,\,m}
\nn
\\
&
+\frac38D^{kl}\chi^{||(1)'}D_{kl}\chi^{||(1)'}
+\frac{1}{2}D^{kl}\chi^{||(1)}D_{kl}\chi^{||(1)''}
+\frac{a'}{a}\chi^{\top(1)kl}D_{kl}\chi^{||(1)'}
+\frac{a'}{a}D^{kl}\chi^{||(1)}\chi^{\top(1)'}_{kl}
\nn\\
&
+\chi^{\top(1)km}D_{ml}\chi^{||(1),\,l}_{,\,k}
-\frac{1}{2}\chi^{\top(1)km}\nabla^2D_{km}\chi^{||(1)}
-\frac{1}{2} D^{km}\chi^{||(1)}\nabla^2\chi^{\top(1)}_{km}
\nn
\\
&
-\frac{3}{4}D^{km}\chi^{||(1),\,l}\chi^{\top(1)}_{km,\,l}
+\frac{1}{2}D^{km}\chi^{||(1),\,l}\chi^{\top(1)}_{kl,\,m}
+\frac{3}{4}\chi^{\top(1)'kl}D_{kl}\chi^{||(1)'}
+\frac{1}{2}\chi^{\top(1)''}_{kl}D^{kl}\chi^{||(1)}
\nn\\
&
+\frac{1}{2}\chi^{\top(1)kl}D_{kl}\chi^{||(1)''}
+\frac{a'}{a}\chi^{\top(1)kl}\chi^{\top(1)'}_{kl}
-\frac{1}{2}\chi^{\top(1)km}\nabla^2\chi^{\top(1)}_{km}
-\frac38\chi^{\top(1)km,\,l}\chi^{\top(1)}_{km,\,l}
\nn\\
&
+\frac{1}{4}\chi^{\top(1)km,\,l}\chi^{\top(1)}_{kl,\,m}
+\frac38\chi^{\top(1)'kl}\chi^{\top(1)'}_{kl}
+\frac{1}{2}\chi^{\top(1)kl}\chi^{\top(1)''}_{kl}
\Big]
\nn\\
&
+\frac{1}{2}\phi^{(2)}_{,\,ij}
+\frac{a'}{2a} D_{ij}\chi^{||(2)'}
+\frac{1}{4}D^k_j\chi^{||(2)}_{,\,ki}
+\frac{1}{4}D^k_i\chi^{||(2)}_{,\,kj}
-\frac{1}{4}\nabla^2D_{ij}\chi^{||(2)}
+\frac{1}{4}D_{ij}\chi^{||(2)''}
\nn\\
&
 +\frac{1}{2}\l[(\frac{a'}a )^2
    -2\frac{a''}a \r]D_{ij}\chi^{||(2)}
+\frac{a'}{2a} \chi^{\perp(2)'}_{ij}
+\frac{1}{4}\chi^{\perp(2)k}_{j,\,ki}
+\frac{1}{4}\chi^{\perp(2)k}_{i,\,kj}
-\frac{1}{4}\nabla^2\chi^{\perp(2)}_{ij}
+\frac{1}{4}\chi^{\perp(2)''}_{ij}
\nn\\
&
+\frac{1}{2}\l[(\frac{a'}a )^2
    -2\frac{a''}a \r]\chi^{\perp(2)}_{ij}
+\frac{a'}{2a} \chi^{\top(2)'}_{ij}
-\frac{1}{4}\nabla^2\chi^{\top(2)}_{ij}
+\frac{1}{4}\chi^{\top(2)''}_{ij}
+\frac{1}{2}\l[(\frac{a'}a )^2
    -2\frac{a''}a \r]\chi^{\top(2)}_{ij}
\nn\\
&
+3\phi^{(1)}_{,\,i}\phi^{(1)}_{,\,j}
+2\phi^{(1)}\phi^{(1)}_{,\,ij}
+\frac{1}{2}\phi^{(1)'}D_{ij}\chi^{||(1)'}
+\frac{1}{2}\phi^{(1)}_{,\,k}D^k_j\chi^{||(1)}_{,\,i}
+\frac{1}{2}\phi^{(1)}_{,\,k}D^k_i\chi^{||(1)}_{,\,j}
\nn\\
&
-\frac32\phi^{(1)}_{,\,k}D_{ij}\chi^{||(1),\,k}
+\phi^{(1)}D^k_j\chi^{||(1)}_{,\,ik}
+\phi^{(1)}D^k_i\chi^{||(1)}_{,\,jk}
-\phi^{(1)}\nabla^2D_{ij}\chi^{||(1)}
-2 D_{ij}\chi^{||(1)}\nabla^2\phi^{(1)}
\nn
\\
&
+\phi^{(1)}_{,\,i}D^k_j\chi^{||(1)}_{,\,k}
+\phi^{(1)}_{,\,j}D^k_i\chi^{||(1)}_{,\,k}
+\phi^{(1)}_{,\,ki}D^k_j\chi^{||(1)}
+\phi^{(1)}_{,\,kj}D^k_i\chi^{||(1)}_{}
+3\phi^{(1)''}D_{ij}\chi^{||(1)}
\nn\\
&
+6\frac{a'}a\phi^{(1)'}D_{ij}\chi^{||(1)}
+\frac{1}{2}\phi^{(1)'}\chi^{\top(1)'}_{ij}
+\frac{1}{2}\phi^{(1)}_{,\,k}\chi^{\top(1)k}_{j,\,i}
+\frac{1}{2}\phi^{(1)}_{,\,k}\chi^{\top(1)k}_{i,\,j}
-\frac32\phi^{(1)}_{,\,k}\chi^{\top(1),\,k}_{ij}
\nn\\
&
-\phi^{(1)}\nabla^2\chi^{\top(1)}_{ij}
-2\chi^{\top(1)}_{ij}\nabla^2\phi^{(1)}
+\phi^{(1)}_{,\,ki}\chi^{\top(1)k}_{j}
+\phi^{(1)}_{,\,kj}\chi^{\top(1)k}_{i}
+3\phi^{(1)''}\chi^{\top(1)}_{ij}
\nn
\\
&
+6\frac{a'}a \phi^{(1)'}\chi^{\top(1)}_{ij}
-\frac{1}{2}D^k_i\chi^{||(1)'}D_{kj}\chi^{||(1)'}
-\frac{1}{2}D^{kl}\chi^{||(1)}_{,\,l}D_{kj}\chi^{||(1)}_{,\,i}
-\frac{1}{2}D^{kl}\chi^{||(1)}_{,\,l}D_{ki}\chi^{||(1)}_{,\,j}
\nn\\
&
+\frac{1}{2}D^{lk}\chi^{||(1)}_{,\,l}D_{ij}\chi^{||(1)}_{,\,k}
-\frac{1}{2}D^{kl}\chi^{||(1)}D_{lj}\chi^{||(1)}_{,\,ik}
-\frac{1}{2}D^{kl}\chi^{||(1)}D_{li}\chi^{||(1)}_{,\,jk}
\nn
\\
&
+\frac{1}{2}D^{kl}\chi^{||(1)}D_{ij}\chi^{||(1)}_{,\,kl}
+\frac{1}{2}D^{kl}\chi^{||(1)}D_{kl}\chi^{||(1)}_{,\,ij}
+\frac{1}{4}D^{kl}\chi^{||(1)}_{,\,i}D_{kl}\chi^{||(1)}_{,\,j}
\nn
\\
&
-\frac{1}{2}D_{kl}\chi^{||(1),\,kl}D_{ij}\chi^{||(1)}
-\frac{1}{2}D^l_i\chi^{||(1)}_{,\,k}D^k_j\chi^{||(1)}_{,\,l}
+\frac{1}{2}D^k_i\chi^{||(1)}_{,\,l}D_{kj}\chi^{||(1),\,l}
-\frac{1}{2}\chi^{\top(1)'}_{kj}D^k_i\chi^{||(1)'}
\nn\\
&
-\frac{1}{2}\chi^{\top(1)'k}_{i}D_{kj}\chi^{||(1)'}
-\frac{1}{2}\chi^{\top(1)}_{kj,\,i}D^{kl}\chi^{||(1)}_{,\,l}
-\frac{1}{2}\chi^{\top(1)}_{ki,\,j}D^{kl}\chi^{||(1)}_{,\,l}
+\frac{1}{2}\chi^{\top(1)}_{ij,\,k}D^{lk}\chi^{||(1)}_{,\,l}
\nn
\\
&
-\frac{1}{2}\chi^{\top(1)}_{lj,\,ik}D^{kl}\chi^{||(1)}
-\frac{1}{2}\chi^{\top(1)kl}D_{lj}\chi^{||(1)}_{,\,ik}
-\frac{1}{2}\chi^{\top(1)}_{li,\,jk}D^{kl}\chi^{||(1)}
-\frac{1}{2}\chi^{\top(1)kl}D_{li}\chi^{||(1)}_{,\,jk}
\nn
\\
&
+\frac{1}{2}\chi^{\top(1)}_{ij,\,kl}D^{kl}\chi^{||(1)}
+\frac{1}{2}\chi^{\top(1)kl}D_{ij}\chi^{||(1)}_{,\,kl}
+\frac{1}{2}\chi^{\top(1)}_{kl,\,ij}D^{kl}\chi^{||(1)}
+\frac{1}{2}\chi^{\top(1)kl}D_{kl}\chi^{||(1)}_{,\,ij}
\nn\\
&
+\frac{1}{4}\chi^{\top(1)}_{kl,\,j}D^{kl}\chi^{||(1)}_{,\,i}
+\frac{1}{4}\chi^{\top(1)kl}_{,\,i}D_{kl}\chi^{||(1)}_{,\,j}
-\frac{1}{2}\chi^{\top(1)}_{ij}D_{kl}\chi^{||(1),\,kl}
-\frac{1}{2}\chi^{\top(1)k}_{j,\,l}D^l_i\chi^{||(1)}_{,\,k}
\nn\\
&
-\frac{1}{2}\chi^{\top(1)l}_{i,\,k}D^k_j\chi^{||(1)}_{,\,l}
+\frac{1}{2}\chi^{\top(1),\,l}_{kj}D^k_i\chi^{||(1)}_{,\,l}
+\frac{1}{2}\chi^{\top(1)k}_{i,\,l}D_{kj}\chi^{||(1),\,l}
-\frac{1}{2}\chi^{\top(1)'k}_{i}\chi^{\top(1)'}_{kj}
\nn
\\
&
-\frac{1}{2}\chi^{\top(1)kl}\chi^{\top(1)}_{lj,\,ik}
-\frac{1}{2}\chi^{\top(1)kl}\chi^{\top(1)}_{li,\,jk}
+\frac{1}{2}\chi^{\top(1)kl}\chi^{\top(1)}_{ij,\,kl}
+\frac{1}{2}\chi^{\top(1)kl}\chi^{\top(1)}_{kl,\,ij}
\nn
\\
&
+\frac{1}{4}\chi^{\top(1)kl}_{,\,i}\chi^{\top(1)}_{kl,\,j}
-\frac{1}{2}\chi^{\top(1)l}_{i,\,k}\chi^{\top(1)k}_{j,\,l}
+\frac{1}{2}\chi^{\top(1)k}_{i,\,l}\chi^{\top(1),\,l}_{kj} \ .
\el
}
The above expressions are valid for a general expansion stage
and corrects  some typos in (B7) of Ref.~\cite{WangZhang2017}
where  four terms of $[2\frac{a''}{a}-(\frac{a'}{a})^2 ]$
were missed and which
was valid only for MD stage.

The 1st-order energy-momentum tensor is
\be\label{T001st}
T^{(1)}_{00}=
(\rho^{(1)}+p^{(1)})\l[U_0 U_0\r]^{(0)}+g^{(0)}_{00}p^{(1)}
= a^2\rho^{(0)}\delta^{(1)},
\ee
\be\label{Ti01st}
T_{0i}^{(1)}=
(\rho^{(0)}+p^{(0)})\l[U_0 U_i\r]^{(1)}
=  -(1+c_s^2)\rho^{(0)}a^2 v^{(1)}_i,
\ee
\be\label{Tij1st}
T^{(1)}_{ij}=
\l[
(\rho+p)U_i U_j+g_{ij}p
\r]^{(1)}
=
a^2\rho^{(0)}\l(
c_s^2\gamma^{(1)}_{ij}
+\delta_{ij}c_L^2\delta^{(1)}
\r) .
\ee

The 2nd-order energy-momentum tensor is
\bl\label{EnMoTensor002nd3}
T^{(2)}_{00}=&
(\frac{1}{2}\rho^{(2)}+\frac{1}{2}p^{(2)})[U_0 U_0]^{(0)}
+(\rho^{(0)}+p^{(0)})[U_0 U_0]^{(2)}
+g^{(0)}_{00}\frac{1}{2}p^{(2)}
\nn\\
=&
\frac{1}{2}a^2\delta^{(2)}\rho^{(0)}
+(1+c_s^2)a^2v^{(1)m}v^{(1)}_{m} \rho^{(0)} ,
\el
\bl\label{T0i2th1}
T_{0i}^{(2)}
=&
(\rho^{(0)}+p^{(0)})\l[U_0\r]^{(0)} \l[U_i\r]^{(2)}
+(\rho^{(1)}+p^{(1)})\l[U_0\r]^{(0)} \l[U_i\r]^{(1)}
\nn\\
=&
-\rho^{(0)}(1+c_s^2)a^2(\gamma_{ij}^{(1)}v^{(1)j}
+\frac{1}{2}v^{(2)}_{i})
-\delta^{(1)}\rho^{(0)}(1+c_L^2)a^2v^{(1)}_i
,
\el
\bl\label{Tij2}
T^{(2)}_{ij}=&
(\rho^{(0)}+p^{(0)})[U_i]^{(1)} [U_j]^{(1)}
+g_{ij}^{(0)}\frac{1}{2}p^{(2)}
+\frac{1}{2}g_{ij}^{(2)}p^{(0)}
+g_{ij}^{(1)}p^{(1)}
\nn\\
=&
\rho^{(0)}(1+c_s^2)a^2 v_i^{(1)}v_j^{(1)}
+a^2\delta_{ij}\frac{1}{2}c_N^2\rho^{(0)}\delta^{(2)}
+\frac{1}{2}a^2\gamma^{(2)}_{ij}c_s^2\rho^{(0)}
\nn\\
&
+a^2\gamma^{(1)}_{ij}c_L^2\rho^{(0)}\delta^{(1)}
.
\el

\section{2nd-order perturbed Einstein equations for a general RW spacetime}

In this Appendix
we  list the perturbed Einstein equations
and the equation of covariant conservation
for a   flat RW spacetime with a general scale factor $a(\tau)$.

The $(00)$ component of 1st-order perturbed Einstein equation,
i.e.,
the 1st-order energy constraint is
\be\label{enCons1st}
-6\frac{a'}{a}\phi^{(1)'}
          +2\nabla^2\phi^{(1) }
          +\frac{1}{3}\nabla^2\nabla^2\chi^{||(1)}
= 3\l(\frac{a'}{a}\r)^2  \delta^{(1)}  .
\ee
The $(0i)$ component of the 1st-order perturbed Einstein equation,
i.e.,
the 1st-order momentum constraint is
\be \label{momentconstr1}
2\phi^{(1)' }_{,i}
+ \frac{1}{2} D_{ij}\chi^{||(1)',j  }
+ \frac{1}{2} \chi^{\perp(1)',j  }_{ij} =
-3(1+c_s^2)\l(\frac{a'}{a}\r)^2 v^{(1)}_i.
\ee
The $(ij)$ component of 1st-order perturbed Einstein equation,
i.e.,
the 1st-order evolution equation is
\bl  \label{evoEq1st}
&
2\phi^{(1)''} \delta_{ij}
+4\frac{a'}{a}\phi^{(1)'}\delta_{ij}
+\phi^{(1) }_{,ij}
-\nabla^2\phi^{(1) }\delta_{ij}
\nn\\
&
+\frac{1}{2} D_{ij}\chi^{||(1)''}
+\frac{a'}{a}D_{ij}\chi^{||(1)'}
+\frac{1}{6}\nabla^2D_{ij}\chi^{||(1)}
-\frac{1}{9}\delta_{ij}\nabla^2\nabla^2\chi^{||(1) }
\nn\\
&
+\frac{1}{2} \chi^{\perp(1)''}_{ij}
+\frac{a'}{a}\chi^{\perp(1)'}_{ij}
\nn \\
&
+\frac{1}{2} \chi^{\top(1)''}_{ij}
+\frac{a'}{a} \chi^{\top(1)'}_{ij}
- \frac{1}{2}\nabla^2\chi^{\top(1) }_{ij}
=
3 c_L^2\l(\frac{a'}{a}\r)^2\delta^{(1)}\delta_{ij}
\ .
\el
The 1st-order energy conservation is
\be\label{enCons1stZero}
\delta^{(1)'}
+\frac{2a''}{a'}\delta^{(1)}
+\frac{3a'}{a}(
c_L^2-\frac{1}{3})\delta^{(1)}
=3(1+c_s^2)\phi^{(1)'}
-(1+c_s^2) \nabla^2v^{||(1)}
.
\ee
The 1st-order momentum conservation is
\be\label{MoConsv7}
c_L^2\delta^{(1),\,i}
+\frac{2a''}{a'}(1+c_s^2)v^{(1)i}
+(1+c_s^2)v^{(1)'\,i}
=0 .
\ee

The   following are
the 2nd-order perturbed equations for the scalar-scalar coupling
of 1st-order perturbations.
The $(00)$ component of 2nd-order perturbed Einstein equation is
\be  \label{2nd00Einstein}
 G^{(2)}_{00}
 = 8\pi G T_{00}^{(2)} ,
\ee
where $G^{(2)}_{00}$ is given by (\ref{G002}) and
 $T_{00}^{(2)}$ is given by (\ref{EnMoTensor002nd3}).
For the scalar-scalar coupling,
this gives the  2nd-order energy constraint
{\allowdisplaybreaks
\bl  \label{Ein2th001}
&\nabla^2\phi^{(2)}_S
-\frac{3a'}{a} \phi^{(2)'}_S
+\frac{1}{4}D^{kl}\chi^{||(2)}_{S,\,kl}
\nn\\
&
-12\frac{a'}a\phi^{(1)'}\phi^{(1)}
+3\phi^{(1) '}\phi^{(1) '}
+3\phi^{(1)}_{,\,k}\phi^{(1),\,k}
+8\phi^{(1)}\nabla^2\phi^{(1)}
+2\phi^{(1)}D^{kl}\chi^{||(1)}_{,\,kl}
\nn\\
    &
-\phi^{(1)}_{,\,kl}D^{kl}\chi^{||(1)}
-\frac{1}{8} D^{kl}\chi^{||(1)'}D_{kl}\chi^{||(1)'}
-\frac{a'}a D^{kl}\chi^{||(1)}D_{kl}\chi^{||(1)'}
            \nn\\
            &
-D_{ml}\chi^{||(1),\,l}_{,\,k}D^{km}\chi^{||(1)}
+\frac{1}{2}D^{km}\chi^{||(1)}\nabla^2D_{km}\chi^{||(1)}
-\frac{1}{2}D^{km}\chi^{||(1)}_{,\,k}D_{ml}\chi^{||(1),\,l}
\nn\\
&
    +\frac{3}{8}D^{km}\chi^{||(1)}_{,\,l}D_{km}\chi^{||(1),\,l}
-\frac{1}{4}D^{km}\chi^{||(1)}_{,\,l}D^l_k\chi^{||(1)}_{,\,m}
\nn\\
=&
8\pi G \l[
\frac{1}{2}a^2\delta^{(2)}_S\rho^{(0)}
+(1+c_s^2)a^2v^{||(1),m}v^{||(1)}_{,m} \rho^{(0)}
\r] ,
\el
}
where  a subscript ``S" in $\phi^{(2)}_S$, etc.,
indicates the case of scalar-scalar coupling,
and the 1st-order scalar perturbations
$\phi^{(1)}$, $\chi^{||(1)}$, $v^{||(1)}$ are already shown in Sec. 3.
Moving the scalar-scalar coupling terms to the rhs
and using the Friedmann equation (\ref{Ein0th00})
give the following 2nd-order energy constraint:
\be  \label{Ein2th003}
-\frac{6a'}{a} \phi^{(2)'}_S
 +2\nabla^2\phi^{(2)}_S
 +\frac{1}{3}\nabla^2\nabla^2\chi^{||(2)}_{S}
=  3\l(\frac{a'}{a}\r)^2\delta^{(2)}_S +E_S ,
\ee
where
\bl\label{ES1}
E_S
=
&
6(1+c_s^2)\l(\frac{a'}{a}\r)^2v^{||(1),\,k}v^{||(1)}_{,\,k}
+24\frac{a'}a\phi^{(1)'}\phi^{(1)}
-6\phi^{(1) '}\phi^{(1) '}
-6\phi^{(1)}_{,\,k}\phi^{(1),\,k}
-16\phi^{(1)}\nabla^2\phi^{(1)}
\nn\\
    &
-\frac{8}{3}\phi^{(1)}\nabla^2\nabla^2\chi^{||(1)}
+2\phi^{(1),\,kl}\chi^{||(1)}_{,\,kl}
-\frac{2}{3}\nabla^2\phi^{(1)}\nabla^2\chi^{||(1)}
+\frac{1}{4} \chi^{||(1)',\,kl}\chi^{||(1)'}_{,\,kl}
            \nn\\
            &
-\frac{1}{12} \nabla^2\chi^{||(1)'}\nabla^2\chi^{||(1)'}
+\frac{2a'}a \chi^{||(1),\,kl}\chi^{||(1)'}_{,\,kl}
-\frac{2a'}{3 a}\nabla^2\chi^{||(1)}\nabla^2\chi^{||(1)'}
+\frac{1}{3}\chi^{||(1),\,kl}\nabla^2\chi^{||(1)}_{,\,kl}
\nn\\
&
-\frac{1}{9}\nabla^2\chi^{||(1)}\nabla^2\nabla^2\chi^{||(1)}
-\frac{1}{4}\chi^{||(1),\,klm}\chi^{||(1)}_{,\,klm}
+\frac{5}{12}\nabla^2\chi^{||(1)}_{,\,k}\nabla^2\chi^{||(1),\,k}
.
\el
The $(0i)$ component of 2nd-order perturbed Einstein equation is
\be \label{Ein2nd0i1}
 G^{(2)}_{0i}=  R^{(2)}_{0i}=8\pi G T_{0i}^{(2)},
\ee
where  $R^{(2) }_{0i}$ is given by  (\ref{cnm13})
and $T^{(2)}_{0i}$ by  (\ref{T0i2th1}).
For the scalar-scalar coupling
it is
\bl\label{MoConstr2ndv1}
&  \phi^{(2)'} _{S,\,i}
+\frac{1}{4}D_{ij}\chi^{||(2)',\,j}_S
+\frac{1}{4}\chi^{\perp(2)',\,j}_{S\,ij}
+4\phi^{(1)'} \phi^{(1) } _{,\,i}+4\phi^{(1) } \phi^{(1) '} _{,\,i}
+\phi^{(1) }D_{ij}\chi^{||(1)',\,j}
\nn\\
&
+\phi^{(1)'}D_{ij}\chi^{||(1) ,\,j}
-\frac{1}{2} \phi^{(1) ,\,j }D_{ij}\chi^{||(1)' }
+\phi^{(1)' ,\,j }D_{ij}\chi^{||(1) }
-\frac{1}{2} D^k_{j}\chi^{||(1),\,j}D_{ik}\chi^{||(1)' }
\nn\\
&
-\frac{1}{2} D^k_{j}\chi^{||(1)  }D_{ik}\chi^{||(1)',\,j}
  + \frac{1}{2} D^{jk}\chi^{||(1)'}_{,\,i}D_{jk}\chi^{||(1)  }
+ \frac{1}{4} D^{jk}\chi^{||(1)' }D_{jk}\chi^{||(1) }_{,\,i}
\nn\\
=&
8\pi G\l[
-a^2\rho^{(0)}(1+c_s^2)(
-2\phi^{(1)}v^{||(1)}_{,i}+v^{||(1),m}D_{im}\chi^{||(1)}
+\frac{1}{2}v^{(2)}_{S\,i})
-a^2\rho^{(0)}(1+c_L^2)\delta^{(1)}v^{||(1)}_{,i}
\r]
.
\el
Moving the couplings to the rhs
and using the Friedmann equation (\ref{Ein0th00}),
one has the 2nd-order momentum constraint
\be\label{MoConstr2ndv3}
2 \phi^{(2)'} _{S,\,i}
+\frac{1}{2}D_{ij}\chi^{||(2)',\,j}_S
+\frac{1}{2}\chi^{\perp(2)',\,j}_{S\,ij}
=
-3(1+c_s^2)\l(\frac{a'}{a}\r)^2v^{(2)}_{S\,i}
+M_{S\,i}
\ee
where
\bl\label{MSi1}
M_{S\,i}
\equiv&
12(1+c_s^2)\l(\frac{a'}{a}\r)^2\phi^{(1)}v^{||(1)}_{,i}
-6(1+c_s^2)\l(\frac{a'}{a}\r)^2v^{||(1),k}\chi^{||(1)}_{,ki}
\nn\\
&
+2(1+c_s^2)\l(\frac{a'}{a}\r)^2v^{||(1)}_{,\,i}\nabla^2\chi^{||(1)}
-6(1+c_L^2)\l(\frac{a'}{a}\r)^2\delta^{(1)}v^{||(1)}_{,i}
\nn\\
&
-8\phi^{(1)'} \phi^{(1) } _{,\,i}
-8\phi^{(1) } \phi^{(1) '} _{,\,i}
-\frac{4}{3}\phi^{(1) }\nabla^2\chi^{||(1)'}_{,\,i}
-\frac{4}{3}\phi^{(1)'}\nabla^2\chi^{||(1)}_{ ,\,i}
\nn\\
&
+ \phi^{(1) ,\,k }\chi^{||(1)' }_{,\,ki}
-\frac{1}{3}\phi^{(1) }_{,\,i }\nabla^2\chi^{||(1)' }
-2\phi^{(1)' ,\,k }\chi^{||(1) }_{,\,ki}
+\frac{2}{3}\phi^{(1)'}_{ ,\,i }\nabla^2\chi^{||(1) }
\nn\\
&
+\frac{2}{3}\chi^{||(1)' }_{,\,ki}\nabla^2\chi^{||(1),\,k}
-\frac{1}{18}\nabla^2\chi^{||(1)}_{,\,i}\nabla^2\chi^{||(1)' }
-\frac{1}{3}\chi^{||(1)  }_{,\,ki}\nabla^2\chi^{||(1)',\,k}
\nn\\
&
+\frac{1}{9}\nabla^2\chi^{||(1)  }\nabla^2\chi^{||(1)'}_{,\,i}
- \frac{1}{2}\chi^{||(1)',\,kl }\chi^{||(1) }_{,\,kli}
.
\el
The longitudinal part of momentum constraint (\ref{MoConstr2ndv3}) is
\be \label{MoCons2ndNoCurl2}
2\phi^{(2)'}_{S}
+\frac{1}{3}\nabla^2\chi^{||(2)'}_S
=
-3(1+c_s^2)\l(\frac{a'}{a}\r)^2v^{||(2)}_{S}
+\nabla^{-2}M_{S\,l}^{,\,l}
,
\ee
and the transverse part is
\be \label{MoCons2ndCurl1}
\frac{1}{2}\chi^{\perp(2)',\,j}_{S\,ij}
=
-3(1+c_s^2)\l(\frac{a'}{a}\r)^2v^{\perp(2)}_{S\,i}
+ \l( M_{S\,i} -\partial_i\nabla^{-2}M_{S\,l}^{,\,l} \r)
\ ,
\ee
where
\bl\label{MSkk}
M_{S\,l}^{,l}
=&
\nabla^2\Big[
-8\phi^{(1)'} \phi^{(1) }
-\frac{1}{3}\phi^{(1) }\nabla^2\chi^{||(1)' }
-\frac{4}{3}\phi^{(1)'}\nabla^2\chi^{||(1)}
-\frac{1}{18}\nabla^2\chi^{||(1)}\nabla^2\chi^{||(1)' }
\nn\\
&
-\frac{1}{6}\chi^{||(1)}_{,\,k}\nabla^2\chi^{||(1)',\,k}\Big]
+12(1+c_s^2)\l(\frac{a'}{a}\r)^2\phi^{(1),k}v^{||(1)}_{,\,k}
+12(1+c_s^2)\l(\frac{a'}{a}\r)^2\phi^{(1)}\nabla^2v^{||(1)}
\nn\\
&
-6(1+c_s^2)\l(\frac{a'}{a}\r)^2v^{||(1),kl}\chi^{||(1)}_{,kl}
-4(1+c_s^2)\l(\frac{a'}{a}\r)^2v^{||(1),k}\nabla^2\chi^{||(1)}_{,k}
\nn\\
&
+2(1+c_s^2)\l(\frac{a'}{a}\r)^2\nabla^2v^{||(1)}\nabla^2\chi^{||(1)}
-6(1+c_L^2)\l(\frac{a'}{a}\r)^2\delta^{(1)}\nabla^2v^{||(1)}
\nn\\
&
-6(1+c_L^2)\l(\frac{a'}{a}\r)^2\delta^{(1)}_{,\,k}v^{||(1),\,k}
+ \phi^{(1) ,\,kl }\chi^{||(1)' }_{,\,kl}
-\phi^{(1) }\nabla^2\nabla^2\chi^{||(1)'}
+2\nabla^2\phi^{(1)'}\nabla^2\chi^{||(1) }
\nn\\
&
-2\phi^{(1)' ,\,kl}\chi^{||(1) }_{,\,kl}
+\frac{1}{6}\chi^{||(1)}_{,\,k}\nabla^2\nabla^2\chi^{||(1)',\,k}
+\frac{1}{6}\nabla^2\chi^{||(1)  }\nabla^2\nabla^2\chi^{||(1)'}
\nn\\
&
+\frac{1}{6}\chi^{||(1)'}_{,\,kl}\nabla^2\chi^{||(1),\,kl}
+\frac{2}{3}\nabla^2\chi^{||(1)' }_{,\,k}\nabla^2\chi^{||(1),\,k}
- \frac{1}{2}\chi^{||(1)',\,klm }\chi^{||(1) }_{,\,klm}
,
\el
and
{\allowdisplaybreaks
\bl\label{MSiCurl}
&
\l( M_{S\,i}-\partial_i\nabla^{-2}M_{S\,l}^{,l} \r)
\nn\\
=&
12(1+c_s^2)\l(\frac{a'}{a}\r)^2\phi^{(1)}v^{||(1)}_{,i}
-6(1+c_s^2)\l(\frac{a'}{a}\r)^2v^{||(1),k}\chi^{||(1)}_{,ki}
\nn\\
&
+2(1+c_s^2)\l(\frac{a'}{a}\r)^2v^{||(1)}_{,\,i}\nabla^2\chi^{||(1)}
-6(1+c_L^2)\l(\frac{a'}{a}\r)^2\delta^{(1)}v^{||(1)}_{,i}
\nn\\
&
-\phi^{(1) }\nabla^2\chi^{||(1)' }_{,\,i}
+2\phi^{(1)'}_{ ,\,i }\nabla^2\chi^{||(1) }
+ \phi^{(1) ,\,k }\chi^{||(1)' }_{,\,ki}
-2\phi^{(1)' ,\,k }\chi^{||(1) }_{,\,ki}
+\frac{2}{3}\chi^{||(1)' }_{,\,ki}\nabla^2\chi^{||(1),\,k}
\nn\\
&
-\frac{1}{6}\chi^{||(1)}_{,\,ki}\nabla^2\chi^{||(1)',\,k}
+\frac{1}{6}\chi^{||(1)}_{,\,k}\nabla^2\chi^{||(1)',\,k}_{,\,i}
+\frac{1}{6}\nabla^2\chi^{||(1)  }\nabla^2\chi^{||(1)'}_{,\,i}
- \frac{1}{2}\chi^{||(1)',\,kl }\chi^{||(1) }_{,\,kli}
\nn\\
&
+\partial_i\nabla^{-2}\Big[
-12(1+c_s^2)\l(\frac{a'}{a}\r)^2\phi^{(1),k}v^{||(1)}_{,\,k}
-12(1+c_s^2)\l(\frac{a'}{a}\r)^2\phi^{(1)}\nabla^2v^{||(1)}
\nn\\
&
+6(1+c_s^2)\l(\frac{a'}{a}\r)^2v^{||(1),kl}\chi^{||(1)}_{,kl}
+4(1+c_s^2)\l(\frac{a'}{a}\r)^2v^{||(1),k}\nabla^2\chi^{||(1)}_{,k}
\nn\\
&
-2(1+c_s^2)\l(\frac{a'}{a}\r)^2\nabla^2v^{||(1)}\nabla^2\chi^{||(1)}
+6(1+c_L^2)\l(\frac{a'}{a}\r)^2\delta^{(1)}\nabla^2v^{||(1)}
\nn\\
&
+6(1+c_L^2)\l(\frac{a'}{a}\r)^2\delta^{(1)}_{,\,k}v^{||(1),\,k}
- \phi^{(1) ,\,kl }\chi^{||(1)' }_{,\,kl}
+\phi^{(1) }\nabla^2\nabla^2\chi^{||(1)'}
-2\nabla^2\phi^{(1)'}\nabla^2\chi^{||(1) }
\nn\\
&
+2\phi^{(1)' ,\,kl}\chi^{||(1) }_{,\,kl}
 -\frac{1}{6}\chi^{||(1)}_{,\,k}\nabla^2\nabla^2\chi^{||(1)',\,k}
-\frac{1}{6}\nabla^2\chi^{||(1)  }\nabla^2\nabla^2\chi^{||(1)'}
-\frac{1}{6}\chi^{||(1)'}_{,\,kl}\nabla^2\chi^{||(1),\,kl}
\nn\\
&
-\frac{2}{3}\nabla^2\chi^{||(1)' }_{,\,k}\nabla^2\chi^{||(1),\,k}
+ \frac{1}{2}\chi^{||(1)',\,klm }\chi^{||(1) }_{,\,klm}
\Big]  .
\el
}

The $(ij)$ component of 2nd-order perturbed Einstein equation is
\be\label{Ein2ndijss1}
G^{(2)}_{ij}
=8\pi G T^{(2)}_{ij} ,
\ee
where  $G^{(2)}_{ij}$ is given in (\ref{EinsteinTensor2f}) and
$T^{(2)}_{ij}$ in (\ref{Tij2}).
For the scalar-scalar coupling,
(\ref{Ein2ndijss1}) gives the 2nd-order evolution equation
{\allowdisplaybreaks
\bl
&
\delta_{ij}\Big[
-\frac{1}{2}\nabla^2\phi^{(2)}_S
+\phi^{(2)''}_S
+2\frac{{a'}}{a}\phi^{(2)'}_S
+\l[2\frac{a''}{a}-(\frac{a'}{a})^2\r]\phi^{(2)}_S
-\frac{1}{4}D^{kl}\chi^{||(2)}_{S,\,kl}
-2\phi^{(1),\,k}\phi^{(1)}_{,\,k}
\nn\\
&
-2\phi^{(1)}\nabla^2\phi^{(1)}
+\l(\phi^{(1)'}\r)^2
-\phi^{(1)}_{,\,k}D^{lk}\chi^{||(1)}_{,\,l}
-\phi^{(1)}D^{kl}\chi^{||(1)}_{,\,kl}
+\frac{a'}{a}D^{kl}\chi^{||(1)}D_{kl}\chi^{||(1)'}
\nn\\
&
+D_{ml}\chi^{||(1),\,l}_{,\,k}D^{km}\chi^{||(1)}
-\frac{1}{2} D^{km}\chi^{||(1)}\nabla^2D_{km}\chi^{||(1)}
-\frac38D^{km}\chi^{||(1),\,l}D_{km}\chi^{||(1)}_{,\,l}
\nn
\\
&
+\frac{1}{4}D^{km}\chi^{||(1),\,l}D_{kl}\chi^{||(1)}_{,\,m}
+\frac{1}{2}D^{kl}\chi^{||(1)}_{,\,l}D^m_k\chi^{||(1)}_{,\,m}
+\frac38D^{kl}\chi^{||(1)'}D_{kl}\chi^{||(1)'}
\nn\\
&
+\frac{1}{2}D^{kl}\chi^{||(1)}D_{kl}\chi^{||(1)''}
\Big]
+\frac{1}{2}\phi^{(2)}_{S,\,ij}
+\frac{a'}{2a} D_{ij}\chi^{||(2)'}_{S\,}
+\frac{1}{4}D^k_j\chi^{||(2)}_{S,\,ki}
+\frac{1}{4}D^k_i\chi^{||(2)}_{S,\,kj}
\nn\\
&
-\frac{1}{4}\nabla^2D_{ij}\chi^{||(2)}_{S}
+\frac{1}{4}D_{ij}\chi^{||(2)''}_{S}
 +\frac{1}{2}\l[(\frac{a'}a )^2
    -2\frac{a''}a \r]D_{ij}\chi^{||(2)}_{S}
+\frac{a'}{2a} \chi^{\perp(2)'}_{S\,ij}
+\frac{1}{4}\chi^{\perp(2)k}_{S\,j,\,ki}
\nn\\
&
+\frac{1}{4}\chi^{\perp(2)k}_{S\,i,\,kj}
-\frac{1}{4}\nabla^2\chi^{\perp(2)}_{S\,ij}
+\frac{1}{4}\chi^{\perp(2)''}_{S\,ij}
+\frac{1}{2}\l[(\frac{a'}a )^2
    -2\frac{a''}a \r]\chi^{\perp(2)}_{S\,ij}
+\frac{a'}{2a} \chi^{\top(2)'}_{S\,ij}
-\frac{1}{4}\nabla^2\chi^{\top(2)}_{S\,ij}
\nn\\
&
+\frac{1}{4}\chi^{\top(2)''}_{S\,ij}
+\frac{1}{2}\l[(\frac{a'}a )^2
    -2\frac{a''}a \r]\chi^{\top(2)}_{S\,ij}
+3\phi^{(1)}_{,\,i}\phi^{(1)}_{,\,j}
+2\phi^{(1)}\phi^{(1)}_{,\,ij}
+\frac{1}{2}\phi^{(1)'}D_{ij}\chi^{||(1)'}
\nn\\
&
+\frac{1}{2}\phi^{(1)}_{,\,k}D^k_j\chi^{||(1)}_{,\,i}
+\frac{1}{2}\phi^{(1)}_{,\,k}D^k_i\chi^{||(1)}_{,\,j}
-\frac32\phi^{(1)}_{,\,k}D_{ij}\chi^{||(1),\,k}
+\phi^{(1)}D^k_j\chi^{||(1)}_{,\,ik}
+\phi^{(1)}D^k_i\chi^{||(1)}_{,\,jk}
\nn
\\
&
-\phi^{(1)}\nabla^2D_{ij}\chi^{||(1)}
-2 D_{ij}\chi^{||(1)}\nabla^2\phi^{(1)}
+\phi^{(1)}_{,\,i}D^k_j\chi^{||(1)}_{,\,k}
+\phi^{(1)}_{,\,j}D^k_i\chi^{||(1)}_{,\,k}
+\phi^{(1)}_{,\,ki}D^k_j\chi^{||(1)}
\nn\\
&
+\phi^{(1)}_{,\,kj}D^k_i\chi^{||(1)}_{}
+3\phi^{(1)''}D_{ij}\chi^{||(1)}
+6\frac{a'}a\phi^{(1)'}D_{ij}\chi^{||(1)}
-\frac{1}{2}D^k_i\chi^{||(1)'}D_{kj}\chi^{||(1)'}
\nn\\
&
-\frac{1}{2}D^{kl}\chi^{||(1)}_{,\,l}D_{kj}\chi^{||(1)}_{,\,i}
-\frac{1}{2}D^{kl}\chi^{||(1)}_{,\,l}D_{ki}\chi^{||(1)}_{,\,j}
+\frac{1}{2}D^{lk}\chi^{||(1)}_{,\,l}D_{ij}\chi^{||(1)}_{,\,k}
-\frac{1}{2}D^{kl}\chi^{||(1)}D_{lj}\chi^{||(1)}_{,\,ik}
\nn\\
&
-\frac{1}{2}D^{kl}\chi^{||(1)}D_{li}\chi^{||(1)}_{,\,jk}
+\frac{1}{2}D^{kl}\chi^{||(1)}D_{ij}\chi^{||(1)}_{,\,kl}
+\frac{1}{2}D^{kl}\chi^{||(1)}D_{kl}\chi^{||(1)}_{,\,ij}
+\frac{1}{4}D^{kl}\chi^{||(1)}_{,\,i}D_{kl}\chi^{||(1)}_{,\,j}
\nn\\
&
-\frac{1}{2}D_{kl}\chi^{||(1),\,kl}D_{ij}\chi^{||(1)}
-\frac{1}{2}D^l_i\chi^{||(1)}_{,\,k}D^k_j\chi^{||(1)}_{,\,l}
+\frac{1}{2}D^k_i\chi^{||(1)}_{,\,l}D_{kj}\chi^{||(1),\,l}
\nn\\
=&8\pi G a^2\rho^{(0)}\Big\{
(1+c_s^2) v^{||(1)}_{,i}v^{||(1)}_{,j}
+\frac{1}{2}c_N^2\delta^{(2)}_{S}\delta_{ij}
+\frac{1}{2} c_s^2\Big[
    -2\phi^{(2)}_{S}\delta_{ij}
    +D_{ij}\chi^{||(2)}_{S}
   \nn\\
&   +\chi^{\perp(2)}_{S\,ij}
    +\chi^{\top(2)}_{S\,ij}
    \Big]
+c_L^2 \Big[  -2\delta^{(1)}\phi^{(1)}\delta_{ij}
     +\delta^{(1)}D_{ij}\chi^{||(1)}     \Big]  \Big\} .
\el
}
Moving the coupling terms to the rhs
and using the Friedmann equation (\ref{Ein0th00})
yield
the following 2nd-order evolution equation:
\bl\label{Evo2ndSs1}
&
2\phi^{(2)''}_S \delta_{ij}
+4\frac{{a'}}{a}\phi^{(2)'}_S \delta_{ij}
+\phi^{(2)}_{S,\,ij}
-\nabla^2\phi^{(2)}_S \delta_{ij}
+\l[4\frac{a''}{a}
    +6(c_s^2-\frac{1}{3})(\frac{a'}{a})^2\r]\phi^{(2)}_S \delta_{ij}
\nn\\
&
+\frac{1}{2}D_{ij}\chi^{||(2)''}_S
+\frac{a'}{a} D_{ij}\chi^{||(2)'}_S
 +\l[(1-3c_s^2)(\frac{a'}{a})^2
    -2\frac{a''}a \r]D_{ij}\chi^{||(2)}_{S}
\nn\\
&
+\frac{1}{6}\nabla^2D_{ij}\chi^{||(2)}_S
-\frac{1}{9}\delta_{ij}\nabla^2\nabla^2\chi^{||(2) }_S
\nn\\
&
+\frac{1}{2}\chi^{\perp(2)''}_{S\,ij}
+\frac{a'}{a} \chi^{\perp(2)'}_{S\,ij}
+\l[(1-3c_s^2)(\frac{a'}a )^2
    -2\frac{a''}a \r]\chi^{\perp(2)}_{S\,ij}
\nn\\
&
+\frac{1}{2}\chi^{\top(2)''}_{S\,ij}
+\frac{a'}{a} \chi^{\top(2)'}_{S\,ij}
+\l[(1-3c_s^2)(\frac{a'}a )^2
    -2\frac{a''}a \r]\chi^{\top(2)}_{S\,ij}
-\frac{1}{2}\nabla^2\chi^{\top(2)}_{S\,ij}
\nn\\
=&
3c_N^2\l(\frac{a'}{a}\r)^2\delta^{(2)}_{S}\delta_{ij}
+S_{S\,ij}
,
\el
where
{\allowdisplaybreaks
\bl\label{Ss2ndij2}
S_{S\,ij}
&=
6(1+c_s^2)\l(\frac{a'}{a}\r)^2 v^{||(1)}_{,i}v^{||(1)}_{,j}
-12c_L^2\l(\frac{a'}{a}\r)^2\delta^{(1)}\phi^{(1)}\delta_{ij}
+6c_L^2\l(\frac{a'}{a}\r)^2\delta^{(1)}\chi^{||(1)}_{,\,ij}
\nn\\
&
-2c_L^2\l(\frac{a'}{a}\r)^2\delta^{(1)}\nabla^2\chi^{||(1)}\delta_{ij}
-6\phi^{(1)}_{,\,i}\phi^{(1)}_{,\,j}
+4\phi^{(1),\,k}\phi^{(1)}_{,\,k}\delta_{ij}
+4\phi^{(1)}\nabla^2\phi^{(1)}\delta_{ij}
-4\phi^{(1)}\phi^{(1)}_{,\,ij}
\nn\\
&
-2\phi^{(1)'}\phi^{(1)'}\delta_{ij}
-12\frac{a'}a\phi^{(1)'}\chi^{||(1)}_{,\,ij}
+4\frac{a'}a\phi^{(1)'}\nabla^2\chi^{||(1)}\delta_{ij}
-\phi^{(1)'}\chi^{||(1)'}_{,\,ij}
+\frac{1}{3}\phi^{(1)'}\nabla^2\chi^{||(1)'}\delta_{ij}
\nn\\
&
-6\phi^{(1)''}\chi^{||(1)}_{,\,ij}
+2\phi^{(1)''}\nabla^2\chi^{||(1)}\delta_{ij}
-\phi^{(1)}_{,\,j}\nabla^2\chi^{||(1)}_{,\,i}
-\phi^{(1)}_{,\,i}\nabla^2\chi^{||(1)}_{,\,j}
+\phi^{(1)}_{,\,k}\chi^{||(1),\,k}_{,\,ij}
\nn\\
&
-\frac{2}{3}\phi^{(1)}\nabla^2\chi^{||(1)}_{,\,ij}
+\frac{1}{3}\phi^{(1)}_{,\,k}\nabla^2\chi^{||(1),\,k}\delta_{ij}
+\frac{2}{3}\phi^{(1)}\nabla^2\nabla^2\chi^{||(1)}\delta_{ij}
+4\chi^{||(1)}_{,\,ij}\nabla^2\phi^{(1)}
\nn\\
&
-\frac{4}{3}\nabla^2\phi^{(1)}\nabla^2\chi^{||(1)}\delta_{ij}
-2\phi^{(1)}_{,\,ki}\chi^{||(1),\,k}_{,\,j}
-2\phi^{(1)}_{,\,kj}\chi^{||(1),\,k}_{,\,i}
+\frac{4}{3}\phi^{(1)}_{,\,ij}\nabla^2\chi^{||(1)}
\nn\\
&
+\frac{1}{4}\chi^{||(1),\,klm}\chi^{||(1)}_{,\,klm}\delta_{ij}
-\frac{11}{36}\nabla^2\chi^{||(1),\,k}\nabla^2\chi^{||(1)}_{,\,k}\delta_{ij}
-\frac{2}{9}\nabla^2\chi^{||(1)}\nabla^2\nabla^2\chi^{||(1)}\delta_{ij}
\nn\\
&
-\frac{1}{6}\nabla^2\chi^{||(1)}_{,\,i}\nabla^2\chi^{||(1)}_{,\,j}
-\frac{1}{3}\chi^{||(1),\,k}_{,\,j}\nabla^2\chi^{||(1)}_{,\,ik}
-\frac{1}{3}\chi^{||(1),\,k}_{,\,i}\nabla^2\chi^{||(1)}_{,\,jk}
+\frac{2}{3}\chi^{||(1)}_{,\,ij}\nabla^2\nabla^2\chi^{||(1)}
\nn\\
&
+\frac{2}{9}\nabla^2\chi^{||(1)}\nabla^2\chi^{||(1)}_{,\,ij}
-\frac{1}{2}\chi^{||(1),\,kl}_{,\,i}\chi^{||(1)}_{,\,klj}
+\frac{2}{3}\chi^{||(1)}_{,\,kij}\nabla^2\chi^{||(1),\,k}
-\frac{2a'}{a}\chi^{||(1),\,kl}\chi^{||(1)'}_{,\,kl}\delta_{ij}
\nn\\
&
+\frac{2a'}{3a}\nabla^2\chi^{||(1)}\nabla^2\chi^{||(1)'}\delta_{ij}
-\frac{3}{4}\chi^{||(1)',\,kl}\chi^{||(1)'}_{,\,kl}\delta_{ij}
-\chi^{||(1),\,kl}\chi^{||(1)''}_{,\,kl}\delta_{ij}
\nn\\
&
+\frac{1}{3}\nabla^2\chi^{||(1)}\nabla^2\chi^{||(1)''}\delta_{ij}
+\frac{13}{36}\nabla^2\chi^{||(1)'}\nabla^2\chi^{||(1)'}\delta_{ij}
+\chi^{||(1)',\,k}_{,\,i}\chi^{||(1)'}_{,\,kj}
-\frac{2}{3}\chi^{||(1)'}_{,\,ij}\nabla^2\chi^{||(1)'}
 .
\el
}
The trace part of 2nd-order evolution equation (\ref{Evo2ndSs1}) is
\bl\label{Evo2ndSsTr2}
&
2\phi^{(2)''}_S
+4\frac{{a'}}{a}\phi^{(2)'}_S
-\frac{2}{3}\nabla^2\phi^{(2)}_S
+\l[4\frac{a''}{a}
    +6(c_s^2-\frac{1}{3})(\frac{a'}{a})^2\r]\phi^{(2)}_S
-\frac{1}{9}\nabla^2\nabla^2\chi^{||(2) }_S
\nn\\
=&
3c_N^2\l(\frac{a'}{a}\r)^2\delta^{(2)}_{S}
+\frac{1}{3}S_{S\,k}^k
\,,
\el
where
\bl\label{SijTrace}
S_{S\,k}^k
=&
6(1+c_s^2)\l(\frac{a'}{a}\r)^2 v^{||(1)}_{,\,k}v^{||(1),\,k}
-36 c_L^2\l(\frac{a'}{a}\r)^2\delta^{(1)}\phi^{(1)}
+6\phi^{(1)}_{,\,k}\phi^{(1),\,k}
+8\phi^{(1)}\nabla^2\phi^{(1)}
\nn\\
&
-6\phi^{(1)'}\phi^{(1)'}
-4\phi^{(1)}_{,\,kl}\chi^{||(1),\,kl}
+\frac{4}{3}\phi^{(1)}\nabla^2\nabla^2\chi^{||(1)}
+\frac{4}{3}\nabla^2\phi^{(1)}\nabla^2\chi^{||(1)}
\nn\\
&
+\frac{2}{9}\nabla^2\chi^{||(1)}\nabla^2\nabla^2\chi^{||(1)}
-\frac{2}{3}\chi^{||(1),\,kl}\nabla^2\chi^{||(1)}_{,\,kl}
-\frac{5}{12}\nabla^2\chi^{||(1),\,k}\nabla^2\chi^{||(1)}_{,\,k}
\nn\\
&
+\frac{1}{4}\chi^{||(1),\,klm}\chi^{||(1)}_{,\,klm}
-\frac{6a'}{a}\chi^{||(1),\,kl}\chi^{||(1)'}_{,\,kl}
+\frac{2a'}{a}\nabla^2\chi^{||(1)}\nabla^2\chi^{||(1)'}
-\frac{5}{4}\chi^{||(1)',\,kl}\chi^{||(1)'}_{,\,kl}
\nn\\
&
-3\chi^{||(1),\,kl}\chi^{||(1)''}_{,\,kl}
+\nabla^2\chi^{||(1)}\nabla^2\chi^{||(1)''}
+\frac{5}{12}\nabla^2\chi^{||(1)'}\nabla^2\chi^{||(1)'}
 .
\el
The traceless part of 2nd-order evolution equation (\ref{Evo2ndSs1}) is
\bl\label{Evo2ndSsNoTr1}
&
D_{ij}\phi^{(2)}_{S}
+\frac{1}{2}D_{ij}\chi^{||(2)''}_S
+\frac{a'}{a} D_{ij}\chi^{||(2)'}_S
 +\l[(1-3c_s^2)(\frac{a'}{a})^2
    -2\frac{a''}a \r]D_{ij}\chi^{||(2)}_{S}
+\frac{1}{6}\nabla^2D_{ij}\chi^{||(2)}_S
\nn\\
&
+\frac{1}{2}\chi^{\perp(2)''}_{S\,ij}
+\frac{a'}{a} \chi^{\perp(2)'}_{S\,ij}
+\l[(1-3c_s^2)(\frac{a'}a )^2
    -2\frac{a''}a \r]\chi^{\perp(2)}_{S\,ij}
\nn\\
&
+\frac{1}{2}\chi^{\top(2)''}_{S\,ij}
+\frac{a'}{a} \chi^{\top(2)'}_{S\,ij}
+\l[(1-3c_s^2)(\frac{a'}a )^2
    -2\frac{a''}a \r]\chi^{\top(2)}_{S\,ij}
-\frac{1}{2}\nabla^2\chi^{\top(2)}_{S\,ij}
=
\bar S_{S\,ij}
,
\el
where
{\allowdisplaybreaks
\bl\label{SijTraceless}
\bar S_{S\,ij}
&\equiv
S_{S\,ij}-\frac{1}{3}S_{S\,k}^k\delta_{ij}
\nn\\
=&
6(1+c_s^2)\l(\frac{a'}{a}\r)^2 v^{||(1)}_{,i}v^{||(1)}_{,j}
-2(1+c_s^2)\l(\frac{a'}{a}\r)^2 v^{||(1)}_{,\,k}v^{||(1),\,k}\delta_{ij}
+6c_L^2\l(\frac{a'}{a}\r)^2\delta^{(1)}\chi^{||(1)}_{,\,ij}
\nn\\
&
-2c_L^2\l(\frac{a'}{a}\r)^2\delta^{(1)}\nabla^2\chi^{||(1)}\delta_{ij}
-6\phi^{(1)}_{,\,i}\phi^{(1)}_{,\,j}
+2\phi^{(1),\,k}\phi^{(1)}_{,\,k}\delta_{ij}
-4\phi^{(1)}\phi^{(1)}_{,\,ij}
+\frac{4}{3}\phi^{(1)}\nabla^2\phi^{(1)}\delta_{ij}
\nn\\
&
-12\frac{a'}a\phi^{(1)'}\chi^{||(1)}_{,\,ij}
+4\frac{a'}a\phi^{(1)'}\nabla^2\chi^{||(1)}\delta_{ij}
-\phi^{(1)'}\chi^{||(1)'}_{,\,ij}
+\frac{1}{3}\phi^{(1)'}\nabla^2\chi^{||(1)'}\delta_{ij}
-6\phi^{(1)''}\chi^{||(1)}_{,\,ij}
\nn\\
&
+2\phi^{(1)''}\nabla^2\chi^{||(1)}\delta_{ij}
-\phi^{(1)}_{,\,j}\nabla^2\chi^{||(1)}_{,\,i}
-\phi^{(1)}_{,\,i}\nabla^2\chi^{||(1)}_{,\,j}
+\phi^{(1)}_{,\,k}\chi^{||(1),\,k}_{,\,ij}
+\frac{1}{3}\phi^{(1)}_{,\,k}\nabla^2\chi^{||(1),\,k}\delta_{ij}
\nn\\
&
-\frac{2}{3}\phi^{(1)}\nabla^2\chi^{||(1)}_{,\,ij}
+\frac{2}{9}\phi^{(1)}\nabla^2\nabla^2\chi^{||(1)}\delta_{ij}
+4\chi^{||(1)}_{,\,ij}\nabla^2\phi^{(1)}
-\frac{4}{3}\nabla^2\phi^{(1)}\nabla^2\chi^{||(1)}\delta_{ij}
\nn\\
&
-2\phi^{(1)}_{,\,ki}\chi^{||(1),\,k}_{,\,j}
-2\phi^{(1)}_{,\,kj}\chi^{||(1),\,k}_{,\,i}
+\frac{4}{3}\phi^{(1)}_{,\,kl}\chi^{||(1),\,kl}\delta_{ij}
+\frac{4}{3}\phi^{(1)}_{,\,ij}\nabla^2\chi^{||(1)}
-\frac{4}{9}\nabla^2\phi^{(1)}\nabla^2\chi^{||(1)}\delta_{ij}
\nn\\
&
+\frac{2}{3}\chi^{||(1)}_{,\,kij}\nabla^2\chi^{||(1),\,k}
-\frac{1}{6}\nabla^2\chi^{||(1)}_{,\,i}\nabla^2\chi^{||(1)}_{,\,j}
-\frac{1}{6}\nabla^2\chi^{||(1),\,k}\nabla^2\chi^{||(1)}_{,\,k}\delta_{ij}
-\frac{1}{3}\chi^{||(1),\,k}_{,\,j}\nabla^2\chi^{||(1)}_{,\,ik}
\nn\\
&
-\frac{1}{3}\chi^{||(1),\,k}_{,\,i}\nabla^2\chi^{||(1)}_{,\,jk}
+\frac{2}{9}\chi^{||(1),\,kl}\nabla^2\chi^{||(1)}_{,\,kl}\delta_{ij}
+\frac{2}{3}\chi^{||(1)}_{,\,ij}\nabla^2\nabla^2\chi^{||(1)}
+\frac{2}{9}\nabla^2\chi^{||(1)}\nabla^2\chi^{||(1)}_{,\,ij}
\nn\\
&
-\frac{8}{27}\nabla^2\chi^{||(1)}\nabla^2\nabla^2\chi^{||(1)}\delta_{ij}
-\frac{1}{2}\chi^{||(1),\,kl}_{,\,i}\chi^{||(1)}_{,\,klj}
+\frac{1}{6}\chi^{||(1),\,klm}\chi^{||(1)}_{,\,klm}\delta_{ij}
+\chi^{||(1)',\,k}_{,\,i}\chi^{||(1)'}_{,\,kj}
\nn\\
&
-\frac{1}{3}\chi^{||(1)',\,kl}\chi^{||(1)'}_{,\,kl}\delta_{ij}
-\frac{2}{3}\chi^{||(1)'}_{,\,ij}\nabla^2\chi^{||(1)'}
+\frac{2}{9}\nabla^2\chi^{||(1)'}\nabla^2\chi^{||(1)'}\delta_{ij}
\ .
\el
}
The scalar part of  (\ref{Evo2ndSsNoTr1}) is
\bl\label{Evo2ndSsChi1}
&
\phi^{(2)}_{S}
+\frac{1}{2}\chi^{||(2)''}_S
+\frac{a'}{a}\chi^{||(2)'}_S
\nn\\
&
 +\l[(1-3c_s^2)(\frac{a'}{a})^2
    -2\frac{a''}a \r]\chi^{||(2)}_{S}
+\frac{1}{6}\nabla^2\chi^{||(2)}_S
=
\frac{3}{2}\nabla^{-2}\nabla^{-2}\bar S_{S\,kl}^{\, ,\,kl}
,
\el
where
{\allowdisplaybreaks
\bl\label{SijPijbar}
\bar S_{S\,ij}^{,\,ij}
=&
\frac{2}{3}\nabla^2\nabla^2\Big[
-\frac{9}{4}\phi^{(1)}\phi^{(1)}
-\frac{2}{3}\phi^{(1)}\nabla^2\chi^{||(1)}
-\frac{1}{18}\nabla^2\chi^{||(1)}\nabla^2\chi^{||(1)}
-\frac{1}{16}\chi^{||(1)}_{,\,kl}\chi^{||(1),\,kl}
\Big]
\nn\\
&
+\nabla^2\Big[
(1+c_s^2)\l(\frac{a'}{a}\r)^2 v^{||(1)}_{,\,k}v^{||(1),\,k}
+4c_L^2\l(\frac{a'}{a}\r)^2\delta^{(1)}\nabla^2\chi^{||(1)}
-\frac{5}{3}\phi^{(1)}\nabla^2\phi^{(1)}
\nn\\
&
-8\frac{a'}a\phi^{(1)'}\nabla^2\chi^{||(1)}
-\frac{2}{3}\phi^{(1)'}\nabla^2\chi^{||(1)'}
-4\phi^{(1)''}\nabla^2\chi^{||(1)}
-\frac{4}{9}\phi^{(1),k}\nabla^2\chi^{||(1)}_{,\,k}
+\frac{4}{3}\phi^{(1)}_{,\,kl}\chi^{||(1),kl}
\nn\\
&
+\frac{1}{54} \nabla^2\chi^{||(1),k}\nabla^2\chi^{||(1)}_{,\,k}
+\frac{11}{36}\chi^{||(1)}_{,\,kl}\nabla^2\chi^{||(1),kl}
+\frac{1}{6}\chi^{||(1)',kl}\chi^{||(1)'}_{,\,kl}
-\frac{1}{9}\nabla^2\chi^{||(1)'}\nabla^2\chi^{||(1)'}
\Big]
\nn\\
&
+6(1+c_s^2)\l(\frac{a'}{a}\r)^2\nabla^2v^{||(1)}\nabla^2v^{||(1)}
+6(1+c_s^2)\l(\frac{a'}{a}\r)^2v^{||(1)}_{,\,k}\nabla^2v^{||(1),\,k}
\nn\\
&
-6c_L^2\l(\frac{a'}{a}\r)^2\nabla^2\delta^{(1)}\nabla^2\chi^{||(1)}
+6c_L^2\l(\frac{a'}{a}\r)^2\delta^{(1),\,kl}\chi^{||(1)}_{,\,kl}
+2\phi^{(1),\,k}\nabla^2\phi^{(1)}_{,\,k}
\nn\\
&
+2\phi^{(1)}\nabla^2\nabla^2\phi^{(1)}
-12\frac{a'}a\phi^{(1)',\,kl}\chi^{||(1)}_{,\,kl}
+12\frac{a'}a\nabla^2\phi^{(1)'}\nabla^2\chi^{||(1)}
-\phi^{(1)',\,kl}\chi^{||(1)'}_{,\,kl}
\nn\\
&
+\nabla^2\phi^{(1)'}\nabla^2\chi^{||(1)'}
-6\phi^{(1)'',\,kl}\chi^{||(1)}_{,\,kl}
+6\nabla^2\phi^{(1)''}\nabla^2\chi^{||(1)}
+\frac{4}{3}\nabla^2\nabla^2\chi^{||(1)}\nabla^2\phi^{(1)}
\nn\\
&
+\frac{11}{3}\nabla^2\chi^{||(1)}_{,\,k}\nabla^2\phi^{(1),k}
+\frac{2}{3}\phi^{(1),\,k}\nabla^2\nabla^2\chi^{||(1)}_{,\,k}
-3\phi^{(1)}_{,\,klm}\chi^{||(1),klm}
\nn\\
&
+\frac{7}{9}\nabla^2\chi^{||(1)}_{,\,k}\nabla^2\nabla^2\chi^{||(1),k}
+\frac{5}{18}\nabla^2\nabla^2\chi^{||(1)}\nabla^2\nabla^2\chi^{||(1)}
-\frac{1}{2}\chi^{||(1)}_{,\,klm}\nabla^2\chi^{||(1),\,klm}
\nn\\
&
+\frac{1}{3}\chi^{||(1)'}_{,\,kl}\nabla^2\chi^{||(1)',\,kl}
+\frac{1}{3}\nabla^2\chi^{||(1)'}_{,\,k}\nabla^2\chi^{||(1)',\,k}
\ .
\el
}
The vector part of the  equation (\ref{Evo2ndSsNoTr1}) is
\bl\label{Evo2ndSsVec2}
&
\frac{1}{2}\chi^{\perp(2)''}_{S\,ij}
+\frac{a'}{a} \chi^{\perp(2)'}_{S\,ij}
+\l[(1-3c_s^2)(\frac{a'}a )^2
    -2\frac{a''}a \r]\chi^{\perp(2)}_{S\,ij}
\nn\\
& =
\big( \nabla^{-2}\bar S_{S\,kj,\,i}^{,\,k}
+\nabla^{-2}\bar S_{S\,ki,j}^{,\,k}
-2\nabla^{-2}\nabla^{-2}\bar S_{S\,kl,\,ij}^{\, ,\,kl} \big) ,
\el
where the rhs of the above is
{\allowdisplaybreaks
\bl\label{SourceCurl1}
&
\big(\nabla^{-2}\bar S_{S\,kj,\,i}^{,\,k}
+\nabla^{-2}\bar S_{S\,ki,j}^{,\,k}
-2\nabla^{-2}\nabla^{-2}\bar S_{S\,kl,\,ij}^{\, ,\,kl} \big)
\nn\\
=&
\partial_i\Big[\frac{1}{3}\phi^{(1)}\nabla^2\chi^{||(1)}_{,\,j}
+2\phi^{(1),\,k}\chi^{||(1)}_{,\,kj}
+\frac{1}{3}\chi^{||(1)}_{,\,kj}\nabla^2\chi^{||(1),\,k}
\Big]
\nn\\
&
-\partial_i\partial_j\nabla^{-2}\Big[
6c_L^2\l(\frac{a'}{a}\r)^2\delta^{(1)}\nabla^2\chi^{||(1)}
-12\frac{a'}a\phi^{(1)'}\nabla^2\chi^{||(1)}
-\phi^{(1)'}\nabla^2\chi^{||(1)'}
-6\phi^{(1)''}\nabla^2\chi^{||(1)}
\nn\\
&
+\frac{1}{3}\phi^{(1)}\nabla^2\nabla^2\chi^{||(1)}
+2\phi^{(1),\,kl}\chi^{||(1)}_{,\,kl}
+\frac{1}{12}\nabla^2\chi^{||(1)}_{,\,k}\nabla^2\chi^{||(1),\,k}
+\frac{1}{3}\chi^{||(1)}_{,\,kl}\nabla^2\chi^{||(1),\,kl}
\Big]
\nn\\
&
+\partial_i\nabla^{-2}\Big[
6(1+c_s^2)\l(\frac{a'}{a}\r)^2v^{||(1)}_{,j}\nabla^2v^{||(1)}
+6c_L^2\l(\frac{a'}{a}\r)^2\delta^{(1)}\nabla^2\chi^{||(1)}_{,\,j}
+6c_L^2\l(\frac{a'}{a}\r)^2\delta^{(1),\,k}\chi^{||(1)}_{,\,kj}
\nn\\
&
+2\phi^{(1)}\nabla^2\phi^{(1)}_{,\,j}
-12\frac{a'}a\phi^{(1)',\,k}\chi^{||(1)}_{,\,kj}
-12\frac{a'}a\phi^{(1)'}\nabla^2\chi^{||(1)}_{,\,j}
-\phi^{(1)'}\nabla^2\chi^{||(1)'}_{,\,j}
-\phi^{(1)',\,k}\chi^{||(1)'}_{,\,kj}
\nn\\
&
-6\phi^{(1)''}\nabla^2\chi^{||(1)}_{,\,j}
-6\phi^{(1)'',\,k}\chi^{||(1)}_{,\,kj}
+\frac{4}{3}\nabla^2\chi^{||(1)}_{,\,j}\nabla^2\phi^{(1)}
-\frac{5}{3}\phi^{(1),\,k}\nabla^2\chi^{||(1)}_{,\,kj}
-3\phi^{(1)}_{,\,kl}\chi^{||(1),\,kl}_{,\,j}
\nn\\
&
+\frac{5}{18}\nabla^2\chi^{||(1)}_{,\,j}\nabla^2\nabla^2\chi^{||(1)}
-\frac{1}{2}\chi^{||(1)}_{,\,klj}\nabla^2\chi^{||(1),\,kl}
+\frac{1}{3}\chi^{||(1)'}_{,\,kj}\nabla^2\chi^{||(1)',\,k}
\Big]
\nn\\
&
-\partial_i\partial_j\nabla^{-2}\nabla^{-2}\Big[
6(1+c_s^2)\l(\frac{a'}{a}\r)^2\nabla^2v^{||(1)}\nabla^2v^{||(1)}
+6(1+c_s^2)\l(\frac{a'}{a}\r)^2v^{||(1)}_{,\,k}\nabla^2v^{||(1),\,k}
\nn\\
&
-6c_L^2\l(\frac{a'}{a}\r)^2\nabla^2\delta^{(1)}\nabla^2\chi^{||(1)}
+6c_L^2\l(\frac{a'}{a}\r)^2\delta^{(1),\,kl}\chi^{||(1)}_{,\,kl}
+2\phi^{(1),\,k}\nabla^2\phi^{(1)}_{,\,k}
+2\phi^{(1)}\nabla^2\nabla^2\phi^{(1)}
\nn\\
&
-12\frac{a'}a\phi^{(1)',\,kl}\chi^{||(1)}_{,\,kl}
+12\frac{a'}a\nabla^2\phi^{(1)'}\nabla^2\chi^{||(1)}
-\phi^{(1)',\,kl}\chi^{||(1)'}_{,\,kl}
+\nabla^2\phi^{(1)'}\nabla^2\chi^{||(1)'}
\nn\\
&
-6\phi^{(1)'',\,kl}\chi^{||(1)}_{,\,kl}
+6\nabla^2\phi^{(1)''}\nabla^2\chi^{||(1)}
+\frac{4}{3}\nabla^2\nabla^2\chi^{||(1)}\nabla^2\phi^{(1)}
+\frac{11}{3}\nabla^2\chi^{||(1)}_{,\,k}\nabla^2\phi^{(1),\,k}
\nn\\
&
+\frac{2}{3}\phi^{(1),\,k}\nabla^2\nabla^2\chi^{||(1)}_{,\,k}
-3\phi^{(1)}_{,\,klm}\chi^{||(1),\,klm}
+\frac{7}{9}\nabla^2\chi^{||(1)}_{,\,k}\nabla^2\nabla^2\chi^{||(1),\,k}
-\frac{1}{2}\chi^{||(1)}_{,\,klm}\nabla^2\chi^{||(1),\,klm}
\nn\\
&
+\frac{5}{18}\nabla^2\nabla^2\chi^{||(1)}\nabla^2\nabla^2\chi^{||(1)}
+\frac{1}{3}\chi^{||(1)'}_{,\,kl}\nabla^2\chi^{||(1)',\,kl}
+\frac{1}{3}\nabla^2\chi^{||(1)'}_{,\,k}\nabla^2\chi^{||(1)',\,k}
\Big]
\nn\\
&
+(i\leftrightarrow j)
\ .
\el
}
The tensor part (GW) of  the equation (\ref{Evo2ndSsNoTr1}) is
\bl \label{Evo2ndSsTen1}
&
\frac{1}{2}\chi^{\top(2)''}_{S\,ij}
+\frac{a'}{a} \chi^{\top(2)'}_{S\,ij}
+\l[(1-3c_s^2)(\frac{a'}a )^2
    -2\frac{a''}a \r]\chi^{\top(2)}_{S\,ij}
-\frac{1}{2}\nabla^2\chi^{\top(2)}_{S\,ij}
\nn\\
=&
\bar S_{S\,ij}
-\frac{3}{2}D_{ij}\nabla^{-2}\nabla^{-2}\bar S_{S\,kl}^{\, ,\,kl}
-\nabla^{-2}\bar S_{S\,kj,\,i}^{,\,k}
-\nabla^{-2}\bar S_{S\,ki,j}^{,\,k}
+2\nabla^{-2}\nabla^{-2}\bar S_{S\,kl,\,ij}^{\, ,\,kl}
\ ,
\el
where the rhs is the effective source for the 2nd-order tensor
{
\allowdisplaybreaks
\bl\label{2ndTensorSource}
&
\big(\bar S_{S\,ij}
-\frac{3}{2}D_{ij}\nabla^{-2}\nabla^{-2}\bar S_{S\,kl}^{\, ,\,kl}
-\nabla^{-2}\bar S_{S\,kj,\,i}^{,\,k}
-\nabla^{-2}\bar S_{S\,ki,j}^{,\,k}
+2\nabla^{-2}\nabla^{-2}\bar S_{S\,kl,\,ij}^{\, ,\,kl} \big)
\nn\\
=&
D_{ij}\Big[
-\frac{3}{4}\phi^{(1)}\phi^{(1)}
-\frac{1}{3}\phi^{(1)}\nabla^2\chi^{||(1)}
-2\phi^{(1)}_{,\,k}\chi^{||(1),\,k}
+\frac{1}{6}\nabla^2\chi^{||(1)}\nabla^2\chi^{||(1)}
+\frac{1}{16}\chi^{||(1)}_{,\,kl}\chi^{||(1),\,kl}
\nn\\
&
-\frac{1}{3}\chi^{||(1)}_{,\,k}\nabla^2\chi^{||(1),\,k}
\Big]
+6(1+c_s^2)\l(\frac{a'}{a}\r)^2 v^{||(1)}_{,i}v^{||(1)}_{,j}
-2(1+c_s^2)\l(\frac{a'}{a}\r)^2 v^{||(1)}_{,\,k}v^{||(1),\,k}\delta_{ij}
\nn\\
&
+6c_L^2\l(\frac{a'}{a}\r)^2\delta^{(1)}\chi^{||(1)}_{,\,ij}
-2c_L^2\l(\frac{a'}{a}\r)^2\delta^{(1)}\nabla^2\chi^{||(1)}\delta_{ij}
+2\phi^{(1)}\phi^{(1)}_{,\,ij}
-\frac{2}{3}\phi^{(1)}\nabla^2\phi^{(1)}\delta_{ij}
\nn\\
&
-12\frac{a'}a\phi^{(1)'}\chi^{||(1)}_{,\,ij}
+4\frac{a'}a\phi^{(1)'}\nabla^2\chi^{||(1)}\delta_{ij}
-\phi^{(1)'}\chi^{||(1)'}_{,\,ij}
+\frac{1}{3}\phi^{(1)'}\nabla^2\chi^{||(1)'}\delta_{ij}
-6\phi^{(1)''}\chi^{||(1)}_{,\,ij}
\nn\\
&
+2\phi^{(1)''}\nabla^2\chi^{||(1)}\delta_{ij}
+\frac{7}{3}\phi^{(1)}_{,\,ij}\nabla^2\chi^{||(1)}
-\frac{7}{9}\nabla^2\phi^{(1)}\nabla^2\chi^{||(1)}\delta_{ij}
+\frac{1}{3}\phi^{(1)}\nabla^2\chi^{||(1)}_{,\,ij}
\nn\\
&
-\frac{1}{9}\phi^{(1)}\nabla^2\nabla^2\chi^{||(1)}\delta_{ij}
+3\phi^{(1)}_{,\,k}\chi^{||(1),\,k}_{,\,ij}
-\phi^{(1)}_{,\,k}\nabla^2\chi^{||(1),\,k}\delta_{ij}
+2\chi^{||(1),\,k}\phi^{(1)}_{,\,kij}
\nn\\
&
-\frac{2}{3}\chi^{||(1),\,k}\nabla^2\phi^{(1)}_{,\,k}\delta_{ij}
+4\chi^{||(1)}_{,\,ij}\nabla^2\phi^{(1)}
-\frac{4}{3}\nabla^2\phi^{(1)}\nabla^2\chi^{||(1)}\delta_{ij}
+\frac{2}{3}\chi^{||(1)}_{,\,ij}\nabla^2\nabla^2\chi^{||(1)}
\nn\\
&
-\frac{2}{9}\nabla^2\chi^{||(1)}\nabla^2\nabla^2\chi^{||(1)}\delta_{ij}
-\frac{7}{18}\nabla^2\chi^{||(1)}_{,\,i}\nabla^2\chi^{||(1)}_{,\,j}
+\chi^{||(1)}_{,\,kij}\nabla^2\chi^{||(1),\,k}
\nn\\
&
-\frac{11}{54}\nabla^2\chi^{||(1),\,k}\nabla^2\chi^{||(1)}_{,\,k}\delta_{ij}
-\frac{1}{9}\chi^{||(1),\,k}\nabla^2\nabla^2\chi^{||(1)}_{,\,k}\delta_{ij}
+\frac{1}{3}\chi^{||(1)}_{,\,k}\nabla^2\chi^{||(1),\,k}_{,\,ij}
\nn\\
&
-\frac{1}{2}\chi^{||(1),\,kl}_{,\,i}\chi^{||(1)}_{,\,klj}
+\frac{1}{6}\chi^{||(1),\,klm}\chi^{||(1)}_{,\,klm}\delta_{ij}
+\chi^{||(1)',\,k}_{,\,i}\chi^{||(1)'}_{,\,kj}
-\frac{1}{3}\chi^{||(1)',\,kl}\chi^{||(1)'}_{,\,kl}\delta_{ij}
\nn\\
&
-\frac{2}{3}\chi^{||(1)'}_{,\,ij}\nabla^2\chi^{||(1)'}
+\frac{2}{9}\nabla^2\chi^{||(1)'}\nabla^2\chi^{||(1)'}\delta_{ij}
    \nn\\
    &
-\frac{3}{2}D_{ij}\nabla^{-2}\Big[
(1+c_s^2)\l(\frac{a'}{a}\r)^2 v^{||(1)}_{,\,k}v^{||(1),\,k}
+4c_L^2\l(\frac{a'}{a}\r)^2\delta^{(1)}\nabla^2\chi^{||(1)}
-\frac{5}{3}\phi^{(1)}\nabla^2\phi^{(1)}
\nn\\
&
-8\frac{a'}a\phi^{(1)'}\nabla^2\chi^{||(1)}
-\frac{2}{3}\phi^{(1)'}\nabla^2\chi^{||(1)'}
-4\phi^{(1)''}\nabla^2\chi^{||(1)}
-\frac{4}{9}\phi^{(1),k}\nabla^2\chi^{||(1)}_{,\,k}
+\frac{4}{3}\phi^{(1)}_{,\,kl}\chi^{||(1),kl}
\nn\\
&
+\frac{1}{54} \nabla^2\chi^{||(1),k}\nabla^2\chi^{||(1)}_{,\,k}
+\frac{11}{36}\chi^{||(1)}_{,\,kl}\nabla^2\chi^{||(1),kl}
+\frac{1}{6}\chi^{||(1)',kl}\chi^{||(1)'}_{,\,kl}
-\frac{1}{9}\nabla^2\chi^{||(1)'}\nabla^2\chi^{||(1)'}
\Big]
\nn\\
&
-\frac{3}{2}D_{ij}\nabla^{-2}\nabla^{-2}
\Big[
6(1+c_s^2)\l(\frac{a'}{a}\r)^2\nabla^2v^{||(1)}\nabla^2v^{||(1)}
+6(1+c_s^2)\l(\frac{a'}{a}\r)^2v^{||(1)}_{,\,k}\nabla^2v^{||(1),\,k}
\nn\\
&
-6c_L^2\l(\frac{a'}{a}\r)^2\nabla^2\delta^{(1)}\nabla^2\chi^{||(1)}
+6c_L^2\l(\frac{a'}{a}\r)^2\delta^{(1),\,kl}\chi^{||(1)}_{,\,kl}
+2\phi^{(1),\,k}\nabla^2\phi^{(1)}_{,\,k}
+2\phi^{(1)}\nabla^2\nabla^2\phi^{(1)}
\nn\\
&
-12\frac{a'}a\phi^{(1)',\,kl}\chi^{||(1)}_{,\,kl}
+12\frac{a'}a\nabla^2\phi^{(1)'}\nabla^2\chi^{||(1)}
-\phi^{(1)',\,kl}\chi^{||(1)'}_{,\,kl}
+\nabla^2\phi^{(1)'}\nabla^2\chi^{||(1)'}
\nn\\
&
-6\phi^{(1)'',\,kl}\chi^{||(1)}_{,\,kl}
+6\nabla^2\phi^{(1)''}\nabla^2\chi^{||(1)}
+\frac{4}{3}\nabla^2\nabla^2\chi^{||(1)}\nabla^2\phi^{(1)}
+\frac{11}{3}\nabla^2\chi^{||(1)}_{,\,k}\nabla^2\phi^{(1),k}
\nn\\
&
+\frac{2}{3}\phi^{(1),k}\nabla^2\nabla^2\chi^{||(1)}_{,\,k}
-3\phi^{(1)}_{,\,klm}\chi^{||(1),klm}
+\frac{7}{9}\nabla^2\chi^{||(1)}_{,\,k}\nabla^2\nabla^2\chi^{||(1),k}
+\frac{5}{18}\nabla^2\nabla^2\chi^{||(1)}\nabla^2\nabla^2\chi^{||(1)}
\nn\\
&
-\frac{1}{2}\chi^{||(1)}_{,\,klm}\nabla^2\chi^{||(1),\,klm}
+\frac{1}{3}\chi^{||(1)'}_{,\,kl}\nabla^2\chi^{||(1)',\,kl}
+\frac{1}{3}\nabla^2\chi^{||(1)'}_{,\,k}\nabla^2\chi^{||(1)',\,k}
\Big]
\nn\\
    &
-\partial_i\Big[
\frac{1}{3}\phi^{(1)}\nabla^2\chi^{||(1)}_{,\,j}
+2\phi^{(1),\,k}\chi^{||(1)}_{,\,kj}
+\frac{1}{3}\chi^{||(1)}_{,\,kj}\nabla^2\chi^{||(1),\,k}
\Big]
\nn\\
&
-\partial_j\Big[
\frac{1}{3}\phi^{(1)}\nabla^2\chi^{||(1)}_{,\,i}
+2\phi^{(1),\,k}\chi^{||(1)}_{,\,ki}
+\frac{1}{3}\chi^{||(1)}_{,\,ki}\nabla^2\chi^{||(1),\,k}
\Big]
\nn\\
&
+\partial_i\partial_j\nabla^{-2}\Big[
12c_L^2\l(\frac{a'}{a}\r)^2\delta^{(1)}\nabla^2\chi^{||(1)}
-24\frac{a'}a\phi^{(1)'}\nabla^2\chi^{||(1)}
-2\phi^{(1)'}\nabla^2\chi^{||(1)'}
-12\phi^{(1)''}\nabla^2\chi^{||(1)}
\nn\\
&
+\frac{2}{3}\phi^{(1)}\nabla^2\nabla^2\chi^{||(1)}
+4\phi^{(1),\,kl}\chi^{||(1)}_{,\,kl}
+\frac{1}{6}\nabla^2\chi^{||(1)}_{,\,k}\nabla^2\chi^{||(1),\,k}
+\frac{2}{3}\chi^{||(1)}_{,\,kl}\nabla^2\chi^{||(1),\,kl}
\Big]
\nn\\
&
-\partial_i\nabla^{-2}\Big[
6(1+c_s^2)\l(\frac{a'}{a}\r)^2v^{||(1)}_{,j}\nabla^2v^{||(1)}
+6c_L^2\l(\frac{a'}{a}\r)^2\delta^{(1)}\nabla^2\chi^{||(1)}_{,\,j}
+6c_L^2\l(\frac{a'}{a}\r)^2\delta^{(1),\,k}\chi^{||(1)}_{,\,kj}
\nn\\
&
+2\phi^{(1)}\nabla^2\phi^{(1)}_{,\,j}
-12\frac{a'}a\phi^{(1)',\,k}\chi^{||(1)}_{,\,kj}
-12\frac{a'}a\phi^{(1)'}\nabla^2\chi^{||(1)}_{,\,j}
-\phi^{(1)'}\nabla^2\chi^{||(1)'}_{,\,j}
-\phi^{(1)',\,k}\chi^{||(1)'}_{,\,kj}
\nn\\
&
-6\phi^{(1)''}\nabla^2\chi^{||(1)}_{,\,j}
-6\phi^{(1)'',\,k}\chi^{||(1)}_{,\,kj}
+\frac{4}{3}\nabla^2\chi^{||(1)}_{,\,j}\nabla^2\phi^{(1)}
-\frac{5}{3}\phi^{(1),\,k}\nabla^2\chi^{||(1)}_{,\,kj}
-3\phi^{(1)}_{,\,kl}\chi^{||(1),\,kl}_{,\,j}
\nn\\
&
+\frac{5}{18}\nabla^2\chi^{||(1)}_{,\,j}\nabla^2\nabla^2\chi^{||(1)}
-\frac{1}{2}\chi^{||(1)}_{,\,klj}\nabla^2\chi^{||(1),\,kl}
+\frac{1}{3}\chi^{||(1)'}_{,\,kj}\nabla^2\chi^{||(1)',\,k}
\Big]
\nn\\
&
-\partial_j\nabla^{-2}\Big[
6(1+c_s^2)\l(\frac{a'}{a}\r)^2v^{||(1)}_{,i}\nabla^2v^{||(1)}
+6c_L^2\l(\frac{a'}{a}\r)^2\delta^{(1)}\nabla^2\chi^{||(1)}_{,\,i}
+6c_L^2\l(\frac{a'}{a}\r)^2\delta^{(1),\,k}\chi^{||(1)}_{,\,ki}
\nn\\
&
+2\phi^{(1)}\nabla^2\phi^{(1)}_{,\,i}
-12\frac{a'}a\phi^{(1)',\,k}\chi^{||(1)}_{,\,ki}
-12\frac{a'}a\phi^{(1)'}\nabla^2\chi^{||(1)}_{,\,i}
-\phi^{(1)'}\nabla^2\chi^{||(1)'}_{,\,i}
-\phi^{(1)',\,k}\chi^{||(1)'}_{,\,ki}
\nn\\
&
-6\phi^{(1)''}\nabla^2\chi^{||(1)}_{,\,i}
-6\phi^{(1)'',\,k}\chi^{||(1)}_{,\,ki}
+\frac{4}{3}\nabla^2\chi^{||(1)}_{,\,i}\nabla^2\phi^{(1)}
-\frac{5}{3}\phi^{(1),\,k}\nabla^2\chi^{||(1)}_{,\,ki}
-3\phi^{(1)}_{,\,kl}\chi^{||(1),\,kl}_{,\,i}
\nn\\
&
+\frac{5}{18}\nabla^2\chi^{||(1)}_{,\,i}\nabla^2\nabla^2\chi^{||(1)}
-\frac{1}{2}\chi^{||(1)}_{,\,kli}\nabla^2\chi^{||(1),\,kl}
+\frac{1}{3}\chi^{||(1)'}_{,\,ki}\nabla^2\chi^{||(1)',\,k}
\Big]
\nn\\
&
+\partial_i\partial_j\nabla^{-2}\nabla^{-2}\Big[
12(1+c_s^2)\l(\frac{a'}{a}\r)^2\nabla^2v^{||(1)}\nabla^2v^{||(1)}
+12(1+c_s^2)\l(\frac{a'}{a}\r)^2v^{||(1)}_{,\,k}\nabla^2v^{||(1),\,k}
\nn\\
&
-12c_L^2\l(\frac{a'}{a}\r)^2\nabla^2\delta^{(1)}\nabla^2\chi^{||(1)}
+12c_L^2\l(\frac{a'}{a}\r)^2\delta^{(1),\,kl}\chi^{||(1)}_{,\,kl}
+4\phi^{(1),\,k}\nabla^2\phi^{(1)}_{,\,k}
+4\phi^{(1)}\nabla^2\nabla^2\phi^{(1)}
\nn\\
&
-24\frac{a'}a\phi^{(1)',\,kl}\chi^{||(1)}_{,\,kl}
+24\frac{a'}a\nabla^2\phi^{(1)'}\nabla^2\chi^{||(1)}
-2\phi^{(1)',\,kl}\chi^{||(1)'}_{,\,kl}
+2\nabla^2\phi^{(1)'}\nabla^2\chi^{||(1)'}
\nn\\
&
-12\phi^{(1)'',\,kl}\chi^{||(1)}_{,\,kl}
+12\nabla^2\phi^{(1)''}\nabla^2\chi^{||(1)}
+\frac{8}{3}\nabla^2\nabla^2\chi^{||(1)}\nabla^2\phi^{(1)}
+\frac{22}{3}\nabla^2\chi^{||(1)}_{,\,k}\nabla^2\phi^{(1),\,k}
\nn\\
&
+\frac{4}{3}\phi^{(1),\,k}\nabla^2\nabla^2\chi^{||(1)}_{,\,k}
-6\phi^{(1)}_{,\,klm}\chi^{||(1),\,klm}
+\frac{14}{9}\nabla^2\chi^{||(1)}_{,\,k}\nabla^2\nabla^2\chi^{||(1),\,k}
-\chi^{||(1)}_{,\,klm}\nabla^2\chi^{||(1),\,klm}
\nn\\
&
+\frac{5}{9}\nabla^2\nabla^2\chi^{||(1)}\nabla^2\nabla^2\chi^{||(1)}
+\frac{2}{3}\chi^{||(1)'}_{,\,kl}\nabla^2\chi^{||(1)',\,kl}
+\frac{2}{3}\nabla^2\chi^{||(1)'}_{,\,k}\nabla^2\chi^{||(1)',\,k}
\Big].
\el
}

The  equations  of covariant conservation of
energy and momentum  are  given by (\ref{EnConsv2}) and (\ref{MoConsv2}).
Substituting $\Gamma^{\alpha}_{\mu\nu}$
in (\ref{Christopher000})--(\ref{Christopherijk})
and $U^{\mu}$ in (\ref{U0element}) and (\ref{Uielement})
into (\ref{EnConsv2}) and (\ref{MoConsv2}) and only keeping
scalar-scalar couplings,
we arrive at  the 2nd-order energy conservation
\bl\label{enCons2ndGeneral}
&
\delta^{(2)'}_S
+2a''(a')^{-1}\delta^{(2)}_S
+(-1+3c_N^2)a'a^{-1}\delta^{(2)}_S
\nn\\
&
+(1+c_s^2)v^{(2)k}_{S\,,k}
-3(1+c_s^2)\phi^{(2)'}_S
+4(1+c_s^2)a''(a')^{-1} v^{||(1),\,k}v^{||(1)}_{,\,k}
\nn\\
&
+4(1+c_s^2)v^{||(1)',\,k}v^{||(1)}_{,\,k}
+2(1+c_L^2)\delta^{(1)}_{,\,k}v^{||(1),\,k}
+2(1+c_L^2)\delta^{(1)}\nabla^2v^{||(1)}
-6(1+c_L^2)\delta^{(1)}\phi^{(1)'}
\nn\\
&
-12(1+c_s^2) \phi^{(1)'}\phi^{(1)}
-(1+c_s^2)D_{kl}\chi^{||(1)'}D^{kl}\chi^{||(1)}
-6(1+c_s^2)\phi^{(1)}_{,\,k}v^{||(1),\,k}
=0
,
\el
and the 2nd-order momentum conservation equation
\bl\label{MoCons2ndGeneral}
&
c_N^2 \delta^{(2),\,i}_S
+2(1+c_s^2)a''(a')^{-1} v^{(2)i}_S
+(1+c_s^2) v^{(2)'i}_S
\nn\\
&
+4(1+c_L^2)a''(a')^{-1} \delta^{(1)}v^{||(1),\,i}
+2(1+c_L^2) \delta^{(1)'}v^{||(1),\,i}
+2(1+c_L^2) \delta^{(1)}v^{||(1)',\,i}
\nn\\
&
+2(1+c_s^2)v^{||(1),\,i}_{,k} v^{||(1),\,k}
+2(1+c_s^2) v^{||(1),\,i} \nabla^2v^{||(1)}
+4c_L^2\delta^{(1),\,i} \phi^{(1)}
\nn\\
&
-10(1+c_s^2)v^{||(1),\,i}\phi^{(1)'}
-2c_L^2 \delta^{(1)}_{,\,k}D^{ik}\chi^{||(1)}
+2(1+c_s^2)v^{||(1),\,k}D^{i}_{k}\chi^{||(1)'}=0
\ ,
\el
which are  for a general RW spacetime.

\section{Gauge transformations  from synchronous to synchronous}

The formulae  for the gauge transform
between two general coordinates  for a flat RW spacetime,
as well as between two synchronous coordinates,
have been analyzed in Appendix C in Ref.~\cite{WangZhang2017}.
But the transformation of the stress tensor of
a relativistic fluid has not been given.
 Here we shall add some new results on $\rho$ and $U^{\mu}$.

A general   coordinate transformation is given by
\cite{Weinberg1972, Matarrese98,WangZhang2017,Bruni97}
\be \label{xmutransf}
x^\mu \rightarrow  \bar x^\mu = x^\mu +\xi^{(1)\mu}
+ \frac{1}{2}\xi^{(1)\mu}_{,\alpha}\xi^{(1)\alpha}
+ \frac{1}{2}\xi ^{(2)\mu},
\ee
where  $\xi^{(1)\mu } $ is a  1st-order vector field,
and  $\xi ^{(2)\mu}$ is a 2nd-order vector field
which is independent of  $\xi^{(1)\mu } $.
They can be denoted    by the  parameters
\be\label{xi_2}
\xi^{(A)0}=\alpha^{(A)},
 \,\,\,\,\,\,\,\,  \xi^{(A)i}=\partial^i\beta^{(A)}+d^{ (A)i} ,
~~~~\text{with}~~~~
A=1,2.
\ee
with the constraint $\partial_i d^{(A)i}=0$.
The transformation rules of a tensor,
such as a metric,
are
$g_{\mu\nu}(x)=
\frac{ \partial \bar x ^{\alpha}}{ \partial  x^\mu}
\frac{ \partial \bar x ^{\beta }}{ \partial  x^\nu}
\bar g_{\alpha \beta }(\bar x)$ \cite{Weinberg1972,Russ1996}.
Writing
$g _{\mu\nu} =  g^{(0)}_{\mu\nu}
+ g^{(1)}_{\mu\nu}
+\frac{1}{2}g_{\mu\nu}^{(2)}$   to the 2nd   order,
and similar for $\bar g _{\mu\nu}$,
one has
\be\label{metricTrans0th}
g^{(0)}_{\mu\nu} (x)= \bar g^{(0)}_{\mu\nu}(x),
\ee
\be  \label{metricTrans1st}
\bar g^{(1)}_{\mu  \nu  }(  x)
= g^{(1)}_{\mu  \nu  }(  x)
- \mathcal{L}_{\xi^{(1)}} g^{(0)}_{\mu\nu }(x)
,
\ee
\be\label{metricTrans2nd3}
\bar{g}^{(2)}_{\mu\nu }(x)
 =
g^{(2)}_{\mu\nu }(x)
-2\mathcal{L}_{\xi^{(1)}} g^{(1)}_{\mu\nu }(x)
+\mathcal{L}_{\xi^{(1)}}\l(\mathcal{L}_{\xi^{(1)}}g^{(0)}_{\mu\nu }(x)\r)
-\mathcal{L}_{\xi^{(2)}} g^{(0)}_{\mu\nu }(x),
\ee
where the Lie derivative along $\xi^{(1)\mu}$ is defined as
$\mathcal{L}_{\xi^{(1)}} g^{(0)}_{\mu\nu }
\equiv
g^{(0)}_{\mu\nu,\alpha }\xi^{(1)\alpha}
+g^{(0)}_{\mu\alpha} \xi^{(1)\alpha}_{\, ,\nu}
+g^{(0)}_{\nu\alpha} \xi^{(1)\alpha}_{\, ,\mu}
\,  $,
and others  are similarly defined.
Under  (\ref{xmutransf}),
a scalar function transforms as $f(x)=\bar f(\bar x)$.
By writing  $f(x)=f^{(0)}(x)+f^{(1)}(x)+\frac{1}{2}f^{(2)}(x)$,
one has
\be\label{f0Trans}
\bar f^{(0)}(x)=f^{(0)}(x),
\ee
\be\label{f1Trans}
\bar f^{(1)}(x)=f^{(1)}(x)-\mathcal L_{\xi^{(1)}} f^{(0)}(x),
\ee
\be\label{f2Trans}
\bar f^{(2)}(x)=
f^{(2)}(x)
-2\mathcal{L}_{\xi^{(1)}} f^{(1)}(x)
+\mathcal{L}_{\xi^{(1)}}\l(\mathcal{L}_{\xi^{(1)}} f^{(0)}(x)\r)
-\mathcal{L}_{\xi^{(2)}}f^{(0)}(x),
\ee
where $\mathcal L_{\xi}f\equiv f_{,\, \alpha}\xi^{\alpha} $.
A 4-vector $Z^\mu$ transforms as
$\bar Z^\mu(\bar x) = \frac{{\partial\bar x^\mu}}{{\partial x^\alpha}} Z^\alpha(x)$.
Writing  $Z^\mu(x)  =Z^{(0)\mu}(x)+Z^{(1)\mu}(x)+\frac{1}{2}Z^{(2)\mu}(x)$,
one has
\be\label{vect0transf}
\bar Z^{(0)\mu}(x)=Z^{(0)\mu}(x),
\ee
\be \label{vect1transf}
\bar Z^{(1)\mu}(x)=Z^{(1)\mu}(x)-\mathcal L_{\xi^{(1)}} Z^{(0)\mu}(x),
\ee
\be \label{vect2transf}
\bar Z^{(2)\mu}(x)=
Z^{(2)\mu}(x)
-2\mathcal{L}_{\xi^{(1)}} Z^{(1)\mu}(x)
+\mathcal{L}_{\xi^{(1)}}\l(\mathcal{L}_{\xi^{(1)}} Z^{(0)\mu}(x)\r)
-\mathcal{L}_{\xi^{(2)}}Z^{(0)\mu}(x),
\ee
where $\mathcal L_{\xi}Z^\mu
\equiv
Z^\mu_{,\alpha}\xi^\alpha-\xi^\mu_{,\alpha}Z^\alpha$
and $\mathcal L_{\xi}Z_\mu \equiv
Z_{\mu{,\alpha}}  \xi^\alpha + \xi_{ \mu{,\alpha}}   Z^\alpha$.

The above transformations are general for any two coordinates.
In this paper
we are only concerned with the ones from synchronous to synchronous coordinates.
Moreover,  for the 2nd-order transformations
(\ref{metricTrans2nd3}), (\ref{f2Trans}), and  (\ref{vect2transf}),
we should distinguish the role of  $\xi^{(2)\mu}$ from that of $\xi^{(1)\mu}$.
The real interesting case of 2nd-order gauge transformations is that
the 2nd-order perturbations are transformed
while the 1st-order perturbations are   fixed.
As we mentioned around Eq.(\ref{effecttrs}),
this is the effective 2nd-order transformations,
which require that $\xi^{(1)\mu}=0$, but $\xi^{(2)\mu} \ne 0$
 in (\ref{metricTrans2nd3}), (\ref{f2Trans}), and (\ref{vect2transf}).

First consider the 1st-order transformation.
In the synchronous coordinate,
requiring $ \bar g_{00}(x) = -a^2(\tau) $ and $ \bar g_{0i}(x) = 0 $ leads to
$\xi^{(1)\mu}$ as the following
\be \label{xi0trans}
\xi^{(1)0}(\tau, {\bf x}) = \frac{A^{(1)}({\bf x})}{a(\tau)},
\ee
\be  \label{gi0}
 \xi^{(1)i} (\tau, {\bf x})= A^{(1)}({\bf x})^{,i} \int^\tau \frac{d\tau' }{a(\tau')} + C^{(1)i}(\bf x)  ,
\ee
where $A^{(1)}$ and $C^{(1)i}$
are small, arbitrary functions depending on $\bf x$ only,
and $C^{(1)i}$ can be decomposed into
\be\label{C1decomp}
C ^{(1)i} ({\bf x}) = C\, ^{ ||(1) ,\,i}({\bf x})
+  C\, ^{ \bot(1) i}(\bf x)
\ee
where the transverse part satisfies
$ \partial_i C^{ \bot(1)\, i}=0$.
By Eq.(\ref{metricTrans1st}),
the  residual gauge transform
of the metric $g_{ij}$ within the synchronous coordinates
is
\be \label{deltagij}
 \bar g^{(1)}_{ij}  =  g^{(1)}_{ij}
 + a^2 \left(
 -2\frac{a'}{a} \delta_{ij} \frac{A^{(1)}}{a}
        - 2 A^{(1)}_{,\, ij} \int^\tau  \frac{d\tau }{a(\tau)}
          -C^{(1)}_{i,j} -C^{(1)}_{j,i}        \right),
\ee
from which
the residual gauge transformations of each mode are identified as the following
\be\label{gaugetrphi}
  \bar\phi^{(1)}  =   \phi^{(1)}
  + \frac{1}{3}\nabla^2 A^{(1)} \int \frac{d\tau }{a(\tau)}
   + \frac{1}{3}\nabla^2 C^{||(1)}
  +  \frac{a'}{a^2}   A^{(1)},
\ee
\be\label{gaugetrchi}
 \bar \chi^{||(1)} =
 \chi^{||(1)}  - 2 A^{(1)} \int \frac{d\tau }{a(\tau)} -2C^{||(1)} ,
\ee
\be\label{gaugePerpchi}
\bar\chi^{\perp(1)}_{ij}=\chi^{\perp(1)}_{ij}
-C^{\perp(1)}_{i,j}
-C^{\perp(1)}_{j,i},
\ee
\be\label{gaugeGW}
\bar\chi^{\top(1)}_{ij}=\chi^{\top(1)}_{ij}.
\ee
By (\ref{f0Trans}), the 0th-order  energy density
transforms as $\bar\rho^{(0)}=\rho^{(0)}$.
By  (\ref{f1Trans}),
the density perturbation transforms as
\be \label{deltarho}
\bar   \rho ^{(1)}
=  \rho ^{(1)} - \rho^{(0)}_{,0}\frac{A^{(1)}}{a}  ,
\ee
which leads to the transform of the density contrast
\be \label{deltarho1Gen}
\bar\delta^{(1)}
=   \delta^{(1)}
 - \l[
2\frac{a''}{a'a}
-4\frac{a'}{a^2}
\r]A^{(1)} \, .
\ee
By (\ref{vect0transf}),
the 0th-order 4-velocity transforms between synchronous coordinates as
$\bar U^{(0)0}=U^{(0)0}=a^{-1}$ and
$\bar U^{(0)i}=U^{(0)i}=0$.
By (\ref{vect1transf}),
the 1st-order velocity transforms as
\be\label{U10TransGe}
\bar U^{(1)0}  = U^{(1)0} =0,
\ee
\be \label{U1tr}
\bar U^{(1)i}
= U^{(1)i}+\frac{A^{(1),i}}{a^{2}}.
\ee
Using Eq.(\ref{Uielement}) and
$\bar U^{(1)i}=a^{-1}\bar v^{(1)i}$,
(\ref{U1tr})  can written as transformation of
the 1st-order 3-velocity
\be\label{v1TransGe}
\bar v^{(1)i}  =  v^{(1)i}  +\frac{A^{(1),i}}{a},
\ee
whose transverse and   longitudinal  parts are
\be\label{vGaugemodeCurlGe}
\bar v^{\perp(1)i}
=  v^{\perp(1)i}  ,
\ee
\be\label{vGaugemodeNonCurlGe}
\bar v^{||(1)}
= v^{||(1)}+\frac{A^{(1)}}{a}.
\ee

Now  we  determine   the 2nd-order
synchronous-to-synchronous  gauge transform
for a general RW spacetime.
First  we  determine   the 2nd-order vector $\xi ^{(2)}$.
By the requirements  $\bar g^{(2)}_{00}(x)=g^{(2)}_{00}(x)=0$,
$\bar g^{(2)}_{0i}(x)=g^{(2)}_{0i}(x)=0$,
and  (\ref{xi0trans}) and (\ref{gi0}),
the formula  (\ref{metricTrans2nd3}) gives
\be  \label{alpha2_3}
\xi^{(2)0}=
\alpha^{(2)}
=\frac{ A^{(2)}(\mathbf x)}{a(\tau)} \, ,
\ee
\bl\label{xi2ge}
\xi^{(2)}_i (\tau,{\bf x})
=&
4 A^{(1)}({\bf x})_{,i}\int^\tau \frac{\phi^{(1)}(\tau',{\bf x})}{a({\tau'})}d\tau'
- 2A^{(1)}({\bf x})^{,k}\int^\tau  \frac{\chi^{(1)}_{ki}(\tau',{\bf x})}{a(\tau')}d\tau'
\nn\\
&
-\frac{1}{a^2}A^{(1)}({\bf x})A^{(1)}({\bf x})_{,i}
+ 2 A^{(1)}({\bf x})^{,k}A^{(1)}({\bf x})_{,ki} \int^\tau    \frac{ d\tau' }{a(\tau')}
    \int^{\tau'} \frac{d\tau'' }{a(\tau'')}
\nn\\
&
+ 2 A^{(1)}({\bf x})^{,k}C^{||(1)}({\bf x})_{,ki}\int^{\tau} \frac{d\tau'}{a(\tau')}
+A^{(2)}({\bf x})_{,i}\int^\tau \frac{d\tau'}{a(\tau')}
+ C^{(2)}_i ({\bf x}),
\el
where  $A^{(2)}$ is an arbitrary function of 2nd order,
$ C^{(2)}_i $  is an arbitrary 3-vector of  2nd order
and can be decomposed into
$C^{(2)}_i   =   C^{||(2)}_{,\,i}   +C^{\perp(2)}_{\,i}$.
We remark that the transformation (\ref{xi2ge}) is general
as $\chi^{(1)}_{ki}$  in the integration term contains the  1st-order  scalar
as well as the 1st-order tensor.
Equation (\ref{xi2ge}) can be also written in terms of the parameters
\bl\label{beta2_1}
\beta^{(2)}=&
\nabla^{-2}\Big[
\nabla^2 A^{(1)} \int^\tau\frac{4\phi^{(1)}(\tau',{\bf x})}{a({\tau'})}d\tau'
+A^{(1)}_{,k}\int^\tau\frac{4\phi^{(1)}(\tau',{\bf x})^{,k}}{a({\tau'})}d\tau'
\nn\\
&
-A^{(1) ,\, ki}\int^\tau\frac{2\chi^{(1)}_{ki}(\tau',{\bf x})}{a(\tau')}d\tau'
-A^{(1) ,\, k}\int^\tau\frac{2\chi^{(1)}_{ki}(\tau',{\bf x})^{,\,i}}{a(\tau')}d\tau'
\nn\\
&
+ 2 A^{(1),\, k i}C^{||(1)}_{,\,k i}\int^\tau \frac{d\tau'}{a(\tau')}
+ 2 A^{(1) ,\, k}\nabla^2C^{||(1)}({\bf x})_{,k} \int^\tau\frac{d\tau'}{a(\tau')}\Big]
\nn\\
&-\frac{1}{2a^2(\tau)}A^{(1)} A^{(1)}
+ A^{(1),\, k} A^{(1)}_{,\, k}
    \int^\tau  \frac{d\tau'}{a(\tau')}
    \int^{\tau'}\frac{d\tau'' }{a(\tau'')}
\nn\\
&
+A^{(2)} \int^\tau \frac{d\tau'}{a(\tau')}
+C^{||(2)}
\,,
\el
{\allowdisplaybreaks
\bl   \label{d2_2}
d^{(2)}_i=&\xi^{(2)}_i-\beta^{(2)}_{,i}
\nn\\
=&
\partial_i\nabla^{-2}\Big[
-\nabla^2 A^{(1)} \,  \int^\tau \frac{4\phi^{(1)}(\tau',{\bf x})}{a({\tau'})}d\tau'
-A^{(1)}_{,k}\int^\tau \frac{4\phi^{(1)}(\tau',{\bf x})^{,k}}{a({\tau'})}d\tau'
\nn\\
&
+2A^{(1) ,\,kl}\int^\tau \frac{ \chi^{(1)}_{kl}(\tau',{\bf x})}{a(\tau')}d\tau'
+2A^{(1),\,k}\int^\tau \frac{\chi^{(1)}_{kl}(\tau',{\bf x})^{,\,l}}{a(\tau')}d\tau'
\nn\\
&
- 2 A^{(1),\,kl}C^{||(1)}({\bf x})_{,\,kl}\int^\tau \frac{d\tau'}{a(\tau')}
- 2 A^{(1),\, k} \nabla^2C^{||(1)}_{,k}  \int^\tau \frac{d\tau'}{a(\tau')}\Big]
\nn\\
&
+ 4A^{(1)}_{,i}\int^\tau  \frac{\phi^{(1)}(\tau',{\bf x})}{a({\tau'})}d\tau'
-2A^{(1)\, ,k}\int^\tau  \frac{\chi^{(1)}_{ki}(\tau',{\bf x})}{a(\tau')}d\tau'
\nn\\
&+ 2 A^{(1)\, ,k}C^{||(1)}_{,ki}\int^\tau \frac{d\tau'}{a(\tau')}
+ C^{\perp(2)}_i  .
\el
}
With this,
we now determine the transformation of 2nd-order metric perturbations
(see also Ref.~\cite{WangZhang2017})
{\allowdisplaybreaks
\bl\label{phi2TransGen}
\bar \phi^{(2)}  = &
\phi^{(2)}
-\frac{a''}{a^3}A^{(1)}A^{(1)}
+\frac{1}{a^2}\bigg[
-\frac{2}{3}A^{(1)}\nabla^2A^{(1)}
-\frac{1}{3}A^{(1),\,l}A^{(1)}_{,\,l}
\bigg]
-4\frac{a'}{a^2}\phi^{(1)}A^{(1)}
\nn\\
&
-\frac{2}{a}\phi^{(1)'}A^{(1)}
+\frac{a'}{a^2}\l[\int^\tau \frac{d\tau' }{a(\tau')}\r]
\bigg[
-\frac{4}{3}A^{(1)} \nabla^2  A^{(1)}
- A^{(1)}_{,\,l}A^{(1),\,l}
\bigg]
\nn\\
&
+ \l[ \int^\tau  \frac{d\tau'}{a(\tau')}
    \int^{\tau'}\frac{d\tau'' }{a(\tau'')}\r]
\bigg[
\frac{2}{3}A^{(1),\,l}\nabla^2A^{(1)}_{,\,l}
+ \frac{2}{3} A^{(1),\,lm}A^{(1)}_{,\,lm}
\bigg]
\nn\\
&
+\l[\int^\tau \frac{d\tau' }{a(\tau')}\r]^2
\bigg[
-\frac{1}{3}A^{(1),\,l}  \nabla^2  A^{(1)}_{,\,l}
-\frac{2}{3}A^{(1)}_{,\,lm} A^{(1),\,lm}
\bigg]
+ \l[ \int^\tau \frac{d\tau'}{a(\tau')} \r]
\bigg[
\frac{2}{3} A^{(1),\,lm}C^{||(1)}_{,\,lm}
\nn\\
&
+ \frac{2}{3}A^{(1),\,l}\nabla^2C^{||(1)}_{,\,l}
-\frac{4}{3} \phi^{(1)} \nabla^2  A^{(1)}
-2\phi^{(1)}_{,\,l}A^{(1),\,l}
+\frac{2}{3}\chi^{(1)}_{\,lm}A^{(1),\,lm}
\bigg]
\nn\\
&
+\nabla^2A^{(1)} \int^\tau\frac{4\phi^{(1)}(\tau',{\bf x})}{3a({\tau'})}d\tau'
+A^{(1)}_{,\,l}\int^\tau\frac{4\phi^{(1)}(\tau',{\bf x})^{,\,l}}{3a({\tau'})}d\tau'
\nn\\
&
-A^{(1),\,lm}\int^\tau\frac{2\chi^{(1)}_{\,lm}(\tau',{\bf x})}{3a(\tau')}d\tau'
-A^{(1),\,l}\int^\tau\frac{2\chi^{(1)}_{\,lm}(\tau',{\bf x})^{,m}}{3a(\tau')}d\tau'
\nn\\
&
+\frac{a'}{a^2}A^{(2)}
+\frac{1}{3}\nabla^2A^{(2)}\int^\tau \frac{d\tau'}{a(\tau')}
+\frac{1}{3}\nabla^2C^{||(2)}
.
\el
}
The residual gauge transforms
of the  $\chi^{||(2)}$, $\chi^{\perp(2)}_{ij}$,
$\chi^{\top(2)}_{ij}$ are given in (C56), (C57), and (C58) of
Ref.~\cite{WangZhang2017},
which are listed as following
{\allowdisplaybreaks
\bl\label{chi||2transF2}
 \bar\chi^{||(2)}
&=
 \chi^{||(2)}
+ \frac{1}{a^2}   \bigg[
A^{(1)}A^{(1)}
+2\nabla^{-2}\big(A^{(1)}\nabla^2A^{(1)}\big)
+3\nabla^{-2}\nabla^{-2}\big(
A^{(1),\,lm}A^{(1)}_{,\,lm}
\nn\\
&
-\nabla^2A^{(1)}\nabla^2A^{(1)}\big)
\bigg]
+\frac{4a'}{a^2}  \l[\int^\tau  \frac{d\tau' }{a(\tau')}\r]
  \bigg[
2\nabla^{-2}\big(A^{(1)}\nabla^2A^{(1)}\big)
+3\nabla^{-2}\nabla^{-2}\big(
A^{(1),\,lm}A^{(1)}_{,\,lm}
\nn\\
&
-\nabla^2A^{(1)}\nabla^2A^{(1)}\big)
\bigg]
+  \frac{a'}{a^2}   \bigg[
8\nabla^{-2}\big(A^{(1)}\nabla^2C^{||(1)}\big)
+6\nabla^{-2}\nabla^{-2}\big(
-\chi^{(1),\,lm}_{lm}A^{(1)}
\nn\\
&
-\chi^{(1)}_{lm}A^{(1),\,lm}
-2\chi^{(1),\,l}_{lm}A^{(1),\,m}
+2A^{(1),\,lm}C^{||(1)}_{,\,lm}
-2\nabla^2A^{(1)}\nabla^2C^{||(1)}\big)
\bigg]
\nn\\
&
-  \frac{3}{a}   \nabla^{-2} \nabla^{-2}\bigg[
\chi^{(1)',\,lm}_{lm}A^{(1)}
+\chi^{(1)'}_{lm}A^{(1),\,lm}
+2\chi^{(1)',\,l}_{lm}A^{(1),\,l} \bigg]
\nn \\
&
-2\l[\int^\tau  \frac{d\tau'}{a(\tau')}\int^{\tau'}\frac{d\tau'' }{a(\tau'')}\r]A^{(1),\,l}A^{(1)}_{,\,l}
+\l[\int^\tau  \frac{d\tau' }{a(\tau')}\r]^2   \bigg[
2A^{(1)}_{,\,l}A^{(1),\,l}
\nn \\
&
-2\nabla^{-2}\big(A^{(1),\,l}\nabla^2A^{(1)}_{,\,l}\big)
+3\nabla^{-2}\nabla^{-2}\big(
\nabla^2A^{(1),\,l}\nabla^2A^{(1)}_{,\,l}
-A^{(1),\,lmn}A^{(1)}_{,\,lmn}\big)
\bigg]
\nn\\
&
+ \l[\int^\tau \frac{d\tau' }{a(\tau')} \r] \bigg[
2A^{(1),\,l}C^{||(1)}_{,\,l}
+2\nabla^{-2}\big(4\phi^{(1)}\nabla^2A^{(1)}
+\chi^{(1)}_{lm}A^{(1),\,lm}
-2A^{(1),\,l}\nabla^2C^{||(1)}_{,\,l}
\big)
\nn\\
&
+3\nabla^{-2}\nabla^{-2}\big(
-\chi^{(1),\,lmn}_{lm}A^{(1)}_{,n}
-3\chi^{(1)}_{lm,n}A^{(1),\,lmn}
-4\chi^{(1),\,l}_{lm,n}A^{(1),\,mn}
-2\chi^{(1),\,l}_{lm}\nabla^2A^{(1),\,m}
\nn\\
&
-2\chi^{(1)}_{lm}\nabla^2A^{(1),\,lm}
+4\phi^{(1),\,lm}A^{(1)}_{,\,lm}
-4\nabla^2\phi^{(1)}\nabla^2A^{(1)}
+2\nabla^2A^{(1),\,l}\nabla^2C^{||(1)}_{,\,l}
\nn\\
&
-2A^{(1),\,lmn}C^{||(1)}_{,\,lmn}
\big)
\bigg]
+   \bigg[ 2C^{||(1)}_{,\,l}C^{||(1),\,l}
+\nabla^{-2}\big(8\phi^{(1)}\nabla^2C^{||(1)}
+2\chi^{(1)}_{lm}C^{||(1),\,lm}
\nn\\
&
-2 C^{||(1),\,l}\nabla^2C^{||(1)}_{,\,l}\big)
+3\nabla^{-2}\nabla^{-2}\big(
4\phi^{(1),\,lm}C^{||(1)}_{,\,lm}
-4\nabla^2\phi^{(1)}\nabla^2C^{||(1)}
-\chi^{(1),\,lmn}_{lm}C^{||(1)}_{,n}
\nn\\
&
-3\chi^{(1)}_{lm,n}C^{||(1),\,lmn}
-4\chi^{(1),\,l}_{lm,n}C^{||(1),\,mn}
-2\chi^{(1),\,l}_{lm}\nabla^2C^{||(1),\,m}
-2\chi^{(1)}_{lm}\nabla^2C^{||(1),\,lm}
\nn \\
&
+\nabla^2C^{||(1),\,l}\nabla^2C^{||(1)}_{,\,l}
-C^{||(1),\,lmn}C^{||(1)}_{,\,lmn}
\big)
\bigg]
-4 \nabla^{-2}\bigg[
\nabla^2A^{(1)} \int^\tau\frac{2\phi^{(1)}(\tau',{\bf x})}{a({\tau}')}d\tau'
\nn\\
&
+A^{(1)}_{,\,l} \int^\tau\frac{2\phi^{(1)}(\tau',{\bf x})^{,\,l}}{a({\tau}')}d\tau'
- A^{(1),\,lm} \int^\tau \frac{\chi^{(1)}_{lm}(\tau',{\bf x})}{a(\tau')}d\tau'
- A^{(1),\,l}  \int^\tau \frac{\chi^{(1)}_{lm}(\tau',{\bf x})^{,m}}{a(\tau')}d\tau'
\bigg]
\nn\\
&
-2 A^{(2)}  \int^\tau \frac{d\tau' }{a(\tau')}
-2C^{||(2)}
    ,
\el
}
and
{
\allowdisplaybreaks
\bl\label{chiPerp2TransF}
\bar\chi^{\perp(2)}_{ij}
&=
\chi^{\perp(2)}_{ij}
+ \frac{1}{a^2} \bigg[
-2\partial_i\nabla^{-2}\big( A^{(1)}_{,j}\nabla^2A^{(1)} \big)
+\partial_i\partial_j\nabla^{-2}\big(
A^{(1),\,l}A^{(1)}_{,\,l}\big)
\nn\\
&
-2\partial_i\partial_j\nabla^{-2}\nabla^{-2}\big(
A^{(1),\,lm}A^{(1)}_{,\,lm}
-\nabla^2A^{(1)}\nabla^2A^{(1)}
\big)
\bigg]
\nn\\
&
+\frac{4a'}{a^2} \l[\int^\tau \frac{d\tau' }{a(\tau')}\r] \bigg[
-2\partial_i\nabla^{-2}\big( A^{(1)}_{,j}\nabla^2A^{(1)} \big)
+\partial_i\partial_j\nabla^{-2}\big(
A^{(1),\,l}A^{(1)}_{,\,l}
\big)
\nn\\
&
-2\partial_i\partial_j\nabla^{-2}\nabla^{-2}\big(
A^{(1),\,lm}A^{(1)}_{,\,lm}
-\nabla^2A^{(1)}\nabla^2A^{(1)}
\big)
\bigg]
+\frac{2a'}{a^2} \bigg[
2\partial_i\nabla^{-2}\big( 2A^{(1),\,l}C^{||(1)}_{,\,lj}
\nn\\
&
-2A^{(1)}_{,j}\nabla^2C^{||(1)} -\chi^{(1),\,l}_{lj}A^{(1)}
-\chi^{(1)}_{lj}A^{(1),\,l} \big)
+2\partial_i\partial_j\nabla^{-2}\nabla^{-2}\big(
2\nabla^2A^{(1)}\nabla^2C^{||(1)}
\nn\\
&
-2A^{(1),\,lm}C^{||(1)}_{,\,lm}
+\chi^{(1),\,lm}_{lm}A^{(1)}
+\chi^{(1)}_{lm}A^{(1),\,lm}
+2\chi^{(1),\,l}_{lm}A^{(1),m}
\big)
\bigg]
- \frac{1}{a} \bigg[
2\partial_i\nabla^{-2}\big( \chi^{(1)',\,l}_{lj}A^{(1)}
\nn \\
&
 +\chi^{(1)'}_{lj}A^{(1),\,l} \big)
-2\partial_i\partial_j\nabla^{-2}\nabla^{-2}\big(
\chi^{(1)',\,lm}_{lm}A^{(1)}
+\chi^{(1)'}_{lm}A^{(1),\,lm}
+2\chi^{(1)',\,l}_{lm}A^{(1),m}
\big)
\bigg]
\nn \\
&
+ \l[\int^\tau \frac{d\tau' }{a(\tau')}\r]^2\bigg[
2\partial_i\nabla^{-2}\big( A^{(1)}_{,\,lj}\nabla^2A^{(1),\,l} \big)
-\partial_i\partial_j\nabla^{-2}\big(
A^{(1),\,lm}A^{(1)}_{,\,lm}
\big)
\nn\\
&
+2\partial_i\partial_j\nabla^{-2}\nabla^{-2}\big(
A^{(1),\,lmn}A^{(1)}_{,\,lmn}
-\nabla^2A^{(1),\,l}\nabla^2A^{(1)}_{,\,l}
\big)
\bigg]
\nn \\
&
- \l[\int^\tau \frac{d\tau' }{a(\tau')} \r]\bigg[
2\partial_i\nabla^{-2}\big(
4\phi^{(1)}_{,j}\nabla^2A^{(1)}
-4\phi^{(1),\,l}A^{(1)}_{,\,lj}
+\chi^{(1),\,l}_{lj,n}A^{(1),n}
+2\chi^{(1)}_{lj,n}A^{(1),\,ln}
\nn\\
&
+\chi^{(1),m}_{lm}A^{(1),\,l}_{,j}
+\chi^{(1)}_{lm}A^{(1),\,lm}_{,j}
+\chi^{(1)}_{lj}\nabla^2A^{(1),\,l}
+2A^{(1),\,ln}C^{||(1)}_{,\,lnj}
-2A^{(1)}_{,\,lj}\nabla^2C^{||(1),\,l}
\big)
\nn\\
&
-2\partial_i\partial_j\nabla^{-2}\nabla^{-2}\big(
4\nabla^2\phi^{(1)}\nabla^2A^{(1)}
-4\phi^{(1),\,lm}A^{(1)}_{,\,lm}
+\chi^{(1),\,lm}_{lm,n}A^{(1),n}
+3\chi^{(1)}_{lm,n}A^{(1),\,lmn}
\nn\\
&
+4\chi^{(1),\,l}_{lm,n}A^{(1),\,mn}
+2\chi^{(1),\,l}_{lm}\nabla^2A^{(1),m}
+2\chi^{(1)}_{lm}\nabla^2A^{(1),\,lm}
+2A^{(1),\,lmn}C^{||(1)}_{,\,lmn}
\nn\\
&
-2\nabla^2A^{(1),\,l}\nabla^2C^{||(1)}_{,\,l}
\big)
\bigg]
+\bigg[
-2\partial_i\nabla^{-2}\big(
4\phi^{(1)}_{,j}\nabla^2C^{||(1)}
-4\phi^{(1),\,l}C^{||(1)}_{,\,lj}
+\chi^{(1),\,l}_{lj,n}C^{||(1),n}
\nn\\
&
+2\chi^{(1)}_{lj,n}C^{||(1),\,ln}
+\chi^{(1),m}_{lm}C^{||(1),\,l}_{,j}
+\chi^{(1)}_{lm}C^{||(1),\,lm}_{,j}
+\chi^{(1)}_{lj}\nabla^2C^{||(1),\,l}
+C^{||(1),\,ln}C^{||(1)}_{,\,lnj}
\nn\\
&
-C^{||(1)}_{,\,lj}\nabla^2C^{||(1),\,l}
\big)
+2\partial_i\partial_j\nabla^{-2}\nabla^{-2}\big(
4\nabla^2\phi^{(1)}\nabla^2C^{||(1)}
-4\phi^{(1),\,lm}C^{||(1)}_{,\,lm}
+\chi^{(1),\,lm}_{lm,n}C^{||(1),n}
\nn\\
&
+4\chi^{(1),\,l}_{lm,n}C^{||(1),\,mn}
+3\chi^{(1)}_{lm,n}C^{||(1),\,lmn}
+2\chi^{(1),\,l}_{lm}\nabla^2C^{||(1),m}
+2\chi^{(1)}_{lm}\nabla^2C^{||(1),\,lm}
\nn\\
&
+C^{||(1),\,lmn}C^{||(1)}_{,\,lmn}
-\nabla^2C^{||(1),\,l}\nabla^2C^{||(1)}_{,\,l}
\big)
\bigg]
+\partial_i\bigg[
A^{(1),\,l}\int^\tau \frac{2\chi^{(1)}_{lj}(\tau',{\bf x})}{a(\tau')}d\tau'
\nn \\
&
-A^{(1)}_{,j}\int^\tau \frac{4\phi^{(1)}(\tau',{\bf x})}{a({\tau'})}d\tau'
\bigg]
+\partial_i\partial_j\nabla^{-2}\bigg[
\nabla^2A^{(1)} \int^\tau \frac{4\phi^{(1)}(\tau',{\bf x})}{a({\tau'})}d\tau'
\nn\\
&
+A^{(1)}_{,\,l}\int^\tau \frac{4\phi^{(1)}(\tau',{\bf x})^{,\,l}}{a(\tau')}d\tau'
-A^{(1),\,lm}\int^\tau \frac{2\chi^{(1)}_{lm}(\tau' ,{\bf x})}{a(\tau' )}d\tau
\nn\\
&
-A^{(1),\,l}\int^\tau \frac{2\chi^{(1)}_{ln}(\tau' ,{\bf x})^{, n}}{a(\tau' )}d\tau' \bigg]
  -C^{\perp(2)}_{i,j}
+(i \leftrightarrow j ) \ ,
\el
}
which shows   that transformation of $ \chi^{\perp(2)}_{ij}$
  depends on   $\xi^{(2)\mu}$ only through $C^{\perp(2)}_{(i,j)}$,
{\allowdisplaybreaks
\bl\label{chiT2transF2}
\bar\chi^{\top(2)}_{ij}
&=
\chi^{\top(2)}_{ij}
+\l[\frac{1}{a^2}\r]\bigg[
\delta_{ij}\nabla^{-2}\big(
A^{(1),\,lm}A^{(1)}_{,\,lm}
-\nabla^2A^{(1)}\nabla^2A^{(1)}
\big)
+4\nabla^{-2}\big(
A^{(1)}_{,ij}\nabla^2A^{(1)}
\nn\\
&
-A^{(1),\,l}_{,i}A^{(1)}_{,\,lj}
\big)
+\partial_i\partial_j\nabla^{-2}\nabla^{-2}\big(
A^{(1),\,lm}A^{(1)}_{,\,lm}
-\nabla^2A^{(1)}\nabla^2A^{(1)}
\big)
\bigg]
\nn\\
&
+\frac{4a'}{a^2}\l[\int^\tau  \frac{d\tau' }{a(\tau')}\r]\bigg[
\delta_{ij}\nabla^{-2}\big(
A^{(1),\,lm}A^{(1)}_{,\,lm}
-\nabla^2A^{(1)}\nabla^2A^{(1)}
\big)
+4\nabla^{-2}\big(
A^{(1)}_{,ij}\nabla^2A^{(1)}
\nn\\
&
-A^{(1),\,l}_{,i}A^{(1)}_{,\,lj}
\big)
+\partial_i\partial_j\nabla^{-2}\nabla^{-2}\big(
A^{(1),\,lm}A^{(1)}_{,\,lm}
-\nabla^2A^{(1)}\nabla^2A^{(1)}
\big)
\bigg]
\nn\\
&
+\frac{2a'}{a^2} \bigg[
-\delta_{ij}\nabla^{-2}\big(
\chi^{(1),\,lm}_{lm}A^{(1)}
+\chi^{(1)}_{lm}A^{(1),\,lm}
+2\chi^{(1),\,l}_{lm}A^{(1),m}
+2\nabla^2A^{(1)}\nabla^2C^{||(1)}
\nn\\
&
-2A^{(1),\,lm}C^{||(1)}_{,\,lm}
\big)
-2\chi^{(1)}_{ij}A^{(1)}
+2\partial_i\nabla^{-2}\big(
\chi^{(1),\,l}_{lj}A^{(1)}
+\chi^{(1)}_{lj}A^{(1),\,l}
\big)
+2\partial_j\nabla^{-2}\big(
\chi^{(1),\,l}_{li}A^{(1)}
\nn\\
&
+\chi^{(1)}_{li}A^{(1),\,l}
\big)
+4\nabla^{-2}\big(
A^{(1)}_{,ij}\nabla^2C^{||(1)}
+C^{||(1)}_{,ij}\nabla^2A^{(1)}
-A^{(1),\,l}_{,i}C^{||(1)}_{,\,lj}
-A^{(1),\,l}_{,j}C^{||(1)}_{,\,li}
\big)
\nn\\
&
-\partial_i\partial_j\nabla^{-2}\nabla^{-2}\big(
\chi^{(1),\,lm}_{lm}A^{(1)}
+\chi^{(1)}_{lm}A^{(1),\,lm}
+2\chi^{(1),\,l}_{lm}A^{(1),m}
+2A^{(1),\,lm}C^{||(1)}_{,\,lm}
\nn\\
&
-2\nabla^2A^{(1)}\nabla^2C^{||(1)}
\big)
\bigg]
- \frac{1}{a} \bigg[
\delta_{ij}\nabla^{-2}\big(\chi^{(1)',\,lm}_{lm}A^{(1)}
+\chi^{(1)'}_{lm}A^{(1),\,lm}
+2\chi^{(1)',\,l}_{lm}A^{(1),m}
\big)
\nn\\
&
+2\chi^{(1)'}_{ij}A^{(1)}
-2\partial_i\nabla^{-2}\big(
\chi^{(1)',\,l}_{lj}A^{(1)}
+\chi^{(1)'}_{lj}A^{(1),\,l}
\big)
-2\partial_j\nabla^{-2}\big(
\chi^{(1)',\,l}_{li}A^{(1)}
+\chi^{(1)'}_{li}A^{(1),\,l}
\big)
\nn\\
&
+\partial_i\partial_j\nabla^{-2}\nabla^{-2}\big(
\chi^{(1)',\,lm}_{lm}A^{(1)}
+\chi^{(1)'}_{lm}A^{(1),\,lm}
+2\chi^{(1)',\,l}_{lm}A^{(1),m}
\big) \bigg]
\nn\\
&
-\l[\int^\tau  \frac{d\tau' }{a(\tau')}\r]^2\bigg[
\delta_{ij}\nabla^{-2}\big(
A^{(1),\,lmn}A^{(1)}_{,\,lmn}
-\nabla^2A^{(1),\,l}\nabla^2A^{(1)}_{,\,l}
\big)
\nn\\
&
+2\nabla^{-2}\big(-2A^{(1),\,lm}_{,i}A^{(1)}_{,\,lmj}
+2A^{(1)}_{,\,lij}\nabla^2A^{(1),\,l}
\big)
+\partial_i\partial_j\nabla^{-2}\nabla^{-2}\big(
A^{(1),\,lmn}A^{(1)}_{,\,lmn}
\nn\\
&
-\nabla^2A^{(1),\,l}\nabla^2A^{(1)}_{,\,l}
\big)
\bigg]
+\l[\int^\tau  \frac{d\tau' }{a(\tau')}\r]\bigg[
\delta_{ij}\nabla^{-2}\big(
4\phi^{(1),\,lm}A^{(1)}_{,\,lm}
-4\nabla^2\phi^{(1)}\nabla^2A^{(1)}
\nn\\
&
-\chi^{(1),\,lm}_{lm,n}A^{(1),n}
-4\chi^{(1),\,l}_{lm,n}A^{(1),mn}
+\chi^{(1)}_{lm,n}A^{(1),\,lmn}
-2\chi^{(1),\,l}_{lm}\nabla^2A^{(1),m}
\nn\\
&
+2 A^{(1),\,lm}\nabla^2\chi^{(1)}_{lm}
-2A^{(1),\,lmn}C^{||(1)}_{,\,lmn}
+2\nabla^2A^{(1),\,l}\nabla^2C^{||(1)}_{,\,l}
\big)
-2\chi^{(1)}_{ij,\,l}A^{(1),\,l}
\nn\\
&
-2\chi^{(1)}_{li}A^{(1),\,l}_{,j}
-2\chi^{(1)}_{lj}A^{(1),\,l}_{,i}
-4\nabla^{-2}\big(
2\phi^{(1),\,l}_{,i}A^{(1)}_{,\,lj}
+2\phi^{(1),\,l}_{,j}A^{(1)}_{,\,li}
-2\phi^{(1)}_{,ij}\nabla^2A^{(1)}
\nn\\
&
-2A^{(1)}_{,ij}\nabla^2\phi^{(1)}
-A^{(1),\,lm}_{,i}C^{||(1)}_{,\,lmj}
-A^{(1),\,lm}_{,j}C^{||(1)}_{,\,lmi}
+A^{(1)}_{,\,lij}\nabla^2C^{||(1),\,l}
+C^{||(1)}_{,\,lij}\nabla^2A^{(1),\,l}
\big)
    \nn\\
    &
+2\partial_i\nabla^{-2}\big(
\chi^{(1),\,l}_{lj,m}A^{(1),m}
+2\chi^{(1)}_{lj,m}A^{(1),\,lm}
+\chi^{(1),m}_{lm}A^{(1),\,l}_{,j}
+\chi^{(1)}_{lm}A^{(1),\,lm}_{,j}
+\chi^{(1)}_{lj}\nabla^2A^{(1),\,l}
\big)
    \nn\\
    &
+2\partial_j\nabla^{-2}\big(
\chi^{(1),\,l}_{li,m}A^{(1),m}
+2\chi^{(1)}_{li,m}A^{(1),\,lm}
+\chi^{(1),m}_{lm}A^{(1),\,l}_{,i}
+\chi^{(1)}_{lm}A^{(1),\,lm}_{,i}
+\chi^{(1)}_{li}\nabla^2A^{(1),\,l}
\big)
\nn\\
&
+\partial_i\partial_j\nabla^{-2}\nabla^{-2}\big(
4\phi^{(1),\,lm}A^{(1)}_{,\,lm}
-4\nabla^2\phi^{(1)}\nabla^2A^{(1)}
-\chi^{(1),\,lm}_{lm,n}A^{(1),n}
-4\chi^{(1),\,l}_{lm,n}A^{(1),mn}
\nn\\
&
-7\chi^{(1)}_{lm,n}A^{(1),\,lmn}
-2\chi^{(1),\,l}_{lm}\nabla^2A^{(1),m}
-4\chi^{(1)}_{lm}\nabla^2A^{(1),\,lm}
-2 A^{(1),\,lm}\nabla^2\chi^{(1)}_{lm}
\nn\\
&
-2A^{(1),\,lmn}C^{||(1)}_{,\,lmn}
+2\nabla^2A^{(1),\,l}\nabla^2C^{||(1)}_{,\,l}
\big)
\bigg]
+\bigg[
\delta_{ij}\nabla^{-2}\big(
4\phi^{(1),\,lm}C^{||(1)}_{,\,lm}
\nn\\
&
-4\nabla^2\phi^{(1)}\nabla^2C^{||(1)}
-\chi^{(1),\,lm}_{lm,n}C^{||(1),n}
-4\chi^{(1),\,l}_{lm,n}C^{||(1),mn}
+\chi^{(1)}_{lm,n}C^{||(1),\,lmn}
\nn\\
&
-2\chi^{(1),\,l}_{lm}\nabla^2C^{||(1),m}
+2C^{||(1),\,lm}\nabla^2\chi^{(1)}_{lm}
+\nabla^2C^{||(1),\,l}\nabla^2C^{||(1)}_{,\,l}
-C^{||(1),\,lmn}C^{||(1)}_{,\,lmn}
\big)
\nn\\
&
-2 \chi^{(1)}_{ij,\,l}C^{||(1),\,l}
-2\chi^{(1)}_{li}C^{||(1),\,l}_{,j}
-2\chi^{(1)}_{lj}C^{||(1),\,l}_{,i}
-4\nabla^{-2}\big(2\phi^{(1),\,l}_{,i}C^{||(1)}_{,\,lj}
+2\phi^{(1),\,l}_{,j}C^{||(1)}_{,\,li}
\nn\\
&
-2\phi^{(1)}_{,ij}\nabla^2C^{||(1)}
-2C^{||(1)}_{,ij}\nabla^2\phi^{(1)}
-C^{||(1),\,lm}_{,i}C^{||(1)}_{,\,lmj}
+C^{||(1)}_{,\,lij}\nabla^2C^{||(1),\,l}
\big)
\nn\\
&
+2 \partial_i\nabla^{-2}\big(
\chi^{(1),\,l}_{lj,m}C^{||(1),m}
+2\chi^{(1)}_{lj,m}C^{||(1),\,lm}
+\chi^{(1),m}_{lm}C^{||(1),\,l}_{,j}
+\chi^{(1)}_{lm}C^{||(1),\,lm}_{,j}
\nn\\
&
+\chi^{(1)}_{lj}\nabla^2C^{||(1),\,l}
\big)
+2 \partial_j\nabla^{-2}\big(
\chi^{(1),\,l}_{li,m}C^{||(1),m}
+2\chi^{(1)}_{li,m}C^{||(1),\,lm}
+\chi^{(1),m}_{lm}C^{||(1),\,l}_{,i}
\nn\\
&
+\chi^{(1)}_{lm}C^{||(1),\,lm}_{,i}
+\chi^{(1)}_{li}\nabla^2C^{||(1),\,l}
\big)
-\partial_i\partial_j\nabla^{-2}\nabla^{-2}\big(
4\nabla^2\phi^{(1)}\nabla^2C^{||(1)}
-4\phi^{(1),\,lm}C^{||(1)}_{,\,lm}
\nn\\
&
+\chi^{(1),\,lm}_{lm,n}C^{||(1),n}
+4\chi^{(1),\,l}_{lm,n}C^{||(1),mn}
+7\chi^{(1)}_{lm,n}C^{||(1),\,lmn}
+2\chi^{(1),\,l}_{lm}\nabla^2C^{||(1),m}
\nn\\
&
+4\chi^{(1)}_{lm}\nabla^2C^{||(1),\,lm}
+2 C^{||(1),\,lm}\nabla^2\chi^{(1)}_{lm}
+C^{||(1),\,lmn}C^{||(1)}_{,\,lmn}
-\nabla^2C^{||(1),\,l}\nabla^2C^{||(1)}_{,\,l}
\big)
\bigg]
   .
\el
}
Equation (\ref{chiT2transF2}) states  that transformation of $ \chi^{\top(2)}_{ij}$
involves the 1st-order transformation vector $\xi^{(1)\mu}$ only,
but not the 2nd-order vector $\xi^{(2)\mu}$.

In (\ref{xi2ge}), (\ref{phi2TransGen}), (\ref{chi||2transF2}),
(\ref{chiPerp2TransF}), and (\ref{chiT2transF2}),
$\chi^{ (1)}_{ij } $ contains  the tensor  $\chi^{\top(1)}_{ij } $,
which belongs to  the  type of  scalar-tensor coupling
and will not be  considered.
For  RD stage and  the scalar-scalar coupling,
the transformation vector (\ref{alpha2_3}), (\ref{beta2_1}),
and (\ref{d2_2})
reduces to the following
\be\label{alpha2RD1}
\alpha^{(2)}
=\frac{ A^{(2)}(\mathbf x)}{\tau} \, ,
\ee
\bl\label{beta2RD1}
\beta^{(2)}=&
\nabla^{-2}\bigg[
 \nabla^2A^{(1)}({\bf x})\, \int^\tau\frac{4\phi^{(1)}(\tau',{\bf x})}{\tau'}d\tau'
+A^{(1)}({\bf x})_{,k}\int^\tau\frac{4\phi^{(1)}(\tau',{\bf x})^{,k}}{\tau'}d\tau'
\nn\\
&
-A^{(1)}({\bf x})^{,\,ki}\int^\tau\frac{2D_{ki}\chi^{||(1)}(\tau',{\bf x})}{\tau'}d\tau'
-A^{(1)}({\bf x})^{,\,k}\int^\tau\frac{4\nabla^2\chi^{||(1)}(\tau',{\bf x})_{,\,k}}{3\tau'}d\tau'
\nn\\
&
+  2\ln\tau  A^{(1)}({\bf x})^{,\,k i}C^{||(1)}({\bf x})_{,\,k i}
+  2\ln\tau  A^{(1)}({\bf x})^{,k}\nabla^2C^{||(1)}({\bf x})_{,k} \bigg]
\nn\\
&-\frac{1}{2\tau^2}A^{(1)}({\bf x})A^{(1)}({\bf x})
+ \frac{(\ln\tau)^2}{2}A^{(1)}({\bf x})^{,k}A^{(1)}({\bf x})_{,k}
\nn\\
&
+  A^{(2)}({\bf x}) \ln\tau
+C^{||(2)} ({\bf x})
\,,
\el
\bl\label{d2RD1}
d^{(2)}_i
=&
\partial_i\nabla^{-2}\bigg[
-\nabla^2A^{(1)}({\bf x}) \, \int^\tau \frac{4\phi^{(1)}(\tau',{\bf x})}{\tau'}d\tau'
-A^{(1)}({\bf x})_{,k}\int^\tau \frac{4\phi^{(1)}(\tau',{\bf x})^{,k}}{\tau'}d\tau'
\nn\\
&
+2A^{(1)}({\bf x})^{,\,kl}\int^\tau \frac{ D_{kl}\chi^{||(1)}(\tau',{\bf x})}{\tau'}d\tau'
+A^{(1)}({\bf x})^{,\,k}\int^\tau \frac{4\nabla^2\chi^{||(1)}(\tau',{\bf x})_{,\,k}}{3\tau'}d\tau'
\nn\\
&
-2\ln\tau  A^{(1)}({\bf x})^{,\,kl}C^{||(1)}({\bf x})_{,\,kl}
- 2\ln\tau  A^{(1)}({\bf x})^{,k} \nabla^2C^{||(1)}({\bf x})_{,k} \bigg]
\nn\\
&
+ 4A^{(1)}({\bf x})_{,i}\int^\tau  \frac{\phi^{(1)}(\tau',{\bf x})}{\tau'}d\tau'
-2A^{(1)}({\bf x})^{,k}\int^\tau  \frac{D_{ki}\chi^{||(1)}(\tau',{\bf x})}{\tau'}d\tau'
\nn\\
&+ 2 A^{(1)}({\bf x})^{,k}C^{||(1)}({\bf x})_{,ki} \ln\tau
+ C^{\perp(2)}_i ({\bf x}),
\el
and the transformations of the 2nd-order   perturbations
are given by
(\ref{phi2TransRD}), (\ref{chi||2transRD}), (\ref{chiPerp2TransRD}),
(\ref{chiT2transRD}),  (\ref{delta2TransRD}),
 (\ref{v||2TransRD}), and (\ref{vperp2TransRD})
 in the context.

Next we give the residual transform
of $\rho^{(2)}$ and $U^{(2)\mu}$ for a general RW spacetime,
which has not been given  in Ref.~\cite{WangZhang2017}.
By using Eqs.~(\ref{f2Trans}), (\ref{xi0trans}),
(\ref{gi0}), and (\ref{alpha2_3}),
and omitting the 1st-order curl vector,
one gets the residual gauge transform
of the 2nd-order density perturbation
\bl\label{rho2Trans}
\bar \rho^{(2)}
=&
\rho^{(2)}
-\frac{a'}{a^3}\rho^{(0)'}A^{(1)}A^{(1)}
+\frac{1}{a^2}\rho^{(0)''}A^{(1)}A^{(1)}
+\frac{1}{a}
\bigg[
-2\rho^{(1)'}A^{(1)}
+\rho^{(0)'} A^{(1)}_{,\, l}C\, ^{ ||(1) ,\,l}
\bigg]
\nn\\
&
+\frac{1}{a}\l[\int^\tau \frac{d\tau' }{a(\tau')}\r]\rho^{(0)'}A^{(1)}_{,\, l}A^{(1),\,l}
-2\l[\int^\tau \frac{d\tau' }{a(\tau')}\r]\rho^{(1)}_{,\, l}A^{(1),\,l}
-2\rho^{(1)}_{,\, l}C\, ^{ ||(1) ,\,l}
\nn\\
&
-\frac{1}{a}\rho^{(0)'}A^{(2)}
\,   ,
\el
which can be written in terms of the 2nd-order density contrast
{\allowdisplaybreaks
\bl\label{delta2TransGe}
\bar\delta^{(2)}
=&
\delta^{(2)}
-22\frac{a''}{a^3}A^{(1)}A^{(1)}
+2\frac{a'''}{a'a^2}A^{(1)}A^{(1)}
+2\frac{a''^{\,2}}{a'^{\,2}a^2}A^{(1)}A^{(1)}
+24\frac{a'^{\,2}}{a^4}A^{(1)}A^{(1)}
\nn\\
&
+\frac{a''}{a'a}
\bigg[
-4\delta^{(1)}A^{(1)}
+2A^{(1)}_{,\, l}C\, ^{ ||(1) ,\,l}
\bigg]
-4\frac{a'}{a^2}\l[\int^\tau \frac{d\tau' }{a(\tau')}\r]A^{(1)}_{,\, l}A^{(1),\,l}
\nn\\
&
+\frac{a'}{a^2}
\bigg[
-4A^{(1)}_{,\, l}C\, ^{ ||(1) ,\,l}
+8\delta^{(1)}A^{(1)}
\bigg]
-\frac{2}{a}\delta^{(1)'}A^{(1)}
-2\delta^{(1)}_{,\, l}C\, ^{ ||(1) ,\,l}
\nn\\
&
-2\l[\int^\tau \frac{d\tau' }{a(\tau')}\r]\delta^{(1)}_{,\, l}A^{(1),\,l}
+2\frac{a''}{a'a}\l[\int^\tau \frac{d\tau' }{a(\tau')}\r]A^{(1)}_{,\, l}A^{(1),\,l}
\nn\\
&
-2\frac{a''}{a'a}A^{(2)}
+4\frac{a'}{a^2}A^{(2)}
.
\el
}

From (\ref{vect2transf}), (\ref{xi0trans}), (\ref{gi0}),
and (\ref{alpha2_3}),
the 0-component of the 4-velocity transforms between synchronous
as the following:
\be\label{U20TransGe}
\bar U^{(2)0}(x)=
\frac{1}{a}
\l( v^{(1)l}+\frac{A^{(1),\,l}}{a}\r)
    \l( v^{(1)}_{l}+\frac{A^{(1)}_{,\,l}}{a}\r)
        =
a^{-1}\bar v^{(1)l}\bar v^{(1)}_{l}
\,,
\ee
From (\ref{v1TransGe}),
since $\bar v^{(1)i}  =  v^{(1)i}  +a^{-1}A^{(1),i}$,
the above is the definition of $\bar U^{(2)0}(x)$
in the new synchronous coordinate.
By using (\ref{U0element}), (\ref{Uielement}),
 (\ref{xi0trans}), (\ref{gi0}), (\ref{alpha2_3}), and (\ref{xi2ge}),
and omitting the 1st-order curl-vectors,
one has the transformation of $i$-component
\bl
\bar U^{(2)i}
=&
U^{(2)i}
+\frac{4a'}{a^4}A^{(1),i}A^{(1)}
+\frac{2 a'}{a^3}v^{||(1),i}A^{(1)}
\nn\\
&
+\frac{1}{a^2}
\bigg[
-A^{(1),i}_{,\,l}C^{||(1),\,l}
+3A^{(1)}_{,\,l}C^{||(1),\,li}
+4 A^{(1),i}\phi^{(1)}
-2 A^{(1)}_{,\,l}\chi^{(1)li}
\bigg]
\nn\\
&
+\frac{1}{a}\bigg[
-2 v^{||(1),i}_{,\,l}C^{||(1),\,l}
+2 v^{||(1),\,l}C^{||(1),i}_{,\,l}
\bigg]
+\frac{2}{a^2}\l[\int^{\tau} \frac{d\tau' }{a(\tau')}\r] A^{(1)}_{,\,l}A^{(1),\,li}
\nn\\
&
+ \frac{1}{a}\l[\int^\tau \frac{d\tau' }{a(\tau')}\r]
\bigg[
-2 v^{||(1),\,i}_{,\,l}A^{(1),\,l}
+2 v^{||(1),\,l}A^{(1),i}_{,\,l}
\bigg]
+\frac{1}{a^2}A^{(2),i}
.
\el
By  $\bar U^{(2)i}=a^{-1}\bar v^{(2)i}$
and $U^{(2)i}=a^{-1}v^{(2)i}$,
the above is written as
\bl\label{v2iTransGe}
\bar v^{(2)i}
=&
v^{(2)i}
+\frac{4a'}{a^3}A^{(1),i}A^{(1)}
+\frac{2 a'}{a^2}v^{||(1),i}A^{(1)}
\nn\\
&
+\frac{1}{a}
\bigg[
-A^{(1),i}_{,\,l}C^{||(1),\,l}
+3A^{(1)}_{,\,l}C^{||(1),\,li}
+4 A^{(1),i}\phi^{(1)}
-2 A^{(1)}_{,\,l}\chi^{(1)li}
\bigg]
\nn\\
&
+\bigg[
-2 v^{||(1),i}_{,\,l}C^{||(1),\,l}
+2 v^{||(1),\,l}C^{||(1),i}_{,\,l}
\bigg]
+\frac{2}{a}\l[\int^{\tau} \frac{d\tau' }{a(\tau')}\r] A^{(1)}_{,\,l}A^{(1),\,li}
\nn\\
&
+ \l[\int^\tau \frac{d\tau' }{a(\tau')}\r]
\bigg[
-2 v^{||(1),i}_{,\,l}A^{(1),\,l}
+2 v^{||(1),\,l}A^{(1),i}_{,\,l}
\bigg]
+\frac{1}{a}A^{(2),i}
.
\el
This can be decomposed further.
Taking $\nabla^{-2}\partial_i$ upon (\ref{v2iTransGe})
to eliminate $v^{\perp(2)}$ and $\bar v^{\perp(2)}$,
one obtains  the   transform of $ v^{||(2)}$ as the following
\bl\label{v||2TransGe}
\bar v^{||(2)}
=&
v^{||(2)}
+\frac{2a'}{a^3} A^{(1)}A^{(1)}
+\frac{2 a'}{a^2}\nabla^{-2}
\bigg[
v^{||(1),\,l}A^{(1)}_{,\,l}
+A^{(1)}\nabla^2v^{||(1)}
\bigg]
+\frac{1}{a}
\bigg[
A^{(1)}_{,\,l}C^{||(1),\,l}
\nn\\
&
+\nabla^{-2}
\big(
-2C^{||(1),\,l}\nabla^2A^{(1)}_{,\,l}
+2A^{(1)}_{,\,l}\nabla^2C^{||(1),\,l}
+4 A^{(1),\,l}\phi^{(1)}_{,\,l}
+4\phi^{(1)}\nabla^2A^{(1)}
\nn\\
&
-2 A^{(1)}_{,\,lm}\chi^{(1)lm}
-2 A^{(1)}_{,\,l}\chi^{(1)lm}_{,m}
\big)
\bigg]
+\bigg[
2 v^{||(1),\,l}C^{||(1)}_{,\,l}
+\nabla^{-2}
\big(
-4 v^{||(1)}_{,\,lm}C^{||(1),\,lm}
\nn\\
&
-4 C^{||(1),\,l}\nabla^2v^{||(1)}_{,\,l}
\big)
\bigg]
+\frac{1}{a}\l[\int^{\tau} \frac{d\tau' }{a(\tau')}\r] A^{(1)}_{,\,l}A^{(1),\,l}
+ \l[\int^\tau \frac{d\tau' }{a(\tau')}\r]
\bigg[
2 v^{||(1),\,l}A^{(1)}_{,\,l}
\nn\\
&
+\nabla^{-2}
\big(
-4 v^{||(1)}_{,\,lm}A^{(1),\,lm}
-4A^{(1),\,l}\nabla^2v^{||(1)}_{,\,l}
\big)
\bigg]
+\frac{A^{(2)}}{a}
.
\el
Then,
[(\ref{v2iTransGe})$-\partial^i$(\ref{v||2TransGe})]
gives the   transform for $v^{\perp(2)i}$ as the following
\bl\label{vperp2TransGe}
\bar v^{\perp(2)i}
=&
v^{\perp(2)i}
+\frac{2 a'}{a^2}
\bigg[
v^{||(1),i}A^{(1)}
+\partial^i\nabla^{-2}
\big(
-v^{||(1),\,l}A^{(1)}_{,\,l}
-A^{(1)}\nabla^2v^{||(1)}
\big)
\bigg]
\nn\\
&
+\frac{1}{a}
\bigg[
-2A^{(1),i}_{,\,l}C^{||(1),\,l}
+2A^{(1)}_{,\,l}C^{||(1),\,li}
+4 A^{(1),i}\phi^{(1)}
-2 A^{(1)}_{,\,l}\chi^{(1)li}
\nn\\
&
+\partial^i\nabla^{-2}
\big(
2C^{||(1),\,l}\nabla^2A^{(1)}_{,\,l}
-2A^{(1)}_{,\,l}\nabla^2C^{||(1),\,l}
-4 A^{(1),\,l}\phi^{(1)}_{,\,l}
-4\phi^{(1)}\nabla^2A^{(1)}
\nn\\
&
+2 A^{(1)}_{,\,lm}\chi^{(1)lm}
+2 A^{(1)}_{,\,l}\chi^{(1)lm}_{,m}
\big)
\bigg]
\nn\\
&
+\bigg[
-4 v^{||(1),i}_{,\,l}C^{||(1),\,l}
+\partial^i\nabla^{-2}
\big(
4 v^{||(1)}_{,\,lm}C^{||(1),\,lm}
+4 C^{||(1),\,l}\nabla^2v^{||(1)}_{,\,l}
\big)
\bigg]
\nn\\
&
+ \l[\int^\tau \frac{d\tau' }{a(\tau')}\r]
\bigg[
-4 v^{||(1),i}_{,\,l}A^{(1),\,l}
+\partial^i\nabla^{-2}
\big(
4 v^{||(1)}_{,\,lm}A^{(1),\,lm}
+4A^{(1),\,l}\nabla^2v^{||(1)}_{,\,l}
\big)
\bigg]
,
\el
which does not depend on $\xi^{(2)\mu}$,
similar to the transformation of $v^{\perp(1)i}$
that does not  depend on $\xi^{(1)\mu}$
in (\ref{vGaugemodeCurl}).

\end{document}